\newif\iffigs\figstrue
\documentclass[a4paper,11pt]{report}
\usepackage[all]{xypic}
\usepackage{latexsym,amssymb,amsmath,lscape,graphics,setspace}
\usepackage{graphicx,tikz}        
\usepackage{longtable}
\usepackage{multirow}
\usepackage{color}
\usepackage{mathptmx}
\usepackage{slashed,epsfig}
\usepackage{amsfonts}
\usepackage{hyperref}

\newcommand{\mathsym}[1]{{}}
\newcommand{\unicode}[1]{{}}
\newtheorem{congettura}{Conjecture}[section]
\newtheorem{definizione}{Definition}[section]
\newtheorem{teorema}{Theorem}[section]

\newtheorem{statement}{Statement}[section]

\newtheorem{proofteo}{Proof}[teorema]

\newcommand{\bd}{\begin{definizione}}
\newcommand{\ed}{\end{definizione}}

\def\IC{\relax\,\hbox{$\inbar\kern-.3em{\rm C}$}}
\def\IG{\relax\,\hbox{$\inbar\kern-.3em{\rm G}$}}
\def\IB{\relax{\rm I\kern-.18em B}}
\def\ID{\relax{\rm I\kern-.18em D}}
\def\IL{\relax{\rm I\kern-.18em L}}
\def\IF{\relax{\rm I\kern-.18em F}}
\def\IH{\relax{\rm I\kern-.18em H}}
\def\II{\relax{\rm I\kern-.17em I}}
\def\IN{\relax{\rm I\kern-.18em N}}
\def\IP{\relax{\rm I\kern-.18em P}}
\def\IQ{\relax\,\hbox{$\inbar\kern-.3em{\rm Q}$}}
\def\bfzero{\relax\,\hbox{$\inbar\kern-.3em{\rm 0}$}}
\def\IK{\relax{\rm I\kern-.18em K}}
\def\IG{\relax\,\hbox{$\inbar\kern-.3em{\rm G}$}}
 \font\cmss=cmss10 \font\cmsss=cmss10 at 7pt
\def\IR{\relax{\rm I\kern-.18em R}}
\def\ZZ{\relax\ifmmode\mathchoice
{\hbox{\cmss Z\kern-.4em Z}}{\hbox{\cmss Z\kern-.4em Z}} {\lower.9pt\hbox{\cmsss Z\kern-.4em Z}}
{\lower1.2pt\hbox{\cmsss Z\kern-.4em Z}}\else{\cmss Z\kern-.4em Z}\fi}
\def\bfone{\relax{\rm 1\kern-.35em 1}}

\def\diag{{\rm diag}}

\def\n010{N^{0,1,0}}
\def\inbar{\vrule height1.5ex width.4pt depth0pt}
\def\bfzero{\relax{\rm I\kern-.18em 0}}
\def\bfone{\relax{\rm 1\kern-.35em 1}}
\def\twomat#1#2#3#4{\left(\begin{array}{cc}
 {#1}&{#2}\nonumber \\ {#3}&{#4}\nonumber \\
\end{array}
\right)}

\def\o#1#2{{{#1}\over{#2}}}
\DeclareFontFamily{U}{rsf}{} \DeclareFontShape{U}{rsf}{m}{n}{
  <5> <6> rsfs5 <7> <8> <9> rsfs7 <10-> rsfs10}{}
\DeclareMathAlphabet\Scr{U}{rsf}{m}{n}

 \def\cO{{\cal O}}

\renewcommand{\arraystretch}{1.3}
\newcommand{\ft}[2]{{\textstyle\frac{#1}{#2}}}
\def\tilde{\widetilde}

\def\1bar{1\hskip -.275cm -}
\def\2bar{2\hskip -.275cm -}
\def\3bar{3\hskip -.275cm -}

\newsavebox{\uuunit}
\sbox{\uuunit}
                 {\setlength{\unitlength}{0.825em}
                      \begin{picture}(0.6,0.7)
                                      \thinlines
                                      \put(0,0){\line(1,0){0.5}}
                                      \put(0.15,0){\line(0,1){0.7}}
                                      \put(0.35,0){\line(0,1){0.8}}
                                     \multiput(0.3,0.8)(-0.04,-0.02){10}{\rule{0.5pt}{0.5pt}}
                      \end {picture}}

\makeatletter \@addtoreset{equation}{section} \makeatother

\def\bfone{\relax{\rm 1\kern-.35em 1}}

\font\cmss=cmss10 \font\cmsss=cmss10 at 7pt

\newcommand{\uu}{\mathfrak{u}}

\def\bfone{\relax{\rm 1\kern-.35em 1}}
\def\inbar{\vrule height1.5ex width.4pt depth0pt}
\def\IC{\relax\,\hbox{$\inbar\kern-.3em{\rm C}$}}
\def\ID{\relax{\rm I\kern-.18em D}}
\def\IF{\relax{\rm I\kern-.18em F}}
\def\IH{\relax{\rm I\kern-.18em H}}
\def\II{\relax{\rm I\kern-.17em I}}
\def\IN{\relax{\rm I\kern-.18em N}}
\def\IP{\relax{\rm I\kern-.18em P}}
\def\IQ{\relax\,\hbox{$\inbar\kern-.3em{\rm Q}$}}
\def\IR{\relax{\rm I\kern-.18em R}}
\font\cmss=cmss10 \font\cmsss=cmss10 at 7pt
\def\ZZ{\relax\ifmmode\mathchoice
{\hbox{\cmss Z\kern-.4em Z}}{\hbox{\cmss Z\kern-.4em Z}} {\lower.9pt\hbox{\cmsss Z\kern-.4em Z}}
{\lower1.2pt\hbox{\cmsss Z\kern-.4em Z}}\else{\cmss Z\kern-.4em Z}\fi}

 \def\cO{{\cal O}}

\def\ni{\noindent}
\def\tilde{\widetilde}
\def\bar{\overline}

\def\hat{\widehat}

\def\Coe#1.#2.{{#1\over #2}}

\def\coe#1.#2.{\relax{\textstyle {#1 \over #2}}\displaystyle}

\def\to{\rightarrow}
\def\notin{\hbox{{$\in$}\kern-.51em\hbox{/}}}

\def\IE{\relax{{\rm I\kern-.18em E}}}

\def\IGam{\relax{{\rm I}\kern-.18em \Gamma}}

\def\IA{\relax{\hbox{{\rm A}\kern-.82em {\rm A}}}}

\def\diag{{\rm diag}}

\begin{document}
\begin{titlepage}
\begin{center}
\vskip 0.2cm
\vskip 0.2cm {\Large Lectures on resolutions \`{a} la Kronheimer $\mathbb{C}^3/\Gamma\rightarrow Y^\Gamma_{[3]}$ of
orbifold singularities,\\[0.2cm]
 McKay quivers for Gauge Theories on D3 branes,  \\[0.2cm]
 and the issue of  Ricci flat metrics on the resolved three-folds $Y^\Gamma_{[3]}$}\\[2cm]
Pietro~Fr\'e, \\[10pt]
{\sl\small Professor Emeritus, Dipartimento di Fisica, Universit\'a di Torino\\
via P. Giuria 1, \ 10125 Torino \ Italy\\}
\emph{e-mail:} \quad {\small {\tt pietro.fre@unito.it}}\\
\end{center}
\vspace{15pt}
\begin{abstract}
The present Lecture Notes have been prepared to back up a series of
a few seminars  given by the author at the Albert Einstein Institute
in Potsdam. These Notes aim at reviewing a research project
conducted over the last six years about a quite interesting and
challenging topic, namely the use of the generalized Kronheimer
construction and the generalized McKay correspondence for the
crepant resolution $Y^\Gamma_{[3]} \to \frac{\mathbb{C}^3}{\Gamma}$
of orbifold singularities, $\Gamma$ being s finite subgroup of
$\mathrm{SU(3)}$, as a strategic tool to construct holographic dual
pairs of $\mathcal{N}=1$ gauge theories in $D=4$ and D3-brane
solutions of type IIB supergravity. The project, developed by Ugo
Bruzzo with the present author, in various co-authorships with
Massimo Bianchi, Anna Fino, Pietro Antonio Grassi, Dimitry
Markushevich, Dario Martelli, Mario Trigiante and Umar Shahzad, is
still going on and there are a lot of interesting unanswered
questions that are illustrated throughout these Lecture Notes and
particularly emphasized in its last chapter. The mathematical roots
of this project are very robust and ramified in differential
geometry, group theory and algebraic geometry.  A particularly
challenging issue is that of constructing Ricci flat metrics on the
total space of line-bundles over compact K\"ahler manifolds of real
dimension four. In that direction the symplectic action/angle
formalism turns out to be a winning weapon. The ultimate goal is
that of deducing all the aspects  of holographic duality from finite
group theory or, to be more precise, from the faithful embedding of
abstract finite groups into $\mathrm{SU(3)}$.
\end{abstract}

\end{titlepage}
\tableofcontents \noindent {}
\newpage
\chapter{Introduction to these Lectures}\label{introibo}
The invitation by my old time friend and once co-author Prof.
Hermann Nicolai to pay a two-week visit at the Albert Einstein
Institut in Potsdam, where I volunteered to give some
seminars/lectures, has offered me a precious and challenging
opportunity  to reconsider in an overall perspective a research
project to which, over the last six years, I dedicated  most of my
energies in constant cooperation with another old time friend of
mine and distinguished scientist, Prof. Ugo Bruzzo, through a series
of papers written in various co-authorships including Professors
Anna Fino, Dmitry Markushevich, Pietro Antonio Grassi, Massimo
Bianchi, Dario Martelli and Mario Trigiante, more recently also Dr.
Umar Shahzad. The expressions of my deepest gratitude go to all of
them for what they taught me within the framework of our common work
and for the nice results that we could derive together, last but not
least, also for the pleasant friendly atmosphere of our constructive
discussions.
\par
The topics to be reviewed in these lectures have a precise location
in the wide mathematical-physics landscape and are significantly
characterized by their articulated roots in different branches of
pure mathematics: algebraic complex geometry, differential geometry,
finite group-theory, homology and cohomology. There is however, at
the basis of the investigations, a  hard core of inspiring ideas of
theoretical physics nature, I might even say of philosophical
nature, which developed within the framework of
\textit{String/Supergravity/Brane Theory} but have by now acquired a
wider formulation, not necessarily point-wise tied to String Theory
as such. These general ideas have three main components:
\begin{description}
  \item[$\mathcal{A}$)] \textbf{Holography and the Gauge/Gravity
  Correspondence}. This is the idea that a quantum field theory living on
  the boundary $\partial \mathcal{M}_{\textit{ST}}$ of some multi-dimensional space time,  for
  instance the gauge theory describing all the non gravitational
  interactions that lives on the four-dimensional boundary of five-dimensional anti de Sitter space $\mathrm{AdS_5}$ might be
  \textit{"solved"} by means of classical geometrical calculations in
  the bulk $\mathcal{M}_{\textit{ST}}$.
  \item[$\mathcal{B}$)] \textbf{The breaking of conformal invariance
  of the gauge theory by the resolution of orbifold singularities.}
  This is the development within Supergravity and String Theory of
  the consequences of the following chain of results piled up in the
  last four decades:
  \begin{enumerate}
  \item a way to break supersymmetry or other continuous symmetries
  is that of considering as substratum of exact solutions of (Super)-Gravity field equations, flat manifolds $\mathcal{M}_{flat}$
  quotiened by the action of a discrete group $\Gamma$, generically
  named orbifolds $\mathcal{M}_{flat}/\Gamma$.
  \item The space $\mathcal{M}_{flat}/\Gamma$ has singularities in
  the fixed points for the action of $\Gamma$ on $\mathcal{M}_{flat}$
  and there are relevant operators that can deform the orbifold
  solution to one on a smooth manifold $\widetilde{\mathcal{M}}_{smooth}$ that, via the exceptional
  divisors introduced by the blowup morphism,
  develops non trivial homology and cohomology.
  \item The size of such
  new homology cycles are dimensionful parameters that break
  conformal invariance and give rise to more realistic holographic
  pairs \textit{gauge-theory/gravitational solution}.
  \end{enumerate}
  \item[$\mathcal{C}$)]\textbf{Tracing back physical theories to
  group structures.} This is the guiding principle, physical and
  philosophical that aims at reducing the Laws of Nature to Symmetry
  Principles. In its declination in the context of holography and
  orbifolds one would like to develop a \textit{robust conceptual setup} in
  which the dual pairs and all of their aspects can be traced back
  to \textit{group structures}.
\end{description}
The \textit{robust conceptual setup} that provides a rich
play-ground for the theoretical physicist's aspirations concisely
described in point $\mathcal{C}$) was created by mathematicians at
the end of 1980.s in particular by Peter Kronheimer in
\cite{kro1,kro2} with an essential input from the genial discoveries
of John McKay \cite{mckay}. The Kronheimer setup for the resolution
of the $\mathbb{C}^2/\Gamma$ singularities where $\Gamma \subset
\mathrm{SU(2)}$, resolution that produces the so named
$ALE$-manifolds was very recent at the beginning of the years 1990
when I was newly appointed full-professor at SISSA and I had
exceptionally brilliant Ph.D. students working under my supervision:
in alphabetical order Damiano Anselmi, Marco Billo, Alberto
Zaffaroni. It was Marco Billo who attracted my attention to
Kronheimer results and the fascinating McKay correspondence between
the ADE classification of simply laced simple Lie algebras and that
of finite subgroups $\Gamma\subset \mathrm{SO(3)}$ of the rotation
group. Together with Anselmi, Zaffaroni and our senior colleague
Luciano Girardello, whom we all are missing very much, we
immediately conceived the idea of applying the Kronheimer setup to
two-dimensional  field theories with $\mathcal{N}=2$ super-conformal
symmetries \cite{mango}. Several other applications were produced in
the same years by other authors, particulary in connection with
gravitational and gauge instantons
\cite{Bianchi:2009bg,Bianchi:1996zj,Bianchi:1995ad,Bianchi:1995xd,Bianchi:1994gi}.
Few years later, when I had come back to Torino University from
SISSA, discussions with my former Ph.D. students, Matteo Bertolini
and Mario Trigiante, motivated us to revise the use of ALE-manifolds
and the Kronheimer construction for the errand of establishing exact
D3-brane solutions of type IIB supergravity, preserving
$\mathcal{N}=2$-supersymmetry. With other authors from G\"{o}teborg
University we published the papers
\cite{Bertolini:2002pr,Bertolini:2001ma} that can be seen as the
ancestors of the research project reported in the present lectures.
Indeed, although for many years I was involved in different
activities and in different research directions, the drive towards
the Kronheimer construction remained always in my mind and I always
meant to come back to it at some point. By the year 2016, when I
approached the end of my appointment as Scientific Counselor of the
Italian Embassy in Moscow, I came to know a little bit about the
several advances that the mathematical community had done,
generalizing the Kronheimer construction to the case of the
resolution of $\mathbb{C}^3/\Gamma$ singularities where $\Gamma
\subset \mathrm{SU(3)}$. I was quite seriously attracted by the
topic and tried to read many papers. I was finally irreducibly
seduced when I discovered the results of Roan \cite{roanno} and I
learned that in the classification of $\mathrm{SU(3)}$ finite
subgroups achieved one century ago \cite{blicfeltus,blicfeltus2},
the order 168 simple group $\mathrm{PSL(2,7)}$ is the maximal non
abelian one and that the resolution of the singularity
$\mathbb{C}^3/\mathrm{PSL(2,7)}$ had already been done many years
before by Markusevich in \cite{marcovaldo}.  The group
$\mathrm{PSL(2,7)}$ had been my concern in the previous two years in
relation with solutions of D=11 supergravity on special
$\mathrm{T}^7$ torii obtained by quotiening $\mathbb{R}^7$ with
lattices admitting $\mathrm{PSL(2,7)}$ as a crystallographic group
\cite{miol168}.
\par
Plunging into the study of the mathematical literature I soon
realized that  my education in algebraic geometry was almost
evanescent and absolutely inadequate to master  what I wanted to
understand and possibly utilize in connection with M2-branes and
D3-branes. So I called my old friend Ugo Bruzzo with whom,
unfortunately, I had not been in contact for several years and I
asked him if he was interested in joining me in an adventure where I
had just only wishful thinking, while he solidly knew the way. Very
generously he agreed and our collaboration started. We involved also
Anna Fino, a brilliant recently appointed Full Professor of
Differential Geometry of  my own University with whom Ugo had
already extensive contacts and, in one year time, we arrived at the
publication of \cite{Bruzzo:2017fwj} which is the first step  in the
conceptual development that in these lectures I mean to review,
entering now in \textit{medias res}. Before doing that I want to
warn my reader that, notwithstanding the achieved advances, which
undoubtedly we did and which the number of pages of these Lecture
Notes might suggest to be extensive, yet the road ahead is still
long and the open questions are many more than the obtained answers.
\section{Entering in \textit{medias res}}\label{cosemediane}
After the preliminary remarks that were provided in order to locate
both historically and in the context of a long time perspective the
problems and issues addressed in the reviewed research project, I
finally enter in \textit{medias res}, namely in a more technically
detailed account of what is the content of these Lecture Notes.
\par
I plan to report on the advances obtained on the following special
aspect of the gauge/gravity correspondence, within the context of
quiver gauge--theories
\cite{Morrison:1998cs,Bianchi:2014qma,Feng:2007ur,Feng:2000mw,Feng:2001xr}:
\textit{the relevance of the generalized Kronheimer
construction}\cite{Bruzzo:2017fwj,noietmarcovaldo,Bianchi_2021,bruzzo2023d3brane}
for the resolution of $\mathbb{C}^3/\Gamma$ singularities.
\par
I begin with the general problem of establishing holographic dual
pairs whose members are:
\begin{description}
  \item[A)] a gauge theory living on a D3-brane world volume,
  \item[B)] a classical D3-brane
solution of type IIB supergravity in D=10 supergravity.
\end{description}
Gauge theories based on quiver diagrams have been extensively
studied in the literature in connection with the problem of
establishing such holographic dual pairs as described above. Indeed
the quiver diagram is a powerful tool which encodes the data of a
K\"ahler  quotient describing the geometry of the six directions
transverse to the brane. Here I will be concerned with the
mathematical aspects of the dual holographic pairs. In particular
the following statement will be thoroughly illustrated:
\begin{statement}
\label{turiddu} There is a one-to-one map between the field-content
and the interaction structure of a $D=4$, $\mathcal{N}=1$ Gauge
Theory on the D3-brane world volume  and the generalized Kronheimer
algorithm of solving quotient singularities $\mathbb{C}^3/\Gamma$
via a K\"ahler quotient based on the McKay correspondence. All items
on both sides of the one-to-one correspondence are completely
determined by the structure of the finite group $\Gamma$ and by its
specific embedding into $\mathrm{SU(3)}$.
\end{statement}
An ultra short summary of the results that I am going to present is
the following. From the field--theoretic side the essential items
defining the theory are:
\begin{enumerate}
  \item The K\"ahler manifold $\mathcal{S}$ spanned by the
  Wess-Zumino multiplets. This is the $3|\Gamma|$ dimensional
  manifold $\mathcal{S}_\Gamma \, = \,
  \mathrm{Hom}_\Gamma\left(\mathcal{Q}\otimes R,R\right)$ where $\mathcal{Q}$
  is the representation of
  $\Gamma$ inside $\mathrm{SU(3)}$ and $R$ denotes the regular representation
  of the discrete group.
  \item The gauge group $\mathcal{F}_{\Gamma}$. This latter is identified
  as the maximal compact subgroup $\mathcal{F}_{\Gamma}$ of
  the complex quiver group $\mathcal{G}_\Gamma$ of complex
  dimension $|\Gamma|-1$, to be discussed later.
  Since the action of $\mathcal{F}_{\Gamma}$ on $\mathcal{S}_\Gamma$
  is defined by construction, the gauge
  interactions of the Wess-Zumino multiplets are also fixed and the associated moment maps are equally
  uniquely determined.
  \item The Fayet-Iliopoulos parameters. These are in a one-to-one association with the $\zeta$ levels of the
  moment maps corresponding  to the center of the gauge Lie algebra
  $\mathfrak{z}\left[\mathbb{F}_{\Gamma}\right]$\footnote{In this paper
  we follow the convention that the names of the Lie groups are denoted with
  calligraphic letters $\mathcal{F},\mathcal{G},\mathcal{U}$, the corresponding Lie
  algebras being denoted by mathbb letters $\mathbb{F},\mathbb{G},\mathbb{U}$.}. The
  dimension of this center is $r$ which is the number of nontrivial conjugacy classes of the discrete group
  $\Gamma$ and also of its nontrivial irreps.
  \item The superpotential $\mathcal{W}_\Gamma$. This latter is a cubic function uniquely associated with a
  quadratic constraint $[A,B]=[B,C]=[C,A]=0$ which characterizes the generalized Kronheimer construction,
  defines a K\"ahler subvariety $\mathbb{V}_{|\Gamma|+2}\subset \mathcal{S}_\Gamma$ and admits a universal
  group theoretical description in terms of  the quiver group $\mathcal{G}_\Gamma$.
  \item In presence of all the above items the manifold of vacua of the gauge theory, namely of extrema at
  zero of its scalar potential, is just the minimal crepant resolution of the singularity $\mathcal{M}_\Gamma
  \rightarrow \frac{\mathbb{C}^3}{\Gamma}$, obtained as K\"ahler quotient of $\mathbb{V}_{|\Gamma|+2}$ with respect to
  the gauge group $\mathcal{F}_{\Gamma}$.
  \item The Dolbeault cohomology of the space of vacua $\mathcal{M}_\Gamma$ is predicted by the finite group
  $\Gamma$ structure in terms of a grading of its conjugacy classes named \textit{age grading}.
  The construction of the homology cycles and exceptional divisors that are Poincar\'e dual to such
  cohomology classes is the most exciting issue in the present list of \textit{geometry}
  $\Longleftrightarrow$ \textit{field theory} correspondences, liable to give rise to many
  interesting physical applications. In this context a central item is the notion of tautological
  bundles associated with the
  nontrivial irreps of $\Gamma$ that we discuss at length in the sequel \cite{degeratu}.
\end{enumerate}
Indeed within the general scope of quivers a special subclass is
that of \textit{McKay quivers} that are group theoretically defined
by the embedding of a finite discrete group $\Gamma$ in  an
$n$-dimensional complex unitary group
\begin{equation}\label{hukko}
    \Gamma \hookrightarrow \mathrm{SU}(n)
\end{equation}
and are associated with the resolution of $\mathbb{C}^n/\Gamma$
quotient singularities by means of a Kronheimer-like construction
\cite{kro1,kro2,mango}.
\par
The case $n=2$ corresponds to the HyperK\"ahler quotient
construction of ALE-manifolds as the resolution of the
$\mathbb{C}^2/\Gamma$ singularities, the discrete group $\Gamma$
being a finite Kleinian subgroup of $\mathrm{SU(2)}$, as given by
the ADE classification\footnote{For a recent review of these matters
see chapter 8 of \cite{advancio}.}.
\par
The case $n=3$ was the target of many interesting and robust
mathematical developments starting from the middle of the nineties
up to the present day
\cite{itoriddo,62,Kin94,crawthesis,CrawIshii,SardoInfirri:1994is,SardoInfirri:1996ga,SardoInfirri:1996gb,degeratu}.
The main and most intriguing result in this context, which
corresponds to a generalization of the Kronheimer construction and
of the McKay correspondence, is the group theoretical prediction of
the cohomology groups
$\mathrm{H}^{(p,q)}\left(Y^\Gamma_{[3]}\right)$ of the crepant
smooth resolution $Y^\Gamma_{[3]}$ of the quotient singularity
$\mathbb{C}^3/\Gamma$. Specifically, the main output of the
generalized Kronheimer construction for the crepant resolution of a
singularity $\mathbb{C}^3/\Gamma$ is a blowdown morphism:
\begin{equation}\label{caspiterina}
    \mathrm{BD}\, : \quad Y^\Gamma_{[3]} \, \longrightarrow \,
    \frac{\mathbb{C}^3}{\Gamma}
\end{equation}
where $Y^\Gamma_{[3]}$ is a noncompact smooth three-fold with
trivial canonical bundle. On such a complex three-fold
 a Ricci-flat K\"ahler metric
\begin{equation}\label{carigno}
   \text{ds}^2_{\mathrm{RFK}}(Y^\Gamma_{[3]})\, = \,
    \mathrm{\mathbf{g}}^{\mathrm{RFK}}_{\alpha\beta^\star} \,
    dy^\alpha \otimes dy^{\beta^\star}
\end{equation}
with asymptotically conical boundary conditions (Quasi-ALE) is
guaranteed to exist (see e.g.~ \cite{Joyce-QALE}, Thm.~3.3), {\it
although it is not necessarily the one obtained by means of the
K\"ahler quotient.} According to the theorem proved by Ito-Reid
\cite{itoriddo,crawthesis,CrawIshii} and based on the concept of age
grading\footnote{For a recent review of these matters within a
general framework of applications to brane gauge theories see
\cite{Bruzzo:2017fwj,noietmarcovaldo}.}, the homology cycles of
$Y_{[3]}^\Gamma$ are all algebraic and its non vanishing cohomology
groups are all even and of type $\mathrm{H}^{(q,q)}$. We actually
have a correspondence between the cohomology classes of type $(q,q)$
and the discrete group conjugacy classes with age-grading $q$,
encoded in the statement:
\begin{eqnarray}\label{vecchioni}
    \mbox{dim} \, \mathrm{H}^{1,1}\left(Y^\Gamma_{[3]}\right)& =
    & \# \, \mbox{ of junior conjugacy classes in $\Gamma$;}\nonumber\\
\mbox{dim} \, \mathrm{H}^{2,2}\left(Y^\Gamma_{[3]}\right)& =
    & \# \, \mbox{ of senior conjugacy classes in $\Gamma$;}
    \nonumber\\
    && \mbox{all other cohomology groups are trivial}
\end{eqnarray}
The absence of harmonic forms of type $(2,1)$ implies that the
three-folds $Y^\Gamma_{[3]}$ admit no infinitesimal deformations of
their complex structure and is also a serious obstacle, as we
discuss in section \ref{cimice} to the construction of supergravity
D3-brane solutions based on $Y^\Gamma_{[3]}$ that have transverse
three-form fluxes.
\subsection{A special class of ${\cal N}=1,D=4$ gauge theories}\label{specN1theo} In the realization of
the one-to-one map advocated in statement \ref{turiddu},
we are interested in theories where the Wess-Zumino
multiplets are identified with the non-vanishing entries of a
triple of matrices, named $A,B,C,$ and the
superpotential takes the following form:
\begin{eqnarray}\label{signorello}
    \mathcal{W}& = &\mathrm{const} \times\, \mbox{Tr}\left(A\left[B,C\right]\right)
    +\mbox{Tr}\left(B\left[C,A\right]\right)+\mbox{Tr}\left(C\left[A,B\right]\right)\nonumber\\
    &=&3 \, \mathrm{const} \times \, \left(\mbox{Tr}\left(A \, B \, C \right)-\mbox{Tr}\left(A \, C \, B
    \right)\right)
\end{eqnarray}
Because of the positive-definiteness of the K\"ahler metric $g^{ij^\star}$  and of the Killing metric
$\mathbf{m}^{\Lambda\Sigma}$ the zero of the potential, namely the vacua, are characterized by the two
conditions:
 \begin{eqnarray}
   \partial_i\mathcal{W} &=& 0 \quad \Rightarrow [A,B]=[B,C]=[C,A]=0 \label{holomorfocost} \\
   \mathcal{P}_\Lambda & = & \zeta_I \,
\mathfrak{C}_\Lambda^I
 \end{eqnarray}
which will have a distinctive interpretation in the K\"ahler
quotient construction \`a la Kronheimer. Notice that
$\mathfrak{C}_\Lambda^I$ denotes the projector of the gauge Lie
algebra onto its center, as we already said.
\chapter{The Generalized Kronheimer construction and McKay
correspondence for $\mathbb{C}^n/\Gamma$
singularities}\label{partone}
In this chapter I review the main aspects of the Generalized
Kronheimer construction just listed in the introductory chapter and
the construction of D3-brane solutions based on it.
\section{On superconformal gauge theories dual to D3-brane classical solutions}
In this short section we collect some issues and hints relative to
the construction of  superconformal gauge theories dual to orbifolds
of the D3-brane transverse space with respect to $\Gamma \subset
\mathrm{SU(3)}$. The most important conclusion is that, as long as
we require the existence of a complex structure of $\mathbb{R}^6$
compatible with $\mathrm{PSL(2,7)}$\footnote{Following the notations
of \cite{miol168} by $\mathrm{PSL(2,7)}$ we denote the simple group
of order 168 that is isomorphic with $\mathrm{PSL(2,7)}$ and which
is also the largest non abelian non solvable finite subgroup of
$\mathrm{SU(3)}$, according with the classification of
\cite{blicfeltus,blicfeltus2}.}, we reduce the singularity to:
\begin{equation}\label{galisco}
\frac{\mathbb{C}^3}{\Gamma}\quad ; \quad \Gamma \subset
\mathrm{SU(3)}
\end{equation}
As it is well known $\mathrm{SU(4)}$-gauged supergravity is obtained
from d=10 type IIB  supergravity compactified on:
\begin{equation}\label{sferasette}
 \mathrm{AdS_5} \times \mathbb{S}^5
\end{equation}
which is the near horizon geometry of a D3-brane with $\mathbb{R}^6$
transverse space. Indeed $\mathbb{R}^6$ is the metric cone on
$\mathbb{S}^5$. The entire Kaluza-Klein spectrum which constitutes
the spectrum of BPS operators of the d=4 theory is organized in
short representations of the supergroup:
\begin{equation}\label{ortosymplo}
 \mathrm{ SU(2,2|4)}
\end{equation}
Our discussion leads to the conclusion that we can consider the compactification of supergravity on orbifolds
of the following type:
\begin{equation}\label{orbildus}
  \mathbb{C_{\mathrm{\Gamma}}S}^{5} \, = \, \frac{\mathbb{S}^5}{\Gamma}  \, \quad ; \quad \Gamma \subset
  \mathrm{SU(3)}
  \subset  \mathrm{SO(6)} \sim \mathrm{SO(4)}
\end{equation}
The corresponding D3-brane solution has the orbifold:
\begin{equation}\label{rorbildo}
   \mathbb{C_{\mathrm{\Gamma}}R}^{6} \, = \, \frac{\mathbb{R}^6}{\Gamma}
\end{equation}
as transverse space.
\par
The massive and massless modes of the Kalauza Klein spectrum are
easily worked out from the $\mathrm{SU(2,2|4)}$ spectrum of the
5-sphere. Indeed since the group $\Gamma$ is embedded by the above
construction into $\mathrm{SU(4)} \subset \mathrm{SU(2,2|4)}$, it
suffices to cut the spectrum to the $\Gamma$ singlets.
\subsection{Quotient singularities and gauge groups} What we said above can be summarized by saying that
we construct models where D3-branes are probing the singularity
(\ref{rorbildo}) and we might be interested in the smooth manifold
obtained by blowing up the latter.
\par
 In order to derive the dual superconformal field theory it is essential
to specify the embedding of $\Gamma$ into the isometry group
$\mathrm{SU(4)}$ of the five-sphere. Different embeddings of the
same discrete group can lead to different gauge theories on the
world volume.
\subsection{Some suggestions from ALE manifolds} Some useful suggestions on this conceptual link
can arise by comparing  with the case of well known singularities
like $\mathbb{C}^2/\Gamma$, the discrete group $\Gamma$ being one of
the finite subgroups of $\mathrm{SU(2)}$ falling into the ADE
classification.  In that case the blowup of the singularities can be
done by means of a hyperK\"ahler quotient according to the
Kronheimer construction \cite{kro1,kro2}. Essentially the gauge
group is
 to be identified with the nonabelian extension
$\mathrm{U(1)} \to \mathrm{U(N)}$ of the group $\mathcal{F}$ one
utilizes in the hyperK\"ahler quotient. The group
$\mathcal{F}_\Gamma$ is a product of $\mathrm{U(1)}$'s as long as
the discrete group $\Gamma$ is  the cyclic group $\mathbb{Z}_k$, yet
it becomes nonabelian with factors $\mathrm{SU(k)}$ when $\Gamma$ is
nonabelian. As far as we know no one has constructed  gauge theories
corresponding to D3-branes that probe singularities of the type
$\mathbb{C}\times \mathbb{C}^2/\Gamma$ with  a  nonabelian $\Gamma$.
This case might be a ground-zero case to investigate.
\subsection{Moduli of the blowup and superconformal operators: inspirations from
geometry} It must be stressed that in the spectrum of the conformal
field theory obtained at the orbifold point (which corresponds to
the Kaluza Klein spectrum in the case of smooth manifolds) there
must be those associated with the moduli of the blowup. By
similarity with the case of superstrings at orbifold points, we
expect that these are \textit{twisted states}, namely states not
visible in the Kaluza Klein spectrum on the orbifold. Yet when the
orbifold is substituted with its smooth counterpart, obtained
blowing  up  the singularity, these states should appear as normal
states in the supergravity Kaluza Klein spectrum.
\subsection{Temporary conclusion}
The above discussion shows that the embedding of $\Gamma$ is
fundamental. We can treat $\mathrm{PSL(2,7)}\equiv
\mathrm{PSL(2,\mathbb{Z}_7})$ and other discrete group $\Gamma$
singularities when $\Gamma$ has a holomorphic action on
$\mathbb{C}^3$. This happens when
\begin{equation}\label{cavicchius}
  \Gamma \subset \mathrm{SU(3) }\subset \mathrm{SU(4)}
\end{equation}
In this case the singularity is just $\mathbb{C}^3/\Gamma$ and
$\mathbb{C}^3/\mathrm{PSL(2,7)}$ is the   blowup described by
Markushevich in  \cite{marcovaldo}. We mention it again in a later
section. The classification of discrete subgroups of
$\mathrm{SL(3,\mathbb{C})}$ was achieved at the dawn of the XXth
century in \cite{blicfeltus,blicfeltus2}. The largest non-abelian
nontrivial group appearing in this classification is the unique
simple group with 168 elements named $PSL(2,7)$ (see \cite{miol168}
and \cite{nuovogruppo} for a thorough discussion). The other
possibilities are provided by a finite list of cyclic and solvable
groups reviewed for instance in \cite{62}.
\subsection{Quotient singularities and D3-branes} We come now to the mathematics which is of greatest interest
to us in order to address the physical problem at stake,
\textit{i.e.}, the construction of D=4 gauge theories dual to
D3-branes that have the metric cone on orbifolds
$\mathbb{S}^5/\Gamma$ as transverse space. The first step is to show
that such metric cone is just $\mathbb{C}^3/\Gamma$. This is a
rather simple fact but it is of the utmost relevance since it
constitutes the very bridge between the mathematics of quotient
singularities, together with their resolutions, and the physics of
gauge theories. The pivot of this bridge is the complex Hopf
fibration of the $5$-sphere. The argument leading to the above
conclusion was provided in the paper \cite{pappo1} and we do not
deem it necessary to repeat it here. We just jump to the conclusion
there reached. The space $\mathbb{C}^3 -\{0\}$ can be regarded as
the total space of the canonical  $ \mathbb{C}^\star$-fibration over
$\mathbb{P}^2$:
\begin{eqnarray}\label{lignotto}
    \pi & : & \mathbb{C}^3 - \{0\} \, \rightarrow \, \mathbb{P}^2\nonumber\\
    \forall y \in\mathbb{P}^2 & : & \pi^{-1}(y) \sim \mathbb{C}^\star
\end{eqnarray}
By restricting to the unit sphere in $ \mathbb{C}^3 $ we obtain
  the  Hopf fibration of the seven sphere:
\begin{eqnarray}\label{offetto7}
    \pi & : & \mathbb{S}^5 \, \rightarrow \, \mathbb{P}^2\nonumber\\
    \forall y \in\mathbb{P}^2& : & \pi^{-1}(y) \sim \mathbb{S}^1
\end{eqnarray}
\par
The consequence of such a  discussion is that if we have a finite
subgroup $\Gamma \subset \mathrm{SU(3)}$, which obviously is an
isometry of $\mathbb{P}^2$, we can consider its action both on
$\mathbb{P}^2$ and on the seven sphere so that  we have:
\begin{equation}\label{raccimolato}
    \mathrm{AdS_5} \times \frac{\mathbb{ S}^5}{\Gamma} \, \to \partial\mathrm{AdS_5} \times
    \frac{\mathbb{C}^3}{\Gamma}
\end{equation}
We are therefore interested in describing the theory of D3-branes
probing the singularity $\frac{\mathbb{C}^3}{\Gamma}$. Hence an
important guiding line in addressing mathematical questions comes
from their final use  in connection with D3-brane solutions of
$D=10$ type IIB supergravity and  with the construction of quantum
gauge theories dual to such D3-solutions of supergravity.
\par
Recalling the results of \cite{pappo1} we start from the following diagram
\begin{equation}\label{caluffo2}
  K_2 \, \stackrel{\pi}{\longleftarrow} \, \mathcal{M}_5 \,  \stackrel{Cone}{ \hookrightarrow} \,  K_3 \,
  \stackrel{\mathcal{A}}{ \hookrightarrow} \, \mathbb{V}_q
\end{equation}
where $\mathcal{M}_5$ is the compact manifold on which D=10 type IIB
 supergravity is compactified and $\mathbb{V}_q$ denotes some
appropriate algebraic variety of complex dimension $q$. It is
required that $\mathcal{M}_5$ should be a Sasakian manifold.
\par
What Sasakian means is visually summarized in the following table.
\begin{center}
\begin{tabular}{|ccccc|}
  \hline
 base  of the fibration &   projection & $5$-manifold & inclusion & metric cone  \\
  $\mathcal{B}_4$ & $\stackrel{\pi}{\longleftarrow}$ &$\mathcal{M}_5$ &$\hookrightarrow $& $\mathcal{C}\left( \mathcal{M}_5\right)$ \\
  $\Updownarrow$ & $\forall p \in \mathcal{B}_4 \quad \pi^{-1}(p) \, \sim \, \mathbb{S}^1 \,$
   &$\Updownarrow$ &\null & $\Updownarrow$ \\
  K\"ahler $K_2$ &$\null$ & Sasakian&\null &K\"ahler Ricci flat $K_3$ \\
  \hline
\end{tabular}
\end{center}
First of all the $\mathcal{M}_5$ manifold must admit an
$\mathbb{S}^1$-fibration over a K\"ahler two-fold $K_2$:
\begin{equation}\label{fibratoS1}
    \pi \quad : \quad \mathcal{M}_5 \stackrel{\mathbb{S}^1}{\longrightarrow} \,
    K_2
\end{equation}
Calling $z^i$ the three complex coordinates of $K_2$ and $\chi$ the
angle spanning $\mathbb{S}^1$, the fibration  means that the metric
of $\mathcal{M}_5$ admits the following representation:
\begin{equation}\label{fibrametricu}
    ds^2_{\mathcal{M}_5}\, = \, \left(d\chi - \mathcal{A}\right)^2 \, + \, g_{ij^\star} \, dz^i \otimes d{\bar z}^{j^\star}
\end{equation}
where the one--form $\mathcal{A}$ is some suitable connection one--form on the $\mathrm{U(1)}$-bundle
(\ref{fibratoS1}).
\par
Secondly the metric cone $\mathcal{C}\left( \mathcal{M}_5\right)$
over the manifold $\mathcal{M}_5$ defined by the direct product
$\mathbb{R}_+\times \mathcal{M}_5$ equipped with the following
metric:
\begin{equation}\label{gustoconetto}
    ds^2_{\mathcal{C}\left( \mathcal{M}_5\right)} \, =\,dr^2 + 4 \,e^2 \, r^2 \, ds^2_{\mathcal{M}_5}
\end{equation}
should also be a Ricci-flat complex K\"ahler $3$-fold. In the above
equation $e$ just denotes a constant scale parameter with the
dimensions of an inverse length $\left[e\right] = \ell^{-1}$.
\par
Altogether the Ricci flat K\"ahler manifold $K_3$, which plays the
role of transverse space to the D3-branes, is a line bundle over the
base manifold $K_2$:
\begin{eqnarray}\label{convecchio}
    \pi &\quad : \quad& K_3 \, \longrightarrow \, K_2 \nonumber\\
     \forall p \in K_2 && \pi^{-1}(p) \, \sim \, \mathbb{C}
\end{eqnarray}
\par
The fundamental geometrical clue to the field content of the
\textit{superconformal gauge theory} on the boundary is provided by
the construction of the K\"ahler manifold $K_3$ as a holomorphic
algebraic variety in some higher dimensional affine or projective
space $\mathbb{V}_{q}$, plus a K\"ahler quotient. The equations
identifying the algebraic locus in $\mathbb{V}_{q}$ are related with
the superpotential $\mathcal{W}$ appearing in the $d=4$ lagrangian,
while the K\"ahler quotient is related with the $D$-terms appearing
in the same lagrangian. The coordinates $u^\alpha$ of the space
$\mathbb{V}_{q}$ are the scalar fields of the \textit{superconformal
gauge theory}, whose vacua, namely the set of extrema of its scalar
potential, should be in a one--to--one correspondence with the
points of $K_3$. Going from one to multiple D3--branes just means
that the coordinates $z^i$ of $\mathbb{V}_{q}$ acquire color indices
under a proper set of color gauge groups and are turned into
matrices. In this way we obtain \textit{quivers}.
\par
This is the main link between the D=4 dual gauge theories and the
geometry of the transverse space to the branes.
\par
The $\mathrm{AdS_5}$ compactification of $D=10$ supergravity is
obtained by utilizing as complementary $5$-dimensional space a
manifold $\mathcal{M}_5$ which occupies the above displayed position
in the inclusion--projection diagram (\ref{caluffo2}). The metric
cone $\mathcal{C}(\mathcal{M}_5)$ enters the game when, instead of
looking at the vacuum:
\begin{equation}\label{vacuetto}
    \mathrm{AdS_5} \otimes \mathcal{M}_5
\end{equation}
we consider the more general D3-brane solutions of D=10
supergravity, where the D=10 metric is of the following form:
\begin{equation}\label{m2branmet}
    ds^2_{10} \, = \, H(y)^{-\ft 12}\,\left(dx^\mu\otimes dx^\nu\eta_{\mu\nu}\right) - H(y)^{\ft 12} \,
    \left(ds^2_{\mathcal{M}_6} \right)
\end{equation}
$\eta_{\mu\nu}$ being the constant Lorentz metric of
$\mathrm{Mink}_{1,2}$ and:
\begin{equation}\label{metric8}
   ds^2_{\mathcal{M}_6} \, = \,dy^I\otimes dy^J \, g_{IJ}(y)
\end{equation}
being a Ricci-flat metric on an asymptotically locally Euclidean
$6$-manifold $\mathcal{M}_6$. In eqn.\,(\ref{m2branmet}) the symbol
$H(y)$ denotes a harmonic function over the manifold
$\mathcal{M}_6$, namely:
\newcommand*\DAlambert{\mathop{}\!\mathbin\Box}
\begin{equation}\label{cicio}
     \DAlambert_{g} H(y) \, = \, 0
\end{equation}
Eq.\,(\ref{cicio}) is the only differential constraint required in
order to satisfy all the field equations of $D=10$ supergravity in
presence of  the standard D3-brain ansatz that we detail later on
(see eqn.s \ref{ansazzo}). In this more general setup the manifold
$\mathcal{M}_6$ is what substitutes the metric cone
$\mathcal{C}(\mathcal{M}_5)$. To see the connection between the two
viewpoints it suffices to introduce the radial coordinate $r(y)$ by
means of the position:
\begin{equation}\label{radiatore}
    H(y) \, = \, 1 \, - \, \frac{1}{r(y)^4}
\end{equation}
The asymptotic region where $\mathcal{M}_6$ is required to be
locally Euclidean is defined by the condition $r(y) \to \infty$. In
this limit the metric (\ref{metric8}) should approach the flat
Euclidean metric of $\mathbb{R}^6\simeq \mathbb{C}^3$. The opposite
limit where $r(y)\to 0$ defines the near horizon region of the
D3-brane solution. In this region the metric (\ref{m2branmet}) it is
tacitly assumed that it should approach that of the space
(\ref{vacuetto}), the manifold $\mathcal{M}_5$ being a codimension
one submanifold of $\mathcal{M}_6$ defined by the limit $r\to 0$.
\par
As we will see in chapter \ref{riccione} this assumption is not
necessarily verified. It can happen that there is no intermediate
Sasaki Einstein manifold $\mathcal{M}_5$ and that in the limit $r\to
0$ the Ricci flat metric on $\mathcal{M}_5$ simply flows to the
4-dimensional metric of the base manifold $\mathcal{M}_4 \sim K_2$,
the $\mathcal{M}_6$ Ricci flat manifold being a line bundle on
$K_2$.
\par
To be mathematically more precise let us consider the harmonic function as a map:
\begin{equation}\label{haccusmap}
    \mathfrak{H}\quad : \quad \mathcal{M}_6 \, \rightarrow \, \mathbb{R}_+
\end{equation}
This view point introduces a foliation of $\mathcal{M}_6$ into a
one-parameter family of $5$-manifolds:
\begin{equation}\label{romualdo}
  \forall h \in \mathbb{R}_+  \quad : \quad \mathcal{M}_5(h) \, \equiv \,
  \mathfrak{H}^{-1}(h) \subset \mathcal{M}_6
\end{equation}
The tacit assumption that is not necessarily verified is that each
manifold $\mathcal{M}_5(h)$ is isometrical to any other of the same
family, meaning that the metric on each $\mathcal{M}_5(h)$ is the
same abstract metric.  This is not necessarily true. For each $h$
the metric can be deformed, having  different Weyl and Ricci tensors
in intrinsic components. We show that in one of the calculable
examples we were able to analyse, this is exactly what happens.
Independently from that, in order to have the possibility of
residual supersymmetries we are interested in cases where the Ricci
flat manifold $\mathcal{M}_6$ is actually a Ricci-flat K\"ahler
$3$-fold.
\par
When an intermediate Sasaki-Einstein manifold $\mathcal{M}_5$ does
exist the appropriate rewriting of eqn.\,(\ref{caluffo2}) is as
follows:
\begin{equation}\label{caluffo3}
K_2 \, \stackrel{\pi}{\longleftarrow}    \, \mathcal{M}_5 \quad
  \stackrel{\mathfrak{H}^{-1}}{ \longleftarrow} \quad  K_3 \quad
  \stackrel{\mathcal{A}}{ \hookrightarrow} \quad \mathbb{V}_q
\end{equation}
Next we outline a general pattern laid down  that will be our
starting point.
\paragraph{The $\mathcal{N}=4$ case with no singularities.} The prototype of the above inclusion--projection
diagram is provided by the case of the D3-brane solution with all
preserved supersymmetries. In this case we have:
\begin{equation}\label{caluffopiatto}
\mathbb{P}^2 \quad \stackrel{\pi}{\longleftarrow}  \quad
\mathbb{S}^5 \quad
  \stackrel{Cone}{ \hookrightarrow} \quad  \mathbb{C}^3 \quad
  \stackrel{\mathcal{A}=\mathrm{Id}}{ \hookrightarrow} \quad
  \mathbb{C}^3
\end{equation}
On the left we just have the projection map of the Hopf fibration of
the $5$-sphere. On the right we have the inclusion map of the $5$
sphere in its metric cone $\mathcal{C}(\mathbb{S}^5)\equiv
\mathbb{R}^6\sim \mathbb{C}^3$. The last algebraic inclusion map is
just the identity map, since the algebraic variety $\mathbb{C}^3$ is
already smooth and flat and needs no extra treatment.
\paragraph{The singular orbifold cases.} The next orbifold cases are those of interest to us in these lectures.
Let $\Gamma \subset \mathrm{SU(3)}$ be a finite discrete subgroup of
$\mathrm{SU(3)}$. Then eqn.\,(\ref{caluffopiatto}) might be replaced
by the following one:
\begin{equation}\label{calufforbo}
\frac{\mathbb{P}^2}{\Gamma} \quad \stackrel{\pi}{\longleftarrow}
\quad \frac{\mathbb{S}^5}{\Gamma} \quad
  \stackrel{Cone}{ \hookrightarrow} \quad  \frac{\mathbb{C}^3}{\Gamma} \quad
  \stackrel{\mathcal{A}=\mbox{?}}{ \hookrightarrow} \quad \mbox{?}
\end{equation}
The consistency of the above quotient is guaranteed by the inclusion
$\mathrm{SU(3)}\subset \mathrm{SU(4)}$. The question marks can be
removed only by separating the two cases:
\begin{description}
  \item[A)] Case: $\Gamma \subset \mathrm{SU(2)} \subset \mathrm{S\left(U(1) \otimes \mathrm{U(2)}\right)}
  \subset \mathrm{SU(3)}$. Here we obtain:
  \begin{equation}\label{cromatorosso}
    \frac{\mathbb{C}^3}{\Gamma} \, \simeq \, \mathbb{C} \times \frac{\mathbb{C}^2}{\Gamma}
  \end{equation}
and everything is under full control for the Kleinian $\frac{\mathbb{C}^2}{\Gamma}$ singularities and their
resolution {\`a} la Kronheimer in terms of  hyperK\"ahler  quotients.
  \item[B)] Case: $\Gamma \subset \mathrm{SU(3)}$. Here we obtain:
  \begin{equation}\label{cromatonero}
    \frac{\mathbb{C}^3}{\Gamma} \, = \, \text{not reducible to a
     product of two varieties}
  \end{equation}
  and the study and resolution of the singularity $\frac{\mathbb{C}^3}{\Gamma}$ in a physicist friendly way
   is the main issue of the present lectures. The comparison of case B) with the well known case A)
   will provide us with many important hints.
\end{description}
Let us begin by erasing the question marks in case A). Here we can write:
\begin{equation}\label{calufforboA}
\frac{\mathbb{P}^2}{\Gamma} \quad \stackrel{\pi}{\longleftarrow}
\quad \frac{\mathbb{S}^5}{\Gamma} \quad
  \stackrel{Cone}{ \hookrightarrow} \quad  \mathbb{C} \times \frac{\mathbb{C}^2}{\Gamma} \quad
  \stackrel{\mathrm{Id}\times\mathcal{A}_W}{ \hookrightarrow} \quad \mathbb{C} \times \mathbb{C}^3
\end{equation}
In the first inclusion map on the right, $\mathrm{Id}$ denotes the
identity map $\mathbb{C}\to \mathbb{C}$ while $\mathcal{A}_W$
denotes the inclusion of the orbifold $\frac{\mathbb{C}^2}{\Gamma}$
as a singular variety in $\mathbb{C}^3$ cut out by a single
polynomial constraint:
\begin{eqnarray}\label{inclusione}
  \mathcal{A}_W& : &  \frac{\mathbb{C}^2}{\Gamma}\rightarrow
  \mathbf{V}(\mathcal{I}^W_\Gamma)
  \subset \mathbb{C}^3\nonumber\\
\mathbb{C}\left[\mathbf{V}(\mathcal{I}_\Gamma )\right] &=& \frac{\mathbb{C}[u,w,z]}{W_\Gamma(u,w,z)}
\end{eqnarray}
where by $\mathbb{C}\left[{\mathbf{V}}(\mathcal{I}_\Gamma )\right]$ we denote the \textit{coordinate ring} of
the algebraic variety $\mathbf{V}$. As we recall in more detail in the next section, the variables $u,w,z$
are polynomial $\Gamma$-invariant functions of the coordinates $z_1,z_2$ on which $\Gamma$ acts linearly. The
unique generator $W_\Gamma(u,w,z)$ of the ideal $\mathcal{I}^W_\Gamma$ which cuts out the singular variety
isomorphic to $\frac{\mathbb{C}^2}{\Gamma}$ is the unique algebraic relation existing among such invariants.
In the next sections we discuss the relation between this algebraic equation and the embedding in higher
dimensional algebraic varieties associated with the McKay quiver and the  hyperK\"ahler  quotient.
\par
Let us now consider the  case B). Up to this level things go in a quite analogous way with respect to case
A). Indeed we can write
\begin{equation}\label{calufforboB}
\frac{\mathbb{P}^2}{\Gamma} \quad \stackrel{\pi}{\longleftarrow}
\quad \frac{\mathbb{S}^5}{\Gamma} \quad
  \stackrel{Cone}{ \hookrightarrow} \quad   \frac{\mathbb{C}^3}{\Gamma} \quad
  \stackrel{\mathcal{A}_\mathcal{W}}{ \hookrightarrow} \quad \mathbb{C}^4
\end{equation}
In the last inclusion map on the right, $\mathcal{A}_\mathcal{W}$
denotes the inclusion of the orbifold $\frac{\mathbb{C}^3}{\Gamma}$
as a singular variety in $\mathbb{C}^4$ cut out by a single
polynomial constraint:
\begin{eqnarray}\label{inclusioneB}
  \mathcal{A}_\mathcal{W}& : &  \frac{\mathbb{C}^3}{\Gamma}\rightarrow
  \mathbf{V}(\mathcal{I}_\Gamma )
  \subset \mathbb{C}^4\nonumber\\
 \mathbb{C}\left [\mathbf{V}\left(\mathcal{I}_\Gamma \right)\right] & \sim &
 \frac{\mathbb{C}[u_1,u_2,u_3,u_4]}{\mathcal{W}_\Gamma(u_1,u_2,u_3,u_4)}
\end{eqnarray}
Indeed as we show in later sections for the case $\Gamma\,=\,
\mathrm{PSL(2,7)}$, discussed by Markushevich, and  for all of its
subgroups\footnote{The group $\mathrm{PSL(2,7)}$ has three maximal
subgroups, up to conjugation, namely two non conjugate copies of the
octahedral group $\mathrm{O_{24}} \sim \mathrm{S_4}$ and one non
abelian group of order $21$, denoted $\mathrm{G_{21}}$ that is
isomorphic to the semidirect product $\mathbb{Z}_3
\ltimes\mathbb{Z}_7$.}, including $\Gamma\,=\,\mathrm{G_{21}}
\subset \mathrm{PSL(2,7)}$, the variables $u_1,u_2,u_3,u_4$ are
polynomial $\Gamma$-invariant functions of the coordinates
$z_1,z_2,z_3$ on which $\Gamma$ acts linearly. The unique generator
$\mathcal{W}_\Gamma(u_1,u_2,u_3,u_4)$ of the ideal
$\mathcal{I}_\Gamma$ which cuts out the singular variety isomorphic
to $\frac{\mathbb{C}^3}{\Gamma}$ is the unique algebraic relation
existing among such invariants. As for the relation of this
algebraic equation with the embedding in higher dimensional
algebraic varieties associated with the McKay quiver, things are now
more complicated.
\par
As already anticipated in the introductory section, in the years
1990s up to 2010s there has been an intense activity in the
mathematical community on the issue of the crepant resolutions of
$\mathbb{C}^3/\Gamma$ (see for
instance\cite{giapumckay,crawthesis,marcovaldo,roanno}) that went on
almost unnoticed by physicists since it was mostly formulated in the
abstract language of algebraic geometry, providing few clues to the
applicability of such results to gauge theories and branes. Yet,
once translated into more explicit terms, by making use of
coordinate patches,  and equipped with some additional ingredients
of Lie group character, these mathematical results turn out to be
not only useful, but rather of outmost relevance for the physics of
D3-branes. In the present lectures I aim at drawing the consistent,
systematic scheme which emerges in this context upon a proper
interpretation of the mathematical constructions (see in particular
\cite{Bruzzo:2017fwj,noietmarcovaldo,Bianchi_2021,bruzzo2023d3brane}).
\par
So let us consider the case of smooth resolutions. In case A) the smooth resolution is provided by a manifold
$ALE_\Gamma$ and we obtain the following diagram:
\begin{equation}\label{caluffoALE}
 \mathcal{M}_5 \quad
  \stackrel{\mathfrak{H}^{-1}}{ \longleftarrow} \quad  \mathbb{C} \times ALE_\Gamma \quad
  \stackrel{\mathrm{Id}\times qK}{ \longleftarrow} \quad \mathbb{C}
  \times\mathbb{V}_{|\Gamma|+1} \quad \stackrel{\mathcal{A}_{\mathcal{P}}}{\hookrightarrow}
  \quad \mathbb{C} \times \mathbb{C}^{2|\Gamma|}
\end{equation}
In the above equation the map $\stackrel{\mathfrak{H}^{-1}}{
\longleftarrow}$ denotes the inverse of the harmonic function map on
$\mathbb{C}\times ALE_\Gamma$ that we have already discussed. The
map $\stackrel{\mathrm{Id}\times qK}{ \longleftarrow}$ is instead
the product of the identity map $\mathrm{Id} \, : \, \mathbb{C} \to
\mathbb{C}$ with the K\"ahler quotient map:
\begin{equation}\label{fraulein}
   qK \quad : \quad \mathbb{V}_{|\Gamma|+1} \, \longrightarrow \,\mathbb{V}_{|\Gamma|+1}\,  \,
  /\!\!/_{\null_K} \mathcal{F}_{|\Gamma|-1} \, \simeq \, ALE_\Gamma
\end{equation}
of an algebraic variety of complex dimension $|\Gamma|+1$ with respect to a suitable Lie group
$\mathcal{F}_{|\Gamma|-1}$ of real dimension $|\Gamma|-1$. Finally the map
$\stackrel{\mathcal{A}_{\mathcal{P}}}{\hookrightarrow}$ denotes the inclusion map of the variety
$\mathbb{V}_{|\Gamma|+1}$ in $\mathbb{C}^{2|\Gamma|}$. Let $y_1,\dots y_{2|\Gamma|}$ be the coordinates of
$\mathbb{C}^{2|\Gamma|}$. The variety $\mathbb{V}_{|\Gamma|+1}$ is defined by an ideal generated by
$|\Gamma|-1$ quadratic generators:
\begin{eqnarray}\label{congo}
    \mathbb{V}_{|\Gamma|+1} & = & \mathbf{V}\left( \mathcal{I}_\Gamma\right)\nonumber\\
\mathbb{C} \left[\mathbf{V}\left( \mathcal{I}_\Gamma\right) \right]&= & \frac{\mathbb{C}\left[y_1,\dots
y_{2|\Gamma|} \right]}{\left(\mathcal{P}_1(y),\mathcal{P}_2(y),\dots ,\mathcal{P}_{|\Gamma|-1}(y)\right)}
\end{eqnarray}
Actually the $|\Gamma|-1$ polynomials $\mathcal{P}_i(y)$ are the holomorphic part of the triholomorphic
moment maps associated with the triholomorphic action of the group $\mathcal{F}_{|\Gamma|-1}$ on
$\mathbb{C}^{2|\Gamma| }$ and the entire procedure from  $\mathbb{C}^{2|\Gamma|}$ to $ALE_\Gamma$ can be seen
as the  hyperK\"ahler  quotient:
\begin{equation}\label{cagliozzo}
    ALE_\Gamma \, = \,\mathbb{C}^{2|\Gamma|}/\!\!/_{\null_{HK}}\mathcal{F}_{|\Gamma|-1}
\end{equation}
yet we have preferred to split the procedure into two steps in order to compare case A) with case B) where
the two steps are necessarily distinct and separated.
\par
Indeed in case B) we can write the following diagram:
\begin{equation}\label{caluffoBalle}
 \mathcal{M}_5 \quad
  \stackrel{\mathfrak{H}^{-1}}{ \longleftarrow} \quad   Y^\Gamma_{[3]} \quad
  \stackrel{ qK}{ \longleftarrow} \quad \mathbb{V}_{|\Gamma|+2} \quad \stackrel{\mathcal{A}_{\mathcal{P}}}{\hookrightarrow}
  \quad  \mathbb{C}^{3|\Gamma|}
\end{equation}
In this case, just as in the previous one,  the intermediate step is provided by the K\"ahler quotient but
the map on the extreme right $\stackrel{\mathcal{A}_{\mathcal{P}}}{\hookrightarrow}$ denotes the inclusion
map of the variety $\mathbb{V}_{|\Gamma|+2}$ in $\mathbb{C}^{3|\Gamma|}$. Let $y_1,\dots y_{3|\Gamma|}$ be
the coordinates of $\mathbb{C}^{3|\Gamma|}$. The variety $\mathbb{V}_{|\Gamma|+2}$ is defined as the
principal branch of a set of quadratic algebraic equations that are group-theoretically defined. Altogether
the mentioned construction singles out  the holomorphic orbit of a certain group action to be discussed in
detail in the sequel. So we anticipate:
\begin{eqnarray}\label{belgone}
\mathbb{V}_{|\Gamma|+2} & = &  \mathcal{D}_\Gamma \equiv \mathrm{Orbit}_{\mathcal{G}_\Gamma}\left(
L_\Gamma\right)
\end{eqnarray}
where both the set $L_\Gamma$ and the complex  group $\mathcal{G}_\Gamma$ are  completely defined by the
discrete group $\Gamma$ defining the quotient singularity.
\section{D3-brane supergravity solutions on resolved $\mathbb{C}^3/\Gamma$
singularities}\label{cimice}
An apparently  general property of the complex three-folds
$Y_{[3]}^\Gamma$ that emerge from the crepant resolution
construction, \textbf{when there is at least one senior class} and
therefore \textbf{at least one compact component of the exceptional
divisor}  is the following. The non-compact $Y_{[3]}^\Gamma$
corresponds to the total space of some line-bundle over a complex
two-dimensional compact base manifold $\mathcal{M}_B$:
\begin{equation}\label{sagnalotto}
    Y_{[3]}^\Gamma \, \stackrel{\pi}{\longrightarrow} \,\mathcal{M}_B
\end{equation}
According with this structure we name $u,v,w$ the three complex
coordinates of $Y_{[3]}^\Gamma$, $u,v$ being the coordinates of the
base manifold $\mathcal{M}_B$ and $w$ being the coordinate spanning
the fibres. We will use the same names also in more general cases
even if the interpretation of $w$ as a fibre coordinate will be
lost. Hence we have:
\begin{equation}\label{ycordi}
   \pmb{y} \equiv y^\alpha \, = \, \left\{u,v,w\right\} \quad;\quad \bar{\pmb{y}} \equiv y^{\bar{\alpha}}\, = \,
   \left\{ \bar{u},\bar{v},\bar{w}\right\}
\end{equation}
An important observation which ought to be done right at the
beginning is that other K\"ahler metrics
$\hat{\mathrm{\mathbf{g}}}_{\alpha\beta^\star}$ do exist on the
three-fold $Y_{[3]}$ that are not Ricci-flat, although the
cohomology class of the associated K\"ahler form $\hat{\mathbf{K}}$
can be the same as the cohomology class of
$\mathbf{K}_{\mathrm{RFK}}$. Within the framework of the generalized
Kronheimer construction, among such K\"ahler (non-Ricci flat)
metrics we have the one determined by the K\"ahler quotient
according to the formula of Hithchin, Karlhede, Lindstr\"om and
Ro\v{c}ek \cite{HKLR}. Indeed, as we show later in explicit
examples, the K\"ahler metric:
\begin{equation}\label{carigno}
     \text{ds} ^2_{\mathrm{HKLR}}(Y_{[3]})\, = \,\mathrm{\mathbf{g}}^{\mathrm{HKLR}}_{\alpha\beta^\star} \,
     dy^\alpha \otimes dy^{\beta^\star}
\end{equation}
which emerges from the mathematical K\"ahler quotient construction
and which is naturally associated with $Y_{[3]}$ when this latter is
interpreted as the \textit{space of classical vacua} of the D3-brane
gauge theory (set of extrema of the scalar potential), is
generically non Ricci-flat.
\par
On the other hand on \textit{the supergravity side} of the dual
D3-brane pair we need the Ricci-flat metric in order to construct a
bona-fide D3-brane solution of type IIB supergravity.
In particular, calling $Y^\Gamma_{[3]}$ the crepant resolution of
the $\mathbb{C}^3/\Gamma$ singularity, admitting a Ricci-flat
metric, we can construct a bona-fide D3 brane solution which is
solely defined by a single real function $H$ on $Y^\Gamma_{[3]}$,
that should be harmonic with respect to the Ricci-flat metric,
namely:
\begin{equation}\label{merlatino}
    \Box_{\mathbf{g}^{\mathrm{RFK}}} \, H \, = \, 0
\end{equation}
Indeed the function $H(\mathbf{y})$ is necessary and sufficient to
introduce a flux of the Ramond $5$-form so as to produce the
splitting of the $10$-dimensional space into a $4$-dimensional world
volume plus a  transverse $6$-dimensional space that is identified
with the three-fold $Y^\Gamma_{[3]}$. This is the very essence of
the D3-picture.
\par
Yet there is another essential item that was pioneered in
\cite{Bertolini:2001ma,Bertolini:2002pr}, namely the consistent
addition of fluxes for the complex $3$-forms $\mathcal{H}_\pm$ that
appear in the field content of type IIB supergravity. These provide
relevant new charges on both sides of the gauge/gravity
correspondence. In \cite{Bertolini:2001ma,Bertolini:2002pr} such
fluxes were constructed explicitely relying on the special kind of
three-fold:
\begin{equation}\label{specialusto}
    Y_{[3]} \, = \, Y_{[1+2]} \, = \, \mathbb{C}\times \mathrm{ALE}_\Gamma
\end{equation}
constructed by Kronheimer \cite{kro1,kro2} and mentioned above as
case \textbf{A}.
\par
As we explain in detail below, the essential geometrical feature of
$Y_{[3]}$,  required to construct consistent fluxes of the complex
$3$-forms $\mathcal{H}_\pm$, is that $Y_{[3]}$ should admit
imaginary \textit{(anti)-self-dual, harmonic $3$-forms}
$\Omega^{(2,1)}$, which means:
\begin{equation}\label{alaguerre}
    \star_{\mathbf{g}^{\mathrm{RFK}}} \Omega^{(2,1)} \, = \, \pm \, {\rm i} \,  \Omega^{(2,1)}
\end{equation}
and simultaneously:
\begin{equation}\label{conspicua}
 \mathrm{d}\Omega^{(2,1)} \, = \, 0 \quad \Rightarrow \quad
 \mathrm{d}\star_{\mathbf{g}^{\mathrm{RFK}}}\Omega^{(2,1)}\, = \, 0
\end{equation}
Since the Hodge-duality operator involves the use of a metric, we
have been careful in specifying that (anti)-self-duality should
occur with respect to the Ricci-flat metric that is the one used in
the rest of the supergravity solution construction.
\par
The reason why the choice (\ref{specialusto}) of the three-fold
allows the existence of harmonic anti-self dual $3$-forms is easily
understood recalling that the $\mathrm{ALE}_\Gamma$-manifold
obtained from the resolution of $\mathbb{C}^2/\Gamma$ has a compact
support cohomology group of type $(1,1)$ of the following dimension:
\begin{equation}\label{creativina}
    \mathrm{dim}\,\mathrm{H}^{(1,1)}_{comp}\left(\mathrm{ALE}_\Gamma \right) \, = \, r \quad
    \mbox{where}
    \quad r \, = \, \# \,\mbox{ of nontrivial conjugacy classes of
    $\Gamma$}
\end{equation}
Naming $z \in \mathbb{C}$ the coordinate on the factor $\mathbb{C}$
of the product (\ref{specialusto}) and $\omega_I^{(1,1)}$ a basis of
harmonic anti-self dual one-forms on $\mathrm{ALE}_\Gamma$, the
ansatz utilized in \cite{Bertolini:2001ma,Bertolini:2002pr} to
construct the required $\Omega^{(2,1)}$ was the following:
\begin{equation}\label{galeno}
     \Omega^{(2,1)} \, \equiv \, \partial_z
    \, \mathfrak{f}^I (z) \, dz \, \wedge \, \omega^{(1,1)}_I
\end{equation}
where  $\mathfrak{f}^I (z)$ is a set of holomorphic functions of
that variable.  As it is well known $r$ is also the  rank of the
corresponding Lie Algebra in the ADE-classification of the
corresponding Kleinian groups  and the $2$-forms $\omega^{(1,1)}_I$
can be chosen dual to a basis of homology cycles $\mathcal{C}_I$
spanning $H_{2}\left(\mathrm{ALE}_\Gamma\right)$, namely we can set:
\begin{equation}\label{samotracia}
    \int_{\mathcal{C}_I} \,\omega^{(1,1)}_J \, = \, \delta_{IJ}
\end{equation}
The form $\Omega^{(2,1)}$ is closed by construction:
\begin{equation}\label{samellus}
   \mathrm{d} \Omega^{(2,1)} \, = \,0
\end{equation}
and it is also anti-selfdual with respect to the Ricci-flat metric:
\begin{equation}\label{carondimonio}
     \text{ds}^2_{Y_{[1+2]}} \, = \, dz\otimes d\bar{z}\, + \,  \text{ds}^2_{\mathrm{ALE}_\Gamma}
\end{equation}
Hence the question whether we can construct sufficiently flexible
D3-solutions of supergravity with both $5$-form and $3$-form fluxes
depends on the nontriviality of the relevant cohomology group:
\begin{equation}\label{ramingo21}
    \mbox{dim} \, \mathrm{H}^{(2,1)}\left(Y_{[3]}\right)\, > \, 0
\end{equation}
and on our ability to find harmonic (anti)-self dual representatives
of its classes (typically not with compact support and hence non
normalizable).
\par
At this level we find a serious difficulty that was already
mentioned in the introductory section \ref{cosemediane}. Indeed it
appears that we are not able to find the required $\Omega^{(2,1)}$
forms on $Y^\Gamma_{[3]}$ and that no D3-brane supergravity solution
with $3$-form fluxes can be constructed dual to the gauge theory
obtained from the Kronheimer construction dictated by $\Gamma
\subset \mathrm{SU(3)}$. Fortunately, the sharp conclusion encoded
in eqn.~(\ref{vecchioni}) follows from a hidden mathematical
assumption that, in physical jargon, amounts to a rigid universal
choice of the holomorphic superpotential $\mathcal{W}(\Phi)$. Under
appropriate conditions that in collaboration with Massimo Bianchi it
is was planned  to explain in a future publication not yet
accomplished and which are detectable at the level of the McKay
quiver diagram, the superpotential can be deformed (\textbf{mass
deformation}), yielding a family of three-folds
$Y^{\Gamma,\mu}_{[3]}$ which flow, for limiting values of the
parameter ($\mu \to \mu_0$) to a three-fold $Y^{\Gamma,\mu_0}_{[3]}$
admitting imaginary anti self-dual harmonic (2,1)-forms. Since the
content and the interactions of the gauge theory are dictated by the
McKay quiver of $\Gamma$  and by its associated Kronheimer
construction, we are entitled to see its mass deformed version and
the exact D3-brane supergravity solution built on
$Y^{\Gamma,\mu_0}_{[3]}$ as dual to each other.
\par
This, as I already said, will be the object of a future work, which
unfortunately for various reasons including the recent pandemia has
been on hold for a couple of years.
\par
This being said, I begin with an accurate mathematical summary of
the construction of D3-brane solutions of type IIB supergravity
using the geometric formulation of the latter (see a pedagogical
summary in \cite{Fre:2013ika} second volume chapter 6, section 6.8)
obtained within the rheonomy framework \cite{castdauriafre}.
\subsection{Geometric formulation of Type IIB supergravity }
\label{type2bsum} \setcounter{equation}{0} 
In order to discuss conveniently the D3 brane solutions of type IIB
that have as transverse space the crepant resolution of a
$\mathbb{C}^3/\Gamma$ singularity, we have to recall the geometric
Free Differential Algebra formulation of the chiral ten dimensional
theory fixing with care all our conventions, which is not only a
matter of notations but also of principles and geometrical insight.
Indeed the formulation of type IIB supergravity as it appears in
string theory textbooks \cite{greenschwarz,Polchinskibook} is
tailored for the comparison with superstring amplitudes and is quite
appropriate to this goal. Yet, from the viewpoint of the general
geometrical set up of supergravity theories this formulation is
somewhat unwieldy. Specifically it neither makes the
$\mathrm{SU(1,1)/U(1)}$ coset structure of the theory manifest, nor
does it  relate the supersymmetry transformation rules  to the
underlying algebraic structure which, as in all other instances of
supergravities, is a simple and well defined {\sl Free Differential
algebra}. The Free Differential Algebra of type IIB supergravity was
singled out many years ago by Castellani in \cite{castella2b} and
the geometric, manifestly $\mathrm{SU(1,1)}$--covariant formulation
of the theory was constructed by Castellani and Pesando in
\cite{igorleo}. Their formulae and  their transcription from a
complex $\mathrm{SU(1,1)}$ basis to a real
$\mathrm{SL(2,\mathbb{R})}$ basis were summarized and thoroughly
explained in the dedicated section of the book \cite{Fre:2013ika}
quoted above which I refer the reader to.
\subsection{The D3-brane solution with a  $Y_{[3]}$ transverse
manifold} \label{3} 
In this section I discuss a D3-brane solution of type IIB
supergravity in which, transverse to the  brane world-manifold, we
place a smooth non compact three-fold $ Y_{[3]}$ endowed with a
Ricci-flat K\"ahler metric.
\par
The ansatz for the D3-brane solution is characterized by two kinds
of flux; in addition to the usual RR 5-form flux, there might be as
we discussed above a non-trivial flux of the supergravity  complex
3-form field strengths $ \mathcal{H}_{\pm}$.
\par
Let us separate the ten coordinates of space-time into the following
subsets:
\begin{equation}
 x^M = \left \{ \begin{array}{rcll}
x^\mu &:& \mu =0,1,2,3& \mbox{coordinates of the 3-brane world volume}   \\
y^\tau &:& \tau=4,5,6,7,8,9 & \mbox{real coordinates of the $Y$
variety}   \
\end{array} \right.
\label{coordisplit}
\end{equation}
\subsubsection{The D3 brane ansatz}
\label{sol3flux}  We make the following ansatz for the metric:
\begin{eqnarray}
\label{ansazzo}
 \text{ds}^2_{[10]}&=&H(\pmb{y},\bar{\pmb{y}})^{-\frac{1}{2}}\left
(-\eta_{\mu\nu}\,dx^\mu\otimes dx^\nu \right
)+H(\pmb{y},\bar{\pmb{y}})^{\frac{1}{2}} \, \left(
\mathrm{\mathbf{g}}^{\mathrm{RFK}}_{\alpha\beta^\star} \, dy^\alpha \otimes dy^{\beta^\star}\right) \, \nonumber\\
 \text{ds}^2_{Y}&=&\mathrm{\mathbf{g}}_{\alpha\beta^\star}^{\mathrm{RFK}} \, dy^\alpha \otimes dy^{\beta^\star}\nonumber\\
{\rm det}(g_{[10]})&=&H(\pmb{y},\bar{\pmb{y}}){\rm det}(\mathrm{\mathbf{g}^{\mathrm{RFK}}})\nonumber\\
\eta_{\mu\nu}&=&{\rm diag}(+,-,-,-)
\end{eqnarray}
where $\mathrm{\mathbf{g}}^{\mathrm{RFK}}$ is the K\"ahler metric of
the $Y_{[3]}$ manifold
\begin{equation}
  \mathrm{\mathbf{g}}_{\alpha \bar{\beta}}^{\mathrm{RFK}} \, = \, \partial_\alpha \,
  \partial_{\bar{\beta}} \,
  \mathcal{K}^{\mathrm{RFK}}\left(\pmb{y},\bar{\pmb{y}}\right)
\label{m6defi}
\end{equation}
the real function
$\mathcal{K}^{\mathrm{RFK}}\left(\pmb{y},\bar{\pmb{y}}\right)$ being
a suitable K\"ahler potential.
\subsubsection{Elaboration of the ansatz}
In terms of vielbein the ansatz (\ref{ansazzo}) corresponds to
\begin{equation}
V^{A}= \left \{\begin{array}{rcll}
  V^a  & = & H(\pmb{y},\bar{\pmb{y}})^{-1/4} \, dx^a & a=0,1,2,3\\
  V^\ell& = & H(\pmb{y},\bar{\pmb{y}})^{1/4} \,
  \mathbf{e}^\ell & \ell \, = \, 4,5,6,7,8,9
\end{array}\right.
\label{splittoviel}
\end{equation}
where   $\mathbf{e}^\ell$ are the \textbf{sechsbein}   $1$-forms of
the manifold $Y_{[3]}$ of real dimension six. The structure
equations of the latter are\footnote{The hats over the spin
connection and the Riemann tensor denote quantities computed without
the warp factor.}:
\begin{eqnarray}
0& = &  d \, \mathbf{e}^i  - \widehat{\omega}^{ij} \, \wedge \, \mathbf{e}^k \, \eta_{jk}\nonumber\\
\widehat{R}^{ij} & = & d \widehat{\omega}^{ij} -
\widehat{\omega}^{ik}\, \wedge \, \widehat{\omega}^{\ell j} \,
\eta_{k\ell} = \widehat{R}^{ij} _{\phantom{ij}\ell m } \,
\mathbf{e}^\ell \,\wedge \, \mathbf{e}^m \label{structeque}
\end{eqnarray}
The relevant property  of the $Y_{[3]}$ metric that we use in
solving Einstein equations is that it is Ricci-flat:
\begin{equation}
 \widehat{R}^{im}_{\phantom{ij}\ell m } = 0 \label{ricciflatto}
\end{equation}
What we need in order to derive our solution and discuss its
supersymmetry properties is the explicit form of the spin connection
for the full $10$-dimensional metric (\ref{ansazzo}) and  the
corresponding Ricci tensor. From the torsion equation one can
uniquely determine the solution:
\begin{eqnarray}
\omega^{ab} & = & 0 \nonumber\\
\omega^{a\ell} & = &  \ft 1 4 \, H^{-3/2} \, dx^a  \eta^{\ell k} \,
\partial_k \, H
\nonumber\\
\omega^{\ell m} & = & \widehat{\omega}^{\ell m} + \Delta
\omega^{\ell m} \quad ; \quad \Delta \omega^{\ell m} =- \ft 1 2 \,
H^{-1} \, \mathbf{e}^{[\ell}  \, \eta^{m]k} \, \partial_k H
\label{spinconnect}
\end{eqnarray}
Inserting this result into the definition of the curvature $2$-form
we obtain\footnote{The reader should be careful with the indices.
Latin indices are always frame indices referring to the vielbein
formalism. Furthermore we distinguish the 4 directions of the brane
volume by using Latin letters from the beginning of the alphabet
while the 6 transversal directions are denoted by Latin letters from
the middle and the end of the alphabet. For the coordinate indices
we utilize Greek letters and we do exactly the reverse. Early Greek
letters $\alpha,\beta,\gamma,\delta,\dots$ refer to the 6 transverse
directions while Greek letters from the second half of the alphabet
$\mu,\nu,\rho,\sigma,\dots$ refer to the D3 brane world volume
directions as it is customary in $D=4$ field theories. }:
\begin{eqnarray}
R^{a}_{b} & = & -  \frac{1}{8}\, \left [ H^{-3/2} \Box_{\mathbf{g}}
\, H - H^{-5/2} \,
\partial_k H\partial^k H \right] \, \delta^a_b \nonumber\\
R^{a}_{\ell} & = & 0\nonumber\\
R_\ell^m   &=& \frac{1}{8}H^{-3/2} \Box_{\mathbf{g}} H\delta_\ell^m
             - \frac{1}{8} H^{-5/2}\partial_s H\partial^s
             H\delta_\ell^m
             +\frac{1}{4} H^{-5/2} \partial_\ell H\partial^m H
\label{riccius}
\end{eqnarray}
where for any function $f\left(\pmb{y},\bar{\pmb{y}}\right)$ with
support on $Y_{[3]}$:
\begin{equation}
\Box_{\mathbf{g}} \, f\left(\pmb{y},\bar{\pmb{y}}\right) \, =
\,\frac{1}{\sqrt{\mathrm{det}\mathbf{g}}}\, \left( \partial_\alpha
\left(\sqrt{\mathrm{det}\mathbf{g}}\,\,
\mathbf{g}^{\alpha\beta^\star} \,
\partial_{\beta^\star} \,f \right) \right) \label{laplacious}
\end{equation}
denotes the action on it of the Laplace--Beltrami operator  with
respect to the metric (\ref{m6defi}) which is the Ricci-flat one: I
have omitted the superscript $\mathrm{RFK}$ just for simplicity.
Indeed on the supergravity side of the correspondence we ought to
use only the Ricci-flat metric and there is no ambiguity.
\subsubsection{Analysis of the field equations in geometrical terms}
The equations of motion for the scalar fields $\varphi$ and
$C_{[0]}$ and for the 3-form field strength $F^{NS}_{[3]}$ and
$F^{RR}_{[3]}$ can be better analyzed using the complex notation.
Defining:
\begin{eqnarray}
{\mathcal{H}}_\pm & = & \pm 2 \,e^{-\varphi/2} F^{NS}_{[3]} + {\rm
i} 2 \,e^{\varphi/2} \,F^{RR}_{[3]}
\label{mcH} \\
P & =& \ft 1 2 \, d\varphi -{\rm i} \ft 12 \, e^\varphi \,
F_{[1]}^{RR} \label{Psc}
\end{eqnarray}
eqn.s (\ref{dstarP})-(\ref{hodge2formeq}) can be  respectively
written as:
\begin{eqnarray}
d(\star P)- {\rm i} e^{\varphi} dC_{[0]}\wedge \star P + \ft 1 {16}
   {\mathcal{H}}_+ \, \wedge \, \star
  {\mathcal{H}}_+=0 \label{dstarP} \\
d \star {\mathcal{H}}_+ - \frac{{\rm i}}{2} e^{\varphi} dC_{[0]}
\wedge \, \star {\mathcal{H}}_+=
   {\rm i} \, {F}_{[5]}^{RR}\,
  \wedge \, {\mathcal{H}}_+  - P \wedge \star
  {\mathcal{H}}_- \label{hodge2formeq}
\end{eqnarray}
while  the equation for the 5-form becomes:
\begin{equation}
d\star F^{RR}_{[5]} = {\rm i} \, \ft 1 {8} \, {\mathcal{H}}_+ \wedge
{\mathcal{H}}_- \label{f5equazia}
\end{equation}
\par
Besides assuming the structure (\ref{ansazzo}) we also assume that
the two scalar fields, namely the dilaton $\varphi$ and the
Ramond-Ramond $0$-form $C_{[0]}$ are constant and vanishing:
\begin{equation}
  \varphi=0 \quad ; \quad C_{[0]}=0
\label{zerodilat}
\end{equation}
As we shall see, this assumption simplifies considerably the
equations of motion, although these two scalar fields can be easily
restored.
\subsubsection{The three-forms}
The basic ansatz  characterizing the  solution and  providing  its
interpretation as a D3-brane with three-form fluxes  is described
below.
\par
The ansatz for the complex three-forms of type IIB supergravity is
given below and is inspired by what was done in
\cite{Bertolini:2002pr,Bertolini:2001ma} in the case where $Y_{[3]}
= \mathbb{C}\times \mathrm{ALE}_\Gamma$:
\begin{equation}
{\mathcal{H}}_+ \, = \, \Omega^{(2,1)}  \label{hpmposiz}
\end{equation}
where $\Omega^{(2,1)}$ is localized on $Y_{[3]}$ and satisfies
eqn.s~(\ref{alaguerre}-\ref{conspicua})
\par
If we insert the ans\"atze (\ref{zerodilat},\ref{hpmposiz}) into the
scalar field equation (\ref{dstarP}) we obtain:
\begin{eqnarray}
{\mathcal{H}}_+ \, \wedge \, \star_{10} {\mathcal{H}}_+=0
\label{holovincol}
\end{eqnarray}
This equation is automatically satisfied by our ansatz for a very
simple reason that I explain next. The form ${\mathcal{H}}_+$ is by
choice a three-form on $Y_{[3]}$ of type $(2,1)$. Let $\Theta^{[3]}$
be any three-form that is localized on the transverse
six-dimensional \footnote{For the sake of the present calculation
and the following ones where we have to calculate a Hodge dual, it
is more convenient to utilize a set of 6 real coordinates $t^I$
($I=1,\dots,6$) for the manifold $Y_{[3]}$. Let $\partial_I \equiv
\frac{\partial}{\partial t^I}$ denote the standard partial
derivatives with respect to such coordinates.} manifold $Y_{[3]}$:
\begin{equation}\label{terzaformina}
    \Theta^{[3]} \, = \, \Theta_{IJK} \, dt^I\wedge dt^J \wedge dt^K
\end{equation}
When we calculate the Hodge dual of $\Theta^{[3]}$ with respect to
the 10-dimensional metric (\ref{ansazzo}) we obtain a $7$-form with
the following structure:
\begin{equation}\label{salamepuzzoso}
    \star_{10}\,\Theta^{[3]} \, = \, H^{-1} \,
    \mbox{Vol}_{\mathbb{R}^{(1,3)}} \, \wedge \, \widetilde{\Theta}^{[3]}
\end{equation}
where:
\begin{equation}\label{cirimetto}
    \mbox{Vol}_{\mathbb{R}^{(1,3)}} \, = \, \ft {1}{4!} \, dx^{\mu} \wedge dx^{\nu}
    \wedge dx^\rho \wedge dx^\sigma \, \epsilon_{\mu\nu\rho\sigma}
\end{equation}
is the volume-form of the flat D3-brane and
\begin{equation}\label{zante}
    \widetilde{\Theta}^{[3]} \, \equiv \, \star_{\mathbf{g}} \, \Theta^{[3]}
\end{equation}
is the dual of the three-form $\Theta^{[3]}$ with respect to the
metric $\mathbf{g}$ defined on $Y_{[3]}$. Let us now specialize the
three-form $\Theta^{[3]}$ to be of type $(2,1)$:
\begin{equation}\label{specialindo}
    \Theta^{[3]} \, = \, \mathrm{Q}^{(2,1)}
\end{equation}
Preservation of supersymmetry requires the complex three-form
${\mathcal{H}}_+$ to obey the condition\footnote{It also requires
${\mathcal{H}}_+$ to be primitive.}
\begin{equation}\label{gerundio}
    \star_{\mathbf{g}} \, \mathrm{Q}^{(2,1)} \, = - {\rm i} \,
   \mathrm{Q}^{(2,1)}
\end{equation}
Hence:
\begin{equation}\label{ciarlatano}
    {\mathcal{H}}_+ \wedge \star_{10}\, {\mathcal{H}}_+ \, = - {\rm i} \,
   \mathrm{Q}^{(2,1)} \, \wedge \,
    \mathrm{Q}^{(2,1)}
     \,\wedge \, H^{-1} \mbox{Vol}_{\mathbb{R}^{(1,3)}}\, = \, 0
\end{equation}
\subsubsection{The self-dual $5$-form} Next we consider the self-dual $5$-form $F_{[5]}^{RR}$
which by definition must satisfy the following Bianchi identity:
\begin{equation}
d \, F_{[5]}^{RR} = {\rm i} \, \ft 1 8 \,{ \mathcal{H}}_+ \, \wedge
\, { \mathcal{H}}_- \label{f5bianchi}
\end{equation}
Our ansatz for $F_{[5]}^{RR}$ is the following:
\begin{eqnarray}
F_{[5]}^{RR} & = & \alpha \left( U + \star_{10}\, U \right)  \label{genova1}\\
U & =  & \mathrm{d} \left( H^{-1} \, \mbox{Vol}_{\mathbb{R}^{(1,3)}}
\right) \label{f5ansaz}
\end{eqnarray}
where $\alpha$ is a constant to be determined later. By construction
$F_{[5]}^{RR}$ is self-dual and its equation of motion is trivially
satisfied. What is not guaranteed is that also the Bianchi identity
(\ref{f5bianchi}) is fulfilled. Imposing it, results into a
differential equation for the function
$H\left(\pmb{y},\bar{\pmb{y}}\right)$. Let us see how this works.
\par
Starting from the ansatz (\ref{f5ansaz}) we obtain:
\begin{eqnarray}
  U &=& -\frac{1}{4!} \, \epsilon_{\mu\nu\rho\sigma} \,dx^\mu\wedge dx^\nu\wedge dx^\rho \wedge dx^\sigma \wedge \frac{dH}{H^2} \\
 U_{\mu\nu\rho\sigma  I}  &=& -\frac{1}{4!}
  \epsilon_{\mu\nu\rho\sigma} \, \frac{\partial_I H}{H^2} \quad ;
  \quad
  \mbox{all other components vanish}
\end{eqnarray}
 Calculating the components
of the dual form $\star_{10}\, U$ we find that they are non
vanishing uniquely in the six transverse directions:
\begin{eqnarray}
   \star_{10} U &=& \tilde{U}_{I_1\dots I_5} \,\,dt^{I_1} \wedge \dots \wedge dt^{I_5} \nonumber\\
   \tilde{U}_{I_1\dots I_5} &=& - \, \frac{\sqrt{\mbox{det} \,
   g_{(10)}}}{5!} \epsilon_{I_1\dots I_5 J}\,
   \epsilon_{\mu\nu\rho\sigma} \, g_{(10)}^{JK} \, g_{10}^{\mu\mu^\prime}\, g_{(10)}^{\nu\nu^\prime}\, g_{(10)}^{\rho\rho^\prime}\,
   g_{(10)}^{\sigma\sigma^\prime} \, U_{\mu^\prime\nu^\prime\rho^\prime\sigma^\prime  J}
   \nonumber\\
   &=& \frac{\sqrt{\mbox{det}\mathbf{g}}}{5!} \,\epsilon_{I_1\dots I_5 J} \,
   \mathbf{g}^{JK} \, \partial_K \,H
\end{eqnarray}
The essential point in the above calculation is that all powers of
the function $H$ exactly cancel so that $\star_{10} U$ is linear in
the $H$-derivatives \footnote{Note that we use $\mathbf{g}_{IJ}$ to
denote the components of the K\"ahler metric (\ref{m6defi}) in the
real coordinate basis $t^I$.}. Next using the same coordinate basis
we obtain:
\begin{eqnarray}\label{scolopio}
    d \, F_{[5]}^{RR}& =  & \alpha \, d\star U \, =
    \,\alpha \,
    \underbrace{\frac{1}{\sqrt{\mbox{det} \, \mathbf{g} }} \, \partial_I \,
    \left( \sqrt{\mbox{det} \, \mathbf{g} } \, \mathbf{g}^{IJ} \, \partial_J
    H\right)}_{\Box_{\mathbf{g}} \, H} \, \times \, \mbox{Vol}_{Y_{[3]}} \nonumber\\
    &=& \alpha \,\Box_{\mathbf{g}} \, H(\pmb{y},\bar{\pmb{y}}) \, \times
    \,\mbox{Vol}_{Y_{[3]}}
\end{eqnarray}
where:
\begin{eqnarray}\label{supercaffelatte}
    \mbox{Vol}_{Y_{[3]}} & \equiv &\sqrt{\mbox{det} \, \mathbf{g} } \,
    \frac{1}{6!} \epsilon_{I_1 \dots I_6} dt^{I_1}\wedge \dots
    \wedge dt^{I_6} \nonumber\\
    & = & \sqrt{\mbox{det} \, \mathbf{g}}  \,
    \frac{1}{(3!)^2}\,
    \epsilon_{\alpha\beta\gamma}\,
    dy^\alpha \wedge dy^\beta \wedge dy^\gamma \, \wedge \epsilon_{\bar{\alpha}\bar{\beta}\bar{\gamma}}\,
    d\bar{y}^{\bar{\alpha}} \wedge d\bar{y}^{\bar{\beta}} \wedge
    d\bar{y}^{\bar{\gamma}}
\end{eqnarray}
is the volume form of the transverse six-dimensional space. Once
derived with the use of real coordinates, the relation
(\ref{scolopio}) can be transcribed in terms of complex coordinates
and the Laplace-Beltrami operator $\Box_\mathbf{g}$  can be written
as in eqn. (\ref{laplacious}). Let us now analyze the source terms
provided by the three-forms. With our ansatz we obtain:
\begin{eqnarray}
\ft 1 8 \,{ \mathcal{H}}_+  \, \wedge \, {
  \mathcal{H}}_- &= & \mathbb{J}\left(\pmb{y},\bar{\pmb{y}}\right)
  \, \times \,
  \mbox{Vol}_{Y_{[3]}} \nonumber\\
\mathbb{J}\left(\pmb{y},\bar{\pmb{y}}\right) & = & - \, \frac{1}{72
\, \sqrt{\mbox{det}\, \mathbf{g}}} \, \, \,
\Omega_{\alpha\beta\bar{\eta}}\,\, \bar{\Omega}_{\bar{\delta}
\bar{\theta}\gamma} \,\, \epsilon^{\alpha\beta\gamma} \,\,
\epsilon^{\bar{\eta}\bar{\delta}\bar{\theta}}\label{tempra}
\end{eqnarray}
we conclude that the Bianchi identity (\ref{f5bianchi}) is satisfied
by our ansatz if:
\begin{equation}
  \Box_\mathbf{g}\,  H = - \frac
  {1}{\alpha}\, \mathbb{J}\left(\pmb{y},\bar{\pmb{y}}\right)
\label{maindiffe}
\end{equation}
This is the main differential equation to which the entire
construction of the D3-brane solution can be reduced to. We are
going to show that the parameter $\alpha$ is determined by
Einstein's equations and fixed to $\alpha=1$.
\subsubsection{The equations for the three--forms}
Let us consider next the field equation for the complex three-form,
namely eqn. (\ref{hodge2formeq}). Since the two scalar fields are
constant the $\mathrm{SU(1,1)/O(2)}$ connection vanishes and  we
have:
\begin{equation}
d \star {\mathcal{H}}_+ = {\rm i} \, F^{RR}_{[5]}\, \wedge \,
{\mathcal{H}}_+ \label{simpeq}
\end{equation}
Using our ansatz we immediately obtain:
\begin{eqnarray}
d \star {\mathcal{H}}_+ = & = &- 2\, {\rm i} H^{-2} \mathrm{}dH \,
\wedge \, \tilde{\Omega}^{(2,1)}, \wedge \,
\Omega_{\mathbb{R}^{1,3}}\, +2 {\rm i} \, H^{-1} \,
d\tilde{\Omega}^{(2,1)} \wedge \, \Omega_{\mathbb{R}^{1,3}}
 \nonumber\\
{\rm i} \, F^{RR}_{[5]}\, \wedge \, {\mathcal{H}}_+ & = & - 2\,
\alpha {\rm i} H^{-2} dH \, \wedge \, \Omega^{(2,1)} \, \wedge \,
\Omega_{\mathbb{R}^{1,3}}\, \label{pagnacco}
\end{eqnarray}
Hence if $\alpha=1$,  the field equations for the three-form reduces
to:
\begin{equation}\label{cannalotto}
    \tilde{\Omega}^{(2,1)} \, \equiv  \, \star_{\mathbf{g}} {\Omega}^{(2,1)} \, = \,
    -\, {\rm i}\,{\Omega}^{(2,1)} \quad ;
    \quad d\star_{\mathbf{g}} {\Omega}^{(2,1)} \, = \,0 \quad ;
    \quad d  {\Omega}^{(2,1)} \, = \,0
\end{equation}
which are nothing else but eqn.s~(\ref{alaguerre}-\ref{conspicua}).
In other words the solution of type IIB supergravity with three-form
fluxes exists if and only if the transverse space admits
\textit{closed and imaginary anti-self-dual forms} $\Omega^{(2,1)}$
as we already stated\footnote{By construction a closed
anti-self-dual form is also coclosed, namely it is harmonic.}.
\par
In order to show that also the Einstein's equation is satisfied by
our ansatz we have to calculate the (trace subtracted) stress energy
tensor of the five and three index field strengths. For this purpose
we need the components of $F_{[5]}^{RR}$. These are easily dealt
with. Relying on the ansatz (\ref{f5ansaz}) and on
eqn.~(\ref{splittoviel}) for the vielbein we immediately get:
\begin{equation}
  F_{A_1 \dots A_5} =\left\{ \begin{array}{ccc}
    F_{i abcd} & = & \frac {1} {5!} \, f_i \, \epsilon_{abcd} \\
    F_{ j_1\dots i_5}  & = & \frac{1}{5!} \epsilon_{i j_1\dots j_5} \, f^i \\
    \mbox{otherwise} & = & 0 \
  \end{array} \right.
\label{F5intrinsic}
\end{equation}
where:
\begin{equation}
  f_i = - \alpha\, H^{-5/4} \, \partial_i H
\label{fidefi}
\end{equation}
Then by straightforward algebra we obtain:
\begin{eqnarray}
  T^{a}_{b}\left[ F_{[5]}^{RR}\right]  & \equiv & -75 \, F^{a \,\cdot\, \cdot\, \cdot \,\cdot} \,
  F_{b\,\cdot\, \cdot\, \cdot \,\cdot} = -  \frac 1 8 \, \delta^{a}_{b}
  \, f_p \, f^p \nonumber\\
  & = & - \alpha^2 \, \frac 1 8 \, \delta^{a}_{b}
  \, H^{-5/2} \partial_p H\, \partial^p H \nonumber\\
  T^{i}_{j}\left[  F_{[5]}^{RR}\right]  & \equiv & -75 \, F^{i \,\cdot\, \cdot\, \cdot \,\cdot} \,
  F_{j\,\cdot\, \cdot\, \cdot \,\cdot} = \frac 1 4\,  f^i \, f_j \, - \, \frac 1 8 \,
  \delta^i_j \, f_p \, f^p \nonumber\\
  & = & \alpha^2 \, \left[ \frac 1 4\, H^{-5/4}  \partial^i H \, \partial_j H  \, - \, \frac 1 8
  \, \delta^i_j \,  H^{-5/4}  \partial^p H \, \partial_p H \right]
\label{TofF5}
\end{eqnarray}
\par
Inserting eqn.s (\ref{TofF5}) and (\ref{riccius}) into Einstein's
equations:
\begin{eqnarray}
  R^a_b & = &  T^{a}_{b}\left[F_{[5]}^{RR} \right]
 \nonumber\\
  R^i_j & = &T^{i}_{j}\left[F_{[5]}^{RR}\right]
\end{eqnarray}
we see that they are satisfied, provided
\begin{equation}
  \alpha= 1
\label{fixed}
\end{equation}
and the master equation (\ref{maindiffe}) is satisfied. This
concludes our proof that an exact D3-brane solution with a $Y_{[3]}$
transverse space does indeed exist.
\section{Generalities on $\frac{\mathbb{C}^3}{\Gamma}$  singularities}
Recalling what we summarized above we conclude that the singularities relevant to our goals are of the form:
\begin{equation}\label{cialtrus}
   X \, = \, \frac{\mathbb{C}^3}{\Gamma}
\end{equation}
where the finite group $\Gamma \subset \mathrm{SU(3)}$  has a holomorphic action on $\mathbb{C}^3$. For this
case, as we mentioned above,  there is a series of general results and procedures developed in algebraic
geometry that we want to summarize in the  perspective of their use in physics.
\par
To begin with let us observe the schematic diagram sketched here below:
\begin{equation}
{\setlength{\unitlength}{1pt}
\begin{picture}(280,150)(-10,-70)
\thicklines \put(100,50){\circle*{20}} \put(100,35){\makebox(0,0){\Large $\Downarrow$}}
\put(100,25){\makebox(0,0){levels }}\put(100,15){\makebox(0,0){$\zeta$ }}\put(100,5){\makebox(0,0){of moment
map }}
 \put(100,50){\line(-1,-1){70}} \put(35,-15){\circle*{20}}
\put(100,50){\line(1,-1){70}}
\put(165,-15){\circle*{20}}\put(35,-15){\line(1,0){125}}
\put(100,65){\makebox(0,0){$r = \mbox{dim}\,
\mathfrak{z}\left[\mathbb{F}_\Gamma\right]$ center of the Lie
Algebra}} \put(225,-20){\makebox(0,0){$r = \# $ of nontrivial}}
\put(225,-30){\makebox(0,0){$\Gamma$ irred. represent.s}}
\put(-30,-20){\makebox(0,0){$r = \# $ of nontrivial}}
\put(-30,-30){\makebox(0,0) {$\Gamma$ conjugacy classes
}}\put(35,-33){\makebox(0,0){\Large
$\Downarrow$}}\put(165,-33){\makebox(0,0){\Large $\Downarrow$}}
\put(-30,-50){\makebox(0,0){age grading,}}
\put(-30,-60){\makebox(0,0) {exceptional divisors}}
\put(225,-50){\makebox(0,0){first Chern classes of}}
\put(225,-60){\makebox(0,0){tautological bundles}}
\put(35,-50){\circle{15}}\put(165,-50){\circle{15}}\put(35,-50){\circle{5}}\put(165,-50){\circle{5}}
\put(100,-51){\makebox(0,0){\LARGE
$\Leftrightarrow$}}\put(35,-50){\line(1,0){130}}
\end{picture}}
\label{salumaio}
\end{equation}
The fascination of the mathematical construction lying behind the
desingularization process, which has a definite counterpart in the
structure of the  gauge theories describing D3-branes at the
$\mathbb{C}^3/\Gamma$ singularity, is the triple interpretation of
the same number $r$ which alternatively yields:
\begin{itemize}
  \item The number of nontrivial conjugacy classes of the
  finite group $\Gamma$,
  \item The number of irreducible representations of the
  finite group $\Gamma$,
  \item The center of the  Lie algebra $\mathfrak{z}\left[\mathbb{F}_\Gamma\right]$ of the compact gauge group
  $\mathcal{F}_\Gamma$, whose structure, as we will see, is:
  \begin{equation}\label{sinuhe}
    \mathcal{F}_\Gamma \, = \, \bigotimes_{i=1}^r \mathrm{U(n_i)}
  \end{equation}
\end{itemize}
The levels $\zeta_I$ of the moment maps are the main ingredient of
the singularity resolution. At level $\zeta^I=0$ we have the
singular orbifold $\mathcal{M}_0 \, = \,
\frac{\mathbb{C}^3}{\Gamma}$, while at $\zeta^i \neq 0$ we obtain a
smooth manifold $\mathcal{M}_\zeta$ which develops a nontrivial
homology and cohomology. In physical parlance the levels $\zeta^I$
are the Fayet-Iliopoulos parameters appearing in the lagrangian,
while $\mathcal{M}_\zeta$ is the manifold of vacua of the theory,
namely of extrema of the potential, as we already emphasized.
\par
Quite generally, we find that each of the gauge factors $\mathrm{U(n_i)}$ is the structural group of a
holomorphic vector bundle of rank $n_i$:
\begin{equation}\label{broccoletti}
    \mathfrak{V}_i \, \stackrel{\pi}{\longrightarrow} \, \mathcal{M}_\zeta
\end{equation}
whose first Chern class is a nontrivial (1,1)-cohomology class of the resolved smooth
manifold:
\begin{equation}\label{cimedirapa}
    c_1\left(\mathfrak{V}_i\right) \, \in \, H^{1,1}\left(\mathcal{M}_\zeta\right)
\end{equation}
On the other hand the already anticipated very deep theorem
originally proved in the nineties by Reid and Ito \cite{giapumckay}
relates the dimensions of the cohomology groups
$H^{q,q}\left(\mathcal{M}_\zeta\right)$ to the conjugacy classes of
$\Gamma$ organized according to the grading named \textit{age}. So
named \textit{junior classes} of \textit{age} = 1 are associated
with $H^{1,1}\left(\mathcal{M}_\zeta\right)$ elements, while the
so-named \textit{senior classes} of \textit{age} = 2 are associated
with $H^{2,2}\left(\mathcal{M}_\zeta\right)$ elements.
\par
The link that pairs irreps with conjugacy classes is provided by the relation, well-known in algebraic
geometry, between \textit{divisors} and \textit{line bundles}. The conjugacy classes of $\gamma$ can be put
into correspondence with the exceptional divisors created in the resolution $\mathcal{M}_\zeta
\stackrel{\zeta\to 0}{\longrightarrow} \frac{\mathbb{C}^3}{\Gamma}$ and each divisor defines a line bundle
whose first Chern class is an element of the $H^{1,1}\left(\mathcal{M}_\zeta\right)$ cohomology group.
\par
These line bundles labeled by conjugacy classes have to be compared with the line bundles created by the
K\"ahler quotient procedure that are instead associated with the irreps, as we have sketched above. In this
way we build the bridge between conjugacy classes and irreps.
\par
Finally there is the question whether the divisor is compact or not. In the first case, by Poincar\'e
duality, we obtain nontrivial $H^{2,2}\left(\mathcal{M}_\zeta\right)$ elements. In the second case we have
no new cohomology classes. The age grading precisely informs us about the compact or noncompact nature of
the divisors. Each senior class
corresponds to a cohomology class of degree 4, thus signaling the existence of a non-trivial closed
(2,2) form, and via Poincar\'e duality, it also corresponds to a compact component of the exceptional
divisor.
\par
The physics--friendly illustration of this general beautiful scheme,
together with the explicit construction of a few concrete examples
is  among the priorities of the present lectures. Let me begin with
the concept of age grading.
\subsection{The concept of age grading for conjugacy classes
of the discrete group $\Gamma$} \label{vecchiardo} According to the
above quoted theorem that we shall explain below, the \textit{age
grading} of $\Gamma$ conjugacy classes allows to predict the
Dolbeault cohomology of the resolved algebraic variety. It goes as
follows.
\par
 Suppose that $\Gamma$ (a finite group) acts in a  linear way on
$\mathbb{C}^n$. Consider an element $\gamma \in \Gamma$ whose action is the following:
\begin{equation}\label{etoso}
  \gamma.\vec{z} \, = \, \underbrace{\left(
                                 \begin{array}{ccc}
                                   \ldots & \ldots &  \ldots \\
                                   \vdots &  \vdots& \vdots \\
                                  \ldots & \ldots &\ldots \\
                                 \end{array}
                               \right)
   }_{\mathcal{Q}(\gamma)}\, \cdot \, \left(\begin{array}{c}
                               z_1 \\
                               \vdots\\
                               z_n
                             \end{array}
    \right)
\end{equation}
Since in a finite group all elements have a finite order,  there exists  $r \in \mathbb{N}$,  such that $\gamma^r\,
= \, \mathbf{1}$. We define the age of an element in the following way. Let us diagonalize $D(\gamma)$,
namely compute its eigenvalues. They will be as follows:
\begin{equation}\label{gioffo}
  \left(\lambda_1 , \dots , \lambda_n\right ) \, = \, \exp\left[\frac{2\pi \, i}{r}
  \, a_i\right]  \quad; \quad r> a_i \in  \mathbb{N} \quad i\, = \, 1, \dots , n
\end{equation}
We define:
\begin{equation}\label{vecchione}
  \mathrm{age}\left(\gamma \right)\, = \, \frac{1}{r} \, \sum_{i=1}^n a_i
\end{equation}
Clearly the age is a property of the conjugacy class of the element, relative to the considered three-dimensional
complex representation.
\subsection{The fundamental theorem}\label{anglonipponico}
 In \cite{giapumckay} Y. Ito and M. Reid proved the following fundamental theorem:
\begin{teorema}
\label{reidmarktheo} Let $Y_{[3]}^\Gamma\rightarrow
\mathbb{C}^3/\Gamma$ be a crepant \footnote{A resolution of
singularities $X \rightarrow Y_{[3]}^\Gamma$  is crepant when the
canonical bundle of $X$ is the pullback of the canonical bundle of
$Y_{[3]}^\Gamma$.} resolution of a Gorenstein \footnote{A variety is
Gorenstein when the canonical divisor is a Cartier divisor, i.e., a
divisor corresponding to a line bundle.} singularity. Then we have
the following relation between the de-Rham cohomology groups of the
resolved smooth variety $Y_{[3]}^\Gamma$ and the ages of $\Gamma$
conjugacy classes:
$$ \mathrm{dim} \, H^{2k} \left(Y_{[3]}^\Gamma\right) \, = \, \# \mbox{ of age $k$ conjugacy classes of  $\Gamma$}$$
\end{teorema}
On the other hand it happens that all odd cohomology groups are trivial:
\begin{equation}\label{oddi}
    \mathrm{dim} \, H^{2k+1} \left(Y_{[3]}^\Gamma\right) \, = \, 0
\end{equation}
This is true also in the case of $\mathbb{C}^2/\Gamma$
singularities, yet in $n=2,3$ the consequences of the same fact are
drastically different. In all complex dimensions $n$  the
deformations of the K\"ahler class are in one-to-one correspondence
with the harmonic forms $\omega^{(1,1)}$, while those of the complex
structure are in correspondence with the harmonic forms
$\omega^{(n-1,1)}$. In $n=2$ the harmonic $\omega^{(1,1)}$ forms
play the double role of K\"ahler class deformations and complex
structure deformations. This is the reason why we can do a
hyperK\"ahler  quotient and we have both moduli parameters in the
K\"ahler potential and in the polynomials cutting out the smooth
variety. Instead in $n=3$ eqn.\,(\ref{oddi}) implies that the
polynomials constraints cutting the singular locus have no
deformation parameters. The parameters of the resolution occur only
at the level of the K\"ahler quotient and are the levels of the
K\"ahlerian moment maps.
\par Given an algebraic representation of the variety $Y_{[3]}^\Gamma$ as the
vanishing locus of certain polynomials $W(x)\, = \, 0$, the
algebraic $2k$-cycles  are the $2k$-cycles that can be
holomorphically embedded in $Y_{[3]}^\Gamma$.   The following
statement in $n=3$ is  elementary:
\begin{statement}
The Poincar\'e dual of any algebraic $2k$-cycle is   of type $(k,k)$.
\end{statement}
Its converse is known as the Hodge conjecture, stating that any cycle of type $(k,k)$
is a linear combination of algebraic cycles. This will hold true for the varieties we shall be considering.
\par
Thus we conclude that the so named \textit{junior conjugacy classes} (age=1) are in a
one-to-one correspondence with $\omega^{(1,1)}$-forms that span ${H^{1,1}}$, while \textit{conjugacy classes
of age 2} are in one-to-one correspondence with $\omega^{(2,2)}$-forms that span ${H^{2,2}}$.
\section{Comparison with ALE manifolds and comments}
 Let us compare the above
predictions  for the case B) of $\mathbb{C}^3/\Gamma$ singularities with the well known case A) of
$\mathbb{C}^2/\Gamma$ where $\Gamma$ is a Kleinian subgroup of $\mathrm{SU(2)}$ and the resolution of the
singularity leads to an ALE manifold \cite{mango,degeratu,kro1,kro2,Bertolini:2002pr,Bertolini:2001ma}.  As
we already stressed above,  this latter can be explicitly constructed by means of a  hyperK\"ahler  quotient,
according to Kronheimer's construction.
\par
In table \ref{kleinale} I summarize some well known facts about
$Y^\Gamma_{[2]}\to X=\mathbb{C}^2/\Gamma$ which are the following.
Here $\chi$ denotes Hirzebruch's signature characteristic of the
resolved manifold.

\begin{table}[h!]\caption{Finite $\mathrm{SU(2)}$ subgroups versus ALE manifold
properties} \label{kleinale}{\small
\begin{center}
\begin{tabular}{||l|c|c|c|c|c||}\hline
$\null$ & $\null$ & $\null$ & $\null$ & $\null$ & $\null$ \\
$\Gamma$. & $W_\Gamma(u,w,z)$ & ${\cal R}=\frac{{\bf C}[u,w,z]}{\partial W}$
&$|{\cal R}|$ & ${\#} c.~c. $& $\tau\equiv\chi -1$  \\
$\null$ & $\null$ & $\null$ & $\null$ & $\null$ & $\null$ \\
\hline \hline
$\null$ & $\null$ & $\null$ & $\null$ & $\null$ & $\null$ \\
$A_k$&$u^2+w^2 - z^{k+1}$&$\{ 1, z,.. $&$k$&$k+1$&$k$ \\
$~$&$~$&  $.., z^{k-1} \}$&$~$&$~$&$~$ \\
$\null$ & $\null$ & $\null$ & $\null$ & $\null$ & $\null$ \\
\hline
$\null$ & $\null$ & $\null$ & $\null$ & $\null$ & $\null$ \\
$D_{k+2}$&$u^2 +w^2 z + z^{k+1}$&$\{ 1, w, z,w^2,$&$k+2$&$k+3$&$k+2$ \\
$~$&$~$&  $ z^2, ..., z^{k-1} \}$&$~$&$~$&$~$ \\
$\null$ & $\null$ & $\null$ & $\null$ & $\null$ & $\null$ \\
\hline
$\null$ & $\null$ & $\null$ & $\null$ & $\null$ & $\null$ \\
$E_6=$&$u^2+w^3 +z^{4}$&$\{ 1, w,  z,$&$6$&$7$&$6$ \\
${\cal T}$&$~$&  $wz, z^2,wz^{2} \}$&$~$&$~$&$~$ \\
$\null$ & $\null$ & $\null$ & $\null$ & $\null$ & $\null$ \\
\hline
$\null$ & $\null$ & $\null$ & $\null$ & $\null$ & $\null$ \\
$E_7=$&$u^2+w^3 +wz^{3}$&$\{ 1, w, z,w^2,$&$7$&$8$&$7$ \\
${\cal O}$&$~$&  $z^2,wz,w^2z \}$&$~$&$~$&$~$ \\
$\null$ & $\null$ & $\null$ & $\null$ & $\null$ & $\null$ \\
\hline
$\null$ & $\null$ & $\null$ & $\null$ & $\null$ & $\null$ \\
$E_8=$&$u^2+w^3 + z^{5}$&$\{ 1,w, z,z^2,wz,$&$8$&$9$&$8$ \\
${\cal I}$&$~$&  $z^3,wz^2,wz^3  \}$&$~$&$~$&$~$ \\
$\null$ & $\null$ & $\null$ & $\null$ & $\null$ & $\null$ \\
\hline
\end{tabular}\end{center}}\end{table}
\vskip 0.2cm
\begin{enumerate}
  \item As an affine variety the singular orbifold $X$ is described by a single polynomial  equation
  $W_\Gamma(u,w,z)\, =\,0$ in $\mathbb{C}^3$. This equation is simply given by a relation existing
  among the invariants of $\Gamma$ as we anticipated in the previous section.  Note that this is  the case also for $X \, = \,
  \frac{\mathbb{C}^3}{\mathrm{PSL(2,7)}}$, as Markushevich has shown. He has found one polynomial constraint
  $W_{\mathrm{PSL(2,7)}}(u_1,u_2,u_3,u_4)\, = \,0$ of degree 10 in $\mathbb{C}^4$ which describes $X$.
  We  were able to find   a similar result for the subgroup $\mathrm{G_{21}} \subset \mathrm{PSL(2,7)}$
  and obviously also for the cases $\mathbb{C}^3/\mathbb{Z}_3$ and $\mathbb{C}^3/\mathbb{Z}_7$.
  In the $\mathrm{G_{21}}$  case the equation is   of order  16.  We will present  these results in a future publication\footnote{As
  I mentioned above there is a series of partial and not fully accomplished results that were discovered in the
  framework of the my and Ugo Bruzzo's collaboration with Massimo Bianchi that remained on hold for a couple of years in the pandemic times
  and that now we plan to work out in collaboration also with Mario Trigiante within the larger scope provided by the other
  results obtained in
  \cite{Bianchi_2021,bruzzo2023d3brane}. I refer the reader to chapter \ref{aperto} for a discussion of the open questions.}.
  \item The resolved locus $Y_{[2]}^\Gamma$ in the case of ALE manifolds is described  by a deformed equation:
  \begin{eqnarray}
{ W}^{ALE}_{\Gamma} \left (  u,w,z ; \, \mathbf{t} \right )&= &W_{\Gamma}(u,w,z)\, + \,
\sum_{i=1}^{r} \ t_i \, {\cal P}^{(i)}(u,w,z) \nonumber\\
r &\equiv& \mbox{dim}\, \mathcal{R}_\Gamma \label{grouptheory21}
\end{eqnarray}
where
\begin{description}
\item[a)] { $W_{\Gamma}(u,w,z)$ is the simple singularity polynomial
corresponding to the finite subgroup $\Gamma\subset \mathrm{SU(2)}$} according to Arnold's classification of isolated critical points of functions \cite{arnolsimplicius}, named simple singularities in the literature.
\item[b)]{ ${\cal P}^{(i)}(u,w,z)$ is a basis spanning the chiral ring
\begin{equation}
\mathcal{R}_\Gamma=\frac{{\mathbb{C}}[u,w,z]}{\partial W_{\Gamma}} \label{grouptheory22}
\end{equation}
of polynomials in $u,w,z$ that do not vanish upon use of the vanishing relations $\partial_u \, W_{\Gamma}
\,= \, \partial_w \, W_{\Gamma} \, = \, \partial_z  \, W_{\Gamma} \, = \, 0$.}
\item[c)]{The complex parameters $t^i$ are the complex structure moduli and they
are in one--to--one correspondence with the set of complex level parameters ${ \ell}^{\bf X}_+$.}
\end{description}
   \item According to the general view put forward in the previous section, for ALE manifolds we have:
\begin{equation}\label{carezza}
    \mbox{dim} H^{1,1} \, = \, r \, \equiv \,\# \mbox{ nontrivial conjugacy classes of $\Gamma$}
\end{equation}
We also have:
\begin{equation}\label{casilinus}
    \mbox{dim} \mathcal{R}_\Gamma \, = \, r
\end{equation}
as one sees from table \ref{kleinale}. From the point of view of complex  geometry this is the
consequence of a special coincidence, already stressed in the previous section, which applies only to the
case of complex dimension $2$. As one knows, for Calabi-Yau $n$-folds complex structure deformations are
associated with $\omega^{n-1,1}\in H^{n-1,1}$ harmonic forms, while K\"ahler structure deformations, for all
$n$, are associated  with $\omega^{1,1}\in H^{1,1}$ harmonic forms. Hence when $n=2$, the $(1,1)$-forms play
a double role as complex structure deformations and as K\"ahler structure deformations. For instance, this is
well known in the case of $K3$. Hence in the $n=2$ case the number of \textit{nontrivial conjugacy classes}
of the group $\Gamma$ coincides both with the number of K\"ahler moduli and with number of complex structure
moduli of the resolved variety.
\item In the case of $Y^\Gamma_{[3]} \to X=\frac{\mathbb{C}^3}{\Gamma}$ the number of $(1,1)$-forms and hence of
K\"ahler moduli is still related with $r \, = \, \# \mbox{ \textit{junior conjugacy classes of }  $\Gamma$}$
but there are no complex structure deformations.
\end{enumerate}
\subsection{The McKay correspondence for $\mathbb{C}^2/\Gamma$} The   table of characters  $\chi^{(\mu)}_i$ of any finite group
$\gamma$ allows to reconstruct the decomposition coefficients of any representation along the irreducible
representations:
\begin{eqnarray}\label{multipillini}
    D&=&\bigoplus_{\mu =1}^{r} \, a_\mu \, D_\mu \nonumber\\
    a_\mu &=& \o{1}{g} \, \sum_{i =1}^{r} \, g_i \, \chi^{(D)}_{i} \, \chi^{(\mu) \, \star}_{i}
\end{eqnarray}
where $\chi^{(D)}$ is the character of $D$. For the finite subgroups
$\Gamma\subset \mathrm{SU(2)}$ a particularly important case is the
decomposition of the tensor product of an irreducible representation
$D_\mu$ with the defining 2-dimensional representation ${\cal Q}$.
It is indeed at the level of this decomposition that the relation
between these groups and the simply laced Dynkin diagrams becomes
explicit and  is named the McKay correspondence \cite{mckay}. This
decomposition plays a crucial role in the explicit construction of
ALE manifolds according to Kronheimer. Setting
\begin{equation}
{\cal Q} \, \otimes \, D_\mu ~=~\bigoplus_{\nu =0}^{r} \, A_{\mu \nu} \, D_\nu \label{grouptheory16}
\end{equation}
where $D_0$ denotes the identity representation, one finds that the matrix ${\bar
c}_{\mu\nu}=2\delta_{\mu\nu}-A_{\mu\nu}$ is the {\it extended Cartan matrix} relative to the {\it  extended
Dynkin diagram} corresponding to the given group. We remind the reader that the extended Dynkin diagram of
any simply laced Lie algebra is obtained by adding to the {\it  dots} representing  the {\it  simple roots}
$\left \{ \, \alpha_1 \, ......\, \alpha_r \,  \right \}$ an {\it  additional dot} (marked black in figs.
\ref{dynfigure1A}, \ref{dynfigure2A}) representing the negative of the highest root $\alpha_0 \, = \,
\sum_{i=1}^{r} \, n_i \, \alpha_i$ ($n_i$ are the Coxeter numbers). Thus we see a correspondence between the
nontrivial conjugacy classes $\mathcal{C}_i$ (or equivalently the nontrivial irreps) of the group
$\Gamma(\mathbb{G})$ and the simple roots of $\mathbb{G}$. In this correspondence the extended Cartan matrix
provides  the Clebsch-Gordon coefficients (\ref{grouptheory16}), while the Coxeter numbers $n_i$ express the
dimensions of the irreducible representations. All these informations are summarized in Figs.
\ref{dynfigure1A}, \ref{dynfigure2A} where the numbers $n_i$ are attached to each of the dots: the number $1$
is attached to the extra dot since it stands for the identity representation.
\begin{figure}[tb]
\begin{center}
{\setlength{\unitlength}{0.7pt}
\begin{picture}(280,110)(-60,-60)
\thicklines \put(25,43){\circle{8}} \put(-25,43){\circle*{8}} \put(-50,0){\circle{8}}
\put(-25,-43){\circle{8}} \put(25,-43){\circle{8}} \put(35,35){\makebox(0,0){$\cdot$}}
\put(43,25){\makebox(0,0){$\cdot$}} \put(48,13){\makebox(0,0){$\cdot$}} \put(50,0){\makebox(0,0){$\cdot$}}
\put(48,-13){\makebox(0,0){$\cdot$}} \put(43,-25){\makebox(0,0){$\cdot$}}
\put(35,-35){\makebox(0,0){$\cdot$}} \put(-21,43){\line(1,0){42}} \put(-47,3){\line(3,5){21}}
\put(-47,-3){\line(3,-5){21}} \put(-21,-43){\line(1,0){42}} \put(28,52){\makebox(0,0){1}}
\put(-28,52){\makebox(0,0){1}} \put(-58,0){\makebox(0,0){1}} \put(-28,-52){\makebox(0,0){1}}
\put(28,-52){\makebox(0,0){1}} \put(0,0){\makebox(0,0){$A_{k}$}} \put(75,-30){
\begin{picture}(150,50)(0,-30)
\thicklines \put(67,30){$D_{k+2}$} \multiput(30,0)(30,0){2}{\circle{8}} \put(30,-10){\makebox(0,0){2}}
\put(60,-10){\makebox(0,0){2}} \put(34,0){\line(1,0){22}} \put(64,0){\line(1,0){12}}
\put(80,0){\makebox(0,0){$\cdots$}} \put(84,0){\line(1,0){12}} \put(100,0){\circle{8}}
\put(100,-10){\makebox(0,0){2}} \put(0,15){\circle*{8}} \put(8,20){\makebox(0,0){1}} \put(0,-15){\circle{8}}
\put(8,-20){\makebox(0,0){1}} \multiput(130,15)(0,-30){2}{\circle{8}} \put(122,20){\makebox(0,0){1}}
\put(122,-20){\makebox(0,0){1}} \put(3,13){\line(2,-1){22}} \put(3,-13){\line(2,1){22}}
\put(103,2){\line(2,1){22}} \put(103,-2){\line(2,-1){22}}
\end{picture}}
\end{picture}
} \caption{\label{dynfigure1A} Extended Dynkin diagrams of the infinite series}
\end{center}
\end{figure}
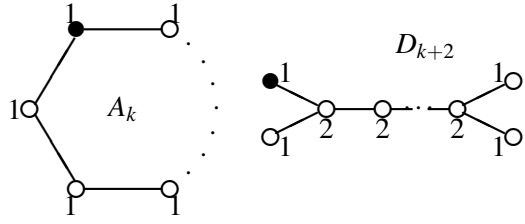
\begin{figure}[tb]
\begin{center}
{\setlength{\unitlength}{0.7pt}
\begin{picture}(300,135)(0,-10)
\thicklines \put(-10,25){$E_8 = {\cal I}$} \multiput(0,0)(30,0){7}{\circle{8}} \put(210,0){\circle*{8}}
\put(0,-10){\makebox(0,0){2}} \put(30,-10){\makebox(0,0){4}} \put(60,-10){\makebox(0,0){6}}
\put(90,-10){\makebox(0,0){5}} \put(120,-10){\makebox(0,0){4}} \put(150,-10){\makebox(0,0){3}}
\put(180,-10){\makebox(0,0){2}} \put(210,-10){\makebox(0,0){1}} \multiput(4,0)(30,0){7}{\line(1,0){22}}
\put(60,4){\line(0,1){22}} \put(60,30){\circle{8}} \put(52,30){\makebox(0,0){3}} \put(140,45){
\begin{picture}(180,50)(0,0)
\thicklines \put(5,25){$E_7= {\cal O}$} \multiput(0,0)(30,0){6}{\circle{8}} \put(180,0){\circle*{8}}
\put(0,-10){\makebox(0,0){1}} \put(30,-10){\makebox(0,0){2}} \put(60,-10){\makebox(0,0){3}}
\put(90,-10){\makebox(0,0){4}} \put(120,-10){\makebox(0,0){3}} \put(150,-10){\makebox(0,0){2}}
\put(180,-10){\makebox(0,0){1}} \multiput(4,0)(30,0){6}{\line(1,0){22}} \put(90,4){\line(0,1){22}}
\put(90,30){\circle{8}} \put(82,30){\makebox(0,0){2}}
\end{picture}}
\put(-20,80){
\begin{picture}(120,50)(0,15)
\thicklines \put(-10,25){$E_6= {\cal T}$} \multiput(0,0)(30,0){5}{\circle{8}} \put(0,-10){\makebox(0,0){1}}
\put(30,-10){\makebox(0,0){2}} \put(60,-10){\makebox(0,0){3}} \put(90,-10){\makebox(0,0){2}}
\put(120,-10){\makebox(0,0){1}} \multiput(4,0)(30,0){4}{\line(1,0){22}} \put(60,4){\line(0,1){22}}
\put(60,30){\circle{8}} \put(52,30){\makebox(0,0){2}} \put(52,60){\makebox(0,0){1}}
\put(60,34){\line(0,1){22}} \put(60,60){\circle*{8}}
\end{picture}}
\end{picture}
} \caption{\label{dynfigure2A} Exceptional extended Dynkin diagrams}
\end{center}
\end{figure}
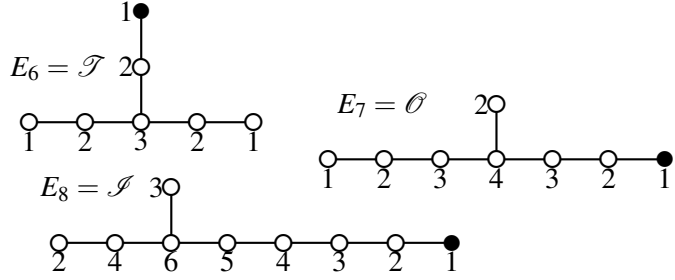
\subsection{Kronheimer's construction} Given any finite subgroup  $\Gamma \subset \mathrm{SU(2)}$, we
consider a space $\mathcal{P}$ whose elements are two-vectors of $|\Gamma |\times |\Gamma |$ complex
matrices: $\left(A, B\right) \in \mathcal{P}$. The action of an element $\gamma\in \Gamma$ on the points of
$\mathcal{P}$ is the following:
\begin{equation}\label{gammaactiononp}
 \left(\begin{array}{c} A \cr B \end{array}\right) \, \stackrel{\gamma}{\longrightarrow}\,
 \left(\begin{array}{cc} u_{\gamma} & i\,{\bar v}_{\gamma} \\
 i\,v_{\gamma}& {\bar u}_{\gamma}\\
 \end{array}\right)\,
 \left( \begin{array}{c} R(\gamma)\,A
\,R(\gamma^{-1}) \\ R(\gamma)\, B \, R(\gamma^{-1}) \\
\end{array}\right)
\end{equation}
where the two-dimensional matrix on the right hand side is the realization of $\gamma$ inside the defining
two-dimensional representation $\mathcal{Q}\subset \mathrm{SU(2)}$, while $R(\gamma)$ is the regular,
$|\Gamma|$-dimensional representation. The basis vectors in $R$ named $e_\gamma$ are in one-to-one
correspondence with the group elements $\gamma \in \Gamma$ and transform as follows:
\begin{equation}
 R(\gamma) \, e_\delta ~=~e_{\gamma \cdot \delta} ~~~~~\forall \, \gamma \, , \, \delta \, \in \,
\Gamma
 \label{grouptheory15}
 \end{equation}
 Intrinsically, the space $\mathcal{P}$ is named as:
 \begin{equation}\label{ommaomma}
    \mathcal{P}\, \simeq \, \mathrm{Hom}\left(R,\mathcal{Q}\otimes R \right)
 \end{equation}
Next we introduce the space $\mathcal{S}$, which by definition is the subspace made of $\Gamma$-invariant
elements in $\mathcal{P}$:
\begin{equation}
\mathcal{S}\equiv\left\{p\in\mathcal{P} / \forall \gamma\in\Gamma, \gamma\cdot p = p\right\}\,\,
\label{carnevalediRio}
\end{equation}
Explicitly the invariance condition  reads as follows:
\begin{equation}\label{invariancecond}
\left(\begin{array}{cc} u_{\gamma} & i\,{\bar v}_{\gamma} \\
 i\,v_{\gamma}& {\bar u}_{\gamma}\\
 \end{array}\right)\, \left(\begin{array}{c} A \cr B \end{array}\right) \, =\,
\left( \begin{array}{c} R(\gamma^{-1})\,A
\,R(\gamma) \\ R(\gamma^{-1})\, B \, R(\gamma) \\
\end{array}\right)
\end{equation}
The decomposition (\ref{grouptheory16}) is very useful in order to determine the $\Gamma$-invariant vector
space (\ref{carnevalediRio}).
\par
A two-vector of matrices can be thought of also as a matrix of
two-vectors: that is, $\mathcal{P}=\mathcal{Q}\otimes{\rm
Hom}(R,R)={\rm Hom}(R,\mathcal{Q}\otimes R)$. Decomposing the
regular representation, $R=\bigoplus_{\nu=0}^{r} n_{\mu} D_{\mu}$
into irreps, using eqn.\,(\ref{grouptheory16}) and Schur's lemma, we
obtain:
\begin{equation}
\mathcal{S}=\bigoplus_{\mu,\nu} A_{\mu,\nu}{\rm Hom}(\mathbb{C}^{n_{\mu}},\mathbb{C}^{n_{\nu}})\,\, .
\label{defmuastratta}
\end{equation}
The dimensions of the irreps,  $n_{\mu}$ are dispayed in figs.
\ref{dynfigure1A},\ref{dynfigure2A}. From
eqn.\,(\ref{defmuastratta}) the real dimension of $\mathcal{S}$
follows immediately: $\dim\, \mathcal{S}=\sum_{\mu,\nu}2 A_{\mu\nu}
n_{\mu}n_{\nu}$ implies, recalling that $A=2\times\mathbf{1}-\bar c$
[see eqn.\,(\ref{grouptheory16})] and that for the  extended Cartan
matrix $\bar c n =0$:
\begin{equation}
\dim_{\mathbb{C}}\, \mathcal{S}=2\sum_{\mu}n_{\mu}^2= 2 |\Gamma |\,\, . \label{dimm}
\end{equation}
Intrinsically, one writes the space $\mathcal{S}$ as:
\begin{equation}\label{carillo}
    \mathcal{S}\, \simeq \, \mathrm{Hom}_{\,\Gamma}\left(R,\mathcal{Q}\otimes R \right)
\end{equation}
So we can summarize the discussion by saying that:
\begin{equation}\label{summillus}
    \mathrm{dim}_\mathbb{C} \, \left[\mathrm{Hom}_{\,\Gamma}\left(R,\mathcal{Q}\otimes R \right)\right] \, =
    \, 2 \, |\Gamma|
\end{equation}
The quaternionic structure of the flat manifolds $\mathcal{P}$ and $\mathcal{S}$ can be seen by simply
writing their elements as follows:
\begin{equation}
 p \,= \,\twomat{A}{iB^{\dagger}}{iB}{A^{\dagger}} \,\in \, \mathrm{Hom}\left(R,\mathcal{Q}\otimes R\right)
 \hskip 1cm A,B\in {\rm End}(R)\,\, .
\label{pquaternions}
\end{equation}
Then the  hyperK\"ahler  forms and the  hyperK\"ahler  metric are defined by the following formulae:
\begin{eqnarray}
  \Theta&=&\mathrm{Tr} (\mathrm{d}p^\dagger\wedge \mathrm{d}p) \,=\, \left(\begin{array}{cc}
                                                   {\rm i} \,\mathbf{K}  &   {\rm i} \overline{\pmb \Omega}  \\
                                                   {\rm i} \, \pmb{\Omega} & -{\rm i} \,\mathbf{K} \\
                                                 \end{array}\right)\nonumber \\
  ds^2 \times \mathbf{1}& = &\mathrm{Tr}(\mathrm{d}p^\dagger \otimes \mathrm{d}p)\label{palestrus}
\end{eqnarray}
In the above equations the trace is taken over the matrices
belonging to ${\rm End}(R)$ in each entry of the quaternion. From
eqn.\,(\ref{palestrus}) we extract the explicit expressions for the
K\"ahler 2-form $\mathbf{K}$ and the holomorphic 2-form
$\pmb{\Omega}$ of the flat  hyperK\"ahler  manifold
$\mathrm{Hom}\left(R,\mathcal{Q}\otimes R\right)$. We have:
\begin{eqnarray}
\mathbf{K} &=& -{\rm i} \left[\mbox{Tr}\left(\mathrm{d}A^\dagger\wedge \mathrm{d}A\right) +
\mbox{Tr}\left(\mathrm{d}B^\dagger\wedge \mathrm{d}B\right) \right]
\, \equiv \, {\rm i} g_{\alpha{\bar \beta}} \, \mathrm{d}q^\alpha \wedge \mathrm{d}q^{\bar \beta} \nonumber\\
ds^2  &=& g_{\alpha{\bar \beta}} \,\mathrm{d}q^\alpha \otimes \mathrm{d}q^{\bar \beta}  \nonumber\\
\pmb{\Omega}& = & 2 \mbox{Tr}\left(\mathrm{d}A\wedge \mathrm{d}B\right) \, \equiv \,\Omega_{\alpha\beta} \,
\mathrm{d}q^\alpha \wedge \mathrm{d}q^{\beta}\label{ermengarda}
\end{eqnarray}
Starting from the above written formulae,  by means of an elementary
calculation one verifies that both the metric and the  hyperK\"ahler
forms are invariant with respect to the action of the discrete group
$\Gamma$ defined in eqn.\,(\ref{gammaactiononp}). Hence one can
consistently reduce the space
$\mathrm{Hom}\left(R,\mathcal{Q}\otimes R\right)$ to the invariant
space $\mathrm{Hom}_{\, \Gamma}\left(R,\mathcal{Q}\otimes R\right)$
defined in eqn.\,(\ref{carnevalediRio}). The  hyperK\"ahler
$2$-forms and the metric of the flat space $\mathcal{S}$, whose real
dimension is $4|\Gamma|$,  are given by eqs.(\ref{ermengarda}) where
the matrices $A,B$ satisfy the invariance condition
eqn.\,(\ref{invariancecond}). I mentioned in the introductory
chapter \ref{introibo} that in the early 1990.s the Kronheimer
construction was utilized by myself in collaboration with Anselmi,
Bill\`{o}, Girardello and Zaffaroni to discuss $\mathcal{N}=2$
conformal field theories in two-dimensions. Furthermore as I already
said, in the same decade ALE-manifolds were the object of several
other applications to theoretical physics in the context of
gravitational and gauge instantons with a particularly active role
played by the group of Rome Tor Vergata
\cite{Bianchi:2009bg,Bianchi:1996zj,Bianchi:1995ad,Bianchi:1995xd,Bianchi:1994gi}.
\subsubsection{Solution of the invariance constraint in the case of the cyclic
groups $A_k$} The space $\mathcal{S}$ can be easily described when
$\Gamma$ is the cyclic group $A_{k}= \mathbb{Z}_{k+1},$ whose
multiplication table can be read off. We can immediately read it off
from the matrices of the regular representation. Obviously, it is
sufficient to consider the representative of the first element
$e_1$, as $R(e_j)=(R(e_1))^j$.
\par
One has:
\begin{equation}
R(e_1)=\left(\begin{array}{ccccc}0&0&\cdots &0&1\cr 1&0&\cdots &0&0\cr 0&1&\cdots &0&0\cr \vdots &\vdots
&\ddots &\vdots &\vdots\cr 0&0&\cdots &1&0\end{array}\right) \label{r1offdiag}
\end{equation}
Actually, the invariance condition eqn.\,(\ref{invariancecond}) is
best solved by changing basis so as to diagonalize the regular
representation, realizing explicitly its decomposition in terms of
the $k$ unidimensional irreps. Let $\nu=e^{2\pi i\over {k+1}}$, be a
$(k+1)$th root of unity  so that $\nu^{k+1}=1$. The looked for
change of basis is performed by means of the matrix:
\begin{eqnarray}\label{canovaccio}
    S_{ij} & = & \frac{1}{\sqrt{k+1}} \, \nu^{ij} \quad; \quad i,j \, = \, 0,1,2,\dots , k \nonumber\\
    \left(S^{-1}\right)_{ij}\, = \, \left(S^\dagger\right)_{ij} & = & \frac{1}{\sqrt{k+1}} \, \nu^{k+1-ij}
\end{eqnarray}
In the new basis we find:
\begin{eqnarray}\label{caramellina}
\widehat{R}(e_0)& \equiv & S^{-1}\, R(e_0)\, S \, = \, \left(\begin{array}{ccccc}
                                                                     1 & 0 & \ldots & 0 & 0 \\
                                                                     0 & 1 & 0 \ldots & 0 & 0 \\
                                                                     \vdots & \vdots & \ddots & \vdots &
                                                                     \vdots \\
                                                                     0 & 0 & \ldots & 1 & 0 \\
                                                                     0 & 0 & \ldots & 0 & 1
                                                                   \end{array}
     \right)\nonumber\\
    \widehat{R}(e_1)& \equiv & S^{-1}\, R(e_1)\, S \, = \, \left(\begin{array}{ccccc}
                                                                     1 & 0 & \ldots & 0 & 0 \\
                                                                     0 & \nu & 0 \ldots & 0 & 0 \\
                                                                     \vdots & \vdots & \ddots & \vdots &
                                                                     \vdots \\
                                                                     0 & 0 & \ldots & \nu^{k-1} & 0 \\
                                                                     0 & 0 & \ldots & 0 & \nu^{k}
                                                                   \end{array}
     \right)
\end{eqnarray}
Eq.\,(\ref{caramellina}) displays on the diagonal the representatives of $e_j$ in the one-dimensional irreps.
\par
In the above basis, the explicit solution of
eqn.\,(\ref{invariancecond}) is given  by
\begin{equation}
A=\left(\begin{array}{ccccc}0&u_0&0&\cdots &0\cr 0&0&u_1&\cdots &0\cr \vdots&\vdots& \vdots&\ddots &\vdots\cr
\vdots &\vdots &\vdots &\null & u_{k-1}\cr u_{k}&0&0&\cdots &0\end{array}\right) \hskip 0.3cm ;\hskip 0.3cm
B=\left(\begin{array}{ccccc}0&0&\cdots &\cdots&v_k\cr v_0&0&\cdots &\cdots &0\cr 0&v_1&\cdots &\cdots &0\cr
\vdots &\vdots &\ddots &\null &\vdots \cr 0&0&\cdots &v_{k-1}&0\end{array}\right) \label{invariantck}
\end{equation}
We see that these matrices are parameterized in terms of $2k+2$
complex, i.e. $4(k+1)=|A_{k}|$ real parameters. In the $D_{k+2}$
case, where the regular representation is $4k$-dimensional, choosing
appropriately a basis, one can solve analogously
eqn.\,(\ref{invariancecond}); the explicit expressions are too
large, so we do not write them. The essential point is that the
matrices $A$ and $B$ no longer correspond to two distinct set of
parameters, the group being nonabelian.
\subsection{The gauge group for the quotient and its moment maps}
The next step in the Kronheimer construction of the ALE manifolds is the determination of the group
$\mathcal{F}$ of triholomorphic isometries with respect to which we will perform the quotient. We borrow from
physics the nomenclature \textit{gauge group} since in a $\mathcal{N}=3,4$ rigid three-dimensional gauge
theory where the space $\mathrm{Hom}_{\, \Gamma}\left(R,\mathcal{Q}\otimes R\right)$ is the flat manifold of
hypermultiplet scalars, the triholomorphic moment maps of $\mathcal{F}$  emerge as scalar dependent nonderivative terms in the hyperino supersymmetry transformation rules generated by the \textit{gauging} of the
group $\mathcal{F}$.
\par
 Consider the action of $\mathrm{SU(|\Gamma |)}$ on $\mathrm{Hom}\left(R,\mathcal{Q}\otimes R\right)$ given,
 using the quaternionic notation
for the elements of $\mathrm{Hom}\left(R,\mathcal{Q}\otimes R\right)$, by
\begin{equation}
\forall g\in \mathrm{SU(|\Gamma |)}\quad , \quad g \quad :\quad
\left(\begin{array}{cc} A & i\,B^{\dagger} \\
i\,B & A^{\dagger}\\ \end{array}\right) \longmapsto
\left(\begin{array}{cc} gAg^{-1}& i \, g\, B^\dagger\,g^{-1}\\
i\, g\, B \, g^{-1} & g\, A^{\dagger}\, g^{-1}\end{array}\right)\label{sunaction}
\end{equation}
It is easy to see that this action is a triholomorphic isometry of $\mathrm{Hom}\left(R,\mathcal{Q}\otimes
R\right)$. Indeed both the  hyperK\"ahler  forms $\Theta$ and the metric $ds^2$ are invariant.
\par
Let $\mathcal{F}\subset \mathrm{SU(|\Gamma |)}$ be the subgroup of
the above group which {\it commutes with the action of $\Gamma$ on
the space $\mathrm{Hom}\left(R,\mathcal{Q}\otimes R\right)$}, action
which was defined in eqn.\,(\ref{gammaactiononp}). Then the action
of $\mathcal{F}$ descends to $\mathrm{Hom}_{\,
\Gamma}\left(R,\mathcal{Q}\otimes
R\right)\subset\mathrm{Hom}\left(R,\mathcal{Q}\otimes R\right)$ to
give a {\it triholomorphic isometry}: indeed the metric and the
hyperK\"ahler  forms on the space $\mathrm{Hom}_{\,
\Gamma}\left(R,\mathcal{Q}\otimes R\right)$ are just the restriction
of those on $\mathrm{Hom}\left(R,\mathcal{Q}\otimes R\right)$.
Therefore one can take the  hyperK\"ahler  quotient of
$\mathrm{Hom}_{\, \Gamma}\left(R,\mathcal{Q}\otimes R\right)$ with
respect to $\mathcal{F}$.
\par
Let $\{f_A\}$ be a basis of generators for $\mathbb{F}$, the Lie algebra of $\mathcal{F}$. Under the
infinitesimal action of $f=\mathbf{1}+\lambda^A f_A\in \mathbb{F}$, the variation of $p\in \mathrm{Hom}_{\,
\Gamma}\left(R,\mathcal{Q}\otimes R\right)$ is $\delta p= \lambda^A\delta_A p$, with
\begin{equation}
\delta_A p = \twomat{[f_A,A]}{i[f_A,B^{\dagger}]}{i[f_A,B]} {[f_A,A^{\dagger}]}\, \label{deltaam}
\end{equation}
The components of the momentum map are then given by
\begin{equation}
\mu_A=\mathrm{Tr}\,( q^\dagger\,\delta_A p)\,\,\,\equiv\,\,\,
\mathrm{Tr}\,\twomat{f_A\,\mu_3(p)}{f_A\,\mu_-(p)}{f_A\,\mu_+(p)}{f_A\,\mu_3(p)} \label{momentummatrix}
\end{equation}
so that the real and holomorphic maps $\mu_3:\mathrm{Hom}_{\, \Gamma}\left(R,\mathcal{Q}\otimes
R\right)\rightarrow\mathbb{F}^*$ and $\mu_+:\mathrm{Hom}_{\, \Gamma}\left(R,\mathcal{Q}\otimes
R\right)\,\rightarrow\,\mathbb{C} \times\mathbb{F}^*$ can be represented as matrix-valued maps:
\begin{eqnarray}
\mu_3(p)&=&-i\left([A,A^{\dagger}]+[B,B^{\dagger}]\right)\nonumber\\
\mu_+(p)&=&\left([A,B]\right)\,\, . \label{momentums}
\end{eqnarray}
In this way we get:
\begin{equation}\label{carontasco}
    \mu_A \, = \, \left(\begin{array}{cc}
                          \mathfrak{P}^3_A & \mathfrak{P}_A^- \\
                          \mathfrak{P}_A^+  & -\mathfrak{P}^3_A
                        \end{array}
     \right)
\end{equation}
where:
\begin{eqnarray}
  \mathfrak{P}^3_A &=& - {\rm i} \, \left[\mbox{Tr} \left(\left[ A \, , \, A^\dagger \right] \, f_A \right)
  + \mbox{Tr} \left(\left[ B^\dagger \, , \, B \right] \, f_A \right)\right]\nonumber \\
  \mathfrak{P}^+_A &=& \mbox{Tr} \left(\left[ A \, , \, B \right] \, f_A \right) \label{cubicolario}
\end{eqnarray}
Let $\mathfrak{Z}^\star$ be the dual of the center of $\mathbb{F}$.
\par
In correspondence with a level $\zeta=\{\zeta^3,\zeta^+\}\in{\mathbb R}^3\otimes\mathfrak{Z}^\star$ we can form
the  hyperK\"ahler  quotient:
\begin{equation}\label{grancassone}
    \mathcal{M}_{\zeta}\equiv\mu^{-1}(\zeta)\, /\!\!/_{\null_{HK}}\, \mathcal{F}
\end{equation}
{\it Varying $\zeta$ and $\Gamma$ all ALE manifolds can be obtained as $\mathcal{M}_{\zeta}$}.
\par
First of all, it is not difficult to check that $\mathcal{M}_{\zeta}$ is four-dimensional. Let us see how
this happens. There is a nice characterization of the group $\mathcal{F}$ in terms of the extended Dynkin
diagram associated with $\Gamma$. We have
\begin{equation}
\mathcal{F}=\bigotimes_{\mu=1}^{r+1} \mathrm{U(n_{\mu})}\,\bigcap \, \mathrm{SU(|\Gamma|)} \label{formofF}
\end{equation}
where the sum is extended to all the irreducible representations of
the group $\Gamma$ and $n_\mu$ are their dimensions. One should also
take into account that the determinant of all the  elements must be
one, since $\mathcal{F}\subset \mathrm{SU(|\Gamma |)}$. Pictorially
the group $\mathcal{F}$ has a $\mathrm{U(n_{\mu})}$ factor for each
dot of the diagram, $n_{\mu}$ being associated with the dots as in
figs. \ref{dynfigure1A},\ref{dynfigure2A}. $\mathcal{F}$ acts on the
various \textit{components} of $\mathrm{Hom}_{\,
\Gamma}\left(R,\mathcal{Q}\otimes R\right)$ that are in
correspondence with the edges of the diagram, see
eqn.\,(\ref{defmuastratta}), as dictated by the
 diagram structure. From eqn.\,(\ref{formofF}) it is immediate to derive:
\begin{equation}\label{cargnolabato}
    \mbox{dim}\, \mathcal{F}=\sum_{\mu} n_{\mu}^2 -1 = |\Gamma |-1
\end{equation}
It follows that
\begin{equation}
\mbox{dim}_\mathbb{R}\,\mathcal{M}_{\zeta}=\dim_\mathbb{R}\, \mathrm{Hom}_{\,
\Gamma}\left(R,\mathcal{Q}\otimes R\right) -4\, \mbox{dim}_\mathbb{R}\, \mathcal{F} = 4|\Gamma | -(4|\Gamma
|-1)=4\, \label{dimxzeta}
\end{equation}
 Analyzing the construction we see that there are two steps.
In the first step, by setting the holomorphic part of the moment map to its level $\zeta$, we define an
algebraic locus in $\mbox{Hom}_\Gamma(\mathcal{Q}\otimes R,R)$. Next the K\"ahler quotient further reduces
such a locus to the necessary complex dimension 2. The two steps are united in one because of the
triholomorphic character of the isometries. As we already stressed in the previous section,
 in complex dimension 3 the isometries are not triholomorphic rather just holomorphic;  hence the holomorphic part
 of the moment map does not exist and the two steps have to be separated.
 There must be  another principle that leads to  impose those constraints that cut  out the algebraic locus
 $\mathbb{V}_{|\Gamma|+2}$ of which
 we perform  the K\"ahler quotient in the next step (see eqn.\,(\ref{belgone})).
 The main question is to spell out such principles. As anticipated,  equation
 $\mathbf{p}\wedge \mathbf{p}\, = \, 0$
 is the one that does the job. We are not able to reduce the $3|\Gamma|^2$ quadrics on $3|\Gamma|$ variables
 to an ideal with $2|\Gamma|-2$ generators, yet we know that such reduction must
 exist. Indeed, by means of another argument
 that utilizes Lie group orbits we can show that  there is a variety of complex dimension 3, named
 $\mathcal{D}^0_\Gamma$ which is in the kernel of the equation $\mathbf{p}\wedge \mathbf{p} \, = \, 0$.
\subsubsection{The triholomorphic moment maps in the $A_k$ case of Kronheimer construction}
\label{CDdiscussia}
 The structure of $\mathcal{F}$ and the momentum map for its action are very simply worked
out in the $A_{k}$ case. An element $f\in \mathcal{F}$ must commute
with the action of $A_{k}$ on $\mathcal{P}$,
eqn.\,(\ref{gammaactiononp}), where the two-dimensional
representation in the l.h.s. is given by:
\begin{equation}
 \Gamma (A_{k})\, \ni \, \gamma_\ell \, = \, {\cal Q}_\ell \, \equiv \, \twomat {e^{2\pi i
\ell/(k+1)}}{0}{0}{e^{-\, 2\pi i \ell /(k+1)}} \quad ;\quad\{ \ell=1,.....,k+1\}\, \label{grouptheory7}
\end{equation}
Then $f$ must have the form
\begin{equation}
f={\rm diag} (e^{i\varphi_0},e^{i\varphi_1},\ldots ,e^{i\varphi_{k}}) \hskip 0.2cm ; \hskip 0.2cm
\sum_{i=0}^k \varphi_{i}=0\,\, . \label{Fforck}
\end{equation}
Thus $\mathbb{F}$ is just the algebra of diagonal traceless $k+1$-dimensional matrices, which is
$k$-dimensional. Choose a basis of generators for $\mathbb{F}$, for instance:
\begin{eqnarray}
  f_1 &=&\diag(1,-1,0,\ldots,0) \nonumber \\
  f_2 &=& \diag (1,0,-1,0,\ldots,0)\nonumber \\
  \dots &=& \dots \nonumber \\
 f_{k}&=&\diag (1,0,0,\ldots,0,-1) \label{lavailmuso}
\end{eqnarray}
From eqn.\,(\ref{cubicolario}) we immediately obtain the components
of the momentum map:
\begin{eqnarray}
\mathfrak{P}^3_A&=&|u^0|^2-|u^k|^2-|v_0|^2+|v_k|^2 +\left( |u^{A-1}|^2-|u^A|^2-|v_{A-1}|^2 +|v_{A}|^2\right)
\nonumber \\
\mathfrak{P}^+_A&=&u^0 v_0 - u^k v_k +\left(u^{A-1}v_{A-1}\, - \, u^A \, v_A\right)\quad ,\quad (A=1,\dots
,k)\label{momentummapck}
\end{eqnarray}
\subsection{Level sets and Weyl chambers} If $\mathcal{F}$ acts freely on $\mu^{-1}(\zeta)$ then
$\mathcal{M}_{\zeta}$ is a smooth manifold. This happens or does not happen depending on the value of
$\zeta$. A simple characterization of $\mathfrak{Z}$ can be given in terms of the simple Lie algebra
$\mathbb{G}$ associated with $\Gamma$. There exists an isomorphism between $\mathfrak{Z}$ and the Cartan
subalgebra $\mathcal{H}_{CSA}\subset \mathbb{G}$. Thus we have
\begin{eqnarray}
\dim\, \mathfrak{Z}=\dim\,\mathcal{H}_{CSA} &=&{\rm rank}\,\mathbb{G}\nonumber\\
&=&\#{\rm \, of\,\,\,nontrivial\,\,\,conj.\,\,\,classes\,\,\,in}\,\,\,\Gamma \,\,\label{golosaidea}
\end{eqnarray}
The space $\mathcal{M}_{\zeta}$ turns out to be singular when, under
the above identification $\mathfrak{Z}\sim\mathcal{H}_{CSA}$, any of
the level components $\zeta^i\in {\mathbb R}^3\otimes \mathfrak{Z}$ lies
on a wall of a Weyl chamber. In particular, as the point
$\zeta^i=0$, ($i=1,\dots,r$) is identified with the origin of the
root space, which lies of course on all the walls of the Weyl
chambers, {\it the space $\mathcal{M}_0$ is singular}. Not too
surprisingly we will see in a moment that $\mathcal{M}_0$
corresponds to the {\it orbifold limit} $\mathbb{C}^2/\Gamma$ of a
family of ALE manifolds with boundary at infinity
$\mathbb{S}^3/\Gamma$.
\par
To verify this statement in general let us choose the natural basis
for the regular representation $R$, in which the basis vectors
$e_{\delta}$ transform as in eqn.\,(\ref{grouptheory15}). Define the
space $L\subset \mathcal{S}$ as follows:
\begin{equation}\label{thespacel}
L \,= \, \left\{\left(\begin{array}{c} C\\ D \end{array}\right)\in\mathcal{S}\,/\,C,D \,\,{\rm
are\,\,diagonal\,\,in\,\,the\,\,basis\,\,}\left\{e_{\delta}\right\} \right\}
\end{equation}
For every element $\gamma\in\Gamma$ there is a pair of numbers
$(c_{\gamma},d_{\gamma})$ given by the corresponding entries of
$C,D$\,:\,\, $C\cdot e_{\gamma}=c_{\gamma}e_{\gamma}$, $D\cdot
e_{\gamma}=d_{\gamma} e_{\gamma}$. Applying the invariance condition
eqn.\,(\ref{invariancecond}), which is valid since
$L\subset\mathcal{S}$, we obtain:
\begin{equation}
\left(\begin{array}{c} c_{\gamma\cdot\delta}\\ d_{\gamma\cdot\delta} \end{array}\right) =
\left(\begin{array}{cc}u_{\gamma}&{i\bar v_{\gamma}}\\ {iv_{\gamma}}&{\bar u_{\gamma}} \end{array} \right)
\left(\begin{array}{c} c_{\delta}\\
d_{\delta} \end{array}\right)\label{orbitofthepair}
\end{equation}
We can identify $L$ with $\mathbb{C}^2$ associating for instance
$(C,D)\in L \longmapsto (c_0,d_0)\in \mathbb{C}^2$. Indeed all the
other pairs $(c_{\gamma}, d_{\gamma})$ are determined in terms of
eqn.\,(\ref{orbitofthepair}) once $(c_0,d_0)$ are given. By
eqn.\,(\ref{orbitofthepair}) the action of $\Gamma$ on $L$ induces
exactly the action of $\Gamma$ on $\mathbb{C}^2$ provided by  its
two-dimensional defining representation inside $\mathrm{SU(2)}$. It
is quite easy to show the following fundamental fact: {\it each
orbit of $\mathcal{F}$ in $\mu^{-1}(0)$ meets $L$ in one orbit of
$\Gamma$}. Because of the above identification between $L$ and
$\mathbb{C}^2$, this leads to conclude that {\it
$\mu^{-1}(0)/\mathcal{F}$ is isometric to $\mathbb{C}^2/\Gamma$}.
Instead of reviewing the formal proofs of these statements as
devised by Kronheimer, we will verify them explicitly in the case of
the cyclic groups, giving a description which sheds some light on
the {\it deformed} situation; that is we show in which way a nonzero
level $\zeta^+$ for the holomorphic momentum map puts
$\mu^{-1}(\zeta)$ in correspondence with the affine hypersurface in
$\mathbb{C}^3$ cut out by the polynomial constraint
(\ref{grouptheory21})  which is a deformation  of that describing
the $\mathbb{C}^2/\Gamma$ orbifold, obtained for $\zeta^+=0$.
\subsubsection{Retrieving the polynomial constraint from the  hyperK\"ahler  quotient in the  $\Gamma$=$A_k$ case.}
We can directly realize $\mathbb{C}^2/\Gamma$ as an affine algebraic
surface in $\mathbb{C}^3$ by expressing the coordinates $x$, $y$ and
$z$ of $\mathbb{C}^3$ in terms of the matrices $(C,D)\in L$. The
explicit parametrization of the matrices in ${\cal S}$ in the
$A_{k}$ case, which was given in eqn.\,(\ref{invariantck}) in the
basis in which the regular representation $R$ is diagonal, can be
conveniently rewritten in the \textit{natural basis}
$\left\{e_{\gamma} \right\}$ via the matrix $S^{-1}$ defined in
eqn.\,(\ref{canovaccio}). The subset $L$ of diagonal matrices
$(C,D)$ is given by:
\begin{equation}
C=c_0\, {\rm diag}(1,\nu,\nu^2,\ldots,\nu^{k}),\hskip 12pt D=d_0\, {\rm
diag}(1,\nu^{k},\nu^{k-1},\ldots,\nu), \label{unouno}
\end{equation}
This is nothing but the fact that $ L \sim \mathbb{C}^2$. The set of pairs $\left(\begin{array}{c}\nu^m c_0\cr
\nu^{k-m}d_0\end{array} \right)$, $m=0,1,\ldots,k$ is an orbit of $\Gamma$ in $\mathbb{C}^2$ and determines
the corresponding orbit of $\Gamma$ in $L$. To describe $\mathbb{C}^2 / A_{k}$ one needs to identify a
suitable set of invariants $(u,w,z)\in \mathbb{C}^3$ such that
\begin{equation}\label{wgammus}
  0\,= \,  W_\Gamma (u,w,z) \, \equiv \, u^2+w^2\, - \, z^{k+1}
\end{equation}
To this effect we define:
\begin{equation}\label{transformillina}
    u= \ft 12 \left(x+y\right) \quad ; \quad w = -{\rm i}\ft 12 \, \left(x-y\right) \quad \Rightarrow
    \quad xy \, = \, u^2 + w^2
\end{equation}
and we make the following ansatz:
\begin{equation}
x=\det\, C \hskip 12pt ;\hskip 12pt y=\det\, D, \hskip 12pt ;\hskip 12pt z=\frac{1}{k+1} \mathrm{Tr} \,CD.
\label{identifyAk}
\end{equation}
This guess is immediately confirmed  by the study of the deformed surface. We know that there is a one-to-one
correspondence between the orbits of $\mathcal{F}$ in $\mu^{-1}(0)$ and those of $\Gamma$ in $L$. Let us
realize this correspondence  explicitly.
\par
Choose the basis where $R$ is diagonal. Then $(A,B) \in {\cal S}$
have the form of eqn.\,(\ref{invariantck}). The relation
$xy=z^{k+1}$  holds also true when, in eqn.\,(\ref{identifyAk}), the
pair $(C,D)\in L$ is replaced by an element $(A,B)\in \mu^{-1}(0)$.
\par
To see this, let us describe the elements $(A,B)\in\mu^{-1}(0)$. We
have to equate the right hand sides of eqn.\,(\ref{momentums}) to
zero. We note that:
\begin{equation}\label{candelinusA1}
    [A,B]=0 \quad \Rightarrow \quad v_i={u_0v_0\over u_i} \quad \forall i
\end{equation}
Secondly,
\begin{equation}\label{candelinusA2}
    [A,A^\dagger]+[B,B^\dagger]=0 \quad \Rightarrow \quad |u_i|=|u_j| \, \mbox{and} \,|v_i|=|v_j| \quad \forall i,j
\end{equation}
From the previous two equations we conclude that:
\begin{equation}\label{candelinusA3}
    u_j=|u_0|{\rm e}^{i\phi_j} \quad ; \quad v_j=|v_0|{\rm e}^{i\psi_j}
\end{equation}
Finally:
\begin{equation}\label{candelinusA4}
    [A,B]=0 \quad \Rightarrow \quad \psi_j=\Phi- \phi_j \quad \forall j
\end{equation}
where $\Phi$ is an arbitrary overall phase.
\par
In this way, we have characterized $\mu^{-1}(0)$ and we immediately
check that the pair $(A,B)\in\mu^{-1}(0)$ satisfies $xy=z^{k+1}$ if
$x=\det\,A$, $y=\det\,B$ and $z= 1/(k+1) \, \mathrm{Tr} \,AB$ as we
have proposed in eqn.\,(\ref{identifyAk}).
\par
After this explicit solution of the momentum map constraint has been implemented we are left with $k+4$
parameters, namely the $k+1$ phases $\phi_j$, $j=0,1,\ldots k$, plus the absolute values $|u_0|$ and $|v_0|$
and the overall phase $\Phi$. So we have:
\begin{equation}\label{pralasso}
    {\rm dim}\,\mu^{-1}(0)={\rm dim}\,{\cal S}-3 \,{\rm dim}\,\mathcal{F}=
4|\Gamma|-3(|\Gamma|-1)=|\Gamma|+3
\end{equation}
where $|\Gamma|=k+1$.
\par
Now we perform the quotient of $\mu^{-1}(0)$ with respect to $ \mathcal{F} $. Given a set of phases $f_i$
such that $\sum_{i=0}^{k}f_i=0\, {\rm mod} \, 2\pi$ and given $f={\rm diag} ({\rm e}^{if_0},{\rm
e}^{if_1},\ldots,{\rm e}^{if_{k}})\in  \mathcal{F} $, the orbit of $ \mathcal{F} $ in $\mu^{-1}(0)$ passing
through  $\left(\begin{array}{c}A\cr B\end{array}\right)$ has the form $\left(\begin{array}{c}fAf^{-1}\cr
fBf^{-1}\end{array}\right)$.
\par
Choosing $f_j=f_0+j\psi+\sum_{n=0}^{j-1}\phi_n$, $j=1,\ldots,k$, with $\psi=-{1\over k}\sum_{n=0}^{k}\phi_n$,
and $f_0$ determined by the condition $\sum_{i=0}^{k}f_i=0\, {\rm mod} \, 2\pi$, one obtains
\begin{equation}
fAf^{-1}=a_0\left(\begin{array}{ccccc} 0 & 1 & 0 & \ldots & 0\cr 0 & 0 & 1 & \ldots & 0\cr
  & \ldots  & & \dots  &  \cr
0 & 0  &  \dots & 0 & 1 \cr 1 & 0 & 0 & \dots & 0 \cr \end{array}\right)\, ,  \hskip 10pt
fBf^{-1}=b_0\left(\begin{array}{ccccc} 0 & 0  &  \dots & 0 & 1 \cr 1 & 0 & 0 & \dots & 0 \cr 0 & 1 & 0 &
\ldots & 0\cr
  & \ldots  & & \dots  &  \cr
0 & \ldots & 0 & 1 & 0\cr \end{array}\right) \label{op}
\end{equation}
where $a_0=|u_0|{\rm e}^{i\psi}$ and $b_0=|v_0|{\rm e}^{i(\Phi-\psi)}$. Since the phases $\phi_j$ are
determined modulo $2\pi$, it follows that $\psi$ is determined modulo $2\pi\over {k+1}$. Thus we can say
$(a_0,b_0)\in {\mathbb{C}^2/\Gamma}$. This is the one-to-one correspondence between $\mu^{-1}(0)/\mathcal{F}$
and $\mathbb{C}^2/\Gamma$.
\par
Next we derive the deformed relation between the invariants $x,y,z$. It fixes the correspondence between the
resolution of the singularity performed in the momentum map approach and the resolution performed on the
hypersurface $xy=z^{k+1}$ in $\mathbb{C}^3$. To this purpose, we focus on the holomorphic part of the
momentum map, i.e.\ on the equation:
\begin{eqnarray}
    [A,B]=\Lambda_0 & = & {\rm
diag}(\lambda_0,\lambda_1,\lambda_2,\ldots,\lambda_{k})\in \mathfrak{Z}\otimes\mathbb{C}\label{cruspilliA}\\
\lambda_{0}&=&-\sum_{i=1}^{k}\lambda_i\label{cruspilliB}
\end{eqnarray}
Let us recall the expression (\ref{invariantck}) for the matrices
$A$ and $B$. Naming $a_i=u_iv_i$, eqn.\,(\ref{cruspilliA}) implies:
\begin{equation}\label{minestrarancida}
    a_i=a_0+\lambda_i \quad ; \quad i=1,\ldots,k
\end{equation}
Let $\Lambda={\rm diag}(\lambda_1,\lambda_2,\ldots,\lambda_{k})$. We have
\begin{equation}
xy=\det A \, \det B= a_0\, \Pi_{i=1}^{k}(a_0+\lambda_i)=a_0^{k+1}\,\det \left(1+{1\over {a_0}}
\Lambda\right)=\sum_{i=0}^{k}a_0^{k+1-i}S_i(\Lambda) \label{def1}
\end{equation}
The $S_i(\Lambda)$ are the symmetric polynomials in the eigenvalues of $\Lambda$. They are defined by the
relation $\det (1+ t \Lambda)= \sum_{i=0}^{k} t^i S_i(\Lambda)$ and are given by:
\begin{equation}\label{symmipolli}
   S_i(\Lambda)=\sum_{j_1<j_2<\cdots<j_i}\lambda_{j_1}\lambda_{j_2} \cdots \lambda_{j_i}
\end{equation}
In particular, $S_0=1$ and $S_1=\sum_{i=1}^{k} \lambda_i$. Define $S_{k+1}(\Lambda)=0$, so that we can
rewrite:
\begin{equation}\label{svekolnyk}
    xy=
\sum_{i=0}^{k+1}a_0^{k+1-i}S_i(\Lambda)
\end{equation}
and note that
\begin{equation}
z={1\over {k+1}}\mathrm{Tr} AB=a_0+{1\over {k+1}}S_1(\Lambda).
\end{equation}
Then the desired deformed relation between $x$, $y$ and $z$ is obtained by substituting $a_0=z-{1\over k}S_1$
in (\ref{def1}), thus obtaining
\begin{eqnarray}
&&xy=\sum_{m=0}^{k+1}\sum_{n=0}^{k+1-m}\left(\begin{array}{c}k+1-m\cr n\end{array}\right) \,z^n
\left(-{1\over {k+1}}S_1\right)^{k+1-m-n}S_m \, z^n=\sum_{n=0}^{k+1}
t_{n+1} \, z^n \nonumber\\
&&\Longrightarrow\hskip 10pt t_{n+1}=\sum_{m=0}^{k+1-n}\left(\begin{array}{c}k+1-m\cr n\end{array}\right)
\left(-{1\over {k+1}}S_1\right)^{k+1-m-n} \label{def2}
\end{eqnarray}
Note in particular that $t_{k+2}=1$ and $t_{k+1}=0$, i.e.
\begin{equation}\label{cromagnolo}
    xy=z^{k+1}+\sum_{n=0}^{k}t_{n+1}z^n
\end{equation}
which means that the deformation proportional to $z^{k}$ is absent. This establishes a clear correspondence
between the momentum map construction and the polynomial ring ${\mathbb{C}[x,y,z]\over \partial W}$ where
$W(x,y,z)=xy-z^{k+1}$. Moreover, note that we have only used one of the
momentum map equations, namely $[A,B]=\Lambda_0$. The equation $[A,A^\dagger]+[B,B^\dagger]=\Sigma$ has been
completely ignored. This means that the deformation of the complex structure is described by the parameters
$\Lambda$, while the parameters $\Sigma$ describe the deformation of the K\"ahler structure. The relation
(\ref{def2}) can also be written in a simple factorized form, namely
\begin{equation}
xy=\Pi_{i=0}^{k}(z-\mu_i),
\end{equation}
where
\begin{eqnarray}
\mu_i&=&{1\over k}(\lambda_1+\lambda_2+\cdots +\lambda_{i-1} -2\lambda_i+\lambda_{i+1}+\cdots+\lambda_k),
\,\,\,\,
i=1,\ldots,k-1\nonumber\\
\mu_0&=&-\sum_{i=1}^k\mu_i={1\over k}S_1.
\end{eqnarray}
\section{Generalization of the correspondence: McKay quivers for $\mathbb{C}^3/\Gamma$ singularities}
\label{generaKronh} One can generalize the extended Dynkin diagrams
obtained in the above way by constructing McKay quivers, according
to the following definition:
\begin{definizione}
Let us consider the quotient $\mathbb{C}^n/\Gamma$, where $\Gamma$ is a finite group that acts on
$\mathbb{C}^n$ by means of the complex representation $\mathcal{Q}$ of dimension $n$ and let $\mathrm{D}_i$,
($i=1,\dots ,r+1$) be the set of irreducible representations of $\Gamma$ having denoted by $r+1$  the number
of conjugacy classes of $\Gamma$. Let the matrix $\mathcal{A}_{ij}$ be defined by:
\begin{equation}\label{quiverro2}
    \mathcal{Q}\otimes \mathrm{D}_i \, = \, \bigoplus_{j=1}^{r+1} \, \mathcal{A}_{ij}\,\mathrm{D}_j
\end{equation}
To such a matrix we associate a quiver diagram in the following way. Every irreducible representation is
denoted by a circle  labeled with a number equal to the dimension of the corresponding irrep. Next we write an
oriented line going from circle $i$ to  circle $j$ if $\mathrm{D}_j$ appears in the decomposition of
$\mathcal{Q}\otimes \mathrm{D}_i$, namely if the matrix element $\mathcal{A}_{ij}$ does not vanish.
\end{definizione}
The analogue of the extended Cartan matrix discussed in the case of $\mathbb{C}^2/\Gamma$ is defined below:
\begin{equation}\label{caspiterina}
    \bar{c}_{ij} \, = \, n \, \delta_{ij} \, - \, \mathcal{A}_{ij}
\end{equation}
and it has the same property, namely,  it admits the vector of irrep dimensions
\begin{equation}\label{dimvecco}
    \mathbf{n}\, \equiv \, \{1,n_1,\dots,n_r\}
\end{equation}
as a null vector:
\begin{equation}\label{nullitone}
    \bar{c}.\mathbf{n}\,= \, \mathbf{0}
\end{equation}
Typically the McKay quivers encode the information determining the
interaction structure of the dual gauge theory on the brane world
volume. Indeed the bridge between Mathematics and Physics is located
precisely at this point. In the case of a single $D3$-brane, the
$n|\Gamma|$ complex coordinates ($n$=2, or 3) of the flat K\"ahler
manifold $Hom_\Gamma(R,Q\otimes R)$ are the scalar fields of the
Wess-Zumino multiplets, the unitary group $\mathcal{F}$ commuting
with the action of $\Gamma$ is the \textit{gauge group}, the moment
maps of $\mathcal{F}$ enter the definition of the potential,
according to standard supersymmetry formulae and the holomorphic
constraints defining the $\mathbf{V}_{|\Gamma|+2}$ variety  have to
be related with the superpotential $\mathfrak{W}$ of the
$\mathcal{N}=1$ theories in d=4 (\textit{i.e.} the $n$=3 case where
the singular space is $\mathbb{C}^3/\Gamma$). In the case of
$\mathcal{N}=2$ theories,  (\textit{i.e.} the $n$=2 case where the
singular space is $\mathbb{C} \times \mathbb{C}^2/\Gamma$), the
holomorphic constraints $\mathcal{P}_i(y)$ are identified with the
holomorphic part of the tri-holomorphic moment map. When one goes to
the case of multiple $D3$-branes the gauge group is enlarged by
color indices. This is another story. The first step is to
understand the case of one $D3$-brane and here the map between
Physics and Mathematics is one-to-one.
\subsection{Representations of the quivers and K\"ahler quotients}
Let us now follow the same steps of the Kronheimer construction and
derive the representations of the $\mathbb{C}^3/\Gamma$ quivers. The
key point is the construction of the analogues of the spaces
$\mathcal{P}_\Gamma$ in eqn.\,(\ref{ommaomma}) and of its invariant
subspace $\mathcal{S}_\Gamma$ in eqn.\,(\ref{carnevalediRio}). To
this effect we introduce three matrices $|\Gamma|\times|\Gamma|$
named $A,B,C$ and set:
\begin{eqnarray}
    p \in \mathcal{P}_\Gamma \, \equiv \, \mbox{Hom}\left(R,\mathcal{Q}\otimes R\right) \, \Rightarrow\,
    p\,=\, \left(\begin{array}{c}
                   A \\
                   B \\
                   C
                 \end{array}
     \right) \label{homqg}
\end{eqnarray}
The action of the discrete group $\Gamma$ on the space $\mathcal{P}_\Gamma$ is defined in full analogy with
the Kronheimer case:
\begin{equation}\label{gammazione}
    \forall \gamma \in \Gamma: \quad \gamma\cdot p \,\equiv\, \mathcal{Q}(\gamma)\,\left(\begin{array}{c}
                  R(\gamma)\, A \, R(\gamma^{-1})\\
                   R(\gamma)\, B \, R(\gamma^{-1}) \\
                  R(\gamma)\, C \, R(\gamma^{-1})
                 \end{array}
     \right)
\end{equation}
where $\mathcal{Q}(\gamma)$ denotes the three-dimensional complex representation of the group element
$\gamma$, while $R(\gamma)$ denotes its $|\Gamma|\times|\Gamma|$-matrix image in the regular representation.
\par
In complete analogy with eqn.\,(\ref{carnevalediRio}) the subspace
$\mathcal{S}_\Gamma$ is obtained by setting:
\begin{equation}
\mathcal{S}_\Gamma \, \equiv \, \mbox{Hom}_\Gamma\left(R,Q\otimes R\right)\, = \,
\left\{p\in\mathcal{P}_\Gamma / \forall \gamma\in\Gamma, \gamma\cdot p = p\right\}\,\,
\label{carnevalediPaulo}
\end{equation}
\par
Just as in the previous case a three-vector of matrices can be
thought  as a matrix of three-vectors: that is,
$\mathcal{P}_\gamma=\mathcal{Q}\otimes{\rm Hom}(R,R)={\rm
Hom}(R,\mathcal{Q}\otimes R)$. Decomposing the regular
representation, $R=\bigoplus_{i=0}^{r} n_i D_i$ into irreps, using
eqn.\,(\ref{quiverro2}) and Schur's lemma, we obtain:
\begin{equation}
\mathcal{S}_\Gamma=\bigoplus_{i,j} A_{i,j}{\rm Hom}(\mathbb{C}^{n_{i}},\mathbb{C}^{n_{j}})\,\,
\label{defmuastrattaG}
\end{equation}
The properties (\ref{caspiterina},\ref{dimvecco},\ref{nullitone}) of
the matrix $A_{ij}$ associated with the quiver diagram guarantee, in
perfect analogy with eqn.\,(\ref{dimm})
\begin{equation} \dim_{\mathbb{C}}\,
\mathcal{S}_\Gamma\simeq \, \mathrm{Hom}_{\,\Gamma}\left(R,\mathcal{Q}\otimes R \right)\,=3\sum_{i}n_{i}^2= 3
|\Gamma |\,\, . \label{dimmus}
\end{equation}
\subsection{The quiver Lie group, its maximal compact subgroup and the K\"ahler quotient} \label{VG2p1}
We address now the most important point, namely the reduction of the
$3|\Gamma|$-dimensional complex manifold
$\mbox{Hom}_\Gamma\left(R,\mathcal{Q}\otimes R\right)$ to a
$|\Gamma|+2$-dimensional subvariety  of which we will perform the
K\"ahler quotient in order to obtain the final 3-dimensional
(de-singularized) smooth manifold that provides the crepant
resolution. The  inspiration about  how this can be done is provided
by comparison with the $\mathbb{C}^2/\Gamma$ case, \textit{mutatis
mutandis}. The key formulae to recall are the following ones:
eqn.\,(\ref{momentums}), (\ref{formofF}) and (\ref{thespacel}).
\par
From eqn.\,(\ref{momentums}) we see that the analytic part of the
triholomorphic moment map is provided by the projection onto the
gauge group generators of the commutator $\left[A\, , \, B\right]$.
When the level parameters are all zero (namely when the locus
equation is not perturbed by the elements of the chiral ring) the
outcome of the moment map equation is simply the condition
$\left[A\, , \, B\right]=0$. In the case of $\mathbb{C}^3/\Gamma$ we
already know that there are no deformations of the complex structure
and that the analogue of the holomorphic moment map constraint has
to be a rigid parameterless condition.  Namely the ideal that cuts
out the $\mathbb{V}_{|\Gamma|+2}$ variety should be generated by a
list of quadric polynomials $\mathcal{P}_i(y)$ fixed once and for
all in a parameterless way. It is reasonable to guess that these
equations should be a  generalization of the condition $\left[A\, ,
\, B\right]=0$. In the $\mathbb{C}^3/\Gamma$ case  we have three
matrices $A,B,C$ and the obvious generalization is given below:
\begin{equation}\label{ganimusco}
  \mathbf{p} \wedge \mathbf{p} \, = \,0
\end{equation}
where:
\begin{eqnarray}
  \mathbf{ p} & = & \left( \begin{array}{c}
                               A \\
                               B \\
                               C
                             \end{array}
  \right) \, \in \, \text{Hom}_{\Gamma}\left(R,\mathcal{Q}\otimes R\right)\nonumber\\
  p_1 &=& A \quad ; \quad  p_2 \, = \, B \quad ; \quad p_3 \, =\, C \label{luciopatta}
\end{eqnarray}
This is a short-hand for the following explicit equations
\begin{eqnarray}
 0 &=& \epsilon^{ijk}\mathbf{ p}_i \cdot \mathbf{p}_j \nonumber\\
 &\Updownarrow& \nonumber\\
  0 &=& \left[ A,B\right] \, = \, \left[ B,C\right] \, = \, \left[ C,A\right] \label{poffarbacchio}
\end{eqnarray}
Eq.\,(\ref{ganimusco}) is the very same equation numbered (1.18) in Craw's doctoral thesis \cite{crawthesis}. We will see in a
moment that it is indeed the correct equation reducing $\mbox{Hom}_\Gamma\left(R,\mathcal{Q}\otimes R\right)$
to a $|\Gamma|+2$-dimensional subvariety. The way to understand it goes once again through a detailed
comparison with the Kronheimer case.
\par
One has to discuss the construction of the gauge group and to recall
the identification of the singular orbifold $\mathbb{C}^2/\Gamma$
with the subspace named $L$  defined by eqn.\,(\ref{thespacel}).
Both constructions have a completely parallel analogue in the
$\mathbb{C}^3/\Gamma$ case and these provide the key to understand
why (\ref{ganimusco}) is the right choice. The field theoretic
mechanism that yields the eqn.s (\ref{poffarbacchio}) is realized in
a gauge theory whose scalar fields span the space
$\mathcal{S}_\Gamma$ for a $\mathbb{C}^3/\Gamma$ singularity, if we
introduce the following superpotential:
\begin{equation}\label{cascurtosco}
  \mathcal{W }\, = \, \mbox{Tr}\left[p_x \,p_y \, p_z\right] \, \epsilon^{xyz}
\end{equation}
With this choice the conditions for the vanishing of the scalar
potential are indeed the K\"ahler moment map equations that we are
going to discuss and eqn.\,(\ref{ganimusco}) or equivalently
(\ref{poffarbacchio}).
\subsubsection{Quiver Lie groups}\label{quiverino} We are
interested in determining the subgroup
\begin{equation}\label{gstorto}
  \mathcal{G}_\Gamma \,\subset \, \mathrm{SL(|\Gamma|,\mathbb{C})}
\end{equation}
made by those elements that commute with the group $\Gamma$.
\begin{equation}\label{carciofillo}
   \mathcal{G}_\Gamma  \, = \, \left\{ g \in \mathrm{SL(|\Gamma|,\mathbb{C})} \quad | \quad \forall
   \gamma \in \Gamma \,\, : \,\,
   \left[ \mathrm{D}_\mathrm{R}\left(\gamma\right)\, , \, \mathrm{D}_{\mathrm{def}}\left(g\right)\,\right]
   \, = \, 0\right\}
\end{equation}
In the above equation $D_R(\null)$ denotes the regular representation while $\mathrm{D}_{\mathrm{def}}$
denotes the defining representation of the complex linear group. The two representations,
by construction, have the same dimension and this is the reason why equation (\ref{carciofillo}) makes sense.
\par
It is sufficient to impose the defining constraint for the generators of the group on a generic matrix
depending on $|\Gamma|^2$ parameters: this reduces it to a specific matrix depending on
$|\Gamma|$-parameters. The further condition that the matrix should have determinant one, reduces the number
of free parameters  to $|\Gamma|-1$.  In more abstract terms we can say that the group $\mathcal{G}_\Gamma$
has the following general structure:
\begin{equation}\label{caliente}
\mathcal{G}_\Gamma \, = \, \bigotimes_{\mu = 1}^{r+1} \mathrm{GL(n_\mu, \mathbb{C})}\bigcap
\mathrm{SL(|\Gamma|,\mathbb{C})}
\end{equation}
This is a perfectly analogous result to that displayed in
eqn.\,(\ref{formofF}) for the Kronheimer case. The difference is
that there we had unitary groups while here we are talking about
general linear complex groups with a holomorphic action on the
quiver coordinates. The reason is that we have not yet introduced a
K\"ahler structure on the quiver space $\mathrm{Hom}_\Gamma\left(R\,
,\,  \mathcal{Q}\otimes R\right)$: we do it presently and we shall
realize that isometries of the constructed K\"ahler metric will be
only those elements of  $\mathcal{G}_\Gamma$ that are contained in
the unitary subgroup mentioned below:
\begin{equation}\label{colasciutto}
\mathcal{F}_\Gamma \,\equiv \,\bigotimes_{\mu = 1}^{r+1} \mathrm{U(n_\mu)}\bigcap \mathrm{SU(|\Gamma|)} \,
\subset \, \mathcal{G}_\Gamma
\end{equation}
\subsubsection{The holomorphic quiver group and the reduction to $V_{|\Gamma|+2}$}
\label{riducocasco} Yet the group $\mathcal{G}_\Gamma$ plays an
important role in understanding the rationale of the holomorphic
constraint (\ref{ganimusco}).  The key item is the coset
$\mathcal{G}_\Gamma/\mathcal{F}_\Gamma$.
\par
Let us introduce some notations. Relaying on eqn.\,(\ref{homqg}) we
define the diagonal  embedding:
\begin{equation}\label{caziloro}
  \mathbb{D} \quad : \quad \mathrm{GL(|\Gamma|,\mathbb{C})} \, \rightarrow \, \mathrm{GL(3|\Gamma|,\mathbb{C})}
\end{equation}
\begin{equation}\label{grossoD}
  \forall M \in \mathrm{GL(|\Gamma|,\mathbb{C})} \quad ; \quad  \mathbb{D}[M] \, \equiv \, \left(
  \begin{array}{c|c|c}
  M & 0 & 0 \\
  \hline
  0 & M & 0 \\
  \hline
  0 & 0 & M \\
  \end{array}
  \right)
\end{equation}
In this notation, the invariance condition that defines $\mathcal{S}_\Gamma \, = \,
\mathrm{Hom}_\Gamma(R,\mathcal{Q}\times R)$ can be rephrased as follows:
\begin{equation}\label{cascalippo}
  \forall \gamma \in \Gamma \quad : \quad \mathcal{Q}[\gamma] \, \mathbf{p} \, = \,
  \mathbb{D}[R^{-1}_\gamma] \,\mathbf{ p} \, \mathbb{D}[R_\gamma]
\end{equation}
It is clear that any  $|\Gamma| \times |\Gamma| $ - matrix $M$ that commutes with $R_\gamma$ realizes an
automorphism of the space $\mathcal{S}_\Gamma$, namely it maps it into itself. The group $\mathcal{G}_\Gamma$
is such an automorphism group. In particular equation (\ref{ganimusco}) or alternatively
(\ref{poffarbacchio}) is invariant under the action of $\mathcal{G}_\Gamma$. Hence the locus:
\begin{eqnarray}
  \mathcal{D}_\Gamma &\subset& \mathcal{S}_\Gamma \nonumber \\
  \mathcal{D}_\Gamma &\equiv& \left\{ \mathbf{p} \in \mathcal{S}_\Gamma \,\, |\,\, \left[A,B\right]\,=\,
  \left[B ,C\right]=\left[C,A\right] \, = \, 0 \right\} \label{granitacaffe}
\end{eqnarray}
is invariant under the action of $\mathcal{G}_\Gamma$. A priori the
locus $\mathcal{D}_\Gamma$ might be empty, but this is not so
because there exists an important solution of the constraint
(\ref{ganimusco}) which is the obvious analogue of the space
$L_\Gamma$ defined for the $\mathbb{C}^2/\Gamma$-case in
eqn.\,(\ref{thespacel}). In full analogy we set:
\begin{equation}\label{thespacellone}
\mathcal{S}_\Gamma\, \supset\, L_\Gamma \,\equiv \, \left\{\left(\begin{array}{c} A_0\\ B_0\\ C_0
\end{array}\right)\in\mathcal{S}_\Gamma \,\,\mid\,\,A_0,B_0,C_0 \,\,{\mbox{
are diagonal in the natural basis of R}\, : \,}\left\{e_{\delta}\right\} \right\}
\end{equation}
Obviously diagonal matrices commute among themselves and they do the same in any other basis where they are
not diagonal, in particular in the \textit{split basis}. By definition we name in this way the basis where
the regular representation R is split into irreducible representations. A general result in finite group
theory tells us that every $n_i$-dimensional irrep $\pmb{D}_i$ appears in R exactly $n_i$-times:
\begin{equation}\label{santachiara}
    R \, = \, \bigoplus_{i=0}^{r} n_i \, \pmb{D}_i \quad ; \quad \mbox{dim}\pmb{D}_i \equiv n_i
\end{equation}
In the split basis every element $\gamma\in \Gamma$ is given by a
block diagonal matrix of the following form:
\begin{equation}\label{mascherone}
 R(\gamma)\,=\,   \left(
       \begin{array}{c|c|c|c|c|c}
         \mathbf{1} & \mathbf{0} & \mathbf{\dots} & \mathbf{\dots} & \mathbf{0} & \mathbf{1} \\
         \hline
         \mathbf{0} &
         \begin{array}{ccc}
         a_{1,1} & \dots & a_{1,n_1} \\
         \vdots & \dots & \vdots \\
         a_{n_1,1} & \dots & a_{n_1,n_1} \\
         \end{array}
          & \mathbf{0} & \dots & \dots & \mathbf{0} \\
         \hline
         \vdots & \dots & \dots & \dots & \dots & \vdots \\
         \hline
         \vdots & \dots & \dots & \dots & \dots & \vdots \\
         \hline
         \mathbf{0} & \dots & \dots & \mathbf{0} & \begin{array}{ccc}
         b_{1,1} & \dots & b_{1,n_{r-1}} \\
         \vdots & \dots & \vdots \\
         b_{n_{r-1},1} & \dots & b_{n_{r-1},n_{r-1}} \\
         \end{array} & \mathbf{0}\\
         \hline
         \mathbf{0} & \dots & \dots & \dots & \mathbf{0} & \begin{array}{ccc}
         c_{1,1} & \dots & c_{1,n_r} \\
         \vdots & \dots & \vdots \\
         c_{n_r,1} & \dots & c_{n_r,n_r} \\
         \end{array} \\
       \end{array}
     \right)
\end{equation}
In analogy to what was noticed for the Kronheimer case,
the space $L_\Gamma$ has complex dimension three (in Kronheimer case it was two):
\begin{equation}\label{pereoliato}
    \mbox{dim}_\mathbb{C} \,L_\Gamma\, = \,3
\end{equation}
Indeed if we fix the first diagonal entry of each of the three matrices, the invariance condition
(\ref{cascalippo}) determines all the other ones uniquely. In any other basis the number of parameters
remains three. Let us call them $(a_0,b_0,c_0)$. Because of the above argument and, once again, in full
analogy with the Kronheimer case, we can conclude that the space $L_\Gamma$ is isomorphic to the singular
orbifold $\mathbb{C}^3/\Gamma$, the $\Gamma$-orbit of a triple $(a_0,b_0,c_0)$ representing a point in
$\mathbb{C}^3/\Gamma$.
\par
The existence of the solution of the constraint
(\ref{ganimusco}) provided by the complex three-dimensional space $L_\Gamma$ shows that we can construct a
variety of dimension $|\Gamma|+2$ which is in the kernel of the constraint (\ref{ganimusco}). This is just
the orbit, under the action of $\mathcal{G}_\Gamma$ of $L_\Gamma$. We set:
\begin{equation}\label{gneccoD}
    \mathcal{D}_\Gamma \,  \equiv \,\mbox{Orbit}_{\mathcal{G}_\Gamma}\left(L_\Gamma\right)
\end{equation}
The counting is easily done.
\begin{enumerate}
  \item A generic point in $L_\Gamma$ has the identity as stability subgroup in $\mathcal{G}_\Gamma$.
  \item The group $\mathcal{G}_\Gamma$ has complex dimension $|\Gamma|-1$, hence we get:
  \begin{equation}\label{licciu}
    \mbox{dim}_{\mathbb{C}}\left(\mathcal{D}_\Gamma\right) \, = \,|\Gamma|-1+3 = |\Gamma|+2
  \end{equation}
\end{enumerate}
\par
In the sequel we define the variety $V_{|\Gamma|+2}$ to be equal to $\mathcal{D}_\Gamma^0$.
\subsubsection{The coset $\mathcal{G}_\Gamma/\mathcal{F}_\Gamma$ and the K\"ahler quotient}
It is now high time to introduce the K\"ahler potential of the original $3|\Gamma|$-dimensional complex flat
manifold $\mathcal{S}_\Gamma$. We set:
\begin{eqnarray}\label{kalerpotent}
    \mathcal{K}_{\mathcal{S}_\Gamma} & \equiv & \mbox{Tr} \left(\mathbf{p}^\dagger \, \mathbf{p}\right)
    \,=\, \mbox{Tr}\left(A^\dagger\, A\right)\, + \,\mbox{Tr}\left(B^\dagger\, B\right)
          \, + \,\mbox{Tr}\left(C^\dagger\, C\right)
\end{eqnarray}
Using the matrix elements of $A,B,C$ as complex coordinates of the  manifold and naming $\lambda_i$ the
independent parameters from which they depend  in a given explicit solution of the invariance constraint, the
K\"ahler metric is defined, as usual, by:
\begin{eqnarray}\label{giorgini}
    ds^2_{\mathcal{S}_\Gamma} & = & g_{\ell\bar{m}}\, d\lambda^\ell \otimes d\bar{\lambda}^{ \bar{m}}
\end{eqnarray}
where:
\begin{equation}\label{cricket}
    g_{\ell\bar{m}} \, = \, \partial_\ell \, \bar{\partial}_{\bar{m}} \,\mathcal{K}
\end{equation}
From eqn.\,(\ref{kalerpotent}) we easily see that the K\"ahler
potential is invariant under the unitary subgroup of the quiver
group defined by:
\begin{equation}\label{unitosubbo}
  \mathcal{F}_\Gamma \, = \, \left\{M \in \mathcal{G}_\Gamma \, \mid \, M\, M^\dagger \, = \, \mathbf{1}\, \right\}
\end{equation}
whose structure was already mentioned in eqn.\,(\ref{colasciutto}).
The center $\mathfrak{z}\left(\mathbb{F}_\Gamma\right)$ of the Lie
algebra $\mathbb{F}_\Gamma$ has dimension $r$, namely the same as
the number of nontrivial conjugacy classes of $\Gamma$ and it has
the following structure:
\begin{equation}\label{centratore}
  \mathfrak{z}\left(\mathbb{F}_\Gamma\right) \, = \, \underbrace{\uu(1)\oplus\uu(1)\oplus\dots\oplus\uu(1)}_{r}
\end{equation}
\par
Since $\mathcal{F}_\Gamma$ acts as a group of isometries on the
space $\mathcal{S}_\Gamma$ we might construct the K\"ahler quotient
of the latter with respect to the former, yet we may do better.
\par
In the case of an abelian $|\Gamma|$ the center $\mathfrak{z}[\mathbb{F}]=\mathbb{F}$ coincides with the
entire gauge algebra.
\par
Let us consider the inclusion map of the variety $\mathcal{D}_\Gamma$ into $\mathcal{S}_\Gamma$:
\begin{equation}\label{includendus}
  \iota \, : \, \mathcal{D}_\Gamma \, \rightarrow \, \mathcal{S}_\Gamma
\end{equation}
and let us define as K\"ahler potential and K\"ahler metric of the locus $\mathcal{D}_\Gamma$  the pull backs
of the K\"ahler potential (\ref{kalerpotent}) and of metric (\ref{giorgini}) of $\mathcal{S}_\Gamma$, namely
let us set:
\begin{eqnarray}
 \mathcal{ K}_{\mathcal{D}_\Gamma} &\equiv&\iota^\star\, \mathcal{ K}_{\mathcal{S}_\Gamma} \label{labio1} \\
  ds^2_{\mathcal{D}_\Gamma}&=& \iota^\star\,  ds^2_{\mathcal{S}_\Gamma} \label{labio2}
\end{eqnarray}
By construction, the isometry group $\mathcal{F}_{\Gamma}$ is inherited by the pullback metric on
$\mathcal{D}_\Gamma$ and we can consider the K\"ahler quotient:
\begin{equation}\label{gioiabella}
  \mathcal{M}_\zeta \, \equiv \, \mathcal{D}_\Gamma /\!\!/^\zeta_{\mathcal{F}_{\Gamma}}
\end{equation}
Let $f_I$ be a basis of generators of  $\mathcal{F}_{\Gamma}$ ($I=1,\dots,|\Gamma|-1$) and
let $Z_i$ ($i=1,\dots,|\Gamma|+2$) be a system of complex coordinates spanning the points of $\mathcal{D}_\Gamma$.
By means of the inclusion map we have:
\begin{equation}\label{craturato}
\forall Z \in  \mathcal{D}_\Gamma \quad : \quad  \iota (Z) \, = \, \mathbf{p}(Z) \, = \,\left(
\begin{array}{c}
A(Z) \\
B(Z) \\
C(Z) \\
\end{array}
\right)
\end{equation}
The action of the \textit{gauge group} $\mathcal{F}_\Gamma$ on $\mathcal{D}_\Gamma$ is implicitly defined by:
\begin{equation}\label{kratinus}
  \mathbf{p}(\delta_I Z) \, = \, \delta_I  \mathbf{p}(Z) \, = \,\left(
  \begin{array}{c}
  \left [f_I \, , \,  A(Z)\right] \\
  \left [f_I \, , \,  B(Z)\right] \\
  \left [f_I \, , \,   C(Z)\right] \\
  \end{array}
  \right)
\end{equation}
and the corresponding real moment maps are easily calculated:
\begin{equation}\label{prosperus}
 \mu_I (Z,\bar{Z})\, = \, \mbox{Tr}\left(f_I \,\left[A(Z),A^\dagger (\bar{Z}) \right]\right)\,
 +\,\mbox{Tr}\left(f_I \,\left[B(Z),B^\dagger (\bar{Z}) \right]\right)\,
 +\,\mbox{Tr}\left(f_I \,\left[C(Z),C^\dagger (\bar{Z}) \right]\right)
\end{equation}
One defines the level sets  by means of the equation:
\begin{equation}\label{carampa}
\mu^{-1}\left( \zeta\right)\, = \, \left\{ Z \, \in \mathcal{D}_\Gamma\, \,  \parallel \, \mu_I (Z,\bar{Z})
\, = \, 0 \quad \mbox{if $f_I \not\in  \mathfrak{Z} $} \quad ; \quad \mu_I (Z,\bar{Z}) \, = \, \zeta_I \quad
\mbox{if $f_I \in  \mathfrak{Z} $}\right\}
\end{equation}
which, by construction, are invariant under the gauge group $\mathcal{F}_\Gamma$ and we can finally set:
\begin{equation}\label{fammicuccia}
\mathcal{M}_\zeta \, \equiv  \, \mu^{-1}\left( \zeta\right)/\!\!/_{\mathcal{F}_\Gamma}\, \equiv \,
\mathcal{D}_\Gamma /\!\!/^\zeta_{\mathcal{F}_{\Gamma}}
\end{equation}
The real and complex dimensions of $\mathcal{M}_\zeta$ are easily calculated. We start from $|\Gamma|+2$
complex dimensions, namely from $2|\Gamma|+4$ real dimensions. The level set equation imposes  $|\Gamma|-1$
real constraints, while the quotiening by the group action takes other $|\Gamma|-1$ parameters  away.
Altogether we remain with $6$ real parameters that can be seen as $3$ complex ones. Hence the manifolds
$\mathcal{M}_\zeta$ are always   complex three-folds that, for generic values of $\zeta$, are smooth:
supposedly the crepant resolutions of the singular orbifold. For  $\zeta \,= \, 0$ the manifold
$\mathcal{M}_0$ degenerates into the singular orbifold $\mathbb{C}^3/\Gamma$, since the solution of the
moment map equation is given by the $\mathcal{F}_\Gamma$ orbit of the locus $L_\Gamma$, namely:
\begin{equation}\label{quadriglia}
  \mu^{-1}\left(0\right) \, = \, \mbox{Orbit}_{\mathcal{F}_\Gamma}\left(L_\Gamma\right)
\end{equation}
Comparing eqn.\,(\ref{gneccoD}) with eqn.\,(\ref{quadriglia}) we are
led to consider the following direct sum decomposition of the Lie
algebra:
\begin{eqnarray}
  \mathbb{G}_\Gamma &=& \mathbb{F}_\Gamma \oplus \mathbb{K}_\Gamma\\
  \left[\mathbb{F}_\Gamma \, , \, \mathbb{F}_\Gamma\right] &\subset & \mathbb{F}_\Gamma \quad ; \quad
\left[\mathbb{F}_\Gamma \, , \, \mathbb{K}_\Gamma\right] \,\subset \, \mathbb{K}_\Gamma \quad ; \quad
\left[\mathbb{K}_\Gamma \, , \, \mathbb{K}_\Gamma\right] \,\subset \, \mathbb{F}_\Gamma \label{salameiolecco}
\end{eqnarray}
where $\mathbb{F}_\Gamma$ is the maximal compact subalgebra and
$\mathbb{K}_\Gamma$ denotes its complementary orthogonal subspace
with respect to the Cartan Killing metric.
\par
A special feature  of all the quiver Groups and Lie Algebras is that $\mathbb{F}_\Gamma$ and
$\mathbb{K}_\Gamma$ have the same real dimension $|\Gamma|-1$ and one can choose a basis of hermitian
generators $T_I$ such that:
\begin{equation}\label{sacherdivuli}
    \begin{array}{ccccccc}
       \forall \pmb{\Phi} \in \mathbb{F}_\Gamma & : & \pmb{\Phi} & = & {\rm i} \times \sum_{I=1}^{|\Gamma|-1} c_I T^I & ; &
       c_I \in \mathbf{R} \\
       \forall \pmb{K} \in \mathbb{K}_\Gamma & : & \pmb{K} & = & \sum_{I=1}^{|\Gamma|-1} b_I T^I & ; &
       b_I \in \mathbf{R} \\
     \end{array}
\end{equation}
Correspondingly a generic element $g\in \mathcal{G}_\Gamma$ can be split as follows:
\begin{equation}\label{consolatio}
   \forall g \in \mathcal G_\Gamma \quad : \quad g=\mathcal{U} \, \mathcal{H} \quad  ; \quad \mathcal{U} \in
   \mathcal{F}_\Gamma \quad ; \quad  \mathcal{H} \in \exp\left[ \mathbb{K}_\Gamma\right]
\end{equation}
Using the above property we arrive at the following parametrization of the space $\mathcal{D}_\Gamma$
\begin{equation}\label{krumiro}
    \mathcal{D}_\Gamma \, = \, \mbox{Orbit}_{\mathcal{F}_\Gamma}\left(\exp\left[
    \mathbb{K}_\Gamma\right]\cdot L_\Gamma\right)
\end{equation}
where, by definition, we have set:
\begin{eqnarray}\label{caresmio}
  p\in \exp\left[
    \mathbb{K}_\Gamma\right]\cdot L_\Gamma  &\Rightarrow & p=\left\{\exp\left[-\pmb{K}\right]\, A_0
    \exp\left[\pmb{K}\right], \, \exp\left[-\pmb{K}\right]\, B_0\,\exp\left[\pmb{K}\right],\, \exp\left[-\pmb{K}\right]\, C_0
    \exp\left[\pmb{K}\right]\right\} \nonumber\\
 \left\{ A_0, \, B_0,\,  C_0\right\} &\in & L_\Gamma \nonumber\\
 \pmb{K} &=& \mathbb{K}_\Gamma
\end{eqnarray}
Relying on this, in the K\"ahler quotient we can invert the order of the operations. First we quotient the
action of the compact gauge group $\mathcal{F}_\Gamma$ and then we implement the moment map constraints. We
have:
\begin{equation}\label{cascapistola}
 \mathcal{D}_{\Gamma}/\!\!/_{\mathcal{F}_\Gamma}\, = \,\exp\left [\mathbb{K}_\Gamma\right]\cdot L_\Gamma
\end{equation}
Calculating the moment maps on $\exp\left [\mathbb{K}_\Gamma\right]\cdot L_\Gamma$ and imposing the moment
map constraint we find:
\begin{equation}\label{carampana}
\mu^{-1}\left( \zeta\right)/\!\!/_{\mathcal{F}_\Gamma}\, = \, \left\{ Z \, \in \exp\left
[\mathbb{K}_\Gamma\right]\cdot L_\Gamma\,  \parallel \, \mu_I (Z,\bar{Z}) \, = \, 0 \quad \mbox{if $f_I
\not\in  \mathfrak{Z} $} \quad ; \quad \mu_I (Z,\bar{Z}) \, = \, \zeta_I
 \quad \mbox{if $f_I \in  \mathfrak{Z} $}\right\}
\end{equation}
Eq.\,(\ref{carampana}) provides an explicit algorithm to calculate the K\"ahler potential of the final
resolved manifold if we are able to solve the constraints in terms of an appropriate triple of complex
coordinates. Furthermore for each level parameter $\zeta_a$ we have to find the appropriate one-parameter
subgroup of $\mathcal{G}_\Gamma$ that lifts the corresponding moment map from the $0$-value to the generic
value $\zeta$. Indeed we recall that the K\"ahler potential of the resolved variety is given by the
celebrated formula:
\begin{equation}\label{celeberro}
  \mathcal{K}_{\mathcal{M}}\, = \, \pi^\star \, \mathcal{K}_{\mathcal{N}} \, + \,\zeta_I \mathfrak{C}^{IJ} \,
  \pmb{\Phi}_J
\end{equation}
where, by definition:
\begin{equation}\label{cirimella}
  \pi \, : \, \mathcal{N} \, \rightarrow \, \mathcal{M}
\end{equation}
is the quotient map and $\exp[\zeta_I  \, \mathfrak{C}^{IJ} \, \pmb{\Phi}_I]\in \exp\left[
    \mathbb{K}_\Gamma\right] \subset\mathcal{G}_{\Gamma}$ is the element of the quiver
group which lifts the moment maps from zero to the values $\zeta_I$,
while $\mathfrak{C}^{IJ}$ is a constant matrix whose definition we
discuss later on. Indeed the rationale behind formula
(\ref{celeberro}) requires a careful discussion, originally due to
Hitchin, Karlhede, Lindstr\"om and Ro\v cek \cite{HKLR} which we
shall review in later sections.
\subsection{Divisors and line bundles} A {\em prime divisor} in a
complex manifold or algebraic variety $X$ is an irreducible closed
codimension one subvariety of $X$. A divisor $\mathfrak{D}$ is a
locally finite formal linear combination
\begin{equation}\label{divisor} \mathfrak{D} = \sum_i a_i\,\mathfrak{D}_i \end{equation}
where the $a_i$ are integers, and the $\mathfrak{D}_i$ are prime divisors. A prime divisor $\mathfrak{D}$ can
be descrived by a collection $\{(U_\alpha,f_\alpha)\}$, where $\{U_\alpha\}$ is an open cover of $X$, and the
$\{f_\alpha\}$ are holomorphic functions on $U_\alpha$ such that $f_\alpha=0$ is the equation of
$\mathfrak{D}\cap U_\alpha$ in $U_\alpha$. As a consequence, the functions $g_{\alpha\beta}
=f_\alpha/f_\beta$ are holomorphic nowhere vanishing functions
$$ g_{\alpha\beta} \colon U_\alpha\cap U_\beta \to \mathbb C^\ast$$
that on triple intersections  $U_\alpha\cap U_\beta \cap U_\gamma $ satisfy the cocycle condition
$$ g_{\alpha\beta} g_{\beta\gamma} = g_{\alpha\gamma} $$
and therefore define a line bundle $\mathcal L(\mathfrak{D})$. If $\mathfrak{D}$ is a divisor as in
\eqref{divisor} then one sets
$$ \mathcal L(\mathfrak{D}) = \bigotimes_i \mathcal L(\mathfrak{D}_i)^{a_i}.$$
The inverse correspondence (from line bundles to divisors) is described as follows. If $s$ is a nonzero
meromorphic section of a line bundle $\mathcal L$, and $V$ is a codimension one subvariety of $X$ over which
$s$ has a zero or a pole, denoted by $\operatorname{ord}_{V}(s)$ the order of the zero, or minus the order of
the pole; then
$$ \mathfrak{D} = \sum_{V} \operatorname{ord}_{V}(s) \cdot V $$
is a divisor, whose associated line bundle $\mathcal L(\mathfrak{D})$ is isomorphic to $\mathcal L$.
\section{The  generalized Kronheimer construction for
$\frac{\mathbb{C}^3}{\Gamma}$ and the Tautological Bundles} In the
present section we aim at extracting a general scheme from the
detailed discussions presented  in the previous sections. Our final
goal is to establish all the algorithmic steps that  give a precise
meaning to each of the lines appearing in the conceptual diagram of
eqn.\,(\ref{salumaio}).
\subsection{Construction of the space $\mathcal{N}_{\zeta }\equiv \mu ^{-1}(\zeta )$ }
Summarizing the points of our construction we have the following
situation. We have considered the moment map
\begin{equation}
\mu : \mathcal{S}_{\Gamma }\longrightarrow  \mathbb{F}_{\Gamma}{}^*
\end{equation}
where \(\mathbb{F}_{\Gamma }{}^*\) is the dual  of the Lie algebra
of the maximal compact subgroup \(\mathcal{F}_{\Gamma }\) of the
quiver group \(\mathcal{G}_{\Gamma }\). Next we have considered the
center of the above Lie algebra $\mathfrak{z}$[\(\mathbb{F}_{\Gamma
}\)]$\subset $\(\mathbb{F}_{\Gamma }\) and its dual
\(\mathfrak{z}\left[\mathbb{F}_{\Gamma } \right]{}^*\). The moment
map can be restricted to the subspace:
\begin{equation}
\mathcal{D}_{\Gamma } \subset  \mathcal{S}_{\Gamma } \quad ; \quad \mathcal{D}_{\Gamma}\equiv
 \left\{\left.p\in  \mathcal{S}_{\Gamma }\right| p\land p = 0\right\}
\end{equation}
which is just the orbit, with respect to the quiver group \(\mathcal{G}_{\Gamma }\), of a locus
$\mathcal{E}_{\Gamma }\subset \mathcal{S}_{\Gamma}$ of complex dimension three obtained in the following way.
\par
Consider the following subspace of \(\mathcal{S}_{\Gamma }^{[0,0]}\)\(\subset \mathcal{S}_{\Gamma }\)
\begin{equation}
\mathcal{S}_{\Gamma }^{[0,0]} =\left\{\left.p\in  \mathcal{S}_{\Gamma } \right| p\land p =0 \quad; \quad \mu
(p)=0\right\}
\end{equation}
whose elements are triples of $|\Gamma |\times |\Gamma |$ complex matrices (A,B,C) satisfying, by the above
definition, in addition to the invariance constraint (\ref{gammazione}-\ref{carnevalediPaulo})  also the
following two ones:
\begin{equation}
[A,B]=[B,C]=[C,A]=0 \quad ; \quad \text{Tr}\left[T_I\,\left( \left[A,A^{\dagger }\right]+\left[B,B^{\dagger
}\right]+\left[C,C^{\dagger }\right]\right)\right]\,=\,0 \quad ; \quad I\,=\, 1,\dots ,\, |\Gamma|-1
\end{equation}
Since the action of the compact group \(\mathcal{F}_{\Gamma }\) leaves both the first and the second
constraint invariant, it follows that it maps the locus { } \(\mathcal{S}_{\Gamma }^{[0,0]}\) into itself
\begin{equation}
\mathcal{F}_{\Gamma } \quad :\quad \mathcal{S}_{\Gamma }^{[0,0]}\, \to\, \mathcal{S}_{\Gamma}^{[0,0]}
\end{equation}
The locus \(\mathcal{E}_{\Gamma }\) is defined as the quotient:
\begin{equation}
\mathcal{E}_{\Gamma } \equiv  \frac{\mathcal{S}_{\Gamma }^{[0,0]}}{\mathcal{F}_{\Gamma}}
\end{equation}
which turns out to be of complex dimension three and to be isomorphic to the singular orbifold :
\begin{equation}
\frac{\mathcal{S}_{\Gamma }^{[0,0]}}{\mathcal{F}_{\Gamma }} \, \simeq\,  \frac{\mathbb{C}^3}{\Gamma}
\end{equation}
Choosing a representative in each equivalence class \(\frac{\mathcal{S}_{\Gamma
}^{[0,0]}}{\mathcal{F}_{\Gamma }}\) simply amounts to a choice of local coordinates on
\(\frac{\mathbb{C}^3}{\Gamma }\) which will be promoted to a system of local coordinates on the manifold
\(\mathcal{M}_{\zeta}\) of the final resolved singularity.
\par We have a canonical algorithm to construct a
canonical coordinate system for { }\(\mathcal{E}_{\Gamma }\) which originates from Kronheimer and from the
1994 paper by Anselmi, Bill\`o, Fr\`e, Girardello and Zaffaroni on ALE manifolds and conformal field theories
\cite{mango}. The construction is the following. We begin with the locus \(L_{\Gamma }\) $\subset
$\(\mathcal{S}_{\Gamma }\) defined as the set of triples (\(A_d\),\(B_d\),\(C_d\)) such that the invariance
constraint (\ref{carnevalediPaulo}) is satisfied with respect to $\Gamma $ and they are diagonal in the
natural basis of the regular representation. We have shown on the basis of several examples that :
\begin{equation}
\mathcal{D}_{\Gamma } = \text{Orbit}_{\mathcal{G}_{\Gamma}}\left(L_{\Gamma }\right) \label{quagliastro}
\end{equation}
We obtain an explicit parameterization of the locus \(\mathcal{E}_{\Gamma }\) by solving the algebraic
equation for the hermitian matrix $\mathcal{V}_0\,\in \, \exp\left[\mathbb{K}_{\Gamma }\right]$, such that
\begin{equation}
\forall p\, \in  \, L_{\Gamma }\quad : \quad \mu \left(\mathcal{V}_0.p\right) = 0
\end{equation}
The important thing is that the solution for the above equation is a constant matrix \(\mathcal{V}_0\),
indipendent from the point p $\in $ \(L_{\Gamma }\). Then we fix the coordinates of our manifold by
parameterizing
\begin{equation}
p \in \mathcal{E}_{\Gamma } \Rightarrow  p= \left(
\begin{array}{c}
 A_0 \\
\begin{array}{c}
 B_0 \\
 C_0 \\
\end{array}
 \\
\end{array}
\right)=\left(
\begin{array}{c}
 \mathcal{V}_0{}^{-1}A_d\mathcal{V}_0 \\
\begin{array}{c}
 \mathcal{V}_0{}^{-1}B_d\mathcal{V}_0 \\
 \mathcal{V}_0{}^{-1}C_d\mathcal{V}_0 \\
\end{array}
 \\
\end{array}
\right)\text{    }\text{where} \left(
\begin{array}{c}
 A_d \\
\begin{array}{c}
 B_d \\
 C_d \\
\end{array}
 \\
\end{array}
\right)\in L_{\Gamma }
\end{equation}
It follows that equation (\ref{quagliastro}) can be substituted by
\begin{equation}
\mathbb{V}_{|\Gamma |+2} \equiv \text{  }\mathcal{D}_{\Gamma } =
\text{Orbit}_{\mathcal{G}_{\Gamma}}\left(\mathcal{E}_{\Gamma }\right)
\end{equation}
We can also introduce a subspace \(\mathcal{D}_{\Gamma }{}^0\)$\subset $\(\mathbb{V}_{|\Gamma |+2}\) which is
the orbit of \(\mathcal{E}_{\Gamma}\) under the compact subgroup \(\mathcal{F}_{\Gamma }\):
\begin{equation}
\mathcal{D}_{\Gamma }{}^0\,=\, \text{Orbit}_{\mathcal{F}_{\Gamma}}\left(\mathcal{E}_{\Gamma }\right)
\end{equation}
This being the case we consider the restriction  of the moment map to \(\mathcal{D}_{\Gamma }\)
\begin{equation}
\mu  : \mathcal{D}_{\Gamma } \longrightarrow  \mathbb{F}_{\Gamma}{}^*
\end{equation}
and given an element
\begin{equation}
\zeta \in \mathfrak{z}\left[\mathbb{F}_{\Gamma }\right]{}^*
\end{equation}
we define:
\begin{equation}
\mathcal{N}_{\zeta }\equiv  \mu ^{-1}(\zeta ) \subset  \mathcal{D}_{\Gamma }\quad:\quad\mathcal{N}_{\zeta} \,
=\, \left\{\left.p\in \mathcal{D}_{\Gamma } \right| \mu (p) = \zeta \right\}
\end{equation}
\\
Obviously we have:
\begin{equation}
\mathcal{N}_0\equiv  \mu ^{-1}(0) =\text{  }\mathcal{D}_{\Gamma}{}^0
\end{equation}
\subsection{The space $\mathcal{N}_{\zeta }$  as a principal fibre bundle}
The space \(\mathcal{N}_{\zeta }\) has a natural structure of an \(\mathcal{F}_{\Gamma }\) principal line
bundle over the quotient \(\mathcal{M}_{\zeta}\):
\begin{equation}
\mathcal{N}_{\zeta }\,\overset{\pi }{\longrightarrow }\,\mathcal{M}_{\zeta }\, \equiv \,\mathcal{N}_{\zeta
}/\!\!/\mathcal{F} _{\Gamma } \label{gualtiero}
\end{equation}
On the tangent space to the total space of the \(\mathcal{F} _{\Gamma }\)--bundle
\(\text{T$\mathcal{N}$}_{\zeta }\) we have a metric induced, as the pullback, by the inclusion map:
\begin{equation}
\iota  : \mathcal{N}_{\zeta } \longrightarrow  \mathcal{S}_{\Gamma}
\end{equation}
of the flat metric $g$ on \(\mathcal{S}_{\Gamma }\)
\begin{equation}
g_{\mathcal{N}} = \iota ^*\left(g_{\mathcal{S}_\Gamma}\right)
\end{equation}
Since the metric \(g_{\mathcal{S}_\Gamma}\) is K\"ahlerian we have a K\"ahler potential
\(\mathcal{K}_{\mathcal{S}_\Gamma}\) from which it derives and we define the function
\begin{equation}
\mathcal{K}_{\mathcal{N}} \equiv \iota ^*\left(\mathcal{K}_{\mathcal{S}_\Gamma}\right)
\end{equation}
This function is not the K\"ahler potential of \(\mathcal{N}_{\zeta
}\) which is not even K\"ahlerian (it has odd dimensions) but it
will be related to the K\"ahler potential of the final quotient
\(\mathcal{M}_{\zeta }\) by means of an argument due to \cite{HKLR},
that we spell out a few lines below. Let us denote by \(p\in
\mathcal{M}_{\zeta }\) a point of the base manifold and by \(\pi
^{-1}(p)\) the \(\mathcal{F} _{\Gamma }\)-fibre over that point.
\subsubsection{The natural connection and the tautological bundles}\label{ciulifischio}
We can determine a natural connection on the principal bundle
(\ref{gualtiero}) through the following steps. As it is observed in
eqn.\,(2.7) of the paper by Degeratu and Walpuski \cite{degeratu},
which agrees with the formulae of the present paper, the quiver
group has always the following form:
\begin{equation}
\mathcal{G}_{\Gamma }=\prod _{I=1}^r \text{GL}\left(\mathbb{C}^{\dim
\left[\pmb{D}_I\right]}\right)
\end{equation}
where \(\pmb{D}_I\) are the nontrivial irreducible representations
of the finite group $\Gamma $, with the exclusion of \(\pmb{D}_0,\)
the identity representation. It also follows that the compact gauge
subgroup \(\mathcal{F}_{\Gamma }\) has the corresponding following
structure
\begin{equation}
\mathcal{F}_{\Gamma }=\prod _{I=1}^r \mathrm{U}\left(\dim
\left[D_I\right]\right)
\end{equation}
Consequently, the principal bundle (\ref{gualtiero}) induces
holomorphic vector bundles of rank \(\dim \left[\pmb{D}_I\right]\)
on which the compact structural group acts non-trivially only with
its component \(\mathrm{U}\left(\dim
\left[\pmb{D}_I\right]\right)\). A natural connection on these
bundles is obtained as it follows
\begin{equation}
\mathbb{A} = \frac{\rm i}{2}\left(\mathcal{H}^{-1}\partial
\mathcal{H} -\mathcal{H}\bar{\partial } \mathcal{H}^{-1}\right) +
g^{-1}d g \in  \underset{I=1}{\overset{r}{\oplus }}
\mathbf{u}\left(\dim \left[D_I\right]\right) \label{rodolfo}
\end{equation}
where $\mathcal{H}$ is a hermitian fibre--metric on the direct sum
of the tautological vector bundles defined below:
\begin{equation}
\mathcal{R} \equiv \underset{I=1}{\overset{r}{\bigoplus }}
\mathcal{R}_I \quad ; \quad \mathcal{R}_I \overset{\pi
}{\longrightarrow } \mathcal{M}_{\zeta }\quad ; \quad \forall p \in
\mathcal{M}_{\zeta } \quad : \quad \pi ^{-1}(p) \simeq
\mathbb{C}^{\dim \left[D_I\right]}
\end{equation}
By definition the matrix $\mathcal{H}$ must be of dimension
\begin{equation}
\dim  [\mathcal{H}]= n\times n \quad\quad\text{where}\quad\quad
n=\sum _{I=1}^r \dim \left[D_I\right]=\sum_{I=1}^r n_I
\end{equation}
In order to find the hermitian matrix $\mathcal{H}$, we argue in the
following way. First we observe that in the regular representation
$R$ each irreducible representation \(\pmb{D}_I\) is contained
exactly \(\dim \left[\pmb{D}_I\right]\) times, so that the form of
the matrix $\mathcal{V}$ corresponding to the hermitian
parametrization of the coset \(\frac{\mathcal{G}_{\Gamma }}{
\mathcal{F}_{\Gamma }}\) has always the following form:
\begin{equation}
\mathcal{V}\, = \,\left(
\begin{array}{|c|c|c|c|c|}
\hline
 \mathfrak{H}_0 & 0 & 0 & \dots  & 0 \\
\hline
 0 & \mathfrak{H}_1\otimes  \pmb{\pmb{1}_{n_1\times  n_1}}  & 0 & \dots &  \vdots \\
\hline
 0 & 0 & \mathfrak{H}_2\otimes  \pmb{\pmb{1}_{n_2\times  n_2}} & \dots &  \vdots \\
\hline
 \vdots & \dots & \dots & \dots\dots &  \vdots \\
\hline
 \vdots & \dots & \dots & \dots &  0 \\
\hline
 0 & \dots & \dots &   0 & \mathfrak{H}_{r}\otimes  \pmb{\pmb{1}_{n_{r}\times  n_{r}}} \\
\hline
\end{array}
\right) \label{cosettusGF}
\end{equation}
where $n_I$ denotes the dimension of the $I$-th nontrivial
representation of the discrete group $\Gamma$ and from this we
extract the block diagonal matrix:
\begin{equation}
\mathcal{H}\text{ }\equiv \text{ }\left(
\begin{array}{|c|c|c|c|c|}
\hline
 \mathfrak{H}_1  & 0 & \dots & \dots & 0 \\
\hline
 0 & \mathfrak{H}_2 & \dots & \dots & \vdots \\
\hline
 \vdots & \dots\dots & \dots\dots & \dots & \vdots \\
\hline
 \vdots & \dots\dots & \dots & \mathfrak{H}_{r-1} & 0 \\
\hline
 0 & \dots\dots & \dots & 0 & \mathfrak{H}_{r} \\
\hline
\end{array}
\right)\label{tautobundmetro}
\end{equation}
The hermitian matrix $\mathcal{H}$ is the fibre metric on the direct
sum:
\begin{equation}\label{direttosummo}
    \mathcal{R}\,=\,\bigoplus_{I=1}^{r} \, \mathcal{R}_I
\end{equation}
of the $r$ tautological bundles that, by construction, are holomorphic vector bundles with rank equal to the
dimension of the $r$ irreducible representations of $\Gamma$:
\begin{equation}\label{tautibundiEach}
    \mathcal{R}_I \, \stackrel{\pi}{\longrightarrow}\, \mathcal{M}_\zeta \quad\quad ;
    \quad \quad\forall p \in \mathcal{M}_\zeta \quad :\quad
    \pi^{-1}(p) \approx \mathbb{C}^{n_I}
\end{equation}
The compatible connection\footnote{Following standard mathematical nomenclature, we call compatible
connection on a holomorphic vector bundle,  one whose $(0,1)$ part is the Cauchy Riemann operator of the
bundle} on the holomorphic vector bundle $\mathcal{R}$ is given by:
\begin{eqnarray}\label{comancio}
    \vartheta & = & \mathcal{H}^{-1} \,\partial\mathcal{H}\, = \,\bigoplus_{I=1}^{r}\, \theta_I \nonumber\\
    \theta_I & = & \mathfrak{H}_I^{-1} \, \partial \mathfrak{H}_I \, \in \, \mathbb{GL}(n_I,\mathbb{C})
\end{eqnarray}
where $\mathbb{GL}(n_I,\mathbb{C})$ is the Lie algebra of
$\mathrm{GL}(n_I,\mathbb{C})$ which is the structural group of the
$I$-th tautological vector bundle. The natural connection of the
$\mathcal{F}_\Gamma$ principal bundle, mentioned in
eqn.\,(\ref{rodolfo}) is just, according to a universal scheme, the
imaginary part of the holomorphic connection $\vartheta$.
\subsubsection{The tautological bundles from the irrep viewpoint and the
K\"ahler potential}\label{realecompdiscus} From the analysis of the
above section we have reached a very elegant conclusion. Once the
matrix $\mathcal{V}$ is calculated as function of the level
parameters $\zeta $ and of the base-manifold coordinates
(\(z_m,\bar{z}_m\)) ($m=1,2,3$), we also have the block diagonal
hermitian matrix $\mathcal{H}$ which encodes the hermitian fibre
metrics \(\mathfrak{H}_i\)\(\left(\zeta ,z,\bar{z}\right)\) on each
of the tautological holomorphic bundles \(\mathfrak{V}_I\) whose
ranks are equal, one by one, { }to the dimensions \(n_I\) of the
irreps of  $\Gamma $. The first Chern classes of these bundles are
represented by the differential $(1,1)$ forms:
\begin{equation}
\omega _I{}^{(1,1)}= \frac{i}{2\pi} \,\bar{\partial }\partial
\text{Log}\left[\text{Det}\left[\mathfrak{H}_{I }\right]\right]
\label{bambolone}
\end{equation}
Let us recall  another remarkable group theoretical fact. The number \textit{r} of nontrivial irreps of
$\Gamma $ is equal to the number \textit{ r} of nontrivial conjugacy classes and to the number \textit{r} of
generators of the center of the compact Lie algebra \(\mathbb{F}_{\Gamma }\), hence also to the number
\textit{ r} of level parameters $\zeta $ and to number \textit{ r} of holomorphic tautological bundles. Now
we are in a position to derive in full generality  the formula for the K\"ahler potential and, hence, for the
K\"ahler metric of the resolved manifold \(\mathcal{M}_{\zeta }\) that we anticipated in (\ref{celeberro}) .
In view of the above discussion, we rewrite the latter as it follows:
\begin{equation}
\mathcal{K}_{\mathcal{M}_{\zeta }}= \mathcal{K}_{\mathcal{S}_{\Gamma
}} \mid _{\mathcal{N}_{\zeta }}+
\zeta^I\mathfrak{C}_{\text{IJ}}\text{Log}\left[\text{Det}\left[\mathfrak{H}_{J
}\right]\right] \label{criceto1}
\end{equation}
where $\mathcal{K}_{\mathcal{S}_{\Gamma }}$ is the K\"ahler
potential of the flat space \(\mathcal{S}_{\Gamma}\) and \(\mid
_{\mathcal{N}_{\zeta }}\) denotes its restriction to the level
surface \(\mathcal{N}_{\zeta }\), while \(\mathfrak{C}_{\text{IJ}}\)
is an r$\times $r constant matrix whose structure we  will define
and determine below. Why the matrix defined there yields the
appropriate K\"ahler potential is what we will now explain starting
from an argument introduced in 1987 by Hitchin, Karlhede,
Lindstr\"om and Ro\v{c}ek.
\paragraph{The HKLR differential equation and its solution} Before explaining the origin
of the matrix \(\mathfrak{C}_{\text{IJ}}\), we would like to stress
that, conceptually it encodes a pairing between the level parameters
( = generators of the Lie algebra center) and the tautological
bundles ( = irreps). If we could understand the relation between
conjugacy classes with their ages and cohomology classes, then we
would have a relation between irreps and conjugacy classes and we
could close the three-sided relation diagram among the center
\(\mathfrak{z}\left[\mathbb{F}_{\Gamma }\right.\)] and the other two
items, irreps and conjugacy classes.  As we are going to show, this
side of the relation is based on the concept of weighted blowup. On
the other hand, understanding the matrix
\(\mathfrak{C}_{\text{IJ}}\), is a pure Lie algebra theory issue,
streaming from the HKLR argument.
\par
Hence,  continuing such an argument, let us  consider the flat K\"ahler manifold \(\mathcal{S}_{\Gamma }\) and
its K\"ahler potential
\begin{equation}
\mathcal{K}= \sum _{i=1}^3 \text{Tr}\left[A_i\, ,\, A_i^{\dagger }
\right]\text{ }\text{where we have defined}\text{  }A_i=\{A,B,C\}
\label{criceto2}
\end{equation}
The exponential of the K\"ahler potential is also, by definition, the hermitian metric on the Hodge line
bundle:
\begin{eqnarray}
&&\mathcal{L}_{\text{Hodge} }\overset{\pi }{\longrightarrow } \mathcal{S}_{\Gamma }\quad \text{ where }\quad
\forall p\in \mathcal{S}_{\Gamma }\quad : \quad \pi ^{-1}(p) \approx \mathbb{C}^* \nonumber\\
&&\quad\left\| W\right\| ^2\equiv e^{\mathcal{K}_{\mathcal{S}}}W\bar{W} \label{criceto3}
\end{eqnarray}
Indeed, the second line of the above equation $\left\| W\right\| ^2$  defines the invariant norm of any
section of $\mathcal{L}_{\text{Hodge}}$.
\par
Let us know consider the action of the quiver group on
\(\mathcal{S}_{\Gamma }\) and its effect on the fibre metric
h=\(e^{\mathcal{K}}\). The maximal compact subgroup
\(\mathcal{F}_{\Gamma }\) is an isometry group for the K\"ahler
metric defined by (\ref{criceto2}). Hence we focus on the orthogonal
(with respect to the Killing form) complement of
\(\mathcal{F}_{\Gamma }\). Let
\begin{equation}
\pmb{\Phi} \in \mathbb{K}_{\Gamma } \label{criceto4}
\end{equation}
be an element of the orthogonal subspace to the maximal compact subalgebra
\begin{equation}
\mathbb{G}_{\Gamma }=\mathbb{F}_{\Gamma }\oplus \mathbb{K}_{\Gamma } \label{criceto5}
\end{equation}
consider the one parameter subgroup generated by this Lie algebra element
\begin{equation}
g(\lambda ) \equiv  e^{\lambda \pmb{\Phi} } \label{criceto6}
\end{equation}
The action of this group on the K\"ahler potential is easily calculated
\begin{equation}
\mathcal{K}_{\mathcal{S}}(\lambda ) =\sum _{i=1}^3 \text{Tr}\left[A_i e^{2\lambda \pmb{\Phi} }A_i^{\dagger }
e^{-2\lambda \pmb{\Phi} }\right]\label{criceto6bis}
\end{equation}
Performing the derivative with respect to $\lambda $ we obtain
\begin{equation}
\partial _{\lambda }\mathcal{K}_{\mathcal{S}}(\lambda )\mid_{\lambda =0}
 =\sum _{i=1}^3 \text{Tr}\left(\pmb{\Phi} \left[A_i, A_i^{\dagger }
\right]\right) \label{criceto7}
\end{equation}
Then we utilize the fact that each element $\pmb{\Phi}\in \mathbb{K}_\Gamma $ is just equal to  ${\rm i}
\times $ \(\pmb{\Phi} _c\) where \(\pmb{\Phi} _c\) denotes an appropriate element of the compact subalgebra.
Hence the above equation becomes
\begin{equation}
\partial _{\lambda }\mathcal{K}_{\mathcal{S}}(\lambda )\mid _{\lambda =0} ={\rm i} \times \sum _{i=1}^3
\text{Tr}\left(\pmb{\Phi} _c\left[A_i, A_i^{\dagger } \right]\right) = i \mathfrak{P}_{\pmb{\Phi} _c}
\label{criceto8}
\end{equation}
Let us decompose the moment map along the standard basis of compact generators. We obtain:
\begin{eqnarray}
\mathfrak{P}_{\Phi }&=&\sum _{I=1}^{|\Gamma| -1} \pmb{\Phi} ^I\text{Tr}\left(\mathfrak{K}_I^c\, \left[A_i,
A_i^{\dagger}
\right]\right)\nonumber\\
&=&{\rm i}\sum _{I=1}^{|\Gamma| -1} \pmb{\Phi} _c^I \mathfrak{P}_I(p) \, = \, \sum _{I=1}^{|\Gamma| -1}
\pmb{\Phi} ^I\mathfrak{P}_I(p)=\sum_{I=1}^{|\Gamma| -1}\pmb{\Phi} ^I\text{Tr}\left(\mathfrak{K}_I\left[A_i,
A_i^{\dagger } \right]\right) \label{criceto9}
\end{eqnarray}
where p $\in $\(\mathcal{D}_{\Gamma } \subset \mathcal{S}_{\Gamma
}\) denotes the arbitrary point in the ambient space described by
the triple of matrices \(A_i\), \(\mathfrak{K}_I\)=i
\(\mathfrak{K}_I^c\) are the $|\Gamma |$-1 noncompact generators of
the quiver group \(\mathcal{G}_{\Gamma }\) that, by construction,
are just as many as the compact generators \(\mathfrak{K}_I^c\) of
the maximal compact subgroup \(\mathcal{F}_{\Gamma }\). Formally
integrating the above differential equation it follows that the
fibre of the metric Hodge line bundle (\ref{criceto3})
\begin{equation}
h(p) \equiv \text{Exp}\left[\mathcal{K}_{\mathcal{S}}(p)\right] \label{criceto10}
\end{equation}
transforms in the following way under the action of the quiver group
\begin{equation}
\forall g\in \mathcal{G}_{\Gamma }\text{           }g : h(p) \longrightarrow  h^g(p) \equiv
 h\left(e^{\text{Log}[g]}p\right)=e^{c(g,p)}h(p)
 \label{criceto11}
\end{equation}
where
\begin{equation}
\text{Log}[g]\in \mathbb{G}_{\Gamma } \label{criceto12}
\end{equation}
is an element of the quiver group Lie algebra and as such can be decomposed along a complete basis of
generators
\begin{equation}
\text{Log}[g]= \sum _{I=1}^7 \pmb{\Phi} ^I\mathfrak{K}_I+\pmb{\Phi} _c{}^I\mathfrak{K}^c{}_I
\label{criceto13}
\end{equation}
and the anomaly \(c(g,p)\)introduced in eqn.\,(\ref{criceto11}) has,
in force of the differential equation discussed above the following
form:
\begin{equation}
c(g,p)=\sum _{I=1}^7  \left(\pmb{\Phi} ^I+i \pmb{\Phi} _c{}^I\right) \mathfrak{P}_I(p) \label{criceto14}
\end{equation}
where \(\mathfrak{P}_I(p)\) are the moment maps at point p.
\par
Next consider the diagram
\begin{equation}
\mathcal{S}_{\Gamma }\overset{\text{       }\iota }{\text{   }\longleftarrow }\text{    }\mathcal{N}_{\zeta
}\text{      }\overset{\pi }{\longrightarrow }\text{    }\mathcal{M}_{\zeta }\text{  }\equiv \text{
}\mathcal{N}_{\zeta }/\mathcal{F} _{\Gamma } \label{criceto15}
\end{equation}
where \(\mathcal{N}_{\zeta }\) is the level surface and \(\mathcal{M}_{\zeta }\) the final K\"ahler threefold
with its associated Hodge line bundle whose curvature is the K\"ahler form \(\mathbf{K} _{\mathcal{M}}\)
\begin{equation}
\mathbf{K} _{\mathcal{M}} \equiv \frac{i}{2\pi }\bar{\partial }\partial  \mathcal{K}_{\mathcal{M}} =
\frac{i}{2\pi }\bar{\partial }\left( \frac{1}{h_{\mathcal{M}}} \partial h_{\mathcal{M}}\right)
\label{criceto19}
\end{equation}
\(\mathcal{K}_{\mathcal{M}}\) being the K\"ahler potential of the resolved variety. Following HKLR, we
require that
\begin{equation}
\pi ^*\mathbf{K} _{\mathcal{M}} = \iota ^*\mathbf{K} _{\mathcal{S}_{\Gamma }} \label{criceto20}
\end{equation}
where \(\mathbf{K} _{\mathcal{S}_{\Gamma }}\) is the K\"ahler form
of the flat K\"ahler manifold \(\mathcal{S}_{\Gamma }\)
=\(\text{Hom}_{\Gamma }\)(Q\(\otimes\)R,R). At the level of fibre
metric on the associated Hodge line bundles, eqn.\,(\ref{criceto20})
amounts to stating that
\begin{equation}
\forall p\in \mathcal{M}_{\zeta }\quad:\quad h_{\mathcal{M}}\text{  }(p) = h_{\mathcal{S}_{\Gamma }}^g (p)
=h_{\mathcal{S}_{\Gamma }} (g.p)=e^{c(g,p)}\text{  }h_{\mathcal{S}_{\Gamma }} (p)\text{    }
\end{equation}
where g is an element of the quiver group that brings the point p
$\in $ \(\mathcal{N}_{\zeta }\) on the level surface of level $\zeta
$ to the reference level surface \(\mathcal{N}_0\) which corresponds
to the singular orbifold \(\frac{\mathbb{C}^3}{\Gamma }\). Applying
this to eqn.\,(\ref{criceto14}) we obtain:
\begin{equation}
c(g,p) =\zeta ^I\pmb{\Phi} _I(p)=\zeta ^I*\text{Tr}\left[\mathfrak{K}_I\text{Log}[g]\right]=\sum _{i=1}^r
\zeta ^I*\text{Tr}\left[\mathfrak{K}_I^{\text{central}}\text{Log}[g]\right]
\end{equation}
since the only non-vanishing levels are located in the Lie Algebra center. On the oher hand we have g =
$\mathcal{H}$ :
\begin{equation}
\text{Tr}\left[\mathfrak{K}_I^{\text{central}}\text{Log}[\mathcal{H}]\right]\equiv\sum _{J=1}^r
\mathfrak{C}_{\text{IJ}}\text{Log}\left[\text{Det}\left[\mathfrak{H}_J\right]\right]
\end{equation}
The above formula defines the constant matrix $\mathfrak{C}_{\text{IJ}}$ and  justifies the final formula
(\ref{criceto1}).
In the case  of cyclic $\Gamma$ the center of the Lie Algebra $\mathbb{F}_\Gamma$ coincides
with the algebra itself and the matrix $\mathfrak{C}_{IJ}$ is just diagonal and essentially trivial.
\paragraph{Expansion to first order} Performing a series expansion
of the 2-forms \(\overset{(1)}{\omega }_I{}^{(1,1)}\) in the $\zeta
$ parameters:
\begin{equation}
\omega _I^{(1,1)} = 0 +\sum _{n=1}^{\infty } \zeta
^n\overset{(n)}{\omega }_I{}^{(1,1)}
\end{equation}
in those few cases where explicit calculations can be performed one
verifies that the order one term is cohomologous to the full form
\(\omega _i^{(1,1)}\). Hence it suffices to solve the moment map
equations at first order in $\zeta $ (which is always possible) and
one obtains  a calculation of the cohomology classes of the resolved
variety according to the above displayed scheme. At the same time
one obtains  a calculation of the K\"ahler potential to the very
same order.
\par
This is a general feature applying to all cases.
\paragraph{Dolbeault cohomology} The objects
we are dealing with are Dolbeault cohomology classes of the final resolved manifold $\mathcal{M}_{\zeta}$
which is K\"ahler as a result of its K\"ahler quotient construction.
\par
When we say that $\omega^{p,q}$ is a harmonic representative of a nontrivial cohomology class in
$H^{1,1}\left(\mathcal{M}_{\zeta}\right)$ we are stating that:
\begin{itemize}
  \item The form is $\partial$-closed and $\overline{\partial}$-closed
  \begin{equation*}
    \partial \omega^{p,q} \, = \,\overline{\partial}\omega^{p,q} \, = \, 0
  \end{equation*}
   \item There do  not exist forms $\phi^{p-1,q}$ and $\phi^{p,q-1}$ such that:
   \begin{equation*}
    \omega^{p,q}\, = \, \partial \phi^{p-1,q} \, = \,\overline{\partial}\phi^{p,q-1}
  \end{equation*}
\end{itemize}
The reason why the $\omega _I^{(1,1)}$ are nontrivial
representatives of $(1,1)$ cohomology classes is that they are
obtained as $\overline{\partial}$ of connection one-forms
$\theta^{(1,0)}$ that are not globally defined.  Indeed if we
introduce the curvatures and the first Chern classes of the
tautological vector bundles  we have the elegant formula anticipated
in eqn.\,(\ref{bambolone}):
\begin{eqnarray}
  \Theta_I &=& \overline{\partial} \theta_I \nonumber\\
  \omega^{(1,1)}_I &\equiv & c_1(\mathcal{R}_I) \, = \, \mathrm{Tr}(\Theta_I) \, = \,\overline{\partial}\,\partial
  \log \left[\mbox{Det} \left(\mathfrak{H}_I\right) \right]
\end{eqnarray}
Comparing now with the definition of Dolbeault cohomology we see
that $\omega^{(1,1)}_I$ are nontrivial cohomology classes because
either
\begin{equation}\label{nobuono}
    \theta^{(1,0)}_I \, \equiv \, \partial \log \left[\mbox{Det} \left(\mathfrak{H}_I\right) \right] \quad \mbox{or}\quad
\theta^{(0,1)}_I \, \equiv \, \overline{\partial} \log
\left[\mbox{Det} \left(\mathfrak{H}_I\right) \right]
\end{equation}
are non-globally defined $1$-forms on the base manifold. This is so because they transform nontrivially from
one local trivialization of the bundle to the next one. The transition functions on the connections are
determined by the transition functions on the metric $\mathcal{H}$, namely on the coset representative. Here
comes the delicate point.
\par
Where from in the Kronheimer--like construction do we know that
there are different local trivializations, otherwise that the
tautological bundles are nontrivial? Computationally we solve the
algebraic equations for $\mathcal{H}$ in terms of the coordinates
$z_i$ $(i=1,2,3)$ parameterizing the locus $L_\Gamma$, which is
equivalent to the singular locus $\frac{\mathbb{C}^3}{\Gamma}$ and
we find $\mathcal{H}=\mathcal{H}(\zeta,z)$ where $\zeta$ are the
level parameters. In order to conclude that the tautological bundle
is nontrivial we should divide the locus $L_\Gamma$ into patches and
find the transition functions of the connections $\theta_I$ from one
patch to the other. Obviously the transition function must be an
element of the the quiver group $g\in \mathcal{G}_{\Gamma}$. At the
first sight it is not clear  how to implement such a program, since
we do not know how we should partition the locus $L_\Gamma$. Clearly
the actual solution of the algebraic equations is complicated and,
as we very well know, we are able to implement it only by means of a
power series in $\zeta$, yet it is obvious that this is not a case
by case study. As everything else in the Kronheimer--like
construction, it must be based on first principles and it is
precisely these first principles that we are going to find out. It
is  at this level that the issue of ages is going to come into play
in an algorithmic way.
\subsection{The exceptional divisor}
According to general lore, the cohomology classes constructed as
first Chern classes of the tautological holomorphic vector bundles
defined by the K\"ahler quotient via  hermitian matrices
$\mathfrak{H}_I$, are naturally associated with  the components of
the exceptional divisor produced by the blow--up of the
singularities. This latter is defined as the vanishing locus of a
global holomorphic section $W(p)$ of a line bundle:
\begin{eqnarray}\label{calendula}
    &&\mathcal{L}_\mathfrak{D} \, \stackrel{\pi}{\longrightarrow} \,\mathcal{M}_\zeta \nonumber\\
    &&\mathfrak{D}\subset \mathcal{M}_\zeta \quad ; \quad \mathfrak{D} \, =\, \left\{
    p \in \mathcal{M}_\zeta \, \mid \, W(p) \, = \, 0 \quad \mbox{where} \quad W \in
    \Gamma\left(\mathcal{L}_\mathfrak{D}\right)
    \right\}
\end{eqnarray}
The line bundle $\mathcal{L}_\mathfrak{D}$ is singled out by the
divisor $\mathfrak{D}$ and for this reason it is labeled by it. Its
first Chern class $\omega_\mathfrak{D}^{(1,1)}$ is certainly a
cohomology class and so it must be a linear combination of the first
Chern classes $\omega_I^{(1,1)}$ created by the K\"ahler quotient
and associated with the hermitian matrices
$\mathfrak{H}_I(\zeta,p)$:
\begin{equation}\label{calzaturificio}
   \left[ \omega_\mathfrak{D}^{(1,1)}\right] \, = \, \mathrm{S}_{\mathfrak{D},I} \,
   \left[ \omega_I^{(1,1)}\right]
\end{equation}
The question is to know which is which and to determine the constant
matrix $\mathrm{S}_{\mathfrak{D},I}$.
\par
Another point is that, at least locally, the entire space
$\mathcal{M}_\zeta$ can be viewed as the total space of a line
bundle over the divisor $\mathfrak{D}$:
\begin{eqnarray}\label{carnitinapotassica}
    && \mathcal{M}_\zeta \, \stackrel{\pi_d}{\longrightarrow} \,\mathfrak{D} \nonumber\\
    && \forall p \in \mathfrak{D}\quad ; \quad \pi_d^{-1}\left(p\right)\, \simeq \, \mathbb{C}^\star
\end{eqnarray}
Furthermore the matrix $\mathfrak{H}_I$ can be viewed as the
invariant norm of a section of the appropriate line bundle:
\begin{equation}\label{cosmitto}
    \mathfrak{H}_I(\zeta,z,\bar{z}) \, = \, H_I(\xi,\bar{\xi}, W,\bar{W}) \, |W|^2
\end{equation}
where $\xi$ denote the two coordinates spanning the divisor
$\mathfrak{D}$ and $W$ spans the vertical fibres  out of the
divisor. The projection $\pi_d$ corresponds to setting $W\to 0$ and
obtaining:
\begin{equation}\label{camalosto}
 \pi_d \, : \,   H(\xi,\bar{\xi}, W,\bar{W}) \longrightarrow h(\xi,\bar{\xi}) \equiv H(\xi,\bar{\xi}, 0,0)
\end{equation}
We expect that the two (1,1)-forms:
\begin{eqnarray}
  \Omega_I &=& \overline{\partial}\partial H_I(\xi,\bar{\xi}, W,\bar{W}) \nonumber\\
  \widehat{\Omega}_I &=& \overline{\partial}\partial h(\xi,\bar{\xi})\label{santeddu}
\end{eqnarray}
should be cohomologous:
\begin{equation}\label{commollo}
    \left[\Omega_I\right] \, = \, \left[\widehat{\Omega}_I\right]
\end{equation}
The form $\widehat{\Omega}_I$ is the first Chern class of the line
bundle (\ref{carnitinapotassica}) while $\Omega_I$ is the first
Chern class of the line bundle (\ref{calendula}) that defines the
divisor.
\paragraph{Divisors and conjugacy classes graded by age.}
Hence the question boils down to the following: \textit{What are the
components of the exceptional divisor of a crepant resolution of the
singularity $\mathbb C^3/\Gamma$,  and how many are they?} The
answer is provided by Theorem \ref{reidmarktheo} (Theorem 1.6 in
\cite{giapumckay}); they are the inverse images via the blowdown
morphism of the irreducible components of the fixed locus of the
action of $\Gamma$ on $\mathbb C^3$, and are in a one-to-one
correspondence with the junior conjugacy classes of $\Gamma$. The
irreducible components of the exceptional divisor may be compact
(corresponding to a component of the fixed locus which is just the
origin of $\mathbb C^3$) or noncompact (corresponding to fixed loci
of higher dimensions, i.e., curves).
\par
Let us consider the case of a cyclic group $\Gamma$, with only the
origin as fixed locus, and choose a generator $\gamma$ of $\Gamma$
of order $r$. As in eqn.\,(\ref{gioffo}), we can write
$\gamma=\frac1r(a_1,a_2,a_3)$. As described in \cite{giapumckay},
sections 2.3 and 2.4, the resolution of singularities is obtained by
iterating the following construction, which uses toric geometry (a
general reference for toric geometry, which in particular explains
how to perform a toric blowup by subdividing the fan of the toric
variety one wants to blowup, is \cite{Fulton-toric}). The fan of the
toric variety $\mathbb C^3$ is the first octant of $\mathbb R^3$,
with all its faces; by adding the ray $\frac1r(a_1,a_2,a_3)$ we
perform a blowup $\mathbb{B}_{[a_1,a_2,a_3]} \to \mathbb C^3$ whose
exceptional divisor  $F$ is the weighted projective space
$\mathbb{WP}[a_1,a_2,a_3]$. The same procedure applied to $\mathbb
C^3/\Gamma$ produces a partial desingularization $ W_\gamma \to
\mathbb C^3/\Gamma$ which is the base of a cyclic covering
$\mathbb{B}_{[a_1,a_2,a_3]} \to W_\gamma$, ramified along the
exceptional divisor $E$ of $ W_\gamma \to \mathbb C^3/\Gamma$. The
situation is summarised by the following diagram
\begin{equation}\label{blowup} \xymatrix@C+40pt{
F \ar@^{(->}[r] \ar[d] & \mathbb{B}_{[a_1,a_2,a_3]} \ar[r]^-{\mbox{\footnotesize weighted blowup}}\ar[d] &  \mathbb C^3 \ar[d] \\
E \ar@^{(->}[r] &W_\gamma \ar[r]^-{\mbox{\footnotesize  crepant resolution}} &  \mathbb C^3/\Gamma
}\,.
\end{equation}

The full desingularization is obtained by performing a multiple toric blowup, adding all rays corresponding to junior conjugacy classes.

\subsection{Steps of a weighted blowup} \label{steppisoffio}
The present section is devoted to give some views on the relation
between the age grading and the cohomology classes of the resolved
variety (to get more extensive information see
\cite{Bruzzo:2017fwj}).
\subsubsection{Weighted projective planes}
Let us define in a pedantic way the weighted blowup of the origin in $\mathbb{C}^3$. To this effect we begin
by recalling the definition of a weighted projective plane $\mathbb{WP}_{[a_1,a_2,a_3]}$, where
$[a_1,a_2,a_3]$ are the weights (good references for weighted projective spaces and line bundles on them are
\cite{Dolgy,Beltrametti-Robbiano,Rossi-Terracini}). We restrict our attention to the case where the weights are integers.
One defines an action of $\mathbb C^\ast$ on $\mathbb C^3-\{0\}$ letting
$$ (y_1,y_2,y_3) \to (y_1 \,\lambda^{a_1},y_2\,\lambda^{a_2},y_3\,\lambda^{a_2}),\qquad \lambda \in \mathbb C^\ast.$$
The weighted projective plane $\mathbb{WP}_{[a_1,a_2,a_3]}$ is the quotient of $\mathbb C^3-\{0\}$ under this action.
It is by construction a complex variety of dimension 2.
\par
To examine the properties of this space it is expedient to assume that the triple $(a_1,a_2,a_3)$ is reduced. One defines
$$ d_ i = \mbox{g.c.d.}\, (a_{i-1},a_{i+1}), \qquad  b_ i = \mbox{l.c.m.} (d_{i-1},d_{i+1}) $$
(where indices in the r.h.s.~are meant mod 3, i.e., $1-1=3$, etc.). The   triple $(a_1,a_2,a_3)$ is reduced
if $(b_1,b_2,b_3) = (1,1,1)$;
otherwise one defines $a'_i=a_i/b_i$. The numbers $a'_i$ are positive integers, the triple $(a'_1,a'_2,a'_3)$
is reduced, and  the weighted projective planes $\mathbb{WP}_{[a_1,a_2,a_3]}$ and $\mathbb{WP}_{[a'_1,a'_2,a'_3]}$
are isomorphic. Henceforth we shall  assume that the triple $(a_1,a_2,a_3)$ is reduced.
It turns out that $\mathbb{WP}_{[a_1,a_2,a_3]}$
is smooth if and only if $(a_1,a_2,a_3) = (1,1,1)$, in which case the weighted projective
plane is just $\mathbb P^2$.
\par
The same construction of line bundles on projective spaces produces
on weighted projective spaces rank one sheaves $\mathcal
O_{\mathbb{WP}_{[a_1,a_2,a_3]}}(i)$, with $i\in \mathbb Z$, that in
general are not locally free (i.e., they are not line bundles), but
only reflexive (i.e., they are isomorphic to their duals). It turns
out that  $\mathcal O_{\mathbb{WP}_{[a_1,a_2,a_3]}}(i)$ is locally
free if and only if $i$ is a multiple of $m = \mbox{l.c.m.}
(a_1,a_2,a_3)$ \cite{Beltrametti-Robbiano}.
\par
The weighted projective plane $\mathbb{WP}_{[a_1,a_2,a_3]}$ is covered by the open sets
$$ U _ i = \{ (y_1,y_2,y_3) \,\vert \, y_i\ne 0\}.$$
On this open cover the  line  bundle  $\mathcal
O_{\mathbb{WP}_{[a_1,a_2,a_3]}}(km)$ has  transition functions
\begin{equation}\label{transitionfunctions}
g_{ij}\colon U_i\cap U_j \to\mathbb C^\ast,\qquad
g_{ij}(y_1,y_2,y_3) =  y_j^{km/a_j}y_i^{-km/a_i}
\end{equation}
where $m = \mbox{l.c.m.}\ (a_1,a_2,a_3)$. In particular, (the
isomorphism class of) $\mathcal O_{\mathbb{WP}_{[a_1,a_2,a_3]}}(m)$
is the (very ample) generator of the Picard group of
$\mathbb{WP}_{[a_1,a_2,a_3]}$, the group of isomorphism classes of
line bundles on $\mathbb{WP}_{[a_1,a_2,a_3]}$, which is isomorphic
to $\mathbb Z$.
We conclude this brief introduction to weighted projective planes by
defining an orbifold K\"ahler metric for the spaces
$\mathbb{WP}_{[a_1,a_2,a_3]}$. Denoting again by $(y_1,y_2,y_3)$ a
set of homogeneous coordinates on $\mathbb{WP}_{[a_1,a_2,a_3]}$, and
$m = \mbox{l.c.m.}\,(a_1,a_2,a_3)$, one can check the 2-form
$$ \omega = \frac{i}{2\pi} \partial\bar\partial\log\sum_{i=1}^3 y_i^{m/a_i}\,\bar y_i^{m/a_i}$$
is invariant under rescaling of the homogeneous coordinates, and
therefore defines a 2-form on the smooth locus  of
$\mathbb{WP}_{[a_1,a_2,a_3]}$; this reduces to the usual
Fubini-Study form when the projective space is smooth.
\subsubsection{The weighted blowup and the tautological bundle}
The weighted blowup of $\mathbb{C}^3$, denoted
$\mathbb{B}_{[a_1,a_2,a_3]}$, is a subvariety
\begin{equation}\label{corleone}
    \mathbb{B}_{[a_1,a_2,a_3]} \, \subset \, \mathbb{C}^3 \times \mathbb{WP}_{[a_1,a_2,a_3]}
\end{equation}
defined by the equations
\begin{equation}\label{eqblowup} z_1y_2^{a_1a_3}=z_2y_1^{a_2a_3},\qquad z_2y_3^{a_1a_2}=z_3y_2^{a_1a_3},\qquad
z_1y_3^{a_1a_2}=z_3y_1^{a_2 a_3}
\end{equation}
where $\{z_1,z_2,z_3\}$ are standard coordinates in $\mathbb C^3$,
and $\{y_1,y_2,y_3\}$ are homogenous coordinates in $
\mathbb{WP}_{[a_1,a_2,a_3]}$. Actually the three equations are not
independent (regarding them as a linear system in the unknowns $z$,
the associated matrix has rank at most 2) and therefore the locus
$\mathbb{B}_{[a_1,a_2,a_3]}$ is 3-dimensional.  The projections of $
\mathbb{C}^3 \times \mathbb{WP}_{[a_1,a_2,a_3]}$ onto its factors
define projections
 \begin{eqnarray}
 p \quad   &:& \quad \mathbb{B}_{[a_1,a_2,a_3]}  \to  \mathbb{C}^3 \nonumber\\
 \pi \quad   &:& \quad \mathbb{B}_{[a_1,a_2,a_3]} \to \mathbb{WP}_{[a_1,a_2,a_3]}
 \nonumber
  \label{portapannolini}
\end{eqnarray}
From eqn.~\eqref{eqblowup} we see that the fibres of $\pi$ are
isomorphic to $\mathbb C$; indeed, by comparing with
eqn.~\eqref{transitionfunctions}, we see that
$\mathbb{B}_{[a_1,a_2,a_3]}$ is the total space of the line bundle
$\mathcal O_{\mathbb{WP}_{[a_1,a_2,a_3]}}(-1)$ over the base
$\mathbb{WP}_{[a_1,a_2,a_3]}$, and $\pi$ is the bundle projection.
On the other hand, the morphism $p$ is   birational, as it is an
isomorphism away from the fibre $p^{-1}(0)$, while the fibre itself
--- the exceptional divisor $F$ of the blowup  ---  is isomorphic to
$\mathbb{WP}_{[a_1,a_2,a_3]}$.
\par
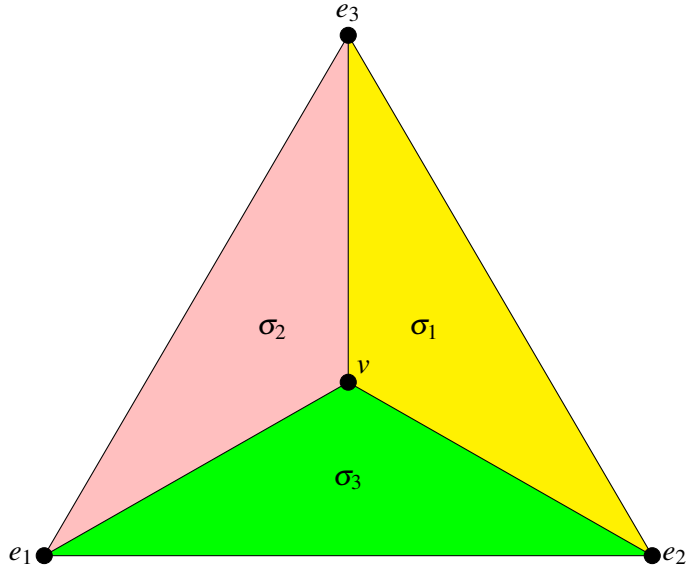
\begin{figure}\label{figure1}
\begin{center}
\begin{tikzpicture}
  \path [fill=pink] (0,0) to (4,2.3) to  (4,6.9) to (0,0) ;
    \path [fill=yellow] (8,0) to (4,2.3) to  (4,6.9) to (8,0) ;
      \path [fill=green] (0,0) to (4,2.3)  to  (8,0) to (0,0);
\draw [fill] (0,0) circle (3pt);
    \draw [fill]  (8,0) circle (3pt);
      \draw [fill] (4,6.9) circle (3pt);
        \draw [fill] (4,2.3) circle (3pt);
\draw (0,0) -- (8,0); \draw (0,0) -- (4,6.9); \draw (8,0) --
(4,6.9); \draw (0,0) -- (4,2.3); \draw (8,0) -- (4,2.3); \draw
(4,6.9) -- (4,2.3); \node at (-0.3,0) {$e_1$}; \node at (8.3,0)
{$e_2$}; \node at (4,7.2) {$e_3$}; \node at (4.2,2.5) {$v$}; \node
at (4,1) {$\sigma_3$}; \node at (3,3) {$\sigma_2$}; \node at (5,3)
{$\sigma_1$};
\end{tikzpicture}
\caption{\label{fan} Representation of the fan of
$\mathbb{B}_{[a_1,a_2,a_3]}$. The vector $v$ \emph{is not} on the
plane singled out by $e_1$, $e_2$, $e_3$.}
\end{center}
\end{figure}

The blowup $\mathbb{B}_{[a_1,a_2,a_3]}$ is nicely described in terms
of toric geometry \cite{Fulton-toric}. Denoting by $\{e_i\}$ the
standard basis of $\mathbb R^3$, the variety
$\mathbb{B}_{[a_1,a_2,a_3]}$ is associated with the fan given by the
one-dimensional cones (rays) generated by
$$ e_1,\qquad  e_2, \qquad e_3, \qquad v = a_1e_1+a_2e_2+a_3e_3.$$
The fan has three 3-dimensional cones $\sigma_i$, corresponding to 3
open affine toric varieties $U_i$ which cover
$\mathbb{B}_{[a_1,a_2,a_3]}$  (see Figure \ref{fan}). It turns out
that $U_i$ is smooth if and only if $a_i=1$, so that
 $\mathbb{B}_{[a_1,a_2,a_3]}$ is smooth if and only if $a_1=a_2=a_3=1$ (in which case the exceptional divisor is a $\mathbb P^2$).
 Moreover, unless again $a_1=a_2=a_3=1$, $F$ is a Weil divisor, so that its associated rank one sheaf (it ideal sheaf,
 i.e., the sheaf of functions  $\mathbb{B}_{[a_1,a_2,a_3]}$ that vanish on $F$), is not locally free, but only reflexive. We shall
denote the dual of this sheaf as $\mathcal
O_{\mathbb{B}_{[a_1,a_2,a_3]}}(F)$. Although this in general is not
locally free, it is still true that its first Chern class is
Poincar\'e dual to $F$.

\par
Let us assume that we are interested in desingularizing a quotient
$\mathbb C^3/\Gamma$, where $\Gamma$ is cyclic, and the
representation of $\Gamma$ on $\mathbb C^3$ has just one junior
class $\frac1r(a_1,a_2,a_3)$, which is compact, i.e., all $a_i$ are
strictly positive. The geometric constructions in this section show
that the orbifold K\"ahler form on $\mathbb{WP}_{[a_1,a_2,a_3]}$
induce by pullback a K\"ahler form on the smooth locus of the blowup
$\mathbb{B}_{[a_1,a_2,a_3]}$. With reference to diagram
\eqref{blowup}, the action of the group $\Gamma$ leaves the
exceptional divisor $F$ pointwise fixed, so that the form descends
to the desingularization $\mathcal M_\zeta$ of $\mathbb C^3/\Gamma$.
However, this K\"ahler form does not appear to coincide with the
K\"ahler form built on $\mathcal M_\zeta$ by means of the K\"ahler
reduction. This comes to no surprise as it is known that there are
several different metrics on a weighted projective space which all
reduce to the standard Fubini-Study metric in the smooth case (see
\cite{Ross-Thomas}).
\subsubsection{Pairing between irreps and conjugacy class in the K\"ahler quotient resolution: open questions}
According to \cite{giapumckay}, in the crepant resolution:
\begin{equation}\label{sandalorotto}
    \mathcal{M}_\zeta \, \longrightarrow \, \frac{\mathbb{C}^3}{\Gamma}
\end{equation}
we obtain a component of the exceptional divisor
$\mathfrak{D}^{(E)}$ for each junior conjugacy class of the group
$\Gamma$, namely we have:
\begin{equation}\label{semifreddocaffe}
    \mathfrak{D}^{(E)} = \bigcup_{i=1}^{ \mbox{$\#$ of junior
    classes}} \,\mathfrak{D}_{[a_1,a_2,a_3]_i}
\end{equation}
When there is just one  junior class,  the procedure
described in previous subsections, which is graphically summarized
in  the diagram (\ref{blowup}), is exhaustive and we easily identify the
exceptional divisor with a single projective plane
$\mathbb{WP}_{[a_1,a_2,a_3]}$. Indeed, the divisor $F$ is the weighted projective plane
$\mathbb{WP}_{[a_1,a_2,a_3]}$  by construction, and the action of $\Gamma$ leaves it pointwise fixed, so
that $E$ is isomorphic to $F$.

Utilizing the correspondence between
line bundles and divisors,
we can conclude that the exceptional divisor
$\mathbb{WP}_{[a_1,a_2,a_3]}$ uniquely identifies a line-bundle,
\textit{i.e.}, the tautological line bundle
$\mathcal{T}_{[a_1,a_2,a_3]}$, whose first Chern class is
necessarily given by:
\begin{equation}\label{crescenzio}
    c_1\left(\mathcal{T}_{[a_1,a_2,a_3]}\right) \, = \, \frac{\rm i}{2\pi} \, \overline{\partial}\,\partial \, \log
    \left(H_{[a_1,a_2,a_3]}\right)
\end{equation}
where $H_{[a_1,a_2,a_3]}$ is a suitable hermitian fibre-metric. The
most  interesting issue is to relate such an invariant fibre metric
and the (1,1)-form $c_1\left(\mathcal{T}_{[a_1,a_2,a_3]}\right)$
with the real functions $\frak{H}_I$ in eqn.(\ref{tautobundmetro})
and the corresponding (1,1)-forms  (\ref{bambolone}), that are
produced by the K\"ahler quotient construction.
\par
In the next section which is the first of the second part such a
relation will be explicitly analyzed in the context of a simple
master example that indeed is characterized by a unique
\textit{junior conjugacy class}.
\par
The construction of the exceptional divisor and the structure of the
blowup in cases with several junior classes is more complicated.
\newpage
\chapter{A  fully calculable master example:
$\mathbb{C}^3/\mathbb{Z}_3$}\label{maestro1}
In this part I focus on the simplest available non-trivial example
of  resolution \`{a} la Kronheimer of a $\mathbb{C}^3/\Gamma$
singularity which, being fully calculable, allows to illustrate all
the so far introduced concepts in a transparent way. An additional
advantage of such a master example is that the Ricci flat metric on
the resolved three-fold $Y^\Gamma_{[3]}$ can also be explicitly
worked out and compared with the HKLR metric emerging from the
K\"ahler quotient.
\section{The master model}
\label{mastello} The master model is provided by quotient
singularity \(\frac{\mathbb{C}^3}{\mathbb{Z}_3}\) in the case where
the generator $Y$ of \(\mathbb{Z}_3\) is of the following  form:
\begin{equation}\label{mastrogiacomo}
Y=\left(
\begin{array}{ccc}
\xi  & 0 & 0 \\
0 & \xi  & 0 \\
0 & 0 & \xi  \\
\end{array}
\right)
\end{equation}
 $\xi $ being a primite cubic root of unity \(\xi ^3\) =1.\\
\section{Analysis of the (1,1)-forms: irreps versus conjugacy classes that is cohomology versus homology }
In the present section following \cite{Bruzzo:2017fwj} I illustrate
in full detail, within the scope of the  one junior class model
defined above by eqn.(\ref{mastrogiacomo}), the relation between the
above extensively discussed $\omega^{(1,1)}_I$ forms ($I = 1,\dots,
r = \#$ of nontrivial irreps), with the exceptional divisors
generated by the blowup of the singularity, together with the other
predictions of the fundamental theorem \ref{reidmarktheo} which
associates cohomology classes of $\mathcal{M}_\zeta$ with conjugacy
classes of $\Gamma$. The number of nontrivial conjugacy classes and
the number of nontrivial irreps are equal to each other so that we
use $r$ in both cases, yet what is the actual pairing is not clear a
priori and it is not intrinsic to group theory, as we have stressed
several times. It is desirable to explore such pairing in an
explicit way and for that we need explicit calculable examples.
These are very few because of the bottleneck constituted by the
solution of the moment map equations, that are algebraic of higher
degree and only seldom admit explicit analytic solutions. For this
reason we introduce here the full-fledged construction of one of
those rare examples, where the moment map equations are solved in
terms of radicals. As anticipated above this model has the
additional nice feature that the number of junior conjugacy classes
is just one. It is our master model for explicit analysis. Such a
master model that we discuss here is based on $\Gamma
=\mathbb{Z}_{3}$  the action of the latter  on $\mathbb{C}^3$ being
degined in eqn.(\ref{mastrogiacomo}). It is also important to stress
that aim of the Kronheimer-like construction is not only the
calculation of cohomology but also the actual determination of the
K\"ahler potential (yielding the K\"ahler metric), which is encoded
in formula (\ref{criceto1}). From this point of view one of the
$\mbox{Det}\mathfrak{H}_i$ may lead to a corresponding
$\omega_I^{(1,1)}=\ft{i}{2\pi}\bar{\partial}\partial\mbox{Det}\mathfrak{H}_I$
that is either exact or cohomologous to another one, yet its
contribution to the K\"ahler potential, which is very important in
physical applications, can not be neglected. It is only the
cohomology class of the K\"ahler 2-form that is affected by the
triviality of one or more of the $\omega_I^{(1,1)}$; the
contributions to the K\"ahler potential that give rise to exact form
deformations of the K\"ahler 2-form are equally important as others.
\par
Having anticipated these general considerations  we turn to the
construction of the master model.
\section{The master model $\frac{\mathbb{C}^3}{\Gamma}$ with generator
$\{\xi,\xi,\xi\}$}\label{masterdegree} In this section we develop
all the calculations for the K\"ahler quotient resolution
corresponding to the $\mathbb{Z}_3$ action defined on $\mathbb{C}^3$
as in eqn.(\ref{mastrogiacomo}). In this case the equation \(p\wedge
p\)=0 which is a set of quadrics has solutions arranged in various
branches. There is a unique, principal branch of the solution that
has maximal dimension \(\mathcal{D}_{\Gamma }^0\) and is indeed
isomorphic to the \(\mathcal{G}_{\Gamma }\) orbit of the singular
locus \(L_{\Gamma }\) . This principal branch is the algebraic
variety \(\mathbb{V}_{|\Gamma |+2}\) mentioned in
eqn.\,(\ref{belgone}), of which we perform the K\"ahler quotient
with respect to the group \(\mathcal{F}_{\Gamma }\)
\begin{equation}
\mathcal{F}_{\Gamma } =\underset{\mu =1}{\overset{r+1}{\otimes }}\mathrm{U}\left(n_{\mu }\right)\cap
\text{SU}(|\Gamma |) = \mathrm{U(1)\otimes U(1)}
\end{equation}
in order to obtain the crepant resolution together with its K\"ahler metric. In the above formula \(n_{\mu
}\) = $\{$1,1,1$\}$ are the dimensions of the irreducible representations of $\Gamma $=\(\mathbb{Z}_3\) and
r+1=3 is the number of conjugacy classes of the group ($r$ is the number of nontrivial
representations). \\
To make a long story short, exactly as in the original Kronheimer
case we are able to retrieve the algebraic equation of the singular
locus from traces and determinants of the quiver matrices restricted
to \(L_{\Gamma }.\) Precisely for the \(\mathbb{Z}_3\) case under
consideration we obtain
\begin{equation}
\mathcal{I}_1=\text{Det}\left[A_o\right] ; \mathcal{I}_2=\text{Det}\left[B_o\right] ;\text{
}\mathcal{I}_3=\text{Det}\left[C_o\right] ; \mathcal{I}_4 =
\frac{1}{3}\text{Tr}\left[A_oB_oC_o\right]\label{osteoporo1}
\end{equation}
and we find the relation
\begin{equation}
\mathcal{I}_1\mathcal{I}_2\mathcal{I}_3=\mathcal{I}_4{}^3
\end{equation}
which reproduces the \(\mathbb{C}^3\) analogue of eqs. (\ref{wgammus}-\ref{identifyAk}) applying to the
\(\mathbb{C}^2\) case of Kronheimer and Arnold.
\par
The main difference, as we have several time observed, is that now
the same equations remain true, with no deformation for the entire
\(\mathcal{G}_{\Gamma }\) = \(\mathbb{C}^*\) \(\times\)
\(\mathbb{C}^*\) orbit of the locus \(L_{\Gamma }\), namely for the
entire { }\(\mathbb{V}_{|\Gamma |+2}\) = \(\mathbb{V}_5\) variety of
which we construct the K\"ahler quotient with respect to the compact
subgroup \(\mathrm{U(1)}\times \mathrm{U(1)}\) $\subset $
\(\mathbb{C}^* \times \mathbb{C}^*\). This is in line with the many
times emphasized feature  that in the $\mathbb{C}^3$ case there is
no deformation of the complex structure.
\subsection{The actual
calculation of the K\"ahler quotient and of the K\"ahler potential}
The calculation of the final form of the K\"ahler potential is
reduced to the solution of a set of two algebraic equations. The
solutions of such equations are accessible in this particular case
since they  reduce to a single cubic for which we have  Cardano{'}s
formula. For this reason the present case is the three-dimensional
analogue of the Eguchi-Hanson space where everything is explicitly
calculable and all theorems admit explicit testing and illustration.
\par
By calculating the ages we determine the number of \(\omega
^{(q,q)}\) harmonic forms (where $q=1,2$). According to  theorem
\ref{reidmarktheo} all these forms (and their dual cycles in
homology) should be in one-to-one correspondence with the \textit{r}
nontrivial conjugacy classes of $\Gamma $. On the other hand the
K\"ahler quotient construction associates one level parameter $\zeta
$ to each generator of the center
$\mathfrak{z}$(\(\mathcal{F}_{\Gamma }\)) of the group
\(\mathcal{F}_{\Gamma }\), two $\zeta $.s in this case, that are in
one-to-one correspondence with the \textit{r} nontrivial irreducible
representation of $\Gamma $. The number is the same, but what is the
pairing between \pmb{irreps} and \pmb{conjugacy classes}? More
precisely how do we see the homology cycles that are created when
each of the \textit{r} level parameters $\zeta $ departs from its
original zero value? Using the explicit expression of the functions
\(\mathfrak{H}_{1,2}\) defined in eqs.
(\ref{cosettusGF}-\ref{bambolone}) we arrive at  the calculation of
the \(\omega ^{(1,1)}{}_{I=1,2}\) forms that encode the first Chern
classes of the two tautological bundles. The expectation from the
age argument is that these two 2-forms should be cohomologous
corresponding to just the unique predicted class of type (1,1) since
\(h^{1,1}\)=1. On the other hand we should be able to construct an
\(\omega ^{(2,2)}\) form representing the unique class that is
Poincar\'e  dual to the exceptional divisor.
\par
In this case we can successfully answer both questions and this is very much illuminating.
\paragraph{Ages.} Indeed taking the explicit generator
\begin{equation}
Y=\left(
\begin{array}{ccc}
 (-1)^{2/3} & 0 & 0 \\
 0 & (-1)^{2/3} & 0 \\
 0 & 0 & (-1)^{2/3} \\
\end{array}
\right)
\end{equation}
 we easily calculate the $\{a_1,a_2,a_3\}$ vectors respectively associated to each of the three conjugacy
 classes and we obtain:
\begin{equation}
a-\text{vectors} =\{\{0,0,0\},\ft 13 \, \{1,1,1\},\ft 13 \,\{2,2,2\}\} \label{fistulario}
\end{equation}
from which we conclude that, apart from the class of the identity, there is just one junior and one senior
class.
\par
Hence we conclude that the Hodge numbers of the resolved variety
should be  \(h^{(0,0) }=1\), \(h^{(1,1) }=1\), \(h^{(2,2) }=1\).  If
we follow the weighted blowup procedure described in section
\ref{steppisoffio}  using the weights of the unique junior class
$\{1,1,1\}$, we see that the projection $\pi$ of
eqn.\,(\ref{portapannolini}) yields
\begin{equation}\label{pisastore}
\pi :\mathbb{B}_{(1,1,1) }\longrightarrow  \mathbb{W}\mathbb{P}_{(1,1,1)}\sim \mathbb{P}^2
\end{equation}
So the blowup replaces the singular point $0 \in \mathbb{C}^3$ with
a $\mathbb{P}^2$, which is compact. As a result, also the
exceptional divisor in the resolution $\mathcal M_\zeta$ is compact.
By Poincar\'e duality this entrains the existence of a harmonic
(2,2)--form associated with the unique senior class.
\subsubsection{The quiver matrix}
In this case, the quiver matrix defined by eqn.\,(\ref{quiverro2})
is the following one :
\begin{equation}
A_{ij} =\left(
\begin{array}{ccc}
 0 & 3 & 0 \\
 0 & 0 & 3 \\
 3 & 0 & 0 \\
\end{array}
\right)
\end{equation}
and it has the graphical representation displayed in fig.
\ref{quiveruMaster}
\begin{figure}
\begin{center}
\begin{tikzpicture}[scale=0.50]
\draw [thick] [fill=yellow] (-5,-5) circle (1.5cm); \node at (-5,-5)
{$\mathrm{U_2(N)}$}; \draw [thick] [fill=green] (5,-5) circle
(1.5cm); \node at (5,-5) {$\mathrm{U_3(N)}$}; \node at (0,0)
{$\mathrm{U_1(N)}$}; \draw [thick] [fill=red] (0,0) circle (1.5cm);
\node at (0,0) {$\mathrm{U_1(N)}$}; \draw [black, line width=0.07cm]
[->] (-6.5,-5) to [out=90,in=-180] (0,1.5); \draw [blue, line
width=0.07cm] [->] (-3.5,-5) to [out=90,in=-180] (0,-1.5); \draw
[red, line width=0.07cm] [->] (-5,-3.5) to [out=90,in=-180](-1.5,0);
\draw [black, line width=0.07cm] [->] (0,1.5) to [out=0,in=90]
(6.5,-5); \draw [blue, line width=0.07cm] [->] (0,-1.5) to
[out=0,in=90] (3.5,-5); \draw [red, line width=0.07cm] [->] (1.5,0)
to [out=0,in=90](5,-3.5);
\draw [blue, line width=0.07cm] [->] (3.5,-5) to (-3.5,-5); 
\draw [red, line width=0.07cm] [->] (5,-6.5) to (-5,-6.5);
\draw [black, line width=0.07cm] [->] (6.5,-5) to [out=-90,in=-90]
(-6.5,-5);
\end{tikzpicture}
\caption{ \label{quiveruMaster} The quiver diagram of the diagonal
embedding of the group $\mathbb{Z}_3 \to \mathrm{SU(3)}$ }
\end{center}
\end{figure}

\subsubsection{The space $\mathcal{S}_{\Gamma } \, = \, \mathrm{Hom}_\Gamma(\mathcal{Q}\otimes R, R)$
in the natural basis} Solving the invariance constraints (\ref{carnevalediPaulo}) in the natural basis of the
regular representation we find the triples of matrices $\{$A,B,C$\}$ spanning the locus \(\mathcal{S}_{\Gamma
}\). They are as follows:
\begin{eqnarray}
A&=&\left(
\begin{array}{ccc}
 \alpha _{1,1} & \alpha _{1,2} & \alpha _{1,3} \\
 (-1)^{2/3} \alpha _{1,3} & (-1)^{2/3} \alpha _{1,1} & (-1)^{2/3} \alpha _{1,2} \\
 -(-1)^{1/3} \alpha _{1,2} & -(-1)^{1/3} \alpha _{1,3} & -(-1)^{1/3} \alpha _{1,1} \\
\end{array}
\right)\nonumber\\
B &=&\left(
\begin{array}{ccc}
 \beta _{1,1} & \beta _{1,2} & \beta _{1,3} \\
 (-1)^{2/3} \beta _{1,3} & (-1)^{2/3} \beta _{1,1} & (-1)^{2/3} \beta _{1,2} \\
 -(-1)^{1/3} \beta _{1,2} & -(-1)^{1/3} \beta _{1,3} & -(-1)^{1/3} \beta _{1,1} \\
\end{array}
\right)\nonumber\\
C&=&\left(
\begin{array}{ccc}
 \gamma _{1,1} & \gamma _{1,2} & \gamma _{1,3} \\
 (-1)^{2/3} \gamma _{1,3} & (-1)^{2/3} \gamma _{1,1} & (-1)^{2/3} \gamma _{1,2} \\
 -(-1)^{1/3} \gamma _{1,2} & -(-1)^{1/3} \gamma _{1,3} & -(-1)^{1/3} \gamma _{1,1} \\
\end{array}
\right) \label{naturaliaz3}
\end{eqnarray}
\paragraph{The locus $L_\Gamma$.} The locus \(L_{\Gamma }\) $\subset $ \(\mathcal{S}_{\Gamma }\) is easily described by the equation:
\begin{eqnarray} A_0 &=&\left(
\begin{array}{ccc}
 \alpha _{1,1} & 0 & 0 \\
 0 & (-1)^{2/3} \alpha _{1,1} & 0 \\
 0 & 0 & -(-1)^{1/3} \alpha _{1,1} \\
\end{array}
\right)\nonumber\\
B_0&=&\left(
\begin{array}{ccc}
 \beta _{1,1} & 0 & 0 \\
 0 & (-1)^{2/3} \beta _{1,1} & 0 \\
 0 & 0 & -(-1)^{1/3} \beta _{1,1} \\
\end{array}
\right)\nonumber\\
C_0&=&\left(
\begin{array}{ccc}
 \gamma _{1,1} & 0 & 0 \\
 0 & (-1)^{2/3} \gamma _{1,1} & 0 \\
 0 & 0 & -(-1)^{1/3} \gamma _{1,1} \\
\end{array}
\right)
\end{eqnarray}
\subsubsection{The space $\mathcal{S}_{\Gamma }$ in the split basis}
Solving the invariance constraints in the split basis of the regular representation we find another
representation of the  triples of matrices $\{A,B,C\}$ that span  the space $\mathcal{S}_{\Gamma }$. They are
as follows:
\begin{eqnarray}\label{sorriento}
A&=&\left(
\begin{array}{ccc}
 0 & 0 & m_{1,3} \\
 m_{2,1} & 0 & 0 \\
 0 & m_{3,2} & 0 \\
\end{array}
\right)\nonumber\\
B&=&\left(
\begin{array}{ccc}
 0 & 0 & n_{1,3} \\
 n_{2,1} & 0 & 0 \\
 0 & n_{3,2} & 0 \\
\end{array}
\right)\nonumber\\
C&=&\left(
\begin{array}{ccc}
 0 & 0 & r_{1,3} \\
 r_{2,1} & 0 & 0 \\
 0 & r_{3,2} & 0 \\
\end{array}
\right)
\end{eqnarray}
\subsubsection{The equation $p \wedge p=0$ and the characterization of the variety
 $\mathbb{V}_5\,=\,\mathcal{D}_{\Gamma }$}
Here we are concerned with the solution of
eqn.\,(\ref{poffarbacchio}) and the characterization of the locus
$\mathcal{D}_{\Gamma}$.
\par
Differently from the more complicated cases of larger groups, in the present abelian case of small order, we
can explicitly solve the quadratic equations provided by the commutator constraints and we discover that
there is a principal branch of the solution, named \(\mathcal{D}_{\Gamma }^0\) that has indeed dimension
5=$|\Gamma |$+2. In addition however there are several other branches with smaller dimension. These branches
describe different components of the locus $\mathcal{D}_\Gamma$. Actually as already pointed out they are all
contained in the \(\mathcal{G}_{\Gamma }\) orbit of the subspace \(L_{\Gamma }\) defined above. The quadratic
equations defining \(\mathcal{D}_{\Gamma }\) have a set of 14 different solutions realized by a number \(n_{i
}\) of constraints on the 9 parameters. Hence there are 14 branches \(\mathcal{D}_{\Gamma }^i\)(i=0,1,...16)
of dimensions:
\begin{equation}
\dim  _{\mathbb{C}} \mathcal{D}_{\Gamma }^i = 9 -n_{i }
\end{equation}
The full dimension table of these branches is displayed below
$$\{5,4,4,4,4,3,3,3,3,3,3,3,2,2\}$$
As we see, there is a unique branch that has the maximal dimension 5 =$|$\(\left.\mathbb{Z}_3\right|+2\).
This is the principal branch \(\mathcal{D}_{\Gamma }^0\). It can be represented by the substitution:
\begin{equation}
n_{2,1}\to \frac{m_{2,1} n_{1,3}}{m_{1,3}},\quad n_{3,2}\to \frac{m_{3,2} n_{1,3}}{m_{1,3}},\quad r_{2,1}\to
\frac{m_{2,1} r_{1,3}}{m_{1,3}},\quad r_{3,2}\to \frac{m_{3,2} r_{1,3}}{m_{1,3}}
\end{equation}
In this way we have reached a complete resolution of the following
problem: how to obtain an explicit parametrization of the variety
\(V_{|\Gamma |+2}\). This variety is described by the following
three matrices depending on the 5 complex variables \(\omega _i\)
(i=1,...,5):
\begin{eqnarray}
A&=&\left(
\begin{array}{ccc}
 0 & 0 & \omega _1 \\
 \omega _2 & 0 & 0 \\
 0 & \omega _3 & 0 \\
\end{array}
\right)\nonumber\\
B&=&\left(
\begin{array}{ccc}
 0 & 0 & \omega _4 \\
 \frac{\omega _2 \omega _4}{\omega _1} & 0 & 0 \\
 0 & \frac{\omega _3 \omega _4}{\omega _1} & 0 \\
\end{array}
\right)\nonumber\\
C&=&\left(
\begin{array}{ccc}
 0 & 0 & \omega _5 \\
 \frac{\omega _2 \omega _5}{\omega _1} & 0 & 0 \\
 0 & \frac{\omega _3 \omega _5}{\omega _1} & 0 \\
\end{array}
\right)
\end{eqnarray}
\subsubsection{The quiver group}
Our next point is the derivation of the group \(\mathcal{G}_{\Gamma }\) defined in eqs. (\ref{gstorto}) and
(\ref{carciofillo}), namely:
\begin{equation}
\mathcal{G}_{\Gamma } = \left\{ g\in  \text{SL}(|\Gamma |,\mathbb{C})\left| \forall \gamma \in \Gamma  :
\left[D_R(\gamma ),D_{\text{def}}(g)\right]\right.=0\right\}
\end{equation}
Let us proceed to this construction. In the diagonal basis of the regular representation this is a very easy
task, since the group is simply given by the diagonal 3$\times $3 matrices with determinant one. We introduce
such matrices
\begin{equation}
\mathfrak{g}\,\in \,\mathcal{G}_{\Gamma}\quad:\quad\mathfrak{g}=\left(
\begin{array}{ccc}
 a_1 & 0 & 0 \\
 0 & a_2 & 0 \\
 0 & 0 & a_3 \\
\end{array}
\right)
\end{equation}
\subsubsection{ $\mathbb{V}_5$ as the orbit under $\mathcal{G}_{\Gamma }$ of the locus $L_{\Gamma }$}
In this section we want to verify and implement
eqn.\,(\ref{belgone}), namely we aim at showing that \(\mathbb{V}_{5
}\)=\(\mathcal{D}_{\Gamma }=\text{Orbit}_{\mathcal{G}_{\Gamma
}}\left(L_{\Gamma }\right)\). To this effect we rewrite the locus
\(L_{\Gamma }\) in the diagonal split basis of the regular
representation. The change of basis is performed by the character
table of the cyclic group \(\mathbb{Z}_3\). The result is displayed
below:
\begin{eqnarray}\label{corridoio}
A_0 &=&\left(
\begin{array}{ccc}
 0 & 0 & \alpha _{1,1} \\
 \alpha _{1,1} & 0 & 0 \\
 0 & \alpha _{1,1} & 0 \\
\end{array}
\right)\nonumber\\
B_0&=&\left(
\begin{array}{ccc}
 0 & 0 & \beta _{1,1} \\
 \beta _{1,1} & 0 & 0 \\
 0 & \beta _{1,1} & 0 \\
\end{array}
\right)\nonumber\\
C_0&=&\left(
\begin{array}{ccc}
 0 & 0 & \gamma _{1,1} \\
 \gamma _{1,1} & 0 & 0 \\
 0 & \gamma _{1,1} & 0 \\
\end{array}
\right)
\end{eqnarray}
Eventually the complex parameters
\begin{equation}
z_1\equiv \alpha _{1,1};  \quad z_2\equiv \beta _{1,1};  \quad  z_3\equiv \gamma _{1,1}
\end{equation}
will be utilized as complex coordinates of the resolved variety when the level parameters \(\zeta _{1,2}\)
are switched on. Starting from the above the orbit is given by:
\begin{equation}\label{colapasta}
    \mbox{Orbit}_{\mathcal{G}_\Gamma} \, \equiv \, \left\{\left\{\mathfrak{g}\, A_0 \,\mathfrak{g}^{-1},
    \mathfrak{g}\, B_0 \,\mathfrak{g}^{-1},\mathfrak{g}\, C_0 \,\mathfrak{g}^{-1} \right\}\quad \mid \quad
    \forall \,\mathfrak{g}\in\mathcal{G}_\Gamma \, , \, \forall \,\{A_0,B_0,C_0\} \in L_\Gamma\,\right\} \,
    \supset
    \, \mathcal{D}_\Gamma^0
\end{equation}
and the correspondence between the parameters of the principal branch \(\mathcal{D}_{\Gamma }^0\) and the
parameters spanning \(\mathcal{G}_{\Gamma }\) and \(L_{\Gamma }\) is provided below:
\begin{equation}
a_1\to \frac{\omega _2^{1/3}}{\omega _1^{1/3}},a_2\to \frac{\omega _3^{1/3}}{\omega _2^{1/3}},a_3\to
\frac{\omega _1^{1/3}}{\omega _3^{1/3}},z_1\to \omega _1^{1/3} \omega _2^{1/3} \omega _3^{1/3},z_2\to
\frac{\omega _2^{1/3} \omega _3^{1/3} \omega _4}{\omega _1^{2/3}},z_3\to \frac{\omega _2^{1/3} \omega
_3^{1/3} \omega _5}{\omega _1^{2/3}}
\end{equation}
Branches of smaller dimension  of the solution are all contained in the
\(\text{Orbit}_{\mathcal{G}_{\Gamma }}\left(L_{\Gamma }\right)\) and
correspond to the orbits of special points of \(L_{\Gamma }\) where
some of the \(z_i\) vanish or satisfy special relations among
themselves. Hence, indeed we have:
$$\mbox{Orbit}_{\mathcal{G}_\Gamma} \, = \,\mathcal{D}_\Gamma$$
\subsubsection{The compact gauge group $\mathcal{F}_{\Gamma } = \mathrm{U(1)}^2$}
We introduce a basis for the generators of the compact subgroup $\mathrm{U(1)}^2 =\mathcal{F}_{\Gamma}
\subset \mathcal{G}_{\Gamma }$ provided by the set of two generators displayed here below
\begin{equation}
T^1=\left(
\begin{array}{|c|c|c|}
\hline
 i & 0 & 0 \\
\hline
 0 & -i & 0 \\
\hline
 0 & 0 & 0 \\
\hline
\end{array}
\right)\quad  ; \quad T^2=\left(
\begin{array}{|c|c|c|}
\hline
 0 & 0 & 0 \\
\hline
 0 & i & 0 \\
\hline
 0 & 0 & -i \\
\hline
\end{array}
\right) \label{birraperoni}
\end{equation}
whose trace-normalization is the \(A_2\) Cartan matrix
\begin{equation}
\text{Tr}\left(T^iT^j\right) =\mathfrak{C}^{\text{ij}}\text{  }= \left(
\begin{array}{cc}
 2 & -1 \\
 -1 & 2 \\
\end{array}
\right)
\end{equation}
\subsubsection{Calculation of the K\"ahler potential and of the moment maps}
Naming \(\Delta _i\) the moduli of the coordinates \(z_i\) and \(\theta _i\) their phases according to
\(z_i\) = \(e^{\text{i$\theta $}_i}\)\(\Delta _i\) and considering a generic element \(\mathfrak{g}_R\) of
the quiver group that is real and hence is a representative of a coset class in \(\frac{\mathcal{G}_{\Gamma
}}{\mathcal{F}_{\Gamma }}\):
\begin{equation}
\mathfrak{g}_R \,= \,\left(
\begin{array}{|c|c|c|}
\hline
 e^{\lambda _1} & 0 & 0 \\
\hline
 0 & e^{-\lambda _1+\lambda _2} & 0 \\
\hline
 0 & 0 & e^{-\lambda _2} \\
\hline
\end{array}
\right)\quad ;\quad\lambda _{1,2}\in \mathbb{R}
\end{equation}
The triple of matrices $\{$A,B,C$\}$=\(\left\{\mathfrak{g}_R A_0 \mathfrak{g}_R{}^{-1}, \mathfrak{g}_R B_0
\mathfrak{g}_R{}^{-1},\mathfrak{g}_R C_0 \mathfrak{g}_R{}^{-1}\right\}\) have the following explicit
appearance:
\begin{eqnarray}
A&=&\left(
\begin{array}{ccc}
 0 & 0 & e^{i \theta _1-\lambda _1-\lambda _2} \Delta _1 \\
 e^{i \theta _1+2 \lambda _1-\lambda _2} \Delta _1 & 0 & 0 \\
 0 & e^{i \theta _1-\lambda _1+2 \lambda _2} \Delta _1 & 0 \\
\end{array}
\right)\nonumber\\
B&=&\left(
\begin{array}{ccc}
 0 & 0 & e^{i \theta _2-\lambda _1-\lambda _2} \Delta _2 \\
 e^{i \theta _2+2 \lambda _1-\lambda _2} \Delta _2 & 0 & 0 \\
 0 & e^{i \theta _2-\lambda _1+2 \lambda _2} \Delta _2 & 0 \\
\end{array}
\right)\nonumber\\
C&=&\left(
\begin{array}{ccc}
 0 & 0 & e^{i \theta _3-\lambda _1-\lambda _2} \Delta _3 \\
 e^{i \theta _3+2 \lambda _1-\lambda _2} \Delta _3 & 0 & 0 \\
 0 & e^{i \theta _3-\lambda _1+2 \lambda _2} \Delta _3 & 0 \\
\end{array}
\right)
\end{eqnarray}
Calculating the K\"ahler potential we find
\begin{equation}\label{carotenebeta}
\mathcal{K}_{\mathcal{S}}|_{\mathcal{D}}\, =
\,\left(\text{Tr}\left[A\, , \, A^\dagger\right]+\text{Tr}\left[B\,
, \, B^\dagger\right]+\text{Tr}\left[C\, , \,
C^\dagger\right]\right) =e^{-2 \left(\lambda _1+\lambda _2\right)}
\left(1+e^{6 \lambda _1}+e^{6 \lambda _2}\right) \left(\Delta
_1^2+\Delta _2^2+\Delta _3^2\right)
\end{equation}
We have used the above notation since \(\text{Tr}\left[A,A^{\dagger
}\right]+\text{Tr}\left[B,B^{\dagger
}\right]+\text{Tr}\left[C,C^{\dagger }\right]\) is the K\"ahler
potential of the  ambient space \(\mathcal{S}_{\Gamma }\) restricted
to the orbit \(\mathcal{D}_{\Gamma }\). Indeed since
\(\mathcal{F}_{\Gamma }\) is an isometry of \(\mathcal{S}_{\Gamma
}\), the dependence in \(\mathcal{K}_{\mathcal{S}}|_{\mathcal{D}}\)
is only on the real part of the quiver group, namely on the real
factors \(\lambda _{1,2}\). Just as it stands,
\(\mathcal{K}_{\mathcal{S}}|_{\mathcal{D}}\) cannot work as K\"ahler
potential of a complex K\"ahler metric. Yet, when the real factors
\(\lambda _{1,2}\) will be turned into functions of the complex
coordinates \(z_i\), then
\(\mathcal{K}_{\mathcal{S}}|_{\mathcal{D}}\) will be enabled to play
the role of a contribution to the K\"ahler potential of the resolved
manifold $\mathcal{M}_\zeta$.
\par
Next we calculate the moment maps according to the formulas:
\begin{eqnarray}
\mathfrak{P}^1 &\equiv& - i\,\text{Tr}\left[ T^1 \left(\left[A,A^{\dagger }\right]+\left[B,B^{\dagger
}\right]+\left[C,C^{\dagger }\right]\right)\right] = e^{-2 \left(\lambda _1+\lambda _2\right)} \left(1-2 e^{6
\lambda _1}+e^{6 \lambda _2}\right)
\left(\Delta _1^2+\Delta _2^2+\Delta _3^2\right)\nonumber\\
\mathfrak{P}^2 & \equiv& - i\,\text{Tr}\left[ T^2
\left(\left[A,A^{\dagger }\right]+\left[B,B^{\dagger
}\right]+\left[C,C^{\dagger }\right]\right)\right] = e^{-2
\left(\lambda _1+\lambda _2\right)} \left(1+e^{6 \lambda _1}-2 e^{6
\lambda _2}\right) \left(\Delta _1^2+\Delta _2^2+\Delta
_3^2\right)\nonumber\\
\end{eqnarray}
\subsubsection{Solution of the moment map equations}
In order to solve the moment map equations it is convenient to introduce the new variables
\begin{equation}
\Upsilon _{1,2} = \exp\left[2 \,\lambda _{1,2}\right]
\end{equation}
and to redefine the moment maps with indices lowered by means of the inverse of the Cartan matrix mentioned
above
\begin{equation}
\mathfrak{P}_I = \left(\mathfrak{C}^{-1}\right){}_{\text{IJ}}
\mathfrak{P}^J
\end{equation}
In this way imposing the level condition
\begin{equation}
\mathfrak{P}_I  = - \zeta _I
\end{equation}
where \(\zeta _{1,2}\) $>$ 0 are the two level parameters, we obtain the final pair of algebraic equations
for the factors \(\Upsilon _{1,2}\)
\begin{equation}
\left\{\frac{\Sigma  \left(-1+\Upsilon _1^3\right)}{\Upsilon _1 \Upsilon _2},\frac{\Sigma \left(-1+\Upsilon
_2^3\right)}{\Upsilon _1 \Upsilon _2}\right\} = \left\{\zeta _{1,}\zeta _2\right\}
\end{equation}
where we have introduced the shorthand notation:
\begin{equation}
\Sigma  = \sum_{i=1}^3 \,|z_i|^2
\end{equation}
The above algebraic system composed of two cubic equations is simple
enough in order to find all of its nine roots by means of Cardano's
formula. The very pleasant property of these solutions is that one
and only one of the nine branches satisfies the correct boundary
conditions, namely provides real  \(\Upsilon _I\)($\zeta $,$\Sigma
$) that are positive for all values of $\Sigma $ and $\zeta $ and
reduce to 1 when $\zeta
$$\rightarrow $0.
\par
The complete solution of the algebraic equations can be written in the following way. For the first factor we
have:
\begin{equation}\label{cannata1}
\Upsilon _1 =\frac{1}{6^{1/3}}\left(\frac{N}{\Sigma ^3\Pi  ^{\frac{1}{3}}}\right)^{\frac{1}{3}}
\end{equation}
where
\begin{eqnarray}
N &=& 2\times 2^{1/3} \zeta _1^3 \zeta _2^2+6 \Sigma ^3\Pi  ^{\frac{1}{3}}+2 \zeta _1^2 \left(3\times 2^{1/3}
\Sigma ^3+\zeta _2\Pi  ^{\frac{1}{3}}\right)+\zeta _1\left(6\times 2^{1/3} \Sigma ^3 \zeta _2+2^{2/3}\Pi
^{\frac{2}{3}}\right)\nonumber\\
\Pi & =& 27 \,\Sigma ^6+9 \,\Sigma ^3 \,\zeta _1^2 \zeta _2+9\, \Sigma ^3 \,\zeta _1 \zeta _2^2+2 \,\zeta
_1^3 \zeta
_2^3+3 \sqrt{3}\, \Sigma ^3\,\mathfrak{R} \nonumber\\
\mathfrak{R} &=&\sqrt{27 \,\Sigma ^6+6 \,\Sigma ^3 \zeta _1 \zeta _2^2-\zeta _1^4 \zeta _2^2-4 \,\Sigma ^3
\zeta _2^3+\zeta _1^3 \left(-4\, \Sigma ^3+2 \zeta _2^3\right)+\zeta _1^2 \left(6\,\Sigma ^3 \zeta _2-\zeta
_2^4\right)}\label{cardanone1}
\end{eqnarray}
For the second factor we have
\begin{equation}
\Upsilon _2= \frac{-\frac{M^{8/3}}{\Sigma ^5}+\frac{18
M^{5/3}}{\Sigma ^2}-72\, M^{2/3} \Sigma +36 \left(\frac{M}{\Sigma
^3}\right)^{2/3} \zeta _1^3-36 \left(\frac{M}{\Sigma
^3}\right)^{2/3} \zeta _1^2 \zeta _2+6 \left(\frac{M}{\Sigma
^3}\right)^{5/3} \zeta _1^2 \zeta _2}{36\times 6^{2/3}\, \Sigma ^2
\zeta _1}\label{cardanone2}
\end{equation}
where
\begin{equation}
M \,= \,\frac{6 \,\Sigma ^3 \Pi ^{1/3}+2^{2/3} \Pi ^{2/3} \zeta
_1+6\times 2^{1/3} \Sigma ^3 \,\zeta _1^2+6\times 2^{1/3} \Sigma ^3
\,\zeta _1 \zeta _2+2 \Pi ^{1/3} \zeta _1^2 \zeta _2+2\times 2^{1/3}
\zeta _1^3 \zeta _2^2}{\Omega ^{1/3}}\label{cardanone3}
\end{equation}
\subsection{Discussion of cohomology in the master model}
Since the two scale factors \(\Upsilon _{1,2}\) are functions only
of $\Sigma $, the two (1,1)-forms, relative to the two tautological
bundles, respectively associated with the first and second
nontrivial irreps of the cyclic group, defined in
eqn.\,(\ref{bambolone}) take the following general appearance:
\begin{eqnarray}
\omega _{1,2}^{(1,1)} &=&\frac{i}{2\pi
}\left(\frac{d}{\text{d$\Sigma $}} \text{Log}\left[\Upsilon
_{1,2}(\Sigma )\right] d\bar{z}^i\wedge {dz}^i
+\frac{d^2}{\text{d$\Sigma $}^2} \text{Log}\left[\Upsilon
_{1,2}(\Sigma )\right] z^j\bar{z}^i \text{d}z^i\wedge d\bar{z}^j\right)\nonumber\\
 &=& \frac{i}{2\pi }\left( f_{1,2} \Theta
\, + \, g_{1,2} \Psi\right)
\end{eqnarray}
where we have introduced the short hand notation
\begin{equation}\label{sciortando}
 \Theta \, = \,  \sum_{i=1}^3 d\bar{z}^i\wedge {dz}^i \quad ;\quad
 \Psi \, = \, \quad \sum_{i,j=1}^3 z^j\, \bar{z}^i {dz}^i\wedge d\bar{z}^j
\end{equation}
Indeed in the present case the fibre metrics $\mathfrak{H}_{1,2}$
are one-dimensional and given by $\mathfrak{H}_{1,2} \, = \,
\sqrt{\Upsilon _{1,2}}$. The most relevant point is that the two
functions $f_{1,2}$ and $g_{1,2}$ being the derivatives (first and
second) of $\Upsilon _{1,2}$ depend only on the variable $\Sigma$.
\par
It follows that a triple wedge product of the two--forms $\omega _{a}^{(1,1)}$ (a=1,2) has always the
following structure:
\begin{equation}\label{strutturone}
    \omega _{a}^{(1,1)}\wedge \omega _{b}^{(1,1)}\wedge \omega _{b}^{(1,1)} \, = \left(\frac{i}{2 \,\pi}\right)^3\,
    \left(f_a\,f_b\,f_c + 2\,
    \Sigma \, \left(g_a \, f_b \, f_c +g_b\,f_c\,f_a +g_c\,f_a\,f_b \right) \right)\times \mbox{Vol}
\end{equation}
where
\begin{equation}\label{voluminoso}
    \mbox{Vol} \, = \, dz_1\wedge dz_2\wedge dz_3\wedge d\bar{z}_1\wedge d\bar{z}_2\wedge d\bar{z}_3
\end{equation}
This structure enables us to calculate intersection integrals of the considered forms very easily. It
suffices to change variables as we explain below. The equations
\begin{equation}\label{varolegione}
    \Sigma \, = \, \sum_{i=1}^3 |z_i|^2 \, = \, \rho^2
\end{equation}
define 5-spheres of radius $\rho$. Introducing the standard Euler angle parametrization of a $5$-sphere, the
volume form (\ref{voluminoso}) reduces to:
\begin{equation}\label{esile}
 \mbox{Vol} \,= \,   8 i \varrho ^5 \cos ^4\left(\theta _1\right) \cos
   ^3\left(\theta _2\right) \cos ^2\left(\theta _3\right)
   \cos \left(\theta _4\right) \, \prod_{i=1}^5 d\theta_i
\end{equation}
The integration on the Euler angles can be easily performed and we obtain:
\begin{equation}\label{integratato}
   \prod_{i}^4 \, \int_{\ft{\pi}{2}}^{\ft{\pi}{2}} \,d\theta_i \, \int_{0}^{2\pi} \, d\theta_5 \, \left(8 i \varrho ^5
   \cos ^4\left(\theta _1\right) \cos^3\left(\theta _2\right) \cos ^2\left(\theta _3\right)
   \cos \left(\theta _4\right) \,\right) \, = \,8 i \pi ^3 \varrho ^5
\end{equation}
Hence defining the intersection integrals:
\begin{equation}\label{interseziunka}
    \mathcal{I}_{abc}\, = \, \int_{\mathcal{M}} \, \omega _{a}^{(1,1)}\wedge \omega _{b}^{(1,1)}\wedge \omega _{b}^{(1,1)}
\end{equation}
we arrive at
\begin{eqnarray}
\mathcal{I}_{\text{abc}}&=&\left(\frac{i}{2\pi }\right)^3\times 8 i \pi ^3 \times \int_0^{\infty } \left(6
\varrho ^5f_af_b f_c+2\varrho ^7 \left(f_b f_c g_a+f_a f_c g_b+f_a f_b g_c\right)\right) \, d\rho\nonumber\\
&=& \int_0^{\infty } \left(6 \varrho ^5f_af_b f_c+2\varrho ^7 \left(f_b f_c g_a+f_a f_c g_b+f_a f_b
g_c\right)\right) \, d\rho
\end{eqnarray}
We have performed the numerical integration of these functions and we have found the following results
\begin{equation}
\begin{array}{lcl}
\left(\zeta _1>0,\zeta _2=0\right)&:&\mathcal{I}_{111} =\frac{1}{8}\\
\left(\zeta _1=0,\zeta _2\geq 0\right)&:&\mathcal{I}_{111} = 0\\
\left(\zeta _1>0,\zeta _2>0\right)&:&\mathcal{I}_{111} = 1\\
\end{array}
\end{equation}
From this we reach the following conclusion. Since the corresponding integral is nonzero it follows that:
\begin{equation}
\omega _S^{(2,2)} \equiv  \omega _1^{(1,1)}\wedge \omega _1^{(1,1)}
\end{equation}
is closed but not exact and by Poincar\'e duality it is the Poincar\'e dual of some cycle S $\in $
\(H_2\)($\mathcal{M}$) such that:
\begin{equation}
\int _S \iota ^*\omega _1^{(1,1)} = \int _{\mathcal{M}}\omega _1^{(1,1)}\wedge \omega _S^{(2,2)}
\end{equation}
where
\begin{equation}
\iota  : S \longrightarrow \mathcal{M}
\end{equation}
is the inclusion map. Since \(H_c^2(\mathcal{M})\)=
\(H^2\)($\mathcal{M}$) and both have dimension 1 it follows that dim
$H_2$($\mathcal{M}$) = 1, so that every nontrivial cycle S is
proportional (as homology class) via some coefficient $\alpha $ to a
single cycle $\mathcal{C}$, namely we have S= $\alpha $
$\mathcal{C}$. Then we can interpret eqn.\,(29) as follows
\begin{equation}
\text{                                              }\int _{\alpha  \mathcal{C}} \iota ^*\omega _1^{(1,1)} =
\alpha \int _{\mathcal{M}}\omega _1^{(1,1)}\wedge \omega _{\mathcal{C}}^{(2,2)}
\end{equation}
If we choose as fundamental cycle, that one for which
\begin{equation}
\int _{\mathcal{C}} \iota ^*\omega _1^{(1,1)}= 1
\end{equation}
we conclude that
\begin{equation}
\alpha  =\left\{
\begin{array}{c|c}
 1 & \text{case} \left\{\zeta _1>0,\zeta _2>0\right\} \\
\hline
 \frac{1}{8} & \text{case} \left\{\zeta _1>0,\zeta _2=0\right\} \\
\end{array}
\right.
\end{equation}
Next we have calculated the intersection integral $\mathcal{I}_{211}$ and we have found:
\begin{equation}
\begin{array}{lcl}
\left(\zeta _1>0,\zeta _2=0\right)&:&\mathcal{I}_{211} =0\\
\left(\zeta _1=0,\zeta _2\geq 0\right)&:&\mathcal{I}_{211} = 0\\
\left(\zeta _1>0,\zeta _2>0\right)&:&\mathcal{I}_{211} = 1\\
\end{array}
\end{equation}
\paragraph{Conclusions on cohomology.}
We have two cases.
\begin{description}
\item[
\(\text{case} \left\{\zeta _1>0,\zeta _2>0\right\}.\)] The the first Chern classes of the two tautological
bundles are cohomologous:
\begin{equation}
\left[\omega _1^{(1,1)} \right] = \left[\omega _2^{(1,1)}\right]=\left[\omega ^{(1,1)}\right]
\end{equation}
\item[\(\text{case} \left\{\zeta _1>0,\zeta _2=0\right\}.\)]
The the first Chern class of the first tautological bundle is nontrivial and generates \,
$H_c^{(1,1)}(\mathcal{M})$ = $H^{1,1}(\mathcal{M})$.
\begin{equation}
\left[\omega _1^{(1,1)} \right] =\text{  }\text{nontrivial}
\end{equation}
The the first Chern class of the second tautological bundle is trivial , namely
\begin{equation}
\omega _2^{(1,1)}\text{  }=\text{  }\text{exact form}
\end{equation}
\end{description}
Obviously since there is symmetry in the exchange of the first and second scale factors, exchanging \(\zeta
_1\)$\leftrightarrow $\(\zeta _2\), the above conclusion is reversed in the case \(\left\{\zeta _1=0,\zeta
_2>0\right\}\).
\par
\begin{figure}
\centering
\includegraphics[height=9cm]{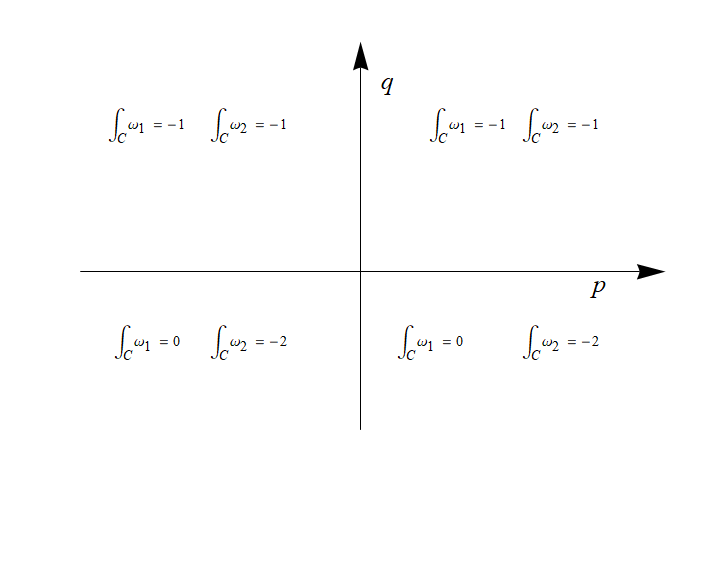}
\caption{ \label{c3periodi} The chamber structure of the resolution
$\mathcal{O}_{\mathbb{P}^2}(-3)\longrightarrow
\frac{\mathbb{C}^3}{\mathbb{Z}_3}$. In the figure we have named
$p=\zeta_1$,$q=\zeta_2$. }
\end{figure}

In passing we have also proved that the unique (2,2)-class is just the square of the unique (1,1)-class
\par
\begin{equation}
\left[\omega ^{(2,2)} \right] =\left[\omega ^{(1,1)}\right]\wedge \left[\omega ^{(1,1)}\right]
\end{equation}
The above results reveal the first instance of the chamber structure
underlying the Kronheimer resolution of $\mathbb{C}^3/\Gamma$
singularities. Indeed the above results are summarized in
fig.\ref{c3periodi} and in the following table:
\begin{equation}\label{trottola}
\begin{array}{||c||c|c||}
\hline \zeta_1\in \mathbb{R} , \,\, \zeta_2 >0 & \int_C \, \omega_1
\, = \, 1 & \int_C \,\omega_2 \, = \, 1
\\
\hline
\zeta_1\in \mathbb{R} , \,\, \zeta_2<0 & \int_C \, \omega_1 \, = \, 0 & \int_C \,\omega_2 \, = \, 2\\
\hline
\zeta_1\in \mathbb{R} , \,\, \zeta_2=0 & \int_C \, \omega_1 \, = \, \frac{1}{2} & \int_C \,\omega_2 \, = \, 2\\
\hline
\zeta_1=0 , \,\, \zeta_2\in \mathbb{R} & \int_C \, \omega_1 \, = \, 0 & \int_C \,\omega_2 \, = \, \frac{1}{2}\\
\hline
\end{array}
\end{equation}
It is evident from table (\ref{trottola}) and from the
fig.\ref{c3periodi} that there are two walls corresponding to the
two coordinate axes $\zeta_1=0$ and $\zeta_2=0$, respectively.
\subsubsection{The exceptional divisor}
Finally let us discuss how we retrieve the exceptional divisor $\mathbb{P}^2$ predicted by the weighted
blowup argument. As we anticipated in eqs. (\ref{cosmitto}-\ref{camalosto}),  replacing the three coordinates
$z_i$ with
\begin{equation}\label{paffuto}
  z_1 \, = \, W \quad ; \quad z_2 \, = \, W \, \xi_1 \quad ; \quad \quad z_3 \, = \, W \, \xi_2
\end{equation}
which is the appropriate change  for one of the three standard open charts of $\mathbb{P}^2$, we obtain
\begin{equation}\label{ripulisti}
\mathfrak{H}_1 (\Sigma)\, = \, \frac{1}{|W|^2 \, H_1(\xi,\bar{\xi},W,\bar{W})}
\end{equation}
where the function $H_1(\xi,\bar{\xi},W,\bar{W})$ has the property that:
\begin{equation}\label{canizzo}
 \lim_{W\to 0} \, \log[H_1(\xi,\bar{\xi},W,\bar{W})]\, = \, - \log[1+|\xi_1|^2 + |\xi_2|^2 ] + \log[\mbox{const}]
\end{equation}
From the above result we conclude that the exceptional divisor
$\mathfrak{D}^{(E)}$ is indeed the locus $W=0$ and that on this
locus the first Chern class of the first tautological bundle reduces
to the K\"ahler 2-form of the Fubini-Study K\"ahler metric on
$\mathbb{P}^2$. Indeed we can write:
\begin{equation}\label{fubinistudia}
  c_1\left(\mathcal{L}_1\right)|_{\mathfrak{D}^{(E)}} \, = \, -\,\frac{i}{2\pi} \,
  \bar{\partial} \, \partial \, \log[1+|\xi_1|^2 + |\xi_2|^2 ]
\end{equation}
From this point of view this master example is the perfect
three-dimensional analogue of the Eguchi-Hanson space, the
$\mathbb{P}^1$ being substituted by a $\mathbb{P}^2$.
\par
\section{Further discussion of the master model as a gauge theory}
\label{parteOP23} From the geometrical viewpoint the  previous
master example of the generalized Kronheimer construction of crepant
resolutions can be summarized by stating that the crepant resolution
is the total space of the following line-bundle:
\begin{equation}\label{rubelli}
    Y_{[3]}^{\mathbb{Z}_3} \, = \, \mathcal{O}_{\mathbb{P}^2}(-3)\, \longrightarrow
    \, \frac{\mathbb{C}^3}{\mathbb{Z}_3}
\end{equation}
Furthermore formulae (\ref{sorriento}) providing the parametrization
of the invariant space $\mathcal{S}_{\mathbb{Z}_3}$ can be written
as follows:
\begin{equation}\label{romil}
 A\, = \,    \left(
\begin{array}{ccc}
 0 & 0 & \Phi^A_{1,3} \\
 \Phi^A_{2,1} & 0 & 0 \\
 0 & \Phi^A_{3,2} & 0 \\
\end{array}
\right) \quad ; \quad B\, = \,    \left(
\begin{array}{ccc}
 0 & 0 & \Phi^B_{1,3} \\
 \Phi^B_{2,1} & 0 & 0 \\
 0 & \Phi^B_{3,2} & 0 \\
\end{array}
\right) \quad ; \quad C\, = \,    \left(
\begin{array}{ccc}
 0 & 0 & \Phi^C_{1,3} \\
 \Phi^C_{2,1} & 0 & 0 \\
 0 & \Phi^C_{3,2} & 0 \\
\end{array}
\right)
\end{equation}
Interpreting the nine complex matrix entries $\Phi^{A,B,C}_{1,3}$,
$\Phi^{A,B,C}_{2,1}$, $\Phi^{A,B,C}_{3,2}$ as scalar fields charged
with respect to the three gauge groups $\mathrm{U_{1,2,3}(N)}$
according with the quiver diagram structure. Indeed with reference
to the quiver diagram of fig.\ref{quiveruMaster}, each of the nine
lines corresponds to one of the nine scalar fields in the
bi-fundamental representation with respect to the two groups
mentioned at its foot. With $\mathrm{N}$ we indicate the number of
D3-branes.
In the case of a single brane (N=1) the quiver group
$\mathcal{G}_{\mathbb{Z}_3}$ has the following structure:
\begin{equation}\label{quiverogruppo}
\mathcal{G}_{\mathbb{Z}_3} \, = \, \mathbb{C}^\star \otimes
\mathbb{C}^\star \, \simeq\, \frac{\mathbb{C}^\star \otimes
\mathbb{C}^\star \otimes
\mathbb{C}^\star}{\mathbb{C}^\star_{central}}
\end{equation}
and its maximal compact subgroup $\mathcal{F}_{\mathbb{Z}_3}\subset
\, \mathcal{G}_{\mathbb{Z}_3}$ is the following:
\begin{equation}\label{gaugogruppo}
\mathcal{F}_{\mathbb{Z}_3} \, = \,\mathrm{ U(1)} \otimes
\mathrm{U(1)} \, \simeq \, \frac{\mathrm{ U(1)} \otimes
\mathrm{U(1)}\otimes \mathrm{U(1)}}{\mathrm{U(1)}_{central}}
\end{equation}
The gauge group $\mathcal{F}_{\mathbb{Z}_3}$ and its
complexification $\mathcal{F}_{\mathbb{Z}_3}$ are embedded into
$\mathrm{SL(3,\mathbb{C})}$ means of the  two generators mentioned
in eqn.(\ref{birraperoni}).
\subsection{The HKLR K\"ahler potential}
The K\"ahler potential of the linear space
$\mathcal{S}_{\mathbb{Z}_3}$, which in the D3-brane gauge theory
provides the kinetic terms of the nine scalar fields
$\Phi^{A,B,C}_{1,2,3}$ is given reinterpreting
eqn.(\ref{carotenebeta}) by:
\begin{equation}\label{simelito}
    \mathcal{K}_0\left(\Phi\right)\, = \, \mbox{Tr}\left(A^\dagger\,A + B^\dagger \, B + C^\dagger C \right)
\end{equation}
where the three matrices $A,B,C$ are those of equation
(\ref{romil}). According with the principles of the Kronheimer
construction, the superpotential is given  by
$\mathcal{W}\left(\Phi\right)\, = \, \text{const}\times
\text{Tr}\left( \left[A,B\right] \, C\right)$. The final HKLR
K\"ahler metric, whose determination  requires two steps of physical
significance:
\begin{enumerate}
  \item Reduction to the critical surface of the superpotential
  \textit{i.e.} $\partial_\Phi \,\mathcal{W} \, = \, 0$
  \item Reduction to the level surfaces of the gauge group moment
  maps by solving the algebraic moment map equations,
\end{enumerate}
is calculated  according with the general theory reported in section
\ref{realecompdiscus}. The final form of the HKLR K\"ahler potential
is provided by:
\begin{eqnarray}\label{HKLRformul}
    \mathcal{K}_{HKLR}(z,\bar{z}, \pmb{\zeta}) & = &\mathcal{K}_0 + \alpha \, \zeta_I \, \mathfrak{C}^{IJ}
    \, \log\left[ \Upsilon_J  \right]\nonumber\\
    & = & \alpha  \Big\{\left(2 \zeta _1-\zeta _2\right) \log \left[\Upsilon _1\right]-
    \left(\zeta _1-2 \zeta _2\right) \log \left[\Upsilon _2\right]\Big\}+\frac{\Sigma  \left(\Upsilon _1^3+\Upsilon
   _2^3+1\right)}{\Upsilon _1 \Upsilon _2}
\end{eqnarray}
where $\Upsilon _{1,2}$ were calculated in eqn.s
(\ref{cannata1},\ref{cardanone1},\ref{cardanone2},\ref{cardanone3}).
In relation with the quoted equations we recall that:
\begin{equation}\label{camillus}
    \Sigma \, \equiv \, |z _1|^2+|z _2|^2+|z _3|^2 \quad ; \quad
    \Upsilon_{1,2} \, = \, \Upsilon _{1,2} \, = \,
    \Lambda_{1,2}\left(\Sigma,\pmb{\zeta}\right)
\end{equation}
where $z_{1,2,3}$ are the three complex coordinates and
$\pmb{\zeta}\, = \, \left\{\zeta_1,\zeta_2\right\}$ the two
Fayet-Iliopoulos parameters.
\subsection{The issue of the Ricci-flat metric}
One main question is whether the metric arising from the K\"ahler
quotient, which is encoded in eqn.~(\ref{HKLRformul}) is Ricci-flat.
A Ricci-flat metric on the crepant resolution of the singularity
$\mathbb{C}^3/\mathbb{Z}_3$, namely on
$\mathcal{O}_{\mathbb{P}^2}(-3)$,  is known in explicit form from
the work of Calabi\footnote{Such metrics were also re-discovered in
the physics literature in \cite{Page:1985bq}.}
\cite{Calabi-Metriques},
 yet it is not a priori
obvious that the metric defined by the K\"ahler potential
(\ref{HKLRformul}) is that one. The true answer is that it is not,
as we show later on. Indeed we are able to construct directly the
K\"ahler potential for the resolution of
$\mathbb{C}^n/\mathbb{Z}_n$, for any $n\geq 2$, in particular
determining the unique Ricci-flat metric on
$\mathcal{O}_{\mathbb{P}^2}(-3)$ with the same isometries as the
metric (\ref{HKLRformul}) and comparing the two we see that they are
different.  Here we stress that the metric defined by
(\ref{HKLRformul}) obviously depends on  the level parameters
$\zeta_1,\zeta_2$ while the Ricci-flat one is unique up to an
overall scale factor. This is an additional reason to understand a
priori that (\ref{HKLRformul}) cannot be the Ricci flat metric.
\par
Actually Calabi in \cite{Calabi-Metriques} found an easy form of the
K\"ahler potential of a Ricci-flat metric on the canonical bundle of
a K\"ahler-Einstein manifold, and that result applies to the cases
of the canonical bundle of $\mathbb P^2$. However, in view of
applications to cases where we shall consider the canonical bundles
of   manifolds which are not K\"ahler-Einstein, in this section we
stick with our strategy of using the metric coming from the K\"ahler
quotient as a starting point.
\par
\subsubsection{The Ricci-flat metric on $Y_{[3]}
=\mathcal{O}_{\mathbb{P}^2}(-3)$} \label{noriccisiricci} As we have
noticed above the HKLR K\"ahler metric defined by the K\"ahler
potential (\ref{HKLRformul}) depends only on the variable $\Sigma$
defined in eqn.~(\ref{camillus}). It follows that the HKLR K\"ahler
metric admits $\mathrm{U(3)}$ as an isometry group, which is the
hidden invariance of $\Sigma$. The already addressed question is
whether the HKLR metric can be Ricci-flat. An almost immediate
result is that a Ricci-flat K\"ahler metric depending only on the
sum of the squared moduli of the complex coordinates is unique (up
to a scale factor) and we can give a general formula for it.
\par
We can present the result in the form of a theorem.
\begin{teorema}
Let $\mathcal{M}_n$ be a non-compact $n$-dimensional K\"ahler
manifold admitting a dense open coordinate patch $z_i$,
$i=1,\dots,n$ which we can identify with the total space of the line
bundle $\mathcal{O}_{\mathbb{P}^{n-1}}(-n)$, the bundle structure
being exposed by the coordinate transformation:
\begin{equation}\label{bundellus}
    z_i\,=\, u_i \, w^{\ft 1 n} \quad , \quad  (i=1,\dots,n-1) \quad \quad ; \quad z_n
    \, = \, w^{\ft 1 n}
\end{equation}
where $u_i$ is a set of inhomogenous coordinates for
$\mathbb{P}^{n-1}$. The K\"ahler potential $\mathcal{K}_n$ of a
$\mathrm{U(n)}$ isometric K\"ahler metric on $\mathcal{M}_n$ must
necessarily be a real function of the unique real variable $\Sigma
\, = \, \sum_{i=1}^n |z_i|^2$. If we require that the metric should
be Ricci-flat, the K\"ahler potential is uniquely defined and it is
the following one:
\begin{equation}\label{kaleropolo}
  \mathcal{K}_n(\Sigma)\, = \, k \, +  \frac{\left(\Sigma ^n+\ell ^n\right)^{-\frac{n-1}{n}}
   \left((n-1) \left(\Sigma ^n+\ell ^n\right)-\ell ^n
   \left(\Sigma ^{-n} \ell ^n+1\right)^{\frac{n-1}{n}} \,
   _2F_1\left(\frac{n-1}{n},\frac{n-1}{n};\frac{2
   n-1}{n};-\ell ^n \Sigma ^{-n}\right)\right)}{n-1}
\end{equation}
where $k$ is an irrelevant additive constant and $\ell>0$ is a
constant that can be reabsorbed by rescaling all the complex
coordinates by a factor $\ell$, namely $z_i\to \ell \tilde{z}_i$.
\end{teorema}
\begin{proofteo}
{\rm The proof of the above statement is rather elementary. It
suffices to recall that the Ricci tensor of any  K\"ahler metric
$\mathbf{g}_{ij^\star}\, = \, \partial_i\partial_{j^\star}
\mathcal{K}(z,\bar{z})$ can always be calculated as follows:
\begin{equation}\label{lombillo}
    \mathrm{Ric}_{ij^\star}[\mathbf{g}] \, = \,
    \partial_i\,\partial_{j^\star} \, \log\left[\mathrm{Det}\left[\mathbf{g}\right]\right]
\end{equation}
In order for the Ricci tensor to be zero it is necessary that
$\mathrm{Det}\left[\mathbf{g}\right]$ be the square modulus of a
holomorphic function $|F(z)|^2$, on the other hand under the
hypotheses of the theorem it is a real function of the real variable
$\Sigma$. Hence it must be a constant. It follows that we have to
impose the equation:
\begin{equation}\label{pomigliano}
    \mathrm{Det}\left[\mathbf{g}\right] \, = \, \ell^2 \, = \,
    \mbox{const}
\end{equation}
Let $\mathcal{K}(\Sigma)$ be the sought for K\"ahler potential,
calculating the K\"ahler metric and its determinant we find:
\begin{equation}\label{bagnasco}
  \mathrm{Det}\left[\mathbf{g}\right] \, = \, \Sigma ^{n-1}  \mathcal{K}(\Sigma)'
  \left(\Sigma ^2\mathcal{K}(\Sigma)''+\Sigma\,   \mathcal{K}(\Sigma)' \right)
\end{equation}
Inserting eqn.~(\ref{bagnasco}) into eqn.~(\ref{pomigliano}) we
obtain a non linear differential equation for $\mathcal{K}(\Sigma)$
of which eqn.~(\ref{kaleropolo}) is the general integral. This
proves the theorem. $\diamondsuit$}
\end{proofteo}
\subsubsection{Particular cases} It is interesting to analyze
particular cases of the general formula (\ref{kaleropolo}).
\paragraph{The case $n=2$: Eguchi-Hanson.} The case $n=2$ yielding a Ricci
flat metric on $\mathcal{O}_{\mathbb{P}^1}(-2)$ is the Eguchi-Hanson
case namely the crepant resolution of the Kleinian singularity
$\mathbb{C}^2/\mathbb{Z}_2$. This is known to be a HyperK\"ahler
manifold and all HyperK\"ahler metrics are Ricci-flat. Hence also
the HKLR metric must be Ricci-flat and identical with the one
defined by eqn.~(\ref{kaleropolo}). Actually we find:
\begin{eqnarray}\label{chicco}
\mathcal{K}_2(\Sigma)&=&\left(\Sigma ^2+\ell
^2\right)^{-\frac{1}{2}}
   \left( \left(\Sigma ^2+\ell ^2\right)-\ell ^2
   \left(\Sigma ^{-2} \ell ^2+1\right)^{\frac{1}{2}} \,
   _2F_1\left(\frac{1}{2},\frac{1}{2};\frac{3}{2};-\ell ^2 \Sigma
   ^{-2}\right)\right)\nonumber\\
   &=&\sqrt{\Sigma ^2+\ell ^2}-\ell  \log
   \left(\sqrt{\Sigma ^2+\ell ^2}+\ell
   \right)+\ell  \log (\Sigma ) \, + \, \mbox{const}
\end{eqnarray}
which follows from the identification of the hypergeometric function
with combinations of elementary transcendental functions occurring
for special values of its indices. The second transcription of the
function is precisely the K\"ahler potential of the Eguchi-Hanson
metric in its HKLR-form as it arises from the Kronheimer
construction.
\paragraph{The case $n=3$: $\mathcal{O}_{\mathbb{P}^2}(-3)$.} The
next case is that of interest for the D3-brane solution. For $n=3$,
setting $\ell=1$, which we can always do by a rescaling of the
coordinates, we find:
\begin{eqnarray}\label{carriolarotta}
 \mathcal{K}_{Rflat}(\Sigma) &=&   \frac{2 \left(\Sigma ^3+1\right)-
   \left(\frac{1}{\Sigma ^3}+1\right)^{2/3} \,
   _2F_1\left(\frac{2}{3},\frac{2}{3};\frac{5}{3};-\frac{
   1}{\Sigma ^3}\right)}{2 \left(\Sigma ^3+1\right)^{2/3}}\nonumber\\
   &=&\frac{2 (\Sigma ^3+1)-\,
   _2F_1\left(\frac{2}{3},1;\frac{5}{3};\frac{1}{\Sigma
   ^3+1}\right)}{2 \left(\Sigma
   ^3+1\right)^{2/3}}
\end{eqnarray}
The second way of writing the K\"ahler potential follows from one of
the standard Kummer relations among hypergeometric functions. There
is a third transcription that also in this case allows to write it
in terms of elementary transcendental functions. Before considering
it we use eqn.~(\ref{carriolarotta}) to study the asymptotic
behavior of the K\"ahler potential for large values of $\Sigma$. We
obtain;
\begin{equation}\label{asintInf}
   \mathcal{K}_{Rflat}(\Sigma) \, \stackrel{\Sigma \to \infty}{\approx} \,  \Sigma-\frac{1 }{6
   \Sigma ^2} +\frac{1}{45 \Sigma ^5} +\mathcal{O}\left(\frac{1}{\Sigma^7}\right)
\end{equation}
Eq.(\ref{asintInf}) shows that the Ricci-flat metric is
asymptotically flat since the K\"ahler potential approaches that of
$\mathbb{C}^3$.
\par
\begin{figure}
\centering
\includegraphics[height=8cm]{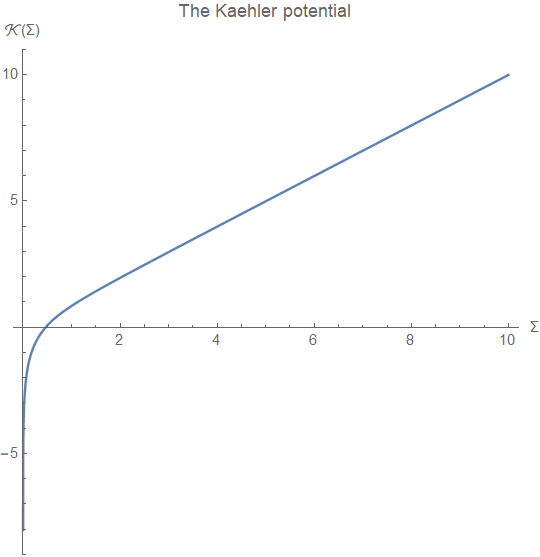}
\caption{ \label{kalleropotto} The plot of the K\"ahler potential
$\mathcal{K}_{Rflat}(\Sigma)$ for the Ricci-flat metric on
$\mathcal{O}_{\mathbb{P}^2}(-3)$. The asymptotic flatness of the
metric is evident from the plot. For large values of $\Sigma$ it
becomes a straight line with angular coefficient $1$. }
\end{figure}

As anticipated, there is an alternative way of writing the K\"ahler
potential (\ref{carriolarotta}) which is the following:
\begin{eqnarray}\label{pagliato}
  \mathcal{K}_{Rflat}(\Sigma)  &=&\frac{\pi }{2
   \sqrt{3}}+ \frac{1}{6} \left(6 \sqrt[3]{\Sigma ^3+1}+2 \log
   \left[\sqrt[3]{\Sigma ^3+1}-1\right]\right.\nonumber\\
   &&\left.-\log
   \left[\left(\Sigma ^3+1\right)^{2/3}+\sqrt[3]{\Sigma
   ^3+1}+1\right]-2 \sqrt{3} \arctan\left[\frac{2
   \sqrt[3]{\Sigma
   ^3+1}+1}{\sqrt{3}}\right]\right)
\end{eqnarray}
The identity of eqn.~(\ref{pagliato}) with
eqn.~(\ref{carriolarotta}) can be worked out with analytic
manipulations that we omit.  The representation (\ref{pagliato}) is
particularly useful to explore the behavior of the K\"ahler
potential at small values of $\Sigma$. We immediately find that:
\begin{equation}\label{sigmuzero}
    \mathcal{K}_{Rflat}(\Sigma) \, \stackrel{\Sigma \to 0}{\approx} \,
    \log[\Sigma]+\frac{\pi }{2
   \sqrt{3}}+\mathcal{O}\left(\Sigma^6\right)
\end{equation}
The behavior of $\mathcal{K}_{Rflat}(\Sigma)$ is displayed in
fig.\ref{kalleropotto}.
\subsection{The harmonic function in the case $Y_{[3]}^{\mathbb{Z}_3}
=\mathcal{O}_{\mathbb{P}^2}(-3)$} Let us now consider the equation
for a harmonic function $H(z,\bar{z})$ on the background of the
Ricci-flat metric of $Y_{[3]}^{\mathbb{Z}_3}$ that we have derived
in the previous sections. Once again we suppose that $H=H(\Sigma)$
is a function only of the real variable $\Sigma$, {\it viz.}
$R=\sqrt{\Sigma}$. For the Ricci-flat metric the Laplacian equation
takes the simplified form: $\partial_i \left( g^{ij^\star} \,
\partial_{j^\star}
    H(\Sigma)\right) \, = \, 0$,
since the determinant of the metric is constant. Using the K\"ahler
metric that follows from the K\"ahler potential $K_{Rflat}(\Sigma)$
defined by eqn.s~(\ref{carriolarotta}),(\ref{pagliato}), we obtain a
differential equation that upon the change of variable
$\Sigma=\sqrt[3]{r}$  takes the following form:
\begin{equation}\label{balzoblu}
     3 r (r+1) C''(r)+(5 r+3) C'(r)\, = \, 0
\end{equation}
The general integral  eqn.~(\ref{balzoblu}) is displayed below:
\begin{equation}\label{integrai}
    C(r)\, = \, \kappa +\lambda  \left(\log
   \left(1-\sqrt[3]{r+1}\right)-\frac{1}{2} \log
   \left((r+1)^{2/3}+\sqrt[3]{r+1}+1\right)-\sqrt{3} \tan
   ^{-1}\left(\frac{2
   \sqrt[3]{r+1}+1}{\sqrt{3}}\right)\right)
\end{equation}
$\kappa,\lambda$ being the two integration constants. We fix these
latter with boundary conditions. We argue in the following way: if
the transverse space to the brane  were the original
$\mathbb{C}^3/\mathbb{Z}_3$ instead of the resolved variety
$\mathcal{O}_{\mathbb{P}^2}(-3)$, then the harmonic function
describing the D3-brane solution would be the following:
\begin{equation}\label{ciumiglio}
    H_{orb}(R) \, = \, 1 + \frac{1}{R^4} \quad ; \quad R \, \equiv \,
    \sqrt{\Sigma} \, = \, \sqrt{\sum_{i}^2 |z_i|^2} \, = \, \sqrt[6]{r}
\end{equation}
The asymptotic identification for $R\to \infty$ of the Minkowski
metric in ten dimension would be guaranteed, while at small values
of $R$ we would find (via dimensional transmutation) the standard
$\mathrm{AdS_5}$-metric times that of $\mathbb{S}^5$ (see the
following eqn.s~(\ref{ciulandario}) and (\ref{adsroba})). In view of
this,  naming $R$ the square root of the variable $\Sigma$, we fix
the coefficients $\kappa,\lambda$ in the harmonic function
$H_{res}(R)$ in such a way that for large values of $R$ it
approaches the harmonic function pertaining to the orbifold case
(\ref{ciumiglio}). The asymptotic expansion of the function: $
H_{res}(R) \, \equiv \, C(r^6) $ is the following one:
\begin{equation}\label{orcoinfi}
   H_{res}(R)\, \stackrel{R\to\infty}{\approx} \,   \left(\lambda -\frac{\pi  \kappa }{2 \sqrt{3}}\right)-\frac{1}{2} \kappa
   \left(\frac{1}{R}\right)^4+O\left(\left(\frac{1}{R}\right)^5\right)
\end{equation}
Hence the function $H_{res}(R)$ approximates the function
$H_{orb}(R)$ if we set  $ \kappa \, = \, 2  \, , \, \lambda =
\frac{\pi}{\sqrt{3}} $. In this way we conclude that:
\begin{eqnarray}\label{resHar}
    H_{res}(R) & = & \frac{1}{3} \left(2 \log \left(\sqrt[3]{R^6+1}-1\right)-\log
   \left(\left(R^6+1\right)^{2/3}+\sqrt[3]{R^6+1}+1\right)\right.\nonumber\\
   &&\left.-2 \sqrt{3} \tan
   ^{-1}\left(\frac{2 \sqrt[3]{R^6+1}+1}{\sqrt{3}}\right)\right)+\frac{\pi}{\sqrt{3}}
\end{eqnarray}
\par
\begin{figure}
\centering
\includegraphics[height=6cm]{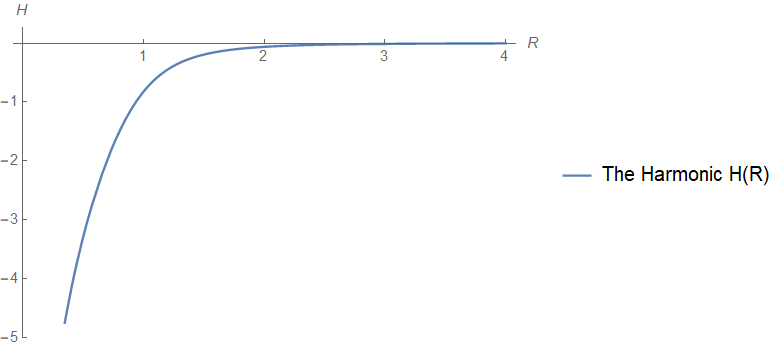}
\caption{ \label{harmoplotto} The plot of the harmonic function
$H_{res}(R)$ for the Ricci-flat metric on
$\mathcal{O}_{\mathbb{P}^2}(-3)$. }
\end{figure}

The overall behavior of the function $H_{res}(R)$ is displayed in
fig.\ref{harmoplotto}.
\subsection{The asymptotic limits of the Ricci-flat metric for the D3-brane solution on $\mathcal{O}_{\mathbb{P}^2}(-3)$
} In the case of a standard D3-brane on $Y_{[3]}\, = \,
\mathbb{C}^3\simeq \mathbb{R}^6$ one writes the same ansatz as in
eqn.~(\ref{ansazzo}) and (\ref{genova1}-\ref{f5ansaz}) where now the
K\"ahler metric is $ \mathbf{g}_{\alpha\beta^\star} \, = \,
\delta_{\alpha\beta^\star} $ Rewriting the complex coordinates in
terms of polar coordinates $ z_1= e^{{\rm i} \varphi _1} R \cos \phi
,\, z_2= e^{{\rm i} \varphi_2} R \cos \chi  \sin \phi ,\, z_3=
e^{{\rm i} \varphi_3} R \sin \chi    \sin \phi$ we obtain that:
\begin{eqnarray}\label{fiasaporita}
   \text{ds}^2_{\mathbb{C}^3} &\equiv& \sum_{i=1}^3 |dz_i|^2\,= \, dR^2+R^2 \, \text{ds}^2_{\mathbb{S}^5}
\end{eqnarray}
where:
\begin{equation}\label{cinquesfera}
  \text{ds}^2_{\mathbb{S}^5} \, = \,    d{\varphi_1}^2 \cos ^2 \phi +\sin ^2\phi
   \left(d{\varphi_2}^2 \cos ^2\chi +d{\varphi_3}^2 \sin ^2\chi +d\chi^2\right)+d\phi^2
\end{equation}
is the $\mathrm{SO(6)}$-invariant metric of a $5$-sphere in polar
coordinates. In other words the Ricci-flat  K\"ahler metric $
\text{ds}^2_{\mathbb{C}^3}$ (which is also Riemann-flat) is that of
the \textit{metric cone} on the Sasaki-Einstein metric of
$\mathbb{S}^5$.  At the same time the $\mathrm{SO(6)}$-invariant
harmonic function on $\mathbb{C}^3$ is given by the already quoted
$H_{orb}(R)$ in (\ref{ciumiglio}), and the complete $10$-dimensional
metric of the D3-brane solution takes the form:
\begin{equation}\label{ciulandario}
    \text{ds}^2_{10|orb}\,
    =\,\frac{1}{\sqrt{1+\frac{1}{R^4}}}\,  \text{ds}^2_{\mathrm{Mink}_{1,3}}
    \, + \,\sqrt{1+\frac{1}{R^4}}\, \left(dR^2+R^2 \,
      \text{ds}^2_{\mathbb{S}^5}\right)
\end{equation}
For $R\to\infty$ the metric (\ref{ciulandario}) approaches the flat
Minkowski metric in $d=10$, while for $R\to 0$ it approaches the
following metric:
\begin{equation}\label{adsroba}
      \text{ds}^2_{10|orb}\, \stackrel{R\to 0}{\approx} \, \underbrace{R^2
    \,\,  \text{ds}^2_{\mathrm{Mink}_{1,3}} \, + \, \frac{dR^2}{R^2}}_{\mathrm{AdS}_5} \, +
    \,\underbrace{  \text{ds}^2_{\mathbb{S}^5}}_{\mathbb{S}^5}
\end{equation}
Let us now consider the asymptotic behavior of the Ricci-flat metric
on $\mathcal{O}_{\mathbb{P}^2}(-3)$. In order to obtain a precise
comparison with the flat orbifold case the main technical point is
provided by the transcription of the $\mathbb{S}^5$-metric in terms
of coordinates well adapted to the Hopf fibration:
\begin{equation}\label{opfo}
    \mathbb{S}^5 \, \stackrel{\pi}{\longrightarrow} \, \mathbb{P}^2
    \quad ; \quad \forall\;  p \in \mathbb{P}^2 \quad \pi^{-1}(p) \sim
    \mathbb{S}^1
\end{equation}
To this effect let ${\mathbf{Y}}=\left\{u,v\right\}$ be a pair of
complex coordinates for $\mathbb{P}^2$ such that the standard
Fubini-Study metric on this compact 2-fold is given by:
\begin{equation}\label{cinesino}
     \text{ds}^2_{\mathbb{P}^2}\, = \, g^{\mathbb{P}^2}_{ij^\star}\, d{\mathbf{Y}}^i \,  d\bar{\mathbf{Y}}^{j^\star}\, \equiv  \,
   \frac{ {d\mathbf{Y}}\cdot d\bar{\mathbf{Y}}}{1+\mathbf{Y}\cdot \bar{\mathbf{Y}}}-\frac{\left(\bar{\mathbf{Y}}\cdot
   d {\mathbf{Y}}\right)\left(\mathbf{Y}\cdot d\bar{\mathbf{Y}}\right)}
    {\left(1+\mathbf{Y}\cdot \bar{\mathbf{Y}}\right)^2}
\end{equation}
the corresponding K\"ahler 2-form being $\mathbb{K}_{\mathbb{P}^2}\,
= \, \frac{{\rm i}}{2\pi} \,g^{\mathbb{P}^2}_{ij^\star}\,
d\mathbf{Y}^i \,\wedge \, d\bar{\mathbf{Y}}^{j^\star}$. Introducing
the one form: $ \Omega \, = \,\frac{{\rm i} \left({\mathbf{Y}}\cdot
d\bar{\mathbf{Y}}-\bar{\mathbf{Y}} \cdot d\mathbf{Y}\right)}{2
\left(1+\mathbf{Y}\cdot \bar{\mathbf{Y}}\right)}$ whose exterior
derivative is the K\"ahler 2-form, $d\Omega \, = \, 2\pi
\,\mathbb{K}_{\mathbb{P}^2} $, the metric of the five-sphere in
terms of these variables is the following one:
\begin{equation}\label{ds5S}
     \text{ds}^2_{\mathbb{S}^5} \, = \,  \text{ds}^2_{\mathbb{P}^2} \, + \,
    \left(d\varphi+ \Omega\right)^2
\end{equation}
where the range of the coordinate $\varphi$ spanning the
$\mathbb{S}^1$ fibre is $\varphi \in \left[0,2\pi\right]$. In this
way the flat metric on the metric cone on $\mathbb{S}^5$, namely
(\ref{fiasaporita}) can be rewritten as follows:
\begin{equation}\label{camillino}
     \text{ds}^2_{\mathbb{C}^3} \, = \, dR^2 + R^2 \,  \text{ds}^2_{\mathbb{P}^2} \,
    + R^2  \left(d\varphi + \Omega\right)^2
\end{equation}
\subsubsection{Comparison of the Ricci-flat metric with the orbifold
metric}\label{Rgrande} In order to compare the exact Ricci-flat
metric streaming from the K\"ahler potential (\ref{carriolarotta})
with the metric (\ref{camillus}) it suffices to turn to toric
coordinates
\begin{equation}\label{toricume}
z_1 = u \sqrt[3]{w}\, , \quad  z_2\, = \,  v \sqrt[3]{w}\, , \quad
z_3 \, = \, \sqrt[3]{w} \quad ; \quad \Sigma \, = \,
\left(1+\varpi\right) \, {\mathfrak{f}}^{1/3} \quad ; \quad
\varpi=|u|^2+|v|^2 \quad ; \quad {\mathfrak{f}} \, = \, |w|^2
\end{equation}
The toric coordinates $\{u,v\}$ span the exceptional divisor
$\mathbb{P}^2$ while $w$ is the fibre coordinate in the bundle.
Setting:
\begin{equation}\label{salmoiraghi}
    w \, = \, e^{{\rm i}\psi}  |w| \, = \, e^{{\rm i}\psi}  \, \left(\frac{R^2}{1 + |u|^2 + |v|^2}\right)^{\ft
    32}
\end{equation}
we obtain:
\begin{eqnarray}\label{caniggia}
     \text{ds}^2_{Rflat} & = & h(R) dR^2 \, + \, f(R) \,  \text{ds}^2_{\mathbb{P}^2}
    \, + \, g(R) \, \left(d\psi + 3 \, \Omega \right)^2 \nonumber\\
      f(R) & = &\sqrt[3]{R^6+1} \quad ; \quad  h(R) \, = \, \frac{R^4}{\left(R^6+1\right)^{2/3}} \quad ;
    \quad g(R) \, = \, \frac{R^6}{9 \left(R^6+1\right)^{2/3}}
    \quad
\end{eqnarray}
From eqn.~(\ref{caniggia}) we derive the asymptotic form of the
metric for large values of $R$, namely:
\begin{equation}\label{sibillata}
   \text{ds}^2_{Rflat}\, \stackrel{R \to \infty}{\approx}\,
   {dR}^2+R^2 \text{ds}_{\mathbb{P}_2}^2+R^2 \left(\frac{\text{d$\psi $}}{3}+\Omega \right)^2
\end{equation}
The only difference between eqn.~(\ref{camillino}) and
eqn.~(\ref{sibillata}) is the range of the angular value $\varphi\,
= \,\frac{\psi}{3}$. Because of the original definition of the angle
$\psi$, the new angle $\varphi\in\left[0,\frac{2\pi}{3}\right]$
takes one third of the values. This means that the asymptotic metric
cone is quotiened by $\mathbb{Z}_3$ as it is natural since we
resolved the singularity $\mathbb{C}^3/\mathbb{Z}_3$.
\subsubsection{Reduction to the exceptional
divisor}\label{ecceziunale} The other important limit of the
Ricci-flat metric is its reduction to the exceptional divisor
$\mathcal{ED}$. In the present case the only fixed point for the
action of $\Gamma=\mathbb{Z}_3$ on $\mathbb{C}^3$ is provided by the
origin $z_{1,2,3}=0$ which, comparing with eqn.~(\ref{toricume}),
means $w=0 \Rightarrow \mathfrak{f}=0$. This is the equation of the
exceptional divisor which is created by the blowup of the unique
singular point.  In the basis of the complex toric coordinates
$\{u,v,w\}$, the K\"ahler metric derived from the K\"ahler potential
(\ref{carriolarotta}) has the following appearance:
\begin{equation}\label{conigliofritto}
 g_{ij^\star}^{Rflat} \, = \,    \left(
\begin{array}{ccc}
 \frac{v \bar{v}+\mathfrak{f} (\varpi +1)^4+1}{(\varpi +1)^2 \left(\mathfrak{f} (\varpi
   +1)^3+1\right)^{2/3}} & -\frac{v \bar{u}}{(\varpi +1)^2 \left(\mathfrak{f} (\varpi
   +1)^3+1\right)^{2/3}} & \frac{w (\varpi +1)^2 \bar{u}}{3 \left(\mathfrak{f} (\varpi
   +1)^3+1\right)^{2/3}} \\
 -\frac{u \bar{v}}{(\varpi +1)^2 \left(\mathfrak{f} (\varpi +1)^3+1\right)^{2/3}} &
   \frac{u \bar{u}+\mathfrak{f} (\varpi +1)^4+1}{(\varpi +1)^2 \left(\mathfrak{f}
   (\varpi +1)^3+1\right)^{2/3}} & \frac{w (\varpi +1)^2 \bar{v}}{3 \left(\mathfrak{f}
   (\varpi +1)^3+1\right)^{2/3}} \\
 \frac{u (\varpi +1)^2 \bar{w}}{3 \left(\mathfrak{f} (\varpi +1)^3+1\right)^{2/3}} &
   \frac{v (\varpi +1)^2 \bar{w}}{3 \left(\mathfrak{f} (\varpi +1)^3+1\right)^{2/3}} &
   \frac{(\varpi +1)^3}{9 \left(\mathfrak{f} (\varpi +1)^3+1\right)^{2/3}} \\
\end{array}
\right)
\end{equation}
where the invariants $\mathfrak{f},\varpi$ are defined in equation
(\ref{toricume}). Hence the reduction of the metric to the
exceptional divisor is obtained by setting
$\mathrm{d}w=\mathrm{d}{\bar w}=0$ in the line element
$\text{ds}^2_{Rflat}\, \equiv \, g_{ij^\star}^{Rflat}  dy^i \,
\,d\bar{y}^{j^\star} $ and performing the limit $\mathfrak{f}\to 0$
on the result. We obtain:
\begin{equation}\label{tersilla}
     \text{ds}^2_{Rflat}\, \stackrel{\mathcal{ED}}{\longrightarrow} \,
    \text{ds}_{\mathbb{P}^2}^2 \, \equiv \,
    \frac{\mathrm{d}v\left( \mathrm{d}{\bar v}+u\,{\bar u} \mathrm{d}{\bar v}-u{\bar v}\,\mathrm{d}{\bar u}\right)+\mathrm{d}u
    \left( \mathrm{d}{\bar u}+ v\,\bar{v}\mathrm{d}{\bar u}-{\bar u}\,v\,\mathrm{d}{\bar v}\right)}{\left(1+u \bar{u}+v \bar{v}\right)^2}
\end{equation}
which is the standard Fubini-Study metric on $\mathbb{P}^2$ obtained
from the K\"ahler potential:
\begin{equation}\label{fubinistudy}
    \mathcal{K}_{\mathbb{P}^2}^{FS}(\varpi)\, = \,  \log\left(1+\varpi\right)
\end{equation}
As we see, the metric on the exceptional divisor obtained from the
Ricci-flat metric has no memory of the Fayet Iliopoulos (or
stability parameters) $p,q$ which characterize instead the HKLR
metric obtained from the Kronheimer construction. This is obvious
since the Ricci-flat metric does not depend on $p,q$. On the other
hand the HKLR metric, that follows from the K\"ahler potential
(\ref{HKLRformul}), strongly depends on the Fayet Iliopoulos
parameters
 \(\zeta_1=p\, , \, \zeta_2=q\) and one naturally expects that the reduction of $\text{ds}^2_{HKLR}$
 to the exceptional divisor will inherit such a dependence. Actually
 this is not the case since the entire dependence from $p,q$ of the HKLR K\"ahler
 potential, once reduced to $\mathcal{ED}$, is localized in an overall
 multiplicative constant and in an irrelevant additive constant. This
 matter of fact is conceptually very important in view of the following conjecture:
\begin{congettura}\label{congetto}
The Ricci-flat metric on $Y^\Gamma_{[3]}$ is completely determined
in terms of a Monge-Amp\`{e}re equation  from the K\"ahler metric on
the exceptional divisor $\mathcal{ED}$, as it is determined by the
 Kronheimer construction.
\end{congettura}
In the present case where, up to a multiplicative constant,
\textit{i.e. a homothety} there is only one Ricci-flat metric on
$\mathcal{O}_{\mathbb{P}^2}(-3)$ with the prescribed isometries, the
above conjecture might be true only if the reduction of the HKLR
metric to the exceptional divisor were unique and
$\zeta_1,\zeta_2$-independent, apart from  overall rescalings. It is
very much reassuring that this is precisely what actually happens.
\subsection{Comment on the near-brane geometry}
\label{commentarius} Let us analyze conceptually the lesson that we
can learn from the present example about the relation between the
Ricci flat metric on the canonical bundle over the exceptional
divisor and the Ricci flat metric of the metric cone over a
$5$-dimensional sasakian manifold. To this effect we need to make a
recollection of formulae and ideas. The exceptional divisor in the
resolution of the considered singularity $\mathbb{C}^3/\mathbb{Z}_3$
is the well known manifold $\mathbb{P}^2$ which happens to be a
coset manifold, but more importantly it is a K\"ahler Einstein
manifold. Hence we can write the following abstract identification:
\begin{equation}\label{idea1}
    \mathbb{P}^2 \, = \, \mathcal{M}^{KE}_B
\end{equation}
where $\mathcal{M}^{KE}_B$ is the  base manifold of a line bundle:
\begin{equation}\label{lineabundello}
    \mathcal{O}_{\mathbb{P}^2}(-3) \, = \, Y^\Gamma_{[3]} \,
    \stackrel{\pi}{\longrightarrow} \,\mathcal{M}^{KE}_B \quad ;
    \quad \forall p \in \mathcal{M}^{KE}_B \quad \pi^{-1}(p) \, \sim
    \, \mathbb{C}
\end{equation}
On the other hand recalling eqn.\eqref{ds5S} we know that there is
another bundle on $\mathcal{M}^{KE}_B=\mathbb{P}^2$. It is a
circle-bundle and its total space is the $5$-sphere, \textit{i.e.} a
Sasaki-Einstein manifold.
\begin{equation}\label{cagnesco}
    \mathbb{S}^5 \, = \, \Upsilon^{SE} \, \stackrel{\tilde{\pi}}{\longrightarrow} \,\mathcal{M}^{KE}_B
    \quad ;
    \quad \forall p \in \mathcal{M}^{KE}_B \quad \tilde{\pi}^{-1}(p) \, \sim
    \, \mathbb{S}^1
\end{equation}
The metric on $\Upsilon^{SE}$ is indeed provided by equation
eqn.\eqref{ds5S} and we can capture the general meaning of the
latter by writing:
\begin{equation}\label{cronicistico}
    \text{ds}^2_{\Upsilon^{SE}} \, = \, \text{ds}^2_{\mathcal{M}^{KE}_B
    } \, + \, \left(d\psi + q\, \Omega_{\mathcal{M}^{KE}_B}\right)^2
\end{equation}
where $q$ is a charge and $\Omega_{\mathcal{M}^{KE}_B}$ is a
K\"ahler $\mathrm{U(1)}$ connection on the K\"ahler Einstein base
manifold, namely one has
\begin{equation}\label{ciargione}
   \mathrm{d}\Omega_{\mathcal{M}^{KE}_B}\, = \, 2\pi \,
   \mathbb{K}_{\mathcal{M}^{KE}_B} \, = \,  \frac{2\pi}{\kappa} \mathbb{R}\mathrm{ic}_{\mathcal{M}^{KE}_B}
\end{equation}
where $\mathbb{K}_{\mathcal{M}^{KE}_B}$ is the K\"ahler $2$-form and
$\mathbb{R}\mathrm{ic}_{\mathcal{M}^{KE}_B}$ is the Ricci $2$-form.
They are proportional to each other because of the K\"ahler Einstein
nature of the metric. If we construct the metric cone on the
Sasakian Einstein manifold $\Upsilon^{SE}$ its metric is Ricci-flat
and we obtain the corresponding D3-brane metric mentioned in
eqn.\eqref{ciulandario} that we can generalize as it follows
\begin{equation}\label{finottera}
    \text{ds}^2_{10}\,
    =\,\frac{1}{\sqrt{1+\frac{1}{R^4}}}\,  \text{ds}^2_{\mathrm{Mink}_{1,3}}
    \, + \,\sqrt{1+\frac{1}{R^4}}\, \left(dR^2+R^2 \,
      \left[\text{ds}^2_{\mathcal{M}^{KE}_B
    } \, + \, \left(d\psi + q\, \Omega_{\mathcal{M}^{KE}_B}\right)^2\right]\right)
\end{equation}
In the near brane limit $R \to 0$ the above metric displays the
mechanism of the dimensional transmutation illustrated in
\eqref{adsroba} that we can generalize by rewriting
\begin{equation}\label{adsroba}
      \text{ds}^2_{[10]}\, \stackrel{R\to 0}{\approx} \, \underbrace{R^2
    \,\,  \text{ds}^2_{\mathrm{Mink}_{1,3}} \, + \, \frac{dR^2}{R^2}}_{\mathrm{AdS}_5} \, +
    \, \text{ds}^2_{\Upsilon^{SE}}
\end{equation}
For the exact D3-brane solution based on the Ricci flat metric on
$Y^{\Gamma}_{[3]}$ eqn.\eqref{finottera} is instead replaced by
\begin{equation}\label{rattoscopa}
    \text{ds}^2_{10}\,
    =\,H_{res}^{-1/2}(R)\,  \text{ds}^2_{\mathrm{Mink}_{1,3}}
    \, + \, H_{res}^{1/2}(R) \left[ h(R) \, dR^2\, +\, f(R) \,
      \text{ds}^2_{\mathcal{M}^{KE}_B} \, + \, g(R) \, \left(d\psi + q\,
\Omega_{\mathcal{M}^{KE}_B}\right)^2\right]
\end{equation}
The functions $H_{res}(R)$, $h(R),f(R),g(R)$ being defined in
eqn.\eqref{resHar} and \eqref{caniggia}. It is then evident that in
the near brane limit $R\to 0$ there is no dimensional transmutation
mechanism and the metric of the Sasakian Einstein manifold
$\Upsilon^{SE} \, = \, \mathbb{S}^5$ does not emerge in any way.
\par
This completely calculable example shows the above mentioned general
features that we will retrieve also in the more complicated cases
studied in chapter \ref{riccione}.
\newpage
\chapter{A second important example with a much richer structure:
$\frac{\mathbb{C}^3}{\mathbb{Z}_4}$} \label{balengusz4}
In this chapter I review the results obtained for the application of
the Kronheimer construction to the case
$\frac{\mathbb{C}^3}{\mathbb{Z}_4}$. For this case, in
\cite{noietmarcovaldo} the resolution was obtained also by utilizing
the independent tools of toric geometry; this allowed a very useful
comparison and a extensive study of the chamber structure. In this
case the compact component of the exceptional divisor is the second
Hirzebruch surface that is known to admit no K\"ahler Einstein
metric. As I will report in chapter \ref{riccione} the absence of
K\"ahler Einstein metrics on the compact exceptional divisor makes
it much more complicated to find the Ricci flat metric on the
resolved three-fold $Y^{\mathbb{Z}_4}_{[3]}$.
\section{The $\mathbb{C}^3/\mathbb{Z}_4$ model, its McKay quiver and
the associated Kronheimer construction} \label{maccaius} The action
of the group $\mathbb{Z}_4$ on $\mathbb{C}^3$ is defined by
introducing the three-dimensional unitary representation
$\mathcal{Q}(A)$ of its abstract generator $A$ that satisfies the
defining relation $A^4 \, = \, \mathbf{e}$. We set:
\begin{equation}\label{generatoreAZ4}
   \mathcal{Q}(A) \, = \, \left(
\begin{array}{ccc}
 i & 0 & 0 \\
 0 & i & 0 \\
 0 & 0 & -1 \\
\end{array}
\right) \quad ; \quad \mathcal{Q}(A)^4 \, = \, \left(
\begin{array}{ccc}
 1 & 0 & 0 \\
 0 & 1 & 0 \\
 0 & 0 & 1 \\
\end{array}
\right)
\end{equation}
Since $\mathbb{Z}_4$ is abelian and cyclic, each of its four
elements corresponds to an entire conjugacy class of which we can
easily calculate the age-vector and the ages according with the
conventions established in section \ref{vecchiardo}. We obtain:
\begin{equation}\label{panefresco}
    \begin{array}{|c||c|c|c|c|}
    \hline
    \text{Conj.
    Class}&\text{Matrix}&\text{age-vector}&\text{age}&\text{name}\\
    \hline
       \mathrm{Id} & \left(
\begin{array}{ccc}
 1 & 0 & 0 \\
 0 & 1 & 0 \\
 0 & 0 & 1 \\
\end{array}
\right) & \ft 14 \, (0,0,0) & 0 & \text{null class} \\
\hline
 \mathcal{Q}(A) & \left(
\begin{array}{ccc}
 i & 0 & 0 \\
 0 & i & 0 \\
 0 & 0 & -1 \\
\end{array}
\right) & \ft 14 \, (1,1,2) & 1 & \text{junior class} \\
\hline \mathcal{Q}(A)^2 & \left(
\begin{array}{ccc}
 -1 & 0 & 0 \\
 0 & -1 & 0 \\
 0 & 0 & 1 \\
\end{array}
\right) & \ft 14 \, (2,2,0) & 1 & \text{junior class} \\
\hline
 \mathcal{Q}(A)^3  & \left(
\begin{array}{ccc}
 -i & 0 & 0 \\
 0 & -i & 0 \\
 0 & 0 & -1 \\
\end{array}
\right) & \ft 14 \, (3,3,2) & 2 & \text{senior class}\\
       \hline
     \end{array}
\end{equation}
Therefore, according with the theorem of Ito and Reid
\cite{itoriddo}, as reviewed in section \ref{anglonipponico}, in the
crepant resolution:
\begin{equation}\label{pirollo}
   Y^{\mathbb{Z}_4}_{[3]} \, \equiv \, \mathcal{M}_\zeta \, \longrightarrow \, \frac{\mathbb{C}^3}{\mathbb{Z}_4}
\end{equation}
the Hodge numbers of the smooth resolved variety are as follows:
\begin{equation}\label{ganimellus}
    h^{(0,0)}\left(\mathcal{M}_\zeta\right) \, = \,1 \quad ; \quad h^{(1,1)}\left(\mathcal{M}_\zeta\right) \, =
    \,2 \quad ; \quad h^{(2,2)}\left(\mathcal{M}_\zeta\right) \, =
    \,1
\end{equation}
Furthermore the existence of a senior class implies that one of the
two generators of $H^{(1,1)}\left(\mathcal{M}_\zeta\right)$
 {can be chosen to have} compact support while the other  {will have} non-compact support. As we
later discuss studying the resolution with the help of toric
geometry, this distinction goes hand in hand with the structure of
the exceptional divisor that has two components, one compact and one
non-compact.

The character table of the $\mathbb{Z}_4$ group is easily calculated
and it foresees four one-dimensional representations that we
respectively name  {$\mathrm{D}_I$}, $(I\, = \, 0,1,2,3)$. The table
is given below.
\begin{equation}\label{caratteruccio}
\begin{array}{c||cccc}
\begin{array}{ccc}
  \text{Irrep}& \setminus & \text{C.C.} \\
\end{array} & \mathbf{e} &A & A^2 & A^3\\
\hline \hline
 {\mathrm{D}_0} &1 & 1 & 1 & 1 \\
 {\mathrm{D}_1}& 1 & i & -1 & -i \\
 {\mathrm{D}_2} & 1 & -1 & 1 & -1 \\
 {\mathrm{D}_3} & 1 & -i & -1 & i \\
\end{array}
\end{equation}
\subsection{The McKay quiver diagram and its representation}
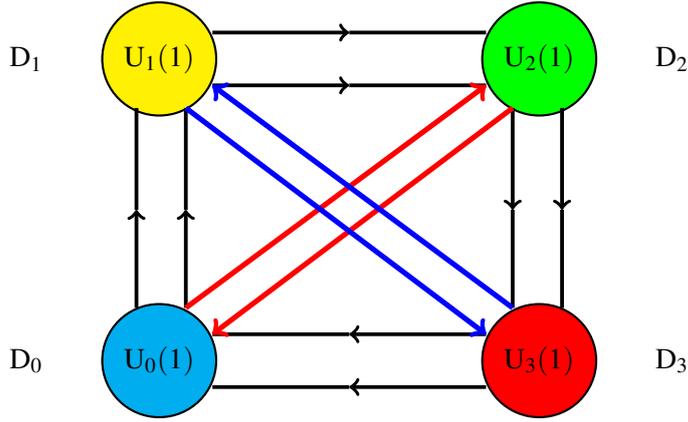
\begin{figure}
\begin{center}
\begin{tikzpicture}[scale=0.50]
\draw [thick] [fill=yellow] (-5,0) circle (1.5 cm) ; \node at (-5,0)
{$\mathrm{U_1(1)}$}; \draw [thick] [fill=green] (5,0) circle (1.5
cm); \node at (5,0) {$\mathrm{U_2(1)}$};  \draw [black,line
width=0.05cm] [->] (-3.6,0.7)-- (0,0.7); \draw [black,line
width=0.05cm] (0,0.7)-- (3.6,0.7); \draw [black,line width=0.05cm]
(3.6,-0.7)-- (0,-0.7); \draw [black,line width=0.05cm][<-]
(0,-0.7)-- (-3.6,-0.7);
 \node at
(8.5,0) {$\mathrm{D}_2$}; \node at (-8.5,0) {$ \mathrm{D}_1$};
\draw [thick] [fill=cyan] (-5,-8) circle (1.5 cm) ; \node at (-5,-8)
{$\mathrm{U_0(1)}$}; \draw [thick] [fill=red] (5,-8) circle (1.5
cm); \node at (5,-8) {$\mathrm{U_3(1)}$};  \draw [black,line
width=0.05cm] (-3.6,-8.7)-- (0,-8.7); \draw [black,line
width=0.05cm][->] (3.6,-7.3)--(0,-7.3); \draw [black,line
width=0.05cm][->] (3.6,-8.7)--(0,-8.7) ; \draw [black,line
width=0.05cm] (-3.6,-7.3)--(0,-7.3) ;
 \node at
(8.5,-8) {$\mathrm{D}_3$}; \node at (-8.5,-8) {$ \mathrm{D}_0$};
\draw [black,line width=0.05cm] [->](-4.3,-6.6)-- (-4.3,-4.0); \draw
[black,line width=0.05cm][->] (-5.6,-6.6)--(-5.6,-4.0); \draw
[black,line width=0.05cm] (-4.3,-4)-- (-4.3,-1.3); \draw [black,line
width=0.05cm](-5.6,-4)--(-5.6,-1.3);
\draw [black,line width=0.05cm] (4.3,-6.6)-- (4.3,-4.0); \draw
[black,line width=0.05cm](5.6,-6.6)--(5.6,-4.0); \draw [black,line
width=0.05cm][->](4.3,-1.3)-- (4.3,-4) ; \draw [black,line
width=0.05cm][->](5.6,-1.3)--(5.6,-4);
\draw [red, line width=0.07cm] [->] (-4.3,-6.6) --(3.6,-0.7);
\draw [red, line width=0.07cm] [->] (4.3,-1.3) --
(-3.6,-7.3); 
\draw [blue, line width=0.07cm] [->] (4.3,-6.6) --(-3.6,-0.7);
\draw [blue, line width=0.07cm] [->] (-4.3,-1.3) --(3.6,-7.3) ;
\end{tikzpicture}
\caption{\label{mckayquivz4} The quiver diagram describing the
$\mathbb{C}^3/\mathbb{Z}_4$ singular quotient and codifying its
resolution via K\"ahler quotient \`{a} la Kronheimer. The same
quiver diagram codifies the construction of the corresponding gauge
theory for a stack of D3-branes. Each node is associated with one of
the 4 irreducible representations of $\mathbb{Z}_4$ and in each node
we located one of the $\mathrm{U_I(1)}$ groups with respect to which
we perform the K\"ahler quotient. This is the case of one
$D3$--brane. For $N$ D3--branes, all gauge groups $\mathrm{U_I(1)}$
are promoted to $\mathrm{U_I(N)}$.}
\end{center}
\end{figure}

The information encoded in eqs. \eqref{panefresco},
\eqref{caratteruccio} is sufficient to calculate the McKay quiver
matrix defined by:
\begin{equation}\label{carteodollo}
    \mathcal{Q} \otimes \mathrm{D}_I\, = \, \bigoplus_{J=0}^3 \,\mathcal{A}_{IJ} \,\mathrm{D}_J
\end{equation}
where $\mathrm{D}_I$ denotes the $4$ irreducible representation of
the group {defined in eqn.} \eqref{caratteruccio}, while
$\mathcal{Q}$ is the representation \eqref{panefresco} describing
the embedding $\mathbb{Z}_4 \hookrightarrow \mathrm{SU(3)}$.
Explicitly we obtain
\begin{equation}\label{quiverroz4}
   \mathcal{A}_{IJ} \, = \, \left(
\begin{array}{cccc}
 0 & 2 & 1 & 0 \\
 0 & 0 & 2 & 1 \\
 1 & 0 & 0 & 2 \\
 2 & 1 & 0 & 0 \\
\end{array}
\right)
\end{equation}
A graphical representation of the quiver matrix \eqref{quiverroz4}
is provided in fig.~\ref{mckayquivz4}.
\par
Every node of the diagram corresponds to an irreducible
representation $\mathcal{D}_I$ and, in the Kronheimer construction,
to a gauge group factor $\mathrm{U}\left(\dim \mathrm{D}_I \right
)$.
\par
As one sees, in every node do enter three lines and go out three
lines. Both the incoming and the outgoing lines are subdivided in a
double line in some direction (or from some direction) and a single
line to some direction, or from some direction.
\par
This information is sufficient to derive the number of Wess-Zumino
multiplets in the corresponding supersymmetric gauge theory and
assign their representations with respect to the three gauge groups
(actually four minus one for the barycentric motion).
\par
From the mathematical viewpoint each line of the diagram corresponds
to an independent parameter appearing in the explicit construction
of the space $\mathcal{S}_{\mathbb{Z}_4}$. This latter is
constructed as follows. Let $R$ denote the regular representation of
$\Gamma$. We consider the space of triplets of $4\times 4$ complex
matrices:
\begin{eqnarray}
    p \in \mathcal{P}_{\mathbb{Z}_4} \, \equiv \, \mbox{Hom}\left(R,\mathcal{Q}\otimes R\right) \, \Rightarrow\,
    p\,=\, \left(\begin{array}{c}
                   A \\
                   B \\
                   C
                 \end{array}
     \right) \label{homqgbis}
\end{eqnarray}
The action of the discrete group $\mathbb{Z}_4$ on the space
$\mathcal{P}_\Gamma$ is defined according with the general scheme
discussed in section \ref{generaKronh} by:
\begin{equation}\label{gammazione}
    \forall \gamma \in \mathbb{Z}_4: \quad \gamma\cdot p \,\equiv\, \mathcal{Q}(\gamma)\,\left(\begin{array}{c}
                  R(\gamma)\, A \, R(\gamma^{-1})\\
                   R(\gamma)\, B \, R(\gamma^{-1}) \\
                  R(\gamma)\, C \, R(\gamma^{-1})
                 \end{array}
     \right)
\end{equation}
where  $R(\gamma)$ denotes its $4 \times 4$-matrix image in the
regular representation.
\par
The subspace $\mathcal{S}_{\mathbb{Z}_4}$ is obtained by setting:
\begin{equation}
\mathcal{S}_{\mathbb{Z}_4} \, \equiv \,
\mbox{Hom}\left(R,\mathcal{Q}\otimes R\right)^{\mathbb{Z}_4}\, = \,
\left\{p\in\mathcal{P}_{\mathbb{Z}_4} / \forall \gamma\in
\mathbb{Z}_4 , \gamma\cdot p = p\right\}\,\,
\label{carnevalediPaulo}
\end{equation}
As we know from the general theory exposed in chapter \ref{partone}
(see section \ref{generaKronh}) the space
$\mathcal{S}_{\mathbb{Z}_4}$ must have complex dimension $3\times
|\mathbb{Z}_4| \, = \, 12$ which is indeed the number of lines in
the quiver diagram of fig. \ref{mckayquivz4}. In the basis where the
regular representation has been diagonalized with the help of the
character Table \eqref{caratteruccio} the general form of the
triplet of matrices composing
$\mbox{Hom}_{\mathbb{Z}_4}\left(R,\mathcal{Q}\otimes R\right)$ and
therefore providing the representation of the quiver diagram of fig.
\ref{mckayquivz4} is the following one:
\begin{equation}\label{pastrugno}
\begin{array}{ccccccc}
       A & = & \left(
\begin{array}{cccc}
 0 & 0 & 0 & \Phi^{(1)}_{0,3} \\
 \Phi^{(1)}_{1,0} & 0 & 0 & 0 \\
 0 & \Phi^{(1)}_{2,1} & 0 & 0 \\
 0 & 0 & \Phi^{(1)}_{3,2} & 0 \\
\end{array}
\right) & ; & B & = & \left(\begin{array}{cccc}
 0 & 0 & 0 & \Phi^{(2)}_{0,3} \\
 \Phi^{(2)}_{1,0} & 0 & 0 & 0 \\
 0 & \Phi^{(2)}_{2,1} & 0 & 0 \\
 0 & 0 & \Phi^{(2)}_{3,2} & 0 \\
\end{array}
\right) \\
\null&\null& \null & \null & \null&\null& \null \\
       C & = & \left(
\begin{array}{cccc}
 0 & 0 & \Phi^{(3)}_{0,2} & 0 \\
 0 & 0 & 0 & \Phi^{(3)}_{1,3} \\
 \Phi^{(3)}_{2,0} & 0 & 0 & 0 \\
 0 & \Phi^{(3)}_{3,1} & 0 & 0 \\
\end{array}
\right) & \null & \null & \null & \null
     \end{array}
\end{equation}
The twelve complex parameters $\Phi^{(J)}_{p,q}$ with $J=1,2,3$,
$p,q=0,1,2,3$, promoted to be functions of the space-time
coordinates $\xi^\mu$:
$$\Phi^{(J)}_{p,q}(\xi)$$
are the complex scalar fields filling the flat K\"ahler manifold of
the Wess-Zumino multiplets in the microscopic lagrangian of the
corresponding gauge theory.
\subsection{The locus $\mathbb{V}_6 \subset \mathcal{S}_{\mathbb{Z}_4}$}
As it was explained in the context of the general framework reviewed
in section \ref{generaKronh}, the $3|\Gamma|$-dimensional flat
K\"ahler manifold $\mathcal{S}_\Gamma$ contains always a
{subvariety} $\mathbb{V}_{|\Gamma|+2} \subset \mathcal{S}_\Gamma$ of
dimension $|\Gamma|+2$ which is singled out by the following set of
quadratic equations:
\begin{equation}\label{ceramicus}
    \left[A\, , \, B\right] \, = \, \left[B\, , \, C\right]\, = \, \left[C\, , \,
    A\right]\, = \, 0
\end{equation}
From the physical point of view, the holomorphic equation
\eqref{ceramicus} occurs as the vanishing of the superpotential
derivatives $\partial_i\mathcal{W}(\Phi) \, = \, 0$ while looking
for the scalar potential extrema, namely for the classical vacua of
the gauge theory. From the mathematical viewpoint the locus
$\mathbb{V}_{|\Gamma|+2}$ is the one we start from in order to
calculate the K\"ahler quotient $\mathcal{M}_\zeta$ which provides
the crepant resolution of the singularity. At the end of the day
$\mathcal{M}_\zeta$ is just the manifold of classical vacua of the
gauge theory. 
As discussed in section \ref{generaKronh}, the vanishing locus
\eqref{ceramicus} {consists of several irreducible components} of
different dimensions,  {and $\mathbb{V}_{|\Gamma|+2}=\mathbb{V}_6$
is the only component of dimension 6. It can be represented in the
form $\mathcal{G}_\Gamma\cdot L_\Gamma$, where $\mathcal{G}_\Gamma$
is the holomorphic quiver group, defined in the next section, and
$L_\Gamma$ is the three dimensional locus that we will shortly
characterize. An open part of this principal component
$\mathbb{V}_6$ can be given by the following explicit equations:}
\begin{eqnarray}\label{bagnomarietta}
  && \Phi^{(2)}_{1,0}\, = \,  \frac{\Phi^{(1)}_{1,0} \Phi^{(2)}_{0,3}}{\Phi^{(1)}_{0,3}},\quad \Phi^{(2)}_{2,1}
   \, = \,  \frac{\Phi^{(1)}_{2,1}
   \Phi^{(2)}_{0,3}}{\Phi^{(1)}_{0,3}},\quad \Phi^{(2)}_{3,2}
   \, = \,  \frac{\Phi^{(1)}_{3,2} \Phi^{(2)}_{0,3}}{\Phi^{(1)}_{0,3}},\nonumber\\
   && \Phi^{(3)}_{1,3}\, = \,
   \frac{\Phi^{(1)}_{1,0} \Phi^{(3)}_{0,2}}{\Phi^{(1)}_{3,2}},\quad
   \Phi^{(3)}_{2,0}\, = \,  \frac{\Phi^{(1)}_{1,0} \Phi^{(1)}_{2,1}
   \Phi^{(3)}_{0,2}}{\Phi^{(1)}_{0,3} \Phi^{(1)}_{3,2}},\quad \Phi^{(3)}_{3,1}\, = \,
   \frac{\Phi^{(1)}_{2,1} \Phi^{(3)}_{0,2}}{\Phi^{(1)}_{0,3}}\end{eqnarray}
\subsection{The holomorphic quiver group $\mathcal{G}_{\mathbb{Z}_4}$ and the gauge group $\mathcal{F}_{\mathbb{Z}_4}$ }
Following the general scheme outlined in \cite{Bruzzo:2017fwj}, we
see that the locus $\mathcal{S}_{\mathbb{Z}_4}$ is mapped into
itself by the action of the \textit{complex quiver group}:
\begin{equation}\label{cromostatico}
    \mathcal{G}_{\mathbb{Z}_4} \, = \, \mathbb{C}^\star \times \mathbb{C}^\star
    \times \mathbb{C}^\star \, \simeq \,\pmb{\Lambda} \, \equiv \, \left(
                                           \begin{array}{c|c|c|c}
                                             \mathbb{C}^\star & 0 & 0 & 0 \\
                                             \hline
                                             0 & \mathbb{C}^\star & 0 & 0 \\
                                             \hline
                                             0 & 0 & \mathbb{C}^\star & 0 \\
                                             \hline
                                             0 & 0 & 0 & \mathbb{C}^\star\\
                                           \end{array}
                                         \right) \quad ; \quad
                                         \mbox{det}\, \pmb{\Lambda}\, =
                                         \, 1
\end{equation}
The gauge group of the final gauge theory is the maximal compact
subgroup of $\mathcal{G}_{\mathbb{Z}_4}$, namely:
\begin{equation}\label{cromodinamico}
    \mathcal{F}_{\mathbb{Z}_4} \, = \, \mathrm{U(1)} \times \mathrm{U(1)}
    \times \mathrm{U(1)} \, \simeq \,\pmb{\Xi} \, \equiv \, \left(
                                           \begin{array}{c|c|c|c}
                                             \mathrm{U(1)} & 0 & 0 & 0 \\
                                             \hline
                                             0 & \mathrm{U(1)} & 0 & 0 \\
                                             \hline
                                             0 & 0 & \mathrm{U(1)} & 0 \\
                                             \hline
                                             0 & 0 & 0 & \mathrm{U(1)}\\
                                           \end{array}
                                         \right) \quad ; \quad
                                         \mbox{det}\, \pmb{\Xi}\, =
                                         \, 1
\end{equation}
The explicit form of the matrices $\pmb{\Lambda}$ and $\pmb{\Xi}$ is
that appropriate to the basis where the regular representation is
diagonalized. In that basis the charge assignments (representation
assignments) of the scalar fields are read off from the
transformation rule:
\begin{equation}\label{caricopendente}
    A(\Phi^\prime) \, = \, \pmb{\Xi}^{-1} \, A(\Phi)\, \pmb{\Xi},
    \quad B(\Phi^\prime) \, = \, \pmb{\Xi}^{-1} \, B(\Phi) \,\pmb{\Xi},
    \quad C(\Phi^\prime) \, = \, \pmb{\Xi}^{-1} \, C(\Phi) \,\pmb{\Xi}
\end{equation}
Then we consider the locus $L_{\mathbb{Z}_4}$ made by those triplets
of matrices $A,B,C$ that belong to $\mathcal{S}_\Gamma$ and are
diagonal in the natural basis of the regular representation. In the
diagonal basis of the regular representation the same matrices
$A,B,C$ have the following form:
\begin{equation}\label{baldovinus}
\begin{array}{ccccccc}
       A_0 & = & \left(
\begin{array}{cccc}
 0 & 0 & 0 & Z^1 \\
 Z^1 & 0 & 0 & 0 \\
 0 & Z^1 & 0 & 0 \\
 0 & 0 & Z^1 & 0 \\
\end{array}
\right) & ; & B_0 & = & \left(\begin{array}{cccc}
 0 & 0 & 0 & Z^2 \\
 Z^2 & 0 & 0 & 0 \\
 0 & Z^2 & 0 & 0 \\
 0 & 0 & Z^2 & 0 \\
\end{array}
\right) \\
\null&\null& \null & \null & \null&\null& \null \\
       C_0 & = & \left(
\begin{array}{cccc}
 0 & 0 & Z^3 & 0 \\
 0 & 0 & 0 & Z^3\\
 Z^3 & 0 & 0 & 0 \\
 0 & Z^3 & 0 & 0 \\
\end{array}
\right) & \null & \null & \null & \null
     \end{array}
\end{equation}
The fields $Z^{1,2,3}$ provide a set of three coordinates spanning
the three-dimensional locus $L_{\mathbb{Z}_4}$. The complete locus
$\mathbb{V}_6$ coincides with the orbit of $L_{\mathbb{Z}_4}$ under
the free action of $\mathcal{G}_{\mathbb{Z}_4}$:
\begin{equation}\label{gestaltus}
    \mathbb{V}_6 \, = \,
{\mathcal{G}_{\mathbb{Z}_4}\cdot L_{\mathbb{Z}_4}}
    \, = \, \left(\begin{array}{c}
                    \pmb{\Lambda}^{-1} \, A_0 \,\,\pmb{\Lambda} \\
                    \pmb{\Lambda}^{-1} \, B_0 \,\,\pmb{\Lambda} \\
                    \pmb{\Lambda}^{-1} \, C_0 \,\,\pmb{\Lambda}
                  \end{array}
     \right)
\end{equation}
\section{The moment map equations} Implementing once more the
general procedure outlined in section \ref{generaKronh} we arrive at
the moment map equations and at the final crepant resolution of the
singularity in the following way.
\par
We refer the reader to section \ref{generaKronh} for the general
definition of the moment map
$$\mu \quad : \quad \mathcal{S}_\Gamma \, \longrightarrow \, \mathbb{F}_\Gamma^\star$$
where $\mathbb{F}_\Gamma^\star$ denotes the dual (as vector spaces)
of the Lie algebra $\mathbb{F}_\Gamma$ of the gauge group. We recall
that the preimage of the level zero moment map is the
$\mathcal{F}_\Gamma$ orbit of the locus $L_\Gamma$:
\begin{equation}\label{quadriglia}
  \mu^{-1}\left(0\right) \, = \,
 {\mathcal{F}_\Gamma\cdot L_\Gamma.}
\end{equation}
Note that $L_\Gamma$   is actually  $\mathbb C^3$, so that the
{image of $L_\Gamma$ in} the K\"ahler quotient of level zero
 coincides with the original singular variety
$\mathbb{C}^3/\Gamma$ see eqn.(\ref{pereoliato}) and following
lines.
\par
Next, as in section \ref{VG2p1}, we  consider the following
decomposition of the Lie quiver group algebra:
\begin{eqnarray}
  \mathbb{G}_{\mathbb{Z}_4} &=& \mathbb{F}_{\mathbb{Z}_4} \oplus
  \mathbb{K}_{\mathbb{Z}_4}\nonumber\\
  \left[\mathbb{F}_{\mathbb{Z}_4} \, , \,
  \mathbb{F}_{\mathbb{Z}_4}\right] &\subset & \mathbb{F}_{\mathbb{Z}_4}
  \quad ; \quad
\left[\mathbb{F}_{\mathbb{Z}_4} \, , \,
\mathbb{K}_{\mathbb{Z}_4}\right] \,\subset \,
\mathbb{K}_{\mathbb{Z}_4} \quad ; \quad
\left[\mathbb{K}_{\mathbb{Z}_4} \, , \,
\mathbb{K}_{\mathbb{Z}_4}\right] \,\subset \,
\mathbb{F}_{\mathbb{Z}_4} \label{salameiolecco}
\end{eqnarray}
where $\mathbb{F}_{\mathbb{Z}_4}$ is the maximal compact subalgebra.
\par
A special feature  of all the quiver  {g}roups and Lie  {a}lgebras
is that $\mathbb{F}_\Gamma$ and $\mathbb{K}_\Gamma$ have the same
real dimension $|\Gamma|-1$ and one can choose a basis of
{H}ermitian generators $T^I$ such that:
\begin{equation}\label{sacherdivuli}
    \begin{array}{ccccccc}
       \forall \pmb{\Phi} \in \mathbb{F}_\Gamma & : & \pmb{\Phi} & = &
       {\rm i} \times \sum_{I=1}^{|\Gamma|-1} c_I T^I & ; &
       c_I \in \mathbb{R} \\
       \forall \pmb{K} \in \mathbb{K}_\Gamma & : & \pmb{K} & = &
       \sum_{I=1}^{|\Gamma|-1} b_I T^I & ; &
       b_I \in \mathbb{R} \\
     \end{array}
\end{equation}
Correspondingly a generic element $g\in \mathcal{G}_{\mathbb{Z}_4}$
can be split as follows:
\begin{equation}\label{consolatio}
   \forall g \in \mathcal G_{\mathbb{Z}_4} \quad : \quad g=\mathcal{U} \,
   \mathcal{H} \quad  ; \quad \mathcal{U} \in
   \mathcal{F}_{\mathbb{Z}_4} \quad ; \quad  \mathcal{H} \in
   \exp\left[ \mathbb{K}_{\mathbb{Z}_4}\right]
\end{equation}
Using the above property we arrive at the following parametrization
of the space $\mathbb{V}_6$
\begin{equation}\label{krumiro}
    \mathbb{V}_6 \, = \,
     {\mathcal{F}_{\mathbb{Z}_4}\cdot}
    \left(\exp\left[
    \mathbb{K}_{\mathbb{Z}_4}\right]\cdot L_{\mathbb{Z}_4}\right)
\end{equation}
where, by definition, we have set:
\begin{eqnarray}
  p\in \exp\left[
    \mathbb{K}_{\mathbb{Z}_4}\right]\cdot L_{\mathbb{Z}_4}  &\Rightarrow &
    p=\left\{\exp\left[-\pmb{K}\right]\, A_0
    \exp\left[\pmb{K}\right], \, \exp\left[-\pmb{K}\right]\, B_0\,
    \exp\left[\pmb{K}\right],\, \exp\left[-\pmb{K}\right]\, C_0
    \exp\left[\pmb{K}\right]\right\} \nonumber\\
 \left\{ A_0, \, B_0,\,  C_0\right\} &\in & L_{\mathbb{Z}_4} \nonumber\\
 { \pmb{K}} & { \in}&  { \mathbb{K}_{\mathbb{Z}_4}} \label{centodiquestigiorni}
\end{eqnarray}
In our case the three generators $T^I$ of the real subspace
 {$\mathbb{K}_{\mathbb{Z}_4}$} have been chosen as follows:
\begin{equation}\label{TIgener}
    \begin{array}{ccccccc}
       T^1 & = & \left(
\begin{array}{cccc}
 1 & 0 & 0 & 0 \\
 0 & -1 & 0 & 0 \\
 0 & 0 & 0 & 0 \\
 0 & 0 & 0 & 0 \\
\end{array}
\right) & ; & T^2 & = & \left(
\begin{array}{cccc}
 0 & 0 & 0 & 0 \\
 0 & 1 & 0 & 0 \\
 0 & 0 & -1 & 0 \\
 0 & 0 & 0 & 0 \\
\end{array}
\right) \\
       \null & \null & \null & \null & \null & \null & \null \\
       T^3 & = & \left(
\begin{array}{cccc}
 0 & 0 & 0 & 0 \\
 0 & 0 & 0 & 0 \\
 0 & 0 & 1 & 0 \\
 0 & 0 & 0 & -1 \\
\end{array}
\right) & \null & \null & \null & \null
     \end{array}
\end{equation}
So that the relevant real group element takes the following form:
\begin{equation}\label{sirenus}
    \exp[\pmb{K}] \, = \, \left(
\begin{array}{cccc}
 \mathfrak{H}_1 & 0 & 0 & 0 \\
 0 & \frac{\mathfrak{H}_2}{\mathfrak{H}_1}
   & 0 & 0 \\
 0 & 0 &
   \frac{\mathfrak{H}_3}{\mathfrak{H}_2} &
   0 \\
 0 & 0 & 0 & \frac{1}{\mathfrak{H}_3} \\
\end{array}
\right)
\end{equation}
where $\mathfrak{H}_I$ are, by definition, real. Relying on this, in
the K\"ahler quotient we can invert the order of the operations.
First we quotient the action of the compact gauge group
$\mathcal{F}_{\mathbb{Z}_4}$ and then we implement the moment map
constraints. We have:
\begin{equation}\label{cascapistola}
 \mathbb{V}_6/\!\!/_{\mathcal{F}_{\mathbb{Z}_4}}\, =
  \, {\left(\exp\left [\mathbb{K}_{\mathbb{Z}_4}\right]\cdot L_{\mathbb{Z}_4}\right)/\mathbb{Z}_4,}
\end{equation}
 {where $\mathbb{Z}_4$ acts on $L_{\mathbb{Z}_4}$ via the action induced by that of
 the stabiliser of $L_{\mathbb{Z}_4}$ in $\mathcal{F}_{\mathbb{Z}_4}$.}
The explicit form of  {the triple} of matrices mentioned in
eqn.~\eqref{centodiquestigiorni} is easily calculated on the basis
of eqs.~\eqref{baldovinus} and \eqref{sirenus}
\begin{equation}\label{caglioccio}
    p\, = \,\left(\begin{array}{c}
                    A \\
                    B \\
                    C
                  \end{array}
     \right) \, \equiv \, \left(\begin{array}{c}
                  \exp[-\pmb{K}] \, A_0 \, \exp[\pmb{K}]\\
                  \exp[-\pmb{K}] \, B_0 \, \exp[\pmb{K}]\\
                  \exp[-\pmb{K}] \, C_0 \, \exp[\pmb{K}]
                  \end{array}
     \right)
\end{equation}
The moment map  {is given by:}
\begin{eqnarray}\label{momentidimappa}
\mu\left(p\right)& = & \left\{ \mathfrak{P}_1,\mathfrak{P}_2,\mathfrak{P}_3 \right\} \nonumber\\
    \mathfrak{P}_I & = &\mathrm{Tr}\left[ T_I \, \left(\left[A\,
    , \, A^\dagger\right]\, +\, \left[B\,
    , \, B^\dagger\right]\, +\, \left[C\,
    , \, C^\dagger\right]\right)\right]
\end{eqnarray}
Imposing the moment map constraint we find:
\begin{equation}\label{carampana}
\mu^{-1}\left( \zeta\right)/\!\!/_{\mathcal{F}_{\mathbb{Z}_4}}\, =
\, \left\{ p\, \in \exp\left [\mathbb{K}_{\mathbb{Z}_4}\right]\cdot
L_{\mathbb{Z}_4}\,
\parallel \, \mathfrak{P}_I (p) \,= \, \zeta_I \right\} {/\mathbb{Z}_4.}
\end{equation}
Eq.\,\eqref{carampana} provides an explicit algorithm to calculate
the K\"ahler potential of the final resolved manifold if we are able
to solve the constraints for $\mathfrak{H}_I$ in terms of the triple
of complex coordinates $Z^i$ ($i = 1,2,3$).  Indeed we recall that
the K\"ahler potential $\mathcal{K}_{\mathcal{M}_\zeta}$ of the
resolved variety $\mathcal M_\zeta = \mathcal
N_\zeta/\mathcal{F}_{\mathbb{Z}_4}$, where $ \mathcal N_\zeta =
\mu^{-1}(\zeta)\subset \mathcal S_\Gamma$, is given by the general
formula \ref{criceto1}:
\begin{equation}\label{celeberro}
  \mathcal{K}_{\mathcal{M}_\zeta}\, =   \mathcal{K}_{\mathcal{N}_\zeta} \,
  + \zeta_I  \, \mathfrak{C}^{IJ} \,
 \log
  { \mathfrak{H}_J^{2\alpha_{\zeta_J}}}
\end{equation}
Here $ \mathcal{K}_{\mathcal{N}_\zeta}$ is the restriction to
$\mathcal{N}_\zeta$ of the  K\"ahler potential of the flat K\"ahler
metric on $\mathcal S_\Gamma$, which is
$\mathcal{F}_{\mathbb{Z}_4}$-invariant and therefore can be regarded
as a function on $\mathcal{M}_\zeta$. The positive rational
constants $\alpha_\zeta$ are to be chosen so that the functions $  {
\mathfrak{H}_J^{2\alpha_{\zeta_J}}}$ are  hermitian fibre metrics on
the three tautological bundles; these constants are completely
determined by the geometry, but it will be easier to fix them later
on, using the fact that Chern characters of the tautological line
bundles are a basis of the cohomology ring of $M_\zeta$
\cite{CrawIshii}. Moreover,
\begin{equation}\label{romualdo}
\mathfrak{C}^{IJ} \, = \, \mbox{Tr}\left(T^I \, T^J\right ) \, = \,
\left(
\begin{array}{ccc}
2 & -1 & 0 \\
-1 & 2 & -1 \\
0 & -1 & 2 \\
\end{array}
\right)
\end{equation}
is the matrix of scalar products of the gauge group generators.
\par
The final outcome of this calculation was already presented in
section 9 of \cite{Bruzzo:2017fwj}. As it was done there, it is
convenient to consider the following linear combinations:
\begin{equation}\label{innominata}
    \left(
\begin{array}{ccc}
 1 & 0 & -1 \\
 1 & -1 & 1 \\
 0 & 1 & 0 \\
\end{array}
\right)\, \left(\begin{array}{c}
                  \mathfrak{P}_1 -\zeta_1\\
                  \mathfrak{P}_2 -\zeta_2\\
                  \mathfrak{P}_3-\zeta_3
                \end{array}
\right) \, = \, 0
\end{equation}
In this way we obtain:
\begin{eqnarray}
\left(
\begin{array}{c}
-\frac{\left(X_1^2-X_3^2\right) \left(X_1 X_3
 \left(\Delta _1^2+\Delta _2^2\right)+\left(1+X_2^2\right)
 \Delta _3^2\right)}{X_1 X_2 X_3} \\
 \frac{\left(X_2+X_2^3-X_1 X_3 \left(X_1^2+X_3^2\right)\right)
 \left(\Delta _1^2+\Delta _2^2\right)}{X_1 X_2 X_3} \\
 -\frac{\left(-1+X_2^2\right) \left(X_2 \left(\Delta _1^2+\Delta _2^2\right)
 +\left(X_1^2+X_3^2\right)
 \Delta _3^2\right)}{X_1 X_2 X_3} \\
\end{array}
\right) & = & \left(
\begin{array}{c}
 \zeta _1-\zeta _3 \\
 \zeta _1-\zeta _2+\zeta _3 \\
 \zeta _2 \\
\end{array}
\right) \label{sakerdivoli}
\end{eqnarray}
where $\Delta_i \, = \, \mid Z^{ i}\mid$ are the moduli of the three
complex coordinates $Z^i$, and $X_J= \mathfrak{H}_J^2$ for
$J=1,2,3$.

Applying the general framework developed in \cite{Bruzzo:2017fwj} we
have
\begin{equation}
\mathcal{H}\text{ }\equiv \text{ }\left(
\begin{array}{|c|c|c|}
\hline
 \mathfrak{H}_1  & 0 & 0  \\
\hline
 0 & \mathfrak{H}_2 & 0 \\
\hline
 0 & 0 &  \mathfrak{H}_{3}  \\
\hline
\end{array}
\right)\label{tautobundmetro}
\end{equation}
and the positive definite hermitian matrix
$\mathcal{H}^{2\alpha_{\zeta_J}}$  is the fibre metric on the direct
sum:
\begin{equation}\label{direttosummo}
    \mathcal{R}\,=\,\bigoplus_{I=1}^{r} \, \mathcal{R}_I
\end{equation}
of the $r=3$ tautological bundles that, by construction, are
holomorphic vector bundles with rank equal to the dimensions of the
three nontrivial irreducible representations of $\Gamma$, which in
this case is always one:
\begin{equation}\label{tautibundiEach}
    \mathcal{R}_I \, \stackrel{\pi}{\longrightarrow}\,
    \mathcal{M}_\zeta\quad\quad ;
    \quad \quad\forall p \in \mathcal{M}_\zeta\quad :\quad
    \pi^{-1}(p) \approx \mathbb{C}^{n_I}
\end{equation}
The compatible connection\footnote{Following standard mathematical
nomenclature, we call compatible connection on a holomorphic vector
bundle,  one whose $(0,1)$ part is the Cauchy-Riemann operator of
the bundle.} on the holomorphic vector bundle
$\mathcal{R}=\bigoplus_I\mathcal R_I$ is given by $\vartheta =
\bigoplus_I\vartheta_I$,  where
\begin{eqnarray}\label{comancio}
    \vartheta_I = \alpha_{\zeta_I} \,\partial \log {X}_I =
    \mathcal{H}^{-2{\alpha_{\zeta_I}}} \,\partial\mathcal{H}^{2\alpha_{\zeta_I}}\end{eqnarray}
which is a 1-form with values in $\mathbb{C}$, the Lie algebra of
the structural group $\mathbb{C}^\ast $ of the $I$-th tautological
vector bundle. The natural connection of the
$\mathcal{F}_{\mathbb{Z}_4}$ principal bundle, mentioned in
eqn.\,\eqref{cromodinamico} is just, according
 {to the} universal scheme, the imaginary part of the
connection $\vartheta$.

In order to solve the system of equations \eqref{sakerdivoli}  it is
convenient to change variables and write:
\begin{eqnarray}\label{lobus}
   && \Sigma \,= \, \Delta_1^2 + \Delta_2^2 \quad ; \quad U \, = \,
   \Delta_3^2
\end{eqnarray}
In this way we obtain:
\begin{equation}
   \left\{ \begin{array}{lcl}
 -U X_2^2 X_1^2-U X_1^2+U X_2^2 X_3^2+U X_3^2-\zeta _1 X_2 X_3 X_1+\zeta _3 X_2 X_3 X_1-\Sigma  X_3 X_1^3
 +\Sigma  X_3^3 X_1 & = &0 \\
 -\zeta _1 X_2 X_3 X_1+\zeta _2 X_2 X_3 X_1-\zeta _3 X_2 X_3 X_1-\Sigma  X_3 X_1^3-\Sigma  X_3^3 X_1+\Sigma  X_2^3
 +\Sigma  X_2 & = &0\\
 -U X_1^2 X_2^2-U X_3^2 X_2^2+U X_1^2+U X_3^2-\zeta _2 X_1 X_3 X_2-\Sigma  X_2^3+\Sigma  X_2& = &0 \\
\end{array}\right. \label{sistemico}
\end{equation}
This is the fundamental algebraic system encoding all information
about the singularity resolution.
\section{Properties of the moment map algebraic system and chamber
structure}\label{GenMap} The resolubility of the system
\eqref{sistemico}, viewed as a set of algebraic equations of higher
order has some very peculiar properties that actually encode the
topology and analytic structure of the resolved manifold
$Y^{\mathbb{Z}_4}_{[3]}$ and of its possible degenerations. The most
relevant property of \eqref{sistemico} is that for generic values
$U>0,\Sigma >0$ it has always one and only one root where all $X_i >
0$ are real positive. This is of course expected from the general
theory, as we know there are a well-defined quotient, and a K\"aher
metric on the quotient,  for every generic choice of the {\em level
parameters} $\zeta$ (these  are called {\em Fayet-Iliopoulos
parameters} in gauge theory, while in algebraic geometry they
correspond to the so-called {\em stability parameters}); but it has
also been verified numerically at an arbitrary large collection of
random points $\zeta \in \mathbb{R}^3$ and for an arbitrary large
collection of points $\{U,\Sigma\} \in \mathbb{R}^2_+$.
\paragraph{The special surface
$\mathcal{S}_2 $.} Although it is not a wall there is in the $\zeta$
space a planar surface defined by the following conditions
\begin{equation}\label{essedue}
 \mathcal{S}_2 \, \equiv \,  \left\{\zeta_1\, =\, \zeta_3\, =\, a
  , \quad\zeta_2\, =\, b \neq 2a \right\}
\end{equation}
where the algebraic system \eqref{sistemico} acquires a more
manageable form without loosing generality. The  strong
simplification is encoded in the following condition:
\begin{equation}\label{simplicius}
    X_1 \, = \, X_2
\end{equation}
Thanks to eqn.\eqref{simplicius}, on the plane $\mathcal{S}_2$ the
moment map system reduces to a system of two rather than three
equations. To understand eqn.\eqref{simplicius} we need to recall
some general properties of the moment map equations for this special
case. Considering the system \eqref{sistemico}, it can be observed
that one of the three functions $X_i$ can always be algebraically
solved in terms of the other two. Indeed we can write:
\begin{eqnarray}\label{caragamba}
    X_3 &=&
   X_1 \sqrt[2]{\frac{\zeta
   _2+\zeta _3 \left(X_2^2-1\right)}{\zeta
   _2+\zeta _1 \left(X_2^2-1\right)}}
\end{eqnarray}
This relation is the algebraic counterpart, in the moment map
equations of the topological result that the homology and cohomology
of the resolved variety $Y^{\mathbb{Z}_4}_{[3]}$ has dimension $2$.
If, inspired by eqn.\eqref{caragamba} we consider the case where all
parameters $\zeta_i$ are different from zero  but two of them,
namely $\zeta_1$ and $\zeta_3$ are equal among themselves,
\text{i.e.} we localize our calculations on the surface
$\mathcal{S}_2$, then eqn.\eqref{simplicius} follows, yielding a
reduced system of moment map equations:
\begin{equation}\label{grumildus}
    \left(
\begin{array}{c}
 -2 a X_2 X_1^2+b X_2 X_1^2-2 \Sigma  X_1^4+\Sigma
   X_2^3+\Sigma  X_2 \\
 -b X_1^2 X_2-2 U X_1^2 X_2^2+2 U X_1^2-\Sigma  X_2^3+\Sigma
   X_2 \\
\end{array}
\right) \, = \, \left(
                  \begin{array}{c}
                    0 \\
                    0 \\
                  \end{array}
                \right)
\end{equation}
\subsection{Generalities on the chamber structure}
\label{classiwall} In general, the space of    parameters $\zeta$
(which are closely related to the {\rm stability parameters} of the
GIT quotient construction)\footnote{Geometric invariant theory,
usually shortened into GIT, is the standard way of taking quotients
in algebraic geometry. The relation between the GIT approach and the
K\"ahler quotient \`a la Kronheimer in the problem at hand is
explored in \cite{degeratu}.} has a chamber structure. Let $C$ be a
chamber in that space, and $\mathcal{W}$   a wall of $C$; denote by
$\mathcal M_C$ the resolution corresponding to a generic $\zeta$ in
$C$ (they are all isomorphic), and by $\mathcal M_{\mathcal{W}}$ the
resolution corresponding to a generic $\zeta$ in ${\mathcal{W}}$.
There is a well-defined morphism $\gamma\,_{\mathcal{W}}\colon
\mathcal M_C \to \mathcal M_{\mathcal{W}}$ (actually one should take
the normalization of the second space, but we skip such details). In
general, the morphism $\gamma\,_{\mathcal{W}}$ contracts curves or
divisors;  in \cite{CrawIshii} the walls are classified according to
the nature of the contractions performed by
$\gamma\,_{\mathcal{W}}$. One says that ${\mathcal{W}}$ is of
\begin{enumerate}\item  type 0 if $\gamma\,_{\mathcal{W}}$ is an isomorphism;
\item type I if  $\gamma\,_{\mathcal{W}}$ contracts a curve to a point;
\item type III if  $\gamma\,_{\mathcal{W}}$ contracts a divisor to a curve.
\end{enumerate}
Walls of type II, that should contract a divisor to a point, do not
actually exist, as shown in \cite{CrawIshii}.
\par
The chamber structure pertaining to our example is analyzed and
reconstructed in detail in section \ref{camerataccademica}. A guide
to the localization of walls and chambers is provided by the
existence of some lines in $\zeta$ space where the system
\eqref{sistemico} becomes solvable by radicals or reduces to a
single algebraic equation. These lines  {either turn out} to be
edges of the convex chambers occuring at   intersections of walls,
or just belong to  walls. We begin by analyzing such solvable lines.
\par
\subsection{The solvable lines  located in  $\zeta$ space} In
the $\zeta$ moduli space there are few subcases where the solution
of the algebraic system \eqref{sistemico} can be reduced to finding
the roots of a single algebraic equation whose order is equal or
less than $6$. As anticipated these solvable cases are located on
walls of the chamber and in most case occur at the intersection of
two walls.
\begin{description}
\item[A)] \textbf{Case Cardano  I, $\zeta_1=0, \, \zeta_2=\zeta_3 =s$}. With this choice the general
solution of the system \eqref{sistemico} is provided by  setting the
ansatz displayed  below and by solving the quartic equation for $X$
contained in the next line:
\begin{eqnarray}
    && X_1\, = \, 1, \quad X_2\, = \,X, \quad X_3\, = \, X \nonumber \\
    &&s X^2-U X^4+U-\Sigma  X^3+\Sigma  X \, = \, 0 \label{equatura}
\end{eqnarray}
Obviously we need to choose a branch of the solution such that $X$
is real and positive. As we discuss later on, this is always
possible for all values of $U$ and $\Sigma$ and the required branch
is unique.

To this effect a simple, but very crucial observation is the
following. The arithmetic square root $\sqrt{|s|}$ of the level
parameter $s$ can be used as length scale of the considered space by
rescaling the coordinates as follows: $Z^i \, \to \, \sqrt{|s|}
\,\tilde{Z}^i$ so that equation \eqref{equatura} can be rewritten as
follows
 \begin{equation}\label{baldop}
   -\tilde{U} X^4+U-\tilde{\Sigma}  X^3+  \mathfrak{s}
   \,X^2+\tilde{\Sigma } X \, = \,0
 \end{equation}
where $\mathfrak{s}$ denotes the sign of the moment map level. This
implies that we have only three cases to be studied, namely:
\begin{equation}\label{xenofonte}
   \mathfrak{s} \, = \, \left \{ \begin{array}{c}
                                   1 \\
                                   0\\
                                   -1
                                 \end{array}\right.
\end{equation}
The second case corresponds to the original singular orbifold while
the first and the third yield one instance of what we name the
Cardano manifold. We will see that it corresponds to one of the
possible degenerations of the full resolution
$Y^{\mathbb{Z}_4}_{[3]}$. In the following we disregard the tildas
and we simply write:
\begin{equation}\label{pirettus}
   -{U} X^4+U-{\Sigma}  X^3 \pm
   \,X^2+{\Sigma } X \, = \,0
 \end{equation}
  \item[B)] \textbf{Case Cardano II $\zeta_3=0, \, \zeta_1=\zeta_2 =s$}. With this choice the general
  solution of the system \eqref{sistemico} is provided by  setting the
  ansatz displayed  below and by solving the quartic equation for
  $X$ contained in the next line:
  \begin{eqnarray}\label{raschiotto}
    && X_1\, = \, X,\quad X_2\, = \,X,  \quad X_3\, = \, 1
    \nonumber\\
    &&-s X^2-U X^4+U-\Sigma  X^3+\Sigma  X\, = \, 0
    \label{konigsberg}
  \end{eqnarray}
As one sees eqn.~\eqref{konigsberg} can be reduced to the form
\eqref{pirettus} by means of a  rescaling similar to that utilized
in the previous case. All previous conclusions apply to this case
upon the exchange of $X_1$ and $X_3$.
\item[C)] \textbf{Case Eguchi-Hanson  $\zeta_2=0, \, \zeta_1=\zeta_3 =s $}.
With this choice the general
  solution of the system \eqref{sistemico} is provided by  setting the
  ansatz displayed  below and by solving the quartic equation for
  $X$ contained in the next line:
  \begin{eqnarray}\label{raschiotto}
    && X_1\, = \, X,\quad X_2\, = \,1\,  \quad X_3\, = \, X
    \nonumber\\
    &&-2 s X^2+2 \Sigma -2 \Sigma  X^4\, = \, 0 \label{ridiculite}
  \end{eqnarray}
The unique real positive branch of the solution to
eqn.~\eqref{ridiculite} is given below:
\begin{equation}\label{segretusquid}
  X\to \frac{\sqrt{\frac{\sqrt{s^2+4 \Sigma ^2}}{\Sigma }-\frac{s}{\Sigma
   }}}{\sqrt{2}}
\end{equation}
We will see in a later section that eqn.\eqref{segretusquid} leads
to a complex three-dimensional manifold that is the tensor product
$\mathrm{EH} \times \mathbb{C}$, having denoted by $\mathrm{EH}$ the
Eguchi-Hanson hyperk\"ahler manifold.
\item[D)] \textbf{Case Kamp\'{e}  $\zeta_2=2s,
\, \zeta_1=\zeta_3 =s$}. With this choice the general solution of
the system \eqref{sistemico} is provided by  setting the ansatz
displayed below and by solving the sextic equation for $X$ contained
in the next line:
\begin{eqnarray}\label{raschiotto}
    && X_1\, = \, \frac{\sqrt[4]{Z^3+Z}}{\sqrt[4]{2}},
    \quad X_2\, = \,Z,  \quad X_3\, = \, \frac{\sqrt[4]{Z^3+Z}}{\sqrt[4]{2}}
    \nonumber\\
    && 2 \left(Z^2+1\right) \left(s Z-U Z^2+U\right)^2-\Sigma ^2 Z
\left(Z^2-1\right)^2\, = \,0
\end{eqnarray}
As in other cases the root of the sextic equation must be chosen
real and positive. Furthermore the absolute value of the parameter
$s$ can be disposed off by means of a rescaling.
\end{description}
\subsection{The K\"ahler potential of the quotient manifolds}
\label{kallusquidam} Before discussing the chamber structure guided
by the discovery of the above mentioned solvable edges $A,B,C,D$ it
is useful to complete the determination of the K\"ahler manifolds
singled out by such edges. This requires considering the explicit
form of the K\"ahler potential for the quotient manifolds. Following
the general rules of the K\"ahler quotient resolution \`a la
{Kronheimer} the restriction of the K\"ahler potential of the linear
space $\mathcal{S}_\Gamma =
{\mbox{Hom}_\Gamma\left(R,\mathcal{Q}\otimes R\right)}$ to the
algebraic locus $\mathcal{D}(L_\Gamma )$ and then to the level
surface $\mathcal{N}_\zeta$ is, for the case under consideration,
the following one:
\begin{equation}\label{KelloN}
   \mathcal{K}_0 \mid_{\mathcal{N}_\zeta} \, = \,\frac{U \left(X_2^2+1\right)
   \left(X_1^2+X_3^2\right)+\Sigma
   \left(X_2^3+X_2+X_1\, X_3
   \left(X_1^2+X_3^2\right)\right)}{X_1
   X_2 X_3}
\end{equation}
The complete K\"ahler potential of the resolved variety is given by:
\begin{equation}\label{caramboletta}
    \mathcal{K}_{\mathcal{M}_\zeta} \, = \, \mathcal{K}_0 \mid_{\mathcal{N}_\zeta}
    \, +  \, \zeta_I  \, \mathfrak{C}^{IJ} \,
    \log\left[X_J\right]
\end{equation}
\par
The main point we need to stress is that, depending on the choices
of the moduli $\zeta_I$ (up to rescalings) we can obtain
substantially different manifolds, both topologically and
metrically.

The generic case which  captures the entire algebraic structure of
the resolved variety, to be discussed in later sections by means of
toric geometry, is provided by
\begin{equation}\label{comancho}
    \zeta_1 \neq 0 \quad, \quad \zeta_2 \neq 0 \quad, \quad \zeta_3 \neq 0
\end{equation}
We name the corresponding K\"ahler manifold $Y$.
\par
For the solvable edges of \textit{moduli space} which we have
classified in the previous section we have instead the following
results
\begin{description}
  \item[A)] \textbf{Cardano case $\mathcal{M}_{0,1,1}$}. We name
  \textit{Cardano manifold} the one emerging from the choice
  $\zeta_1=0,\, \zeta_2=1, \, \zeta_3 \, = \, 1$ where the
  solution of the moment map equations is reduced to the solution
  of the quartic algebraic equation \eqref{pirettus}. Choosing the
  sign plus in that equation and performing the substitution $X_1 =
  1,\, X_2= X, \, X_3=X$ the K\"ahler potential of the
  Cardano manifold $\mathcal{M}_{0,1,1}$ takes the form:
\begin{equation}\label{KpotCardan}
    \mathcal{K}_{\mathcal{M}_{0,1,1}} \, = \, 2\,  \log{X} \, + \,
    \frac{\left(X^2+1\right) \left(U \left(X^2+1\right)+2 \Sigma  X\right)}{X^2}
\end{equation}
where, depending on the $\Sigma,U$ region, $X$ is the positive real
root of the quartic equation
\begin{equation}\label{baldoppo}
   -U X^4+U-\Sigma  X^3+ X^2+\Sigma  X \, = \,0
\end{equation}
We already argued that this exists and is unique in all regions of
the $\Sigma,U$ plane.
\item[B)] \textbf{Cardano case $\mathcal{M}_{1,1,0}$}. This turns
out to be an identical copy of the previous Cardano manifold. It
emerges from the choice $\zeta_1=1,\, \zeta_2=1, \, \zeta_3 \, = \,
0$ for which the solution of the moment map equations are also
reduced to the solution of the quartic algebraic equation
\eqref{pirettus}. Performing the substitution $X_1 = X,\, X_2= X, \,
X_3=1$ the K\"ahler potential of the Cardano manifold
$\mathcal{M}_{1,1,0}$ takes the form:
\begin{equation}\label{croccus}
\mathcal{K}_{\mathcal{M}_{1,1,0}} \, = \, 2 \, \log
(X)+\frac{\left(X^2+1\right) \left(U X^2+U+2 \Sigma X\right)}{X^2}
\end{equation}
which is identical with eqn. \eqref{KpotCardan} and $X$ is once
again the positive real root of the quartic equation
\eqref{baldoppo}.
\par
I do not discuss the explicit solution of the quartic equation
(\ref{baldoppo}) which is reported in the original paper
\cite{noietmarcovaldo}.
\item[C)] \textbf{Eguchi Hanson case $\mathcal{M}_{s,0,s}$}.
When we choose $\zeta_1=s,\, \zeta_2=0, \, \zeta_3 \, = \,  s$ the
moment map system reduces to eqs.  \eqref{ridiculite}. Performing
the substitution $X_1 = X,\, X_2= 1, \, X_3=X$ and using
eqn.~\eqref{segretusquid} the K\"ahler potential of the manifold
$\mathcal{M}_{s,0,s}$ takes the form:
  \begin{equation}\label{canebirbo}
  \mathcal{K}_{\mathcal{M}_{s,0,s}}\, =\underbrace{4\,  s \, \log [X] \,+2 \Sigma  X^2+\frac{2 \Sigma
  }{X^2}}_{\mathcal{K}_2}\, +\,4 U
  \end{equation}
What we immediately observe from eqn.~\eqref{canebirbo} is that the
K\"ahler potential is of the form:
\begin{equation}\label{cancellino}
    \mathcal{K}_{\mathcal{M}_{s,0,s}}\, =
    \,\underbrace{\mathcal{K}_2\left(Z_1,Z_2,\bar{Z}_1,\bar{Z}_2\right)}_{\text{K\"ahler potential of a two-fold}}\, + \,
    4\times|Z_3|^2
\end{equation}
Hence the manifold $\mathcal{M}_{s,0,s}$ is a direct product:
\begin{equation}\label{sapientulo}
    \mathcal{M}_{s,0,s} \, = \, \mathcal{M}_B \, \times \, \mathbb{C}
\end{equation}
It is not difficult to realize that the manifold $\mathcal{M}_B$ is
just the Eguchi-Hanson space $EH \, \equiv \, ALE_{\mathbb{Z}_2}$.
To this effect it suffices to set $s= \frac{\ell}{2}$ and rescale
the coordinates $Z^{1,2} \, \to \, \frac{\check{Z}^{1,2}}{2}$. This
implies $\Sigma = \ft 14 |\check{\mathbf{Z}}|^2$ and the K\"ahler
potential $\mathcal{K}_2\left(Z_1,Z_2,\bar{Z}_1,\bar{Z}_2\right)$
turns out to be
\begin{equation}\label{certamino}
    \mathcal{K}_2 \, = \, \sqrt{\ell^2 +|\check{\mathbf{Z}}|^4} \, +
    \, \ell \, \log
    \left[\frac{-\ell + \sqrt{\ell^2 +|\check{\mathbf{Z}}|^4}}{|\check{\mathbf{Z}}|^2}\right]
    \,
\end{equation}
which is essentially equivalent to the form of the Eguchi-Hanson
K\"ahler potential given  in eqn. (7.22) of \cite{Bruzzo:2017fwj}.
This might be already conclusive, yet for later purposes it is
convenient to consider the further development of the result
\eqref{certamino} since the known and fully computable case of the
Eguchi-Hanson space allows us to calibrate  the general formula for
the K\"ahler potential \eqref{celeberro}. To this effect recalling
the topological structure of the Eguchi Hanson space that is the
total space of the line bundle $\mathcal{O}_{\mathbb{P}^1}(-2)$ we
perform the change of variables:
\begin{equation}\label{ciucius}
    \check{Z}_1 \, = \, u \, \sqrt{v} \quad ; \quad \check{Z}_2 \, = \,\sqrt{v}
\end{equation}
where $u$ is the complex coordinate of the compact base
$\mathbb{P}^1$, while $v$ is the complex coordinate spanning the
non-compact fibre. Upon such a change the K\"ahler potential
\eqref{certamino} becomes:
\begin{equation}\label{cabacchio}
  \mathcal{K}_2 \, = \,  \sqrt{|v|^2   (|u|^2  +1)^2+\ell ^2}\,+\,  {\alpha_\zeta} \,\ell  \log \left(\frac{\sqrt{|v|^2
     (|u|^2  +1)^2+\ell ^2}-\ell }{(|u|^2  +1) \sqrt{|v|^2
    }}\right)
\end{equation}
\par
A further important information can be extracted from the present
case. Setting $v=0$ we perform the reduction to the exceptional
divisor of this partial resolution which is just the base manifold
$\mathbb{P}^1$ of Eguchi-Hanson space. The reduction of the K\"ahler
2-form to this divisor is very simple and it is the following one:
\begin{equation}\label{lampsuco}
\mathbb{K} \mid_{\mathbb{P}^1} \, = \,   \frac{\rho  \sqrt{\ell ^2}
d\rho\wedge d\theta}{\pi \left(\rho
   ^2+1\right)^2}
\end{equation}
where we have set $u=\rho \, \exp[i\,\theta]$. It follows that the
period integral of the K\"ahler 2-form on the unique homology cycle
$C_1$ of the partial resolution $\mathrm{EH}\times \mathbb{C}$ which
is the above mentioned $\mathbb{P}^1$ is:
\begin{equation}\label{califragilisti}
    \int_{C_1} \,\mathbb{K} \, = \, 2\pi \, \int_0^\infty
    \frac{\rho  \sqrt{\ell ^2}
d\rho}{\pi \left(\rho
   ^2+1\right)^2} \, = \, \sqrt{\ell ^2}
\end{equation}
Equation \eqref{califragilisti} sends us two important messages:
\begin{itemize}
  \item Whether the level parameter $s=\ell/2$ is positive or
  negative does not matter.
  \item The absolute value $|s|$ encodes the size of the homology
  cycle in the exceptional divisor. When it vanishes the homology
  cycle shrinks to a point and we have a further degeneration.
\end{itemize}
\item[D)] \textbf{Sextic case or the \textit{Kamp\'{e} manifold}\footnote{Since
it is generally stated that the roots of a general sextic equation
can be written in terms of Kamp\'{e} de  {F\'{e}riet} functions,
although explicit generic formulae are difficult to be found, we
have decided to call  $\mathcal{M}_{s,2s,s}$ the \textit{Kamp\'{e}
manifold} with the same logic that led us to name
$\mathcal{M}_{0,s,s}$ the Cardano manifold.}
 $\mathcal{M}_{s,2s,s}$}.
When we choose $\zeta_1=s,\, \zeta_2=2s, \, \zeta_3 \, = \, s$ the
moment map system reduces to eqs.  \eqref{raschiotto}. With the same
positions used there, the K\"ahler potential of the
$\mathcal{M}_{s,2s,s}$-manifold turns out to be the following one:
\begin{equation}\label{rodriguez}
   \mathcal{K}_{\mathcal{M}_{1,2,1}} \, = \,  2\, s\,
   \log (Z)\, +\, \frac{\sqrt{2}
   \sqrt{Z^3+Z} \left(\sqrt{2} U \sqrt{Z^3+Z}+2 \Sigma
   Z\right)}{Z^2}
\end{equation}
where the function $Z(\Sigma,U)$ of the complex coordinates is the
positive real root, depending on the $\Sigma,U$ region of the sextic
equation:
\begin{equation}\label{sesticina}
     2\left(Z^2+1\right) \left(s Z-U Z^2+U\right)^2-\Sigma ^2 Z
\left(Z^2-1\right)^2\, = \,0
\end{equation}
\end{description}
In the next section we discuss the geometry of the crepant
resolution of the singularity $\frac{\mathbb{C}^3}{\mathbb{Z}_4}$
utilizing toric geometry. Then we return to the formulae for the
K\"ahler potential displayed in the present section in order to see
how the K\"ahler geometry of the entire space and in particular of
the various components of the exceptional divisor is realized in the
various corners of the moduli-space. This will allow us to discuss
the Chamber Structure of this particular instance of K\"ahler
quotient resolution ${\grave{a}}$ la Kronheimer.
\par
As we are going to see, all the four cases analyzed in the present
section, the two Cardano cases, the Eguchi-Hanson case and the
Kamp\'e case correspond to partial resolutions of the orbifold
singularity and indeed they are located on walls or even on edges
where some homology cycles shrink to zero.
\section{Toric geometry description of the crepant resolution}
\label{toricresolution} As announced above in the present section we
study the full and partial resolutions of the singularity
$\mathbb{C}^3/\mathbb{Z}_4$ in terms of toric geometry. Both
resolutions turn out to be the total space of the canonical line
bundle over an algebraic surface, respectively the second Hirzebruch
surface $\mathbb F_2$ and the weighted projective plane
$\mathbb{WP}_{[1,1,2]}$. The main output of this study is provided
by two informations:
\begin{enumerate}
  \item The identification as algebraic varieties of the irreducible components of the
  exceptional divisor $\mathcal{D}_E$ introduced by the resolution.
  \item The explicit form of the atlas of coordinate patches that describe the
  resolved manifold and the coordinate transformation from the
  original $Z_i$ to the new $u,v,w$ (appropriate to each patch) that constitute
  the blowup of the singularities.
\end{enumerate}
The second information of the above list and in particular the
equation of the exceptional divisor in each patch is the main tool
that allows to connect the K\"ahler quotient description outlined in
the previous sections with the algebraic description. In particular
by this token we arrive at the determination of the K\"ahler metric
of the exceptional divisor components induced by the Kronheimer
construction.
\par
The material contained in the following subsections, which is taken
from the original paper \cite{noietmarcovaldo}, involves advanced
concepts of algebraic geometry and might be hard to be grasped by
those readers who are deprived of sufficient background knowledge in
that field. These readers  can skip all the details of the proofs
and of the arguments and just digest the information mentioned in
the above two points. The complete resolution of the singularity
turns out to be the total space of the canonical bundle of the
second Hirzebruch surface:
\begin{equation}\label{risolutissimo}
    Y^{\mathbb{Z}_4}_{[3]} \, = \, \operatorname{tot}\left(K\left[\mathbb{F}_2\right]\right)
\end{equation}
An atlas of coordinate patches on $Y^{\mathbb{Z}_4}_{[3]}$ is
provided in eqn.s(\eqref{settorio},\eqref{subberulle}) torically
characterized in tables \ref{coordinates} and \ref{changes}.
\par
Furthermore the partial resolution
${\tilde{Y}}^{\mathbb{Z}_4}_{[3]}$ that occurs on certain walls is
the total space of the canonical bundle on the weighted projective
space $\mathbb{WP}_{1,1,2}$:
\begin{equation}\label{risolutino}
    {\tilde{Y}}^{\mathbb{Z}_4}_{[3]} \, = \, \operatorname{tot}\left(K\left[\mathbb{WP}_{1,1,2}\right]\right)
\end{equation}
\subsection{The singular variety $ Y^{\mathbb{Z}_4}_0=\mathbb C^3/\mathbb Z_4$}
We shall denote by $(x,y,z)$ the coordinates of $\mathbb{C}^3$, by
$\{\mathbf e_i\}$ the standard basis of $\mathbb R ^3$ and by $\{
\epsilon^i\}$ the dual basis. The action of $\mathbb Z_4$ is given
by
$$(x,y,z) \mapsto (\omega x, \omega y, \omega^2 z )$$
with $\omega^4=1$. It is easy to find a basis for the space of
invariant Laurent polynomials:
\begin{equation}\label{formulauno}
    \mathcal{I}_1 \, = \,x \, y^{-1}, \quad \mathcal{I}_2 \, =
    \,y^2\,z^{-1}, \quad \mathcal{I}_3 \, = \, z^2
\end{equation}
so that the three vectors
$$ \mathbf u^1 = \epsilon^1 -  \epsilon^2, \qquad
\mathbf u^2 =2 \epsilon^2 -  \epsilon^3, \qquad \mathbf u^3 =
2\epsilon^3
$$
generate the lattice $M$ of invariants, which is a sublattice of the
standard (dual) lattice $M_0$. The lattice $N$ dual to $M$ is a
superlattice of the standard lattice $N_0$, and is generated by the
vectors
\begin{equation}\label{doppievu}
\mathbf w_1 = \mathbf e_1, \qquad \mathbf w_2 =\tfrac12 \mathbf
e_1+\tfrac12 \mathbf e_2, \qquad \mathbf w_3 =\tfrac14 \mathbf
e_1+\tfrac14 \mathbf e_2+\tfrac12 \mathbf e_3.
\end{equation}
The generators of the rays giving the cone associated with the
variety $Y^{\mathbb{Z}_4}_0$ are obtained by inverting these
relations, i.e.,
$$\mathbf v_1 = \mathbf w_1, \qquad
\mathbf v_2 = 2  \mathbf w_2 - \mathbf w_1, \qquad \mathbf v_3 = -
\mathbf w_2+2  \mathbf w_3.
$$
From now on, unless differently stated,  coordinate expressions will
always refer to this basis $\{\mathbf w_i\}$ of $N$. So the rays of
the fan $\Sigma_0$ of $Y^{\mathbb{Z}_4}_0$ are
$$ \mathbf v_1 = (1,0,0),\qquad \mathbf v_2=(-1,2,0),\qquad \mathbf v_3=(0,-1,2)$$
They do not form a basis of $N$, according to the fact that
$Y^{\mathbb{Z}_4}_0$ is singular (note indeed that that
$N/\sum_i\mathbb{Z}\mathbf v_i=\mathbb{Z}_4$).
\par
\subsection{The full resolution $Y^{\mathbb{Z}_4}_{[3]}$ of  $Y^{\mathbb{Z}_4}_0=\mathbb C^3/\mathbb Z_4$}
In this section we study the full (smooth) resolution
$Y^{\mathbb{Z}_4}_{[3]}$ of the singular quotient
$Y^{\mathbb{Z}_4}_0$, describing its torus-invariant divisors and
curves and the natural coordinate systems on its affine patches.
\par
\subsubsection{The fan}
The  fan of $Y^{\mathbb{Z}_4}_{[3]}$ is obtained by adding to the
fan $\Sigma_0$ the rays generated by the lattice points lying on the
triangle with vertices $\{\mathbf v_i\}$. These are
$$\mathbf w_2 = (0,1,0),\qquad \mathbf w_3 = (0,0,1).$$
The  torus invariant divisors corresponding to the two new rays of
the  fan are the components of the exceptional divisor. Since
$\mathbf w_3$ is in the interior of the triangle, the corresponding
component of the exceptional divisor is compact, while the component
corresponding to $\mathbf w_2$, which lies on the border,  is
noncompact. Note indeed that, according to the equations
\eqref{doppievu}, $\mathbf w_2$ and $\mathbf w_3$ correspond to the
junior classes $\frac12(1,1,0)$ (noncompact) and $\frac14(1,1,2)$
(compact) associated with the given representation of $\mathbb Z_4$.
\par
\begin{figure}
\begin{center}
\begin{tikzpicture}[scale=1.50]
\draw [thick,->] (0,0) -- (-0.5,1.5) ; \draw [thick,->] (0,0) --
(1.5,0.5) ; \draw [thick,->] (0,0) -- (-0.8,-0.8) ; \draw
[thick,->,green] (0,0) -- (1,-0.2); \node at
(1.2,-0.4){
{\footnotesize $\mathbf w_2= (0,1,0)$}}; \node at
(-1,-1) {\footnotesize$\mathbf v_1=(1,0,0)$}; \node at (2.7,0.5)
{\footnotesize$\mathbf v_2=(-1,2,0)$}; \node at (-0.8,1.7)
{\footnotesize$\mathbf v_3=(0,-1,2)$}; \draw [thick,->,green] (0,0)
-- (0,1); \node at (1,1.1){
{\footnotesize $\mathbf w_3=
(0,0,1)$}}; \draw [fill] (0,0) circle (1.5pt) ;
\end{tikzpicture}\hskip1cm
\begin{tikzpicture}[scale=0.50]
  \path [fill=pink] (0,0) to (4,2.3) to  (4,6.9) to (0,0) ;
    \path [fill=yellow] (8,0) to (4,2.3) to  (4,6.9) to (8,0) ;
      \path [fill=green] (0,0) to (4,2.3)  to  (4,0) to (0,0);
      \path [fill=brown] (4,0) to (4,2.3)  to  (8,0) to (4,0);
  \draw [fill] (0,0) circle (3pt);
    \draw [fill]  (8,0) circle (3pt);
      \draw [fill] (4,6.9) circle (3pt);
        \draw [fill] (4,2.3) circle (3pt);
\draw (0,0) -- (8,0); \draw (0,0) -- (4,6.9); \draw (8,0) --
(4,6.9); \draw (0,0) -- (4,2.3); \draw (8,0) -- (4,2.3); \draw
(4,6.9) -- (4,2.3); \node at (-0.5,0) {$\mathbf v_1$}; \node at
(8.6,0) {$\mathbf v_2$}; \node at (4.2,7.4) {$\mathbf v_3$}; \node
at
(4.7,2.6) {$\mathbf w_3$}; \node at (3,3) {$\sigma_4$}; \node at
(5,3.7) {$\sigma_3$}; \node at (4,-0.5) {$\mathbf w_2$}; \draw
[fill]  (4,0) circle (3pt); \draw (4,0) -- (4,4); \node at (5,1)
{$\sigma_2$}; \node at (3,1) {$\sigma_1$};
\end{tikzpicture}
\caption{\label{SigmaY}  \small The fan $\Sigma_Y$ of the resolution
$Y^{\mathbb{Z}_4}_{[3]}$ and the associated planar graph}
\end{center}
\end{figure}
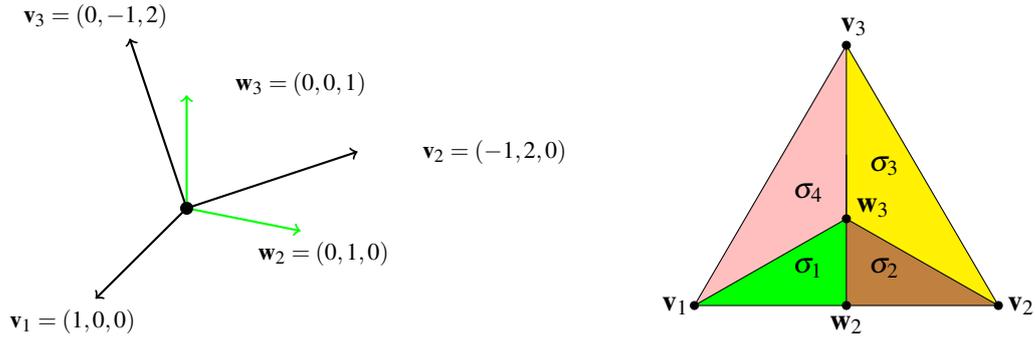
\begin{figure}
\centering
\includegraphics[height=8cm]{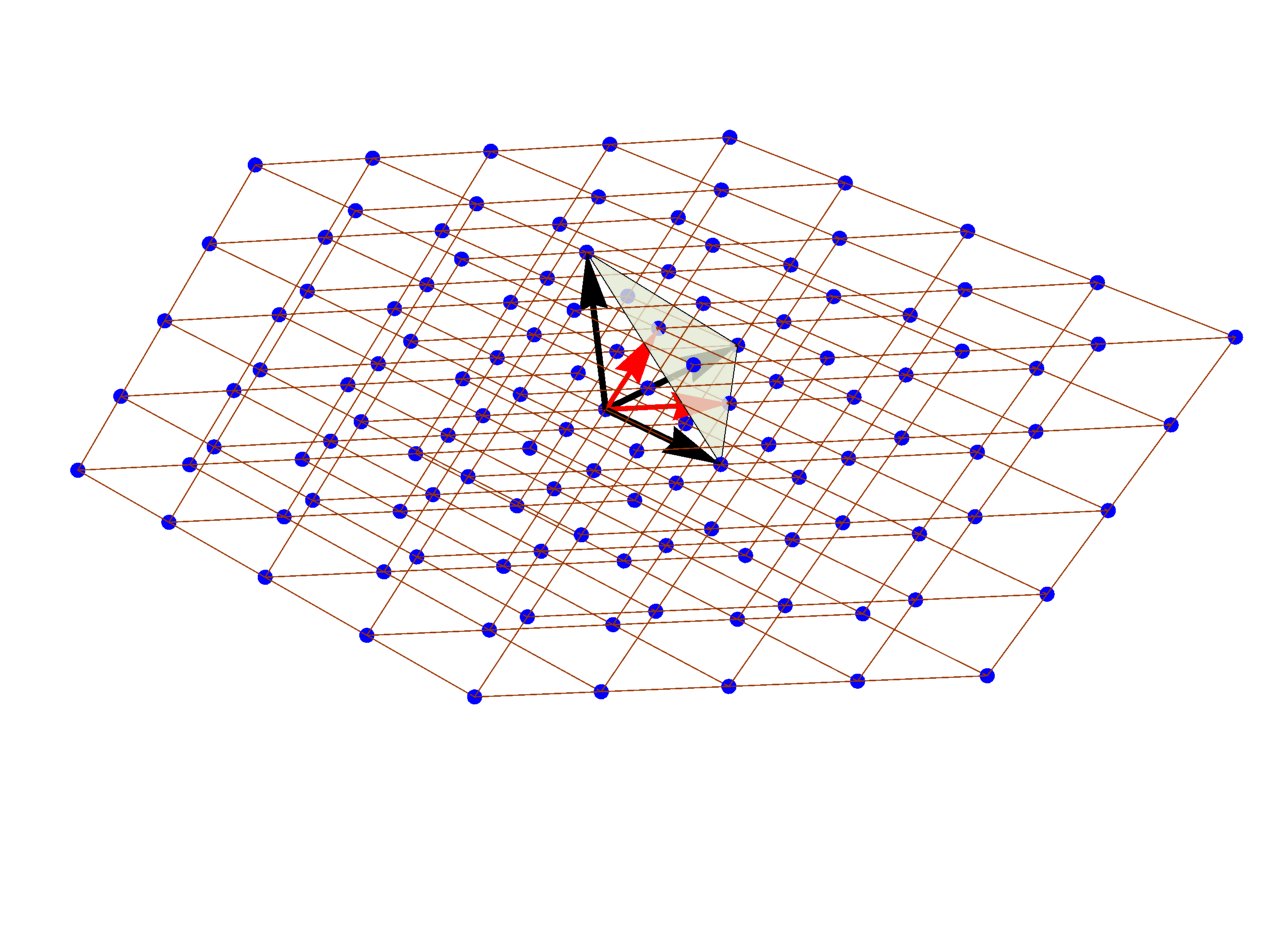}
\vskip -2cm \caption{ \label{z4trian} The figure displays a finite
portion of the lattice $N$ dual {to} the lattice $M$ of
$\mathbb{Z}_4$-invariants and the generators of the cone $\sigma$
describing the singular quotient  $\mathbb{C}^3/\mathbb Z_4$ marked
as fat dark arrows. The extremal points of the three generators
single out a triangle, which  intersects  the lattice $N$ in two
additional points, namely the extremal points of the vector $\mathbf
w_3$  and of the vector $\mathbf w_2$, marked as lighter arrows in
the figure. These vectors have to be added to the fan and divide
the original cone into four maximal cones, corresponding to as many
open charts of the resolved smooth toric variety.}
\end{figure}

We shall denote by $Y^{\mathbb{Z}_4}_{[3]}$ this resolution of
singularities. Figure \ref{SigmaY} shows the fan of
$Y^{\mathbb{Z}_4}_{[3]}$ and the associated planar graph. The planar
graph is obtained by projecting the generators of the rays onto the
triangle formed by the three original vertices; this is shown in a
3-dimensional perspective in Figure \ref{z4trian}.
\par
One can explicitly check that all cones of $\Sigma_Y$ are smooth, so
that $Y^{\mathbb{Z}_4}_{[3]}$ is indeed smooth. Note that all cones
of $\Sigma_Y$ are contained in the cones of $\Sigma_0$, which
corresponds to the existence of a morphism $Y^{\mathbb{Z}_4}_{[3]}
\to Y^{\mathbb{Z}_4}_0$.
\par
If we first make the blowup corresponding to the ray $\mathbf w_3$,
i.e., to the junior class $\frac14(1,1,2)$, according to the general
theory the exceptional divisor is a copy of the weighted projective
plane $\mathbb{WP}_{[1,1,2]}$. When we make the second blowup, i.e.
we blow up the $z$ axis, we also blowup $\mathbb{WP}_{[1,1,2]}$ at
its singular point, so that the compact component of the exceptional
divisor of the resolution of $Y^{\mathbb{Z}_4}_0$ is a copy of the
second Hirzebruch surface $\mathbb F_2$. Moreover, the noncompact
component of the exceptional divisor is isomorphic to $\mathbb
P^1\times\mathbb C$ (which, by the way, turns out to be the weighted
projective space $\mathbb{WP}_{[1,1,0]}$). This will be shown in
more detail in the next sections   (in particular, the compact
exceptional divisor will be characterized as the Hirzebruch surface
$\mathbb F_2$ by computing its fan).
\par
By general theory \cite{itoriddo} (see also \cite{Bruzzo:2017fwj})
we know
$$h_2(Y^{\mathbb{Z}_4}_{[3]},\mathbb{Q}) = 2, \qquad h^2(Y^{\mathbb{Z}_4}_{[3]},\mathbb{Q})=2,
\qquad h^2_c(Y^{\mathbb{Z}_4}_{[3]},\mathbb{Q})=1, \qquad h^4(Y^{\mathbb{Z}_4}_{[3]},\mathbb{Q})=1.$$
\par
\par
\subsubsection{Divisors}  
\begin{figure}
\begin{center}
\begin{tikzpicture}
\draw [thick,->] (0,0) -- (1,0) ;  \node at (1.2,0) {\small $E_3$};
\draw [thick,->] (0,0) -- (-1,2) ;  \node at (-1.1,2.2) {\small
$E_1$}; \draw [thick,->] (0,0) -- (0,-1) ;  \node at (0,-1.4)
{\small $E_4$}; \draw [thick,->] (0,0) -- (0,1);  \node at (0,1.2)
{\small $E_2$};
  \draw [fill] (0,0) circle (1.5pt) ;
  \node at (0,-2) {$D_c$};
  \node at (0.5,0.5) {$\tau_2$} ;
    \node at (0.5,-0.5) {$\tau_3$} ;
        \node at (-0.7,-0.2) {$\tau_4$} ;
            \node at (-0.4,1.5) {$\tau_1$} ;
\end{tikzpicture}
\hskip5mm
\begin{tikzpicture}
\draw [thick,->] (0,0) -- (1,0) ; \draw [thick,->] (0,0) -- (-1,0) ;
\draw [thick,->] (0,0) -- (0,1);
  \draw [fill] (0,0) circle (1.5pt) ;
  \node at (0,-1.3) {$D_{nc}$};
\end{tikzpicture}
\hskip5mm
\begin{tikzpicture}
\draw [thick,->] (0,0) -- (1,0) ; \draw [thick,->] (0,0) -- (-1,2) ;
\draw [thick,->] (0,0) -- (0,1);
  \draw [fill] (0,0) circle (1.5pt) ;
  \node at (0,-1.3) {$D_{EH}$, $D'_{EH}$};
\end{tikzpicture}
\hskip5mm
\begin{tikzpicture}
\draw [thick,->] (0,0) -- (1,0) ; \draw [thick,->] (0,0) -- (-1,4) ;
\draw [thick,->] (0,0) -- (0,1);
  \draw [fill] (0,0) circle (1.5pt) ;
  \node at (0,-1.3) {$D_4$};
\end{tikzpicture}
\caption{\label{DivY}  \small The fans of the 5 toric divisors of
$Y^{\mathbb{Z}_4}_{[3]}$. In the fan of $D_c$ we have labelled the
rays with the names of the corresponding divisors; the $\tau_i$'s
are the maximal cones. }
\end{center}
\end{figure}
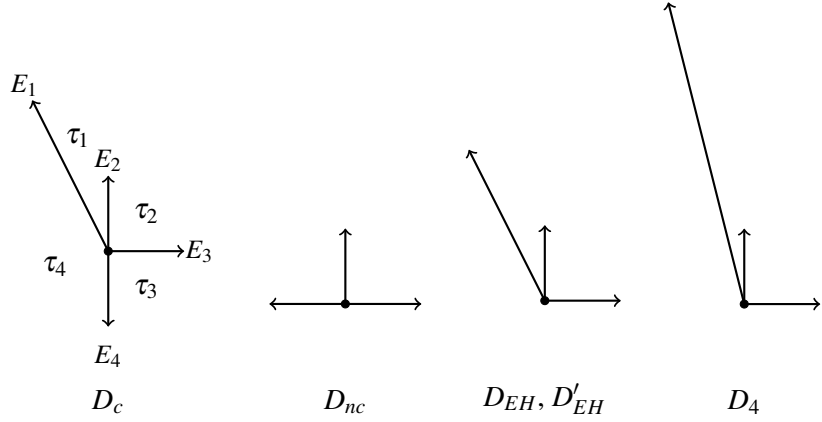

We analyze the toric divisors of $Y^{\mathbb{Z}_4}_{[3]}$; they are summarized in Table \ref{tableDivY}.
Each of these is associated with a ray of the fan $\Sigma_Y$. The
divisors corresponding to $\mathbf w_3$, $\mathbf w_2$, $\mathbf
v_1$, $\mathbf v_3$, $\mathbf v_2$ will be denoted $D_c$, $D_{nc}$,
$D_{EH}$, $D_{4}$, $D'_{EH}$ respectively. Since
$Y^{\mathbb{Z}_4}_{[3]}$ is smooth all of them are Cartier. Table
\ref{divisors} shows the fans of these divisors and what variety
they are as intrinsic varieties. The fans are depicted in Figure
\ref{DivY}.\footnote{In the cases when the ray associated to the
divisor is not a coordinate axis we made a change of basis.} The fan
of $D_c$ is generated by the rays $\mathbf v_2$, $\mathbf w_2$,
$\mathbf v_1$, $\mathbf v_3$, which shows that $D_c$ is the second
Hirzebruch surface $\mathbb F_2$.  The corresponding curves in $D_c$
have been denoted $E_1$, $E_2$, $E_3$, $E_4$ respectively. From the
self-intersections of these curves (in $D_c$)
$$E_1^2=0, \qquad   E_2^2 = -2,\qquad  E_3^2 = 0, \qquad  E_4^2 = 2  $$
we see that $E_2$ is the section of $\mathbb F_2 \to \mathbb P^1$
which squares to $-2$, i.e., the exceptional divisor of the blowup
$\mathbb F_2 \to \mathbb{WP}_{[1,1,2]}$, while $E_4$ is the section
that squares to 2, and $E_1$, $E_3$ are the toric fibres of $\mathbb
F_2 \to \mathbb P^1$.
\begin{table}[ht]
\renewcommand{\arraystretch}{1.50}
\caption{Toric Divisors in $Y^{\mathbb{Z}_4}_{[3]}$. The last column
shows the components of the divisor class on the basis given by
$(D_{nc}, D_c)$.   The variety $\mbox{ALE}_{A_1}$  is the
Eguchi-Hanson space, the crepant resolution of the singular space
$\mathbb C^2/\mathbb{Z}_2$. \label{tableDivY}} \vskip10pt
\centering 
\begin{tabular}{|c |c| c| c| c |} 
 \hline 
Ray & Divisor   & Fan  & Variety & Components \\ [1ex] 
\hline 
$\mathbf w_3$ & $D_c$  & \small $(1,0),\ (-1,2),\ (0,-1),\ (0,1)$ & $\mathbb F_2$ & $ (0,1)$\\   \hline 
$\mathbf w_2$ & $D_{nc}$ & \small  $(1,0),\ (-1,0),\ (0,1)$ &
$\mathbb P^1\times\mathbb C$ & $(1,0)$ \\  \hline $\mathbf v_1$ &
$D_{EH}$  & $ (1,0),\ (-1,2),\ (0,1) $&  $\mbox{ALE}_{A_1}$ &
$(-\frac12,-\frac14)$  \\  \hline $\mathbf v_3$ & $D_{4}$ &  $
(1,0),\ (-1,4),\ (0,1) $ & \small tot  $(\mathcal O(-4) \to \mathbb
P^1$) & $(0,-\frac12)$\\  \hline
$\mathbf v_2$ & $D'_{EH}$ & $ (1,0),\ (-1,2),\ (0,1) $ &  $\mbox{ALE}_{A_1}$  & $(-\frac12,-\frac14)$ \\
  [1ex] 
\hline 
\end{tabular}
\label{divisors} 
\end{table}
\par
Among the 5 divisors only 2 are independent in cohomology.  We
consider  the exact sequence:
\begin{equation}\label{Pic}
0 \to M \xrightarrow{A} \operatorname{Div}_{\mathbb
T}(Y^{\mathbb{Z}_4}_{[3]}) \xrightarrow{B}
\operatorname{Pic}(Y^{\mathbb{Z}_4}_{[3]}) \to  0
\end{equation}
where $M$ is the dual lattice, $ \operatorname{Div}_{\mathbb
T}(Y^{\mathbb{Z}_4}_{[3]}) $ is the group of torus-invariant
divisors, and $\operatorname{Pic}(Y^{\mathbb{Z}_4}_{[3]}) $ is the
Picard group\footnote{The Picard group $\operatorname{Pic}(X)$ of a
complex variety $X$ is the group of isomorphism classes of
holomorphic line bundles on $X$. Using \v Cech cohomology it can be
represented as the cohomology group $H^1(X,\cO_X^\ast)$, where $
\cO_X^\ast$ is the sheaf of nowhere vanishing holomorphic functions
on $X$.} of the full resolution $Y^{\mathbb{Z}_4}_{[3]}$. The
morphism $B$ simply takes the class of a divisor in the Picard
group, while for every $m\in M$, $A(m)$ is the divisor associated
with the rational function defined by $m$. Moreover we know that the
classes of the divisors $D_{nc}$ and $D_{c}$ generate
$\operatorname{Pic}(Y^{\mathbb{Z}_4}_{[3]}) $ over $\mathbb{Q}$
\cite{itoriddo}. With that choice of basis in
$\operatorname{Pic}(Y^{\mathbb{Z}_4}_{[3]}) \otimes\mathbb{Q}$, with
the basis given by the 5 divisors in $\operatorname{Div}_{\mathbb
T}(Y^{\mathbb{Z}_4}_{[3]}) \otimes\mathbb{Q}$, and the basis in
$M\otimes\mathbb{Q}$ given by the duals of the $\{\mathbf w_i\}$,
the morphisms $A$ and $B$ are represented over the rationals by the
matrices
$$ A =\begin{pmatrix} 1 & 0 & 0 \\ -1 & 2 & 0 \\ 0 & -1 & 2 \\ 0 & 1 & 0 \\ 0 & 0 & 1 \end{pmatrix},\qquad
B =\begin{pmatrix}  -\tfrac12 &  -\tfrac12 & 0 & 1 & 0 \\
 -\tfrac14 &  -\tfrac14 &  -\tfrac12 & 0 & 1
\end{pmatrix}
$$
We deduce that the relations among the classes of the 5 toric
divisors in the Picard group are\footnote{The notation $[D]$ means
the class in the Picard group of the line bundle $\cO_X(D)$.}
\begin{gather} [D_{EH}]  =  [D'_{EH}]=  -\tfrac12 [D_{nc}] - \tfrac14 [D_c] \label{Div1} \\
{[}D_{4} ]   =  -\tfrac12 [D_c] \label{eqDiv2}
\end{gather}
Since the canonical divisor can be written as minus the sum of the
torus-invariant divisors, one has
$$ \left[ K_{Y^{\mathbb{Z}_4}_{[3]}} \right] = - [D_{nc}] - [D_c] - [D_{EH}] - [D'_{EH}] - [D_4] = 0 $$
consistently with the fact that the resolution
$Y^{\mathbb{Z}_4}_{[3]} \to Y^{\mathbb{Z}_4}_0$ is crepant. (Note
that $\operatorname{Pic}(Y^{\mathbb{Z}_4}_{[3]})$ is free over
$\mathbb{Z}$, so that the equality $ \left[
K_{Y^{\mathbb{Z}_4}_{[3]}} \right] =0$ in
$\operatorname{Pic}(Y^{\mathbb{Z}_4}_{[3]}) \otimes\mathbb{Q}$ also
implies that $ \left[ K_{Y^{\mathbb{Z}_4}_{[3]}} \right] =0$  in
$\operatorname{Pic}(Y^{\mathbb{Z}_4}_{[3]})$).
\par
The matrix $B$ can also be chosen as
$$ B = \begin{pmatrix}  1 & 1 & 0 & -2 & 0 \\
0 & 0 &  1 & 1 & -2 \end{pmatrix} $$ which corresponds to taking the
classes of $D_{EH}$ and $D_4$ as basis of the Picard group. In this
way the matrix $B$ is integral, which means that $D_{EH}$ and $D_4$
generate $\operatorname{Pic}(Y^{\mathbb{Z}_4}_{[3]})$ over the
integers.
\par
\subsubsection{Toric curves and intersections}  \label{curvesY}
The planar graph in Figure \ref{SigmaY} shows that
$Y^{\mathbb{Z}_4}_{[3]}$ has 4 compact toric curves, corresponding
to the inner edges of the graph. The intersection between a smooth
irreducible curve $C$ and a Cartier divisor $D$ is defined as
$$ C\cdot D = \deg (f^\ast \cO_Y(D))$$
where $f\colon C \to Y^{\mathbb{Z}_4}_{[3]}$ is the embedding, and
$\cO_Y(D)$ is the line bundle associated to the divisor
\footnote{This also allows one to compute the intersection between a
curve $C$ and a Weil divisor $D$. Indeed simplicial toric varieties
are $\mathbb{Q}$-factorial, i.e., every Weil divisor has a multiple
that is Cartier. So if $mD$ is Cartier, one defines
$$ C \cdot D = \tfrac1m \,C\cdot(mD).$$ This may be a rational number. \label{nota}}.
Inspection of the fan allows one to detect when the intersection
is transversal (in which case the intersection mumber is 1), empty
(intersection number 0), or the curve is inside the divisor. The
intersections are shown in Table \ref{tableintersec}.
\par
\begin{table}[ht]
\caption{Intersections among the   toric curves and divisors in the
full resolution $Y^{\mathbb{Z}_4}_{[3]}$. For the curves that are
inside $D_c$ the last two columns also show the identifications with
the curves corresponding to the rays in the fan of $D_c$ of Figure
\ref{DivY}, and what they are inside the second Hirzebruch surface.
Note that $C_1$ is the intersection between the two components of
the exceptional divisor. The basis of the fibration $D_c \to \mathbb
P^1$ may be identified with $C_1$, while $C_2$, $C_4$ are the fibres
over two toric points, which correspond to the cones $\sigma_1$ and
$\sigma_2$.
\label{tableintersec}} 
\vskip10pt
\centering 
\begin{tabular}{|c|c |c| c| c| c| c| c| c| } 
 \hline 
 Edge/face & Curve & $D_c$ & $D_{nc}$   & $D_{EH}$   & $D'_{EH}$ & $D_4$ & Inside $D_c$ &   \\ \hline
 $(\mathbf w_2\mathbf w_3)$ & $C_1$ & 0 & - 2 & 1 & 1 & 0 & $ E_2$ & $-2$-section \\ \hline
 $(\mathbf v_1\mathbf w_3)$ & $C_2$ & -2 & 1 & 0 & 0 & 1 & $E_3$ & fibre \\ \hline
  $(\mathbf v_2\mathbf w_3)$ & $C_4$ & -2 & 1 & 0 & 0 & 1 & $E_1$ & fibre \\ \hline
   $(\mathbf v_3\mathbf w_3)$ & $C_5$ & -4 & 0 & 1 & 1 & 2 & $E_4$ & 2-section\\ \hline
    $(\mathbf v_1\mathbf w_2)$ & $C_3$ & 1 & 0 &  0  & 0 &  0 & &  \\
 \hline 
\end{tabular}
\label{intersections} 
\end{table}
\par
The intersection numbers of $D_{EH}$ and $D_4$ with $C_1$, $C_2$
show that the latter are a basis of
$H_2(Y^{\mathbb{Z}_4}_{[3]},\mathbb{Z})$ dual to
$\{[D_4],[D_{EH}]\}$.
\par
\subsubsection{Coordinate systems and curves}\label{coorcurves}
The four 3-dimensional cones in the fan of $Y^{\mathbb{Z}_4}_{[3]}$
correspond to four affine open varieties, and since all cones are
smooth (basic), they are copies of $\mathbb C^2$. The variables
attached to the rays generating a cone provide a coordinate system
on the corresponding affine set. A face between two 3-dimensional
cones corresponds to the intersection between the two open sets.
Note that all charts have a common intersection, as they all contain
the 3-dimensional torus corresponding to the origin of the fan.
Table \ref{coordinates} shows the association among cones, rays,
coordinates and coordinate expressions of toric  curves. Below we
provide a list of  {coordinate systems}, with all transition
functions between them, and the expressions of the toric curves in
the coordinate systems of the charts they belong to. Table
\ref{changes} displays the coordinate transformations among the four
coordinate systems. We have denoted $C_i$, $i=1\dots 5$ as before,
and moreover $C_6$, $C_7$, $C_8$ are the noncompact toric curves in
the charts 2, 3 and 4 (analogously, $C_3$ was the noncompact curve
in the chart 1). The column ``Dual gen.'' displays the generators of
the dual cone.
\par
In each chart, the coordinates $(u,v,w)$ are related to the
invariants  \eqref{formulauno} as follows:
\begin{equation}\label{settorio}
    \begin{array}{lclcrcccccl}
\left\{u,v,w\right\}_1 &=&\{&\frac{x}{y} &,& \frac{y^2}{z} &,&
   z^2&\} \\
\left\{u,v,w\right\}_2 &=&\{&\frac{y}{x} &,& \frac{x^2}{z} &,&
   z^2&\} \\
\left\{u,v,w\right\}_3&=&\{& \frac{y}{x} &,& \frac{z}{x^2} &,&
   x^4 &\}\\
\left\{u,v,w\right\}_4 &=&\{&\frac{x}{y} &,& y^4 &,&
   \frac{z}{y^2}&\} \\
\end{array}
\end{equation}
Eqs. ~\eqref{settorio} can be easily inverted and one obtains:
\begin{equation}\label{subberulle}
\begin{array}{clclcl}
\mbox{Chart } X_{\sigma_1}& x\to u \sqrt{v} \sqrt[4]{w} &,&
   y\to \sqrt{v} \sqrt[4]{w} &,&
   z\to \sqrt{w} \\
\mbox{Chart } X_{\sigma_2}& x\to \sqrt{v} \sqrt[4]{w} &,&
   y\to u \sqrt{v} \sqrt[4]{w}
   &,& z\to \sqrt{w} \\
\mbox{Chart } X_{\sigma_3}& x\to \sqrt[4]{w} &,& y\to u
   \sqrt[4]{w} &,& z\to v
   \sqrt{w} \\
\mbox{Chart } X_{\sigma_4}& x\to u \sqrt[4]{v} &,& y\to
   \sqrt[4]{v} &,& z\to \sqrt{v}
   w \\
\end{array}
\end{equation}
The irrational coordinate transformations \eqref{subberulle} derived
from the toric construction are the essential tool to relate the
results of the K\"ahler quotient construction with the geometry of
the exceptional divisor as identified by the toric resolution of the
singularity.
The coordinates $x,y,z$ in the above equation are to be identified
with the $Z^{1,2,3}$ that parameterize the locus $L_{\mathbb{Z}_4}$
composed by the matrices $A_0,B_0,C_0$ of eqn.~\eqref{baldovinus}.
As we know  this locus is lifted to the resolved variety
$Y^{\mathbb{Z}_4}_{[3]}$ by the action of the quiver group element
$\exp [\pmb{\Phi}]$, whose corresponding Lie algebra element
$\pmb{\Phi}$ satisfies the moment map equations \eqref{sakerdivoli}.
\newcommand{\eps}[1]{\epsilon_{#1}}
\begin{table}[ht] \small
\caption{For each cone the table assigns a name to the coordinates
associated to the rays. The third column lists the generators of the
dual cones. The $\epsilon$'s here are the dual basis to the $\mathbf
w$. We also write the equations of the toric curves in these
coordinates.} \vskip10pt
\centering 
\begin{tabular}{|c|c |c| c| c| } 
 \hline 
 Cone & Rays & Dual gen. & Coordinates & Curves \\ \hline
 $\sigma_1$ &$ \mathbf v_1  \mathbf w_2  \mathbf  w_3 $ & $\epsilon_1,\epsilon_2,\epsilon_3 $ &
 $u_1,v_1,w_1$ & $C_1: v_1=w_1=0, \ C_2: u_1=w_1=0,\ C_3 = u_1 = v_1 =0 $\\ \hline
 $\sigma_2$ & $ \mathbf v_2  \mathbf w_2  \mathbf  w_3 $  &$ -\eps1,\eps2+2\eps1,\eps3$ &
 $u_2,v_2,w_2$ & $ C_1: v_2=w_2=0, \  C_4: u_2=v_2=0, \ C_6 : u_2 = w_2=0 $\\
 \hline
 $\sigma_3$ &  $ \mathbf v_2  \mathbf v_3  \mathbf  w_3 $  &$-\eps1,-2\eps1-\eps2,\eps3+2\eps2+4\eps1$ &
 $u_3,v_3,w_3$ & $C_4: u_3=w_3=0, \ C_5: v_3=w_3=0, C_7: u_3=v_3=0 $ \\
 \hline
 $\sigma_4$ & $ \mathbf v_1  \mathbf v_3  \mathbf  w_3 $ &$\eps1,2\eps2+\eps3,-\eps2$  &
 $u_4,v_4,w_4$ & $C_2: u_4=w_4=0,\ C_5: v_4=w_4=0, C_8: u_4=v_4=0$ \\
 \hline
\end{tabular}
\label{coordinates} 
\end{table}
\begin{table}[ht] \footnotesize
\hskip-2mm\parbox{\textwidth}{ \caption{Coordinate changes among the
charts described in Table \ref{coordinates}} \vskip10pt
\centering 
\begin{tabular}{|c|c |c| c| c| } 
 \hline 
 & $\sigma_1$  & $\sigma_2$ & $\sigma_3$ & $\sigma_4$ \\ \hline
$\sigma_1$ & id  & $ u_2=\frac1{u_1}, v_2 = u_1^2v_1, w_2=w_1 $  & $
u_3=\frac1{u_1}, v_3 = \frac{1}{u_1^2v_1}, w_3 = u_1v_1^2w_1 $ &
$u_4=u_1, \ v_4 = v_1^2w_1,\ w_4=\frac1{v_1}$ \\ \hline $\sigma_2$ &
$u_1 = \frac1{u_2},v_1=u_2^2v_2,w_1=w_2 $ & id &
$u_3=u_2,v_3=\frac1{v_2},w_3=v_2^2w_2$ &
$u_4=\frac1{u_2},v_4=u_2^4v_2^2w_2,w_4=\frac1{u_2^2v_2}$\\ \hline
$\sigma_3$ &$ u_1=\frac1{u_3},v_1=\frac{u_3^2}{v_3},w_1=v_3^2w_3 $ &
$u_2=u_3,v_2=\frac1{v_3},w_2={v_3^2}{w_3}$& id &
$u_4=\frac1{u_3},v_4=u_3^4w_3,w_4=\frac{v_3}{u_3^2}$\\ \hline
$\sigma_4$ & $ u_1=u_4,v_1=\frac1{w_4},w_1=v_4w_4^2$&
$u_2=\frac1{u_4},v_2=\frac{u_2^4}{w_4},w_2=v_4w_4^2$&$u_3=\frac1{u_4},v_3=
\frac{w_4}{u_4^2},w_3=u_4^4v_4$& id \\ \hline
\end{tabular}
\label{changes} 
}
\end{table}
\subsubsection{$Y^{Z_4}_{[3]}$ as a line bundle on $\mathbb F_2$}\label{Ylinebundle}
The full resolution $Y^{\mathbb{Z}_4}_{[3]}$ is the total space of
the canonical bundle of the second Hirzebruch surface $\mathbb F_2$;
this is quite clear from the blowup procedure. Here we give a toric
description of this fact. The canonical bundle of $\mathbb F_2$ is
the line bundle $\cO_{\mathbb F_2}(-2H)$, where $H$ is the section
of $\mathbb F_2\to\mathbb P^1$ squaring to 2. We regard $H$ as the
toric divisor $E_4$, see Figure \ref{DivY}. To each of the cones
$\tau_i$ of the fan of $\mathbb F_2$ one associates a 3-dimensional
cone $\tilde\sigma_i$, obtaining
\begin{eqnarray*}
\tilde\sigma_1 &=& \mbox{Cone}((0,0,1),(0,1,0),(1,0,0)) \\
\tilde\sigma_2 &=& \mbox{Cone}((0,0,1),(-1,2,0),(0,1,0)) \\
\tilde\sigma_3 &=& \mbox{Cone}((0,0,1),(0,-1,2),(-1,2,0)) \\
\tilde\sigma_4 &=& \mbox{Cone}((0,0,1),(1,0,0),(0,-1,2))
\end{eqnarray*}
This is the fan of $Y^{Z_4}_{[3]}$. So, $ \mbox{tot}(\cO_{\mathbb
F_2}(-2H)) \simeq Y^{\mathbb{Z}_4}_{[3]}$, i.e.,
$Y^{\mathbb{Z}_4}_{[3]}$ is the total space of the canonical bundle
of $\mathbb F_2$. Thus, the canonical bundle of
$Y^{\mathbb{Z}_4}_{[3]}$ is trivial.
\subsection{The partial resolution $\tilde{Y}^{\mathbb{Z}_4}_{[3]}$} \label{Y3sezia}
The full resolution $Y^{\mathbb{Z}_4}_{[3]}$ is obtained by adding
two rays to the fan of $Y^{\mathbb{Z}_4}_0=\mathbb
C^3/\mathbb{Z}_4$. If we add just one we obtain a partial
resolution. Here we examine the partial resolution that will occur
in correspondence of some walls of the stability parameters space.
\subsubsection{The fan}
\begin{figure}
\begin{center}
\begin{tikzpicture}[scale=1.50]
\draw [thick,->] (0,0) -- (-0.5,1.5) ; \draw [thick,->] (0,0) --
(1.5,0.5) ; \draw [thick,->] (0,0) -- (-0.8,-0.8) ; \draw
[thick,->,green] (0,0) -- (0,1); \node at (1,1.1){
{\footnotesize
$\mathbf w_3= (0,0,1)$}}; \node at (-1,-1) {\footnotesize$\mathbf
v_1=(1,0,0)$}; \node at (2.7,0.5) {\footnotesize$\mathbf
v_2=(-1,2,0)$}; \node at (-0.8,1.7) {\footnotesize$\mathbf
v_3=(0,-1,2)$}; \draw [fill] (0,0) circle (1.5pt) ;
\end{tikzpicture}\hskip1cm
\begin{tikzpicture}[scale=0.50]
  \path [fill=pink] (0,0) to (4,2.3) to  (4,6.9) to (0,0) ;
    \path [fill=yellow] (8,0) to (4,2.3) to  (4,6.9) to (8,0) ;
      \path [fill=green] (0,0) to (4,2.3)  to  (4,0) to (0,0);
      \path [fill=green] (4,0) to (4,2.3)  to  (8,0) to (4,0);
  \draw [fill] (0,0) circle (3pt);
    \draw [fill]  (8,0) circle (3pt);
      \draw [fill] (4,6.9) circle (3pt);
        \draw [fill] (4,2.3) circle (3pt);
\draw (0,0) -- (8,0); \draw (0,0) -- (4,6.9); \draw (8,0) --
(4,6.9); \draw (0,0) -- (4,2.3); \draw (8,0) -- (4,2.3); \draw
(4,6.9) -- (4,2.3); \node at (-0.5,0) {$\mathbf v_1$}; \node at
(8.6,0) {$\mathbf v_2$}; \node at (4.2,7.4) {$\mathbf v_3$}; \node
at
(4.7,2.6) {$\mathbf w_3$}; \node at (3,3) {$\sigma_3$}; \node at
(5,3.5) {$\sigma_2$}; \node at (4,1) {$\sigma_1$};
\end{tikzpicture}
\caption{\label{Sigma3}  \small The fan $\Sigma_3$ of the partial
resolution $\tilde{Y}^{\mathbb{Z}_4}_{[3]}$ and the associated
planar graph}
\end{center}
\end{figure}
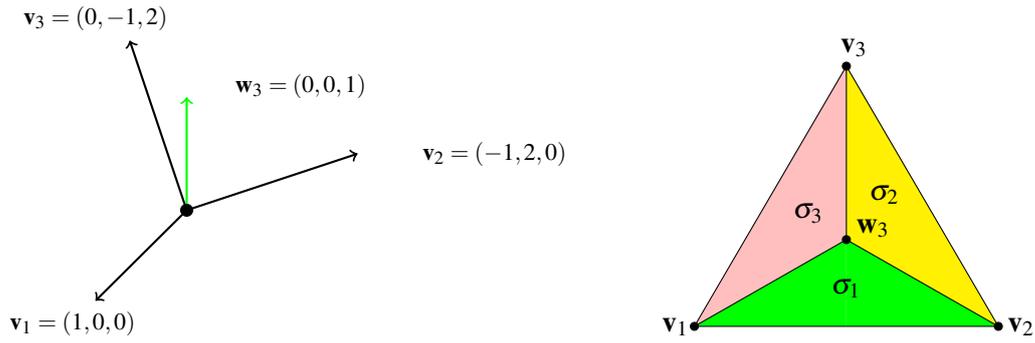

We consider the toric 3-fold $\tilde{Y}^{\mathbb{Z}_4}_{[3]}$ whose
fan $\Sigma_3$ is generated by the 4 rays $\mathbf v_1$,  $\mathbf
v_2$, $\mathbf v_3$, $\mathbf w_3$, which is a partial resolution of
$Y^{\mathbb{Z}_4}_0$.  This will appear as the partial
desingularization occuring at some of the walls of the $\zeta$
parameter space (space of stability conditions). The fan and the
associated planar graph are shown in Figure \ref{Sigma3}. The cone
$\sigma_1$ is singular, while $\sigma_2$ and $\sigma_3$ are smooth,
i.e., $\tilde{Y}^{\mathbb{Z}_4}_{[3]}$ has one singular toric point.
By general theory we know that
$$ h_2(\tilde{Y}^{\mathbb{Z}_4}_{[3]},\mathbb{Q}) = 2, \qquad h^2(\tilde{Y}^{\mathbb{Z}_4}_{[3]},\mathbb{Q}) = h^2_c(\tilde{Y}^{\mathbb{Z}_4}_{[3]},\mathbb{Q})
 = h^4(\tilde{Y}^{\mathbb{Z}_4}_{[3]},\mathbb{Q}) = h^2(\tilde{Y}^{\mathbb{Z}_4}_{[3]},\mathbb{Q}) =1.$$
\subsubsection{Divisors}  
\begin{figure}
\begin{center}
\begin{tikzpicture}
\draw [thick,->] (0,0) -- (1,0) ; \draw [thick,->] (0,0) -- (-1,2) ;
\draw [thick,->] (0,0) -- (0,-1) ;
  \draw [fill] (0,0) circle (1.5pt) ;
  \node at (0,-1.6) {$D_c$};
  \node at (0.4,-1.1) {\small $D_3$};
    \node at (1.4,0) {\small $D_1$};
    \node at (-1.4,2) {\small $D_2$};
\end{tikzpicture}
\hskip10mm
\begin{tikzpicture}
\draw [thick,->] (0,0) -- (1,0) ; \draw [thick,->] (0,0) -- (-1,2) ;
\draw [thick,->] (0,0) -- (0,1);
  \draw [fill] (0,0) circle (1.5pt) ;
  \node at (0,-1.3) {$D_{EH}$, $D'_{EH}$};
\end{tikzpicture}
\hskip10mm
\begin{tikzpicture}
\draw [thick,->] (0,0) -- (1,0) ; \draw [thick,->] (0,0) -- (-1,4) ;
\draw [thick,->] (0,0) -- (0,1);
  \draw [fill] (0,0) circle (1.5pt) ;
  \node at (0,-1.3) {$D_4$};
\end{tikzpicture}
\caption{\label{Div3}  \small The fans of the 4 toric divisors of
$\tilde{Y}^{\mathbb{Z}_4}_{[3]}$}
\end{center}
\end{figure}
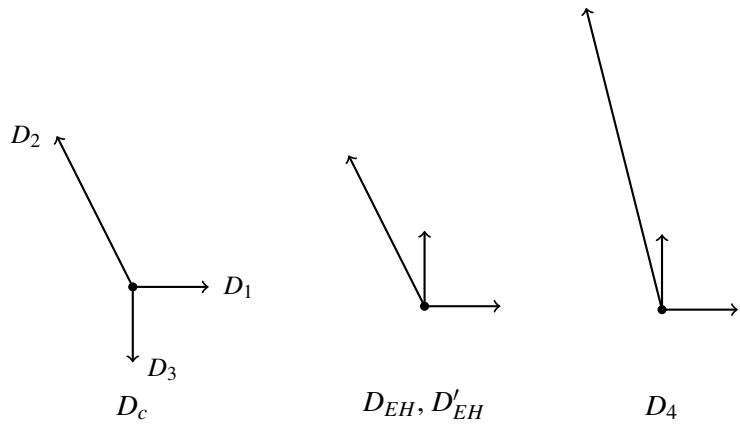

The divisors corresponding
to $\mathbf w_3$, $\mathbf v_1$, $\mathbf v_3$, $\mathbf v_2$ will
be denoted $D_c$,  $D_{EH}$, $D'_{EH}$, $D_{4}$, respectively.  They
are described in Table \ref{divisors3}. The corresponding fans are
shown in Figure \ref{Div3}.
\par
For the variety $\tilde{Y}^{\mathbb{Z}_4}_{[3]}$, which is not
smooth, the Picard group in the sequence \eqref{Pic} must be
replaced by the class group
$\operatorname{Cl}(\tilde{Y}^{\mathbb{Z}_4}_{[3]})$,
 however after tensoring by the rationals the two groups coincide, so that we may ignore this fact.
 The group $\operatorname{Pic}(\tilde{Y}^{\mathbb{Z}_4}_{[3]})\otimes \mathbb{Q}$ is generated by the class of $D_c$.
 The matrices $A$ and $B$ are now
  $$ A =\begin{pmatrix} 1 & 0 & 0 \\ -1 & 2 & 0 \\ 0 & -1 & 2  \\ 0 & 0 & 1 \end{pmatrix},\qquad
B =\begin{pmatrix}
 -\tfrac14 &  -\tfrac14 &  -\tfrac12  & 1
\end{pmatrix}
$$
(the order of the toric divisors is $D_{EH}$,  $D'_{EH}$, $D_4$,
$D_c$). The relations we get among the divisor classes are
$$[D_{EH}]=[D'_{EH}] = -\tfrac14 [D_c],\qquad [D_4] = -\tfrac12 [D_c].$$
Again, this implies
$\left[K_{\tilde{Y}^{\mathbb{Z}_4}_{[3]}}\right]=0$. The integral
generator of $\operatorname{Pic}(\tilde{Y}^{\mathbb{Z}_4}_{[3]})$ is
$D_4$. $D_{EH}$ and $D'_{EH}$ are linearly equivalent, and both
generate the class group.
\par
\begin{table}[ht]\renewcommand{\arraystretch}{1.50}
\caption{Toric Divisors in $\tilde{Y}^{\mathbb{Z}_4}_{[3]}$ \label{divisors3}} 
\vskip10pt
\centering 
\begin{tabular}{|c |c| c| c| c| c| } 
 \hline 
Ray & Divisor   & Fan  & Variety & Type & Component  \\ [1ex] 
\hline 
$\mathbf w_3$ & $D_c$  & \small $(1,0),\ (-1,2),\ (0,-1) $ & $\mathbb{WP}_{[1,1,2]}$ & Cartier & 1 \\
\hline 
$\mathbf v_1$ & $D_{EH}$  & \small$ (1,0),\ (-1,2),\ (0,1) $&
$\mbox{ALE}_{A_1}$  & Weil  &  $-\frac14$ \\   \hline $\mathbf v_2$
& $D'_{EH}$ & \small$ (1,0),\ (-1,2),\ (0,1) $ &  $\mbox{ALE}_{A_1}$
& Weil  &  $-\frac14$ \\   \hline $\mathbf v_3$ & $D_{4}$ &  \small$
(1,0),\ (-1,4),\ (0,1) $ & \small tot  $(\mathcal O(-4)
\to \mathbb P^1$)& Cartier &  $-\frac12$ \\   [1ex] 
\hline 
\end{tabular}
\end{table}

\par
\subsubsection{Toric curves and intersections}
Inspection of the planar graph in Figure \ref{Sigma3} shows that
$\tilde{Y}^{\mathbb{Z}_4}_{[3]}$ has 3 toric compact curves, however
we know that there is only one independent class in
$H_2(\tilde{Y}^{\mathbb{Z}_4}_{[3]},\mathbb Q)$.
 The intersection numbers of the curves with the 4 toric divisors are shown in Table \ref{intersections3}.
The curves $C_1$, $C_2$, $C_4$ are the 3 compact curves. $C_3$ is a
noncompact curve.

\begin{table}[ht]
\renewcommand{\arraystretch}{1.50}
\caption{Intersections among the toric curves and divisors in
$\tilde{Y}^{\mathbb{Z}_4}_{[3]}$. The toric curves are images of
curves in $Y^{\mathbb{Z}_4}_{[3]}$ via the natural map
$Y^{\mathbb{Z}_4}_{[3]}\to \tilde{Y}^{\mathbb{Z}_4}_{[3]}$, and we
have used the same notation for a curve in $Y^{\mathbb{Z}_4}_{[3]}$
and its image in $\tilde{Y}^{\mathbb{Z}_4}_{[3]}$.  The curve $C_8$
is singular (it is actually the only singular curve among the 6
toric curves of $\tilde{Y}^{\mathbb{Z}_4}_{[3]}$). Note that $C_8$
is indeed the strict transform of the $z$ axis, whose points have
nontrivial isotropy, and $\tilde{Y}^{\mathbb{Z}_4}_{[3]}$ does not
resolve this singularity as $D_{nc}$ has been shrunk onto $C_8$. The
curve $C_1$ of $Y^{\mathbb{Z}_4}_{[3]}$ has shrunk to a point. This
can be explicitly checked  by noting that the period of the K\"ahler
form on $C_1$ goes to zero under the blow-down morphism
$Y^{\mathbb{Z}_4}_{[3]}\to \tilde{Y}^{\mathbb{Z}_4}_{[3]}$.
} 
\vskip10pt
\centering 
\begin{tabular}{|c|c |c| c| c| c| } 
 \hline 
 Face  & Curve & $D_c$    & $D_{EH}$   & $D'_{EH}$ & $D_4$  \\ \hline
$(\mathbf v_1\mathbf w_3)$ & $C_2$ & $-2 $ & $\frac12$  &  $\frac12$
& 1   \\ \hline $(\mathbf v_2\mathbf w_3)$ & $C_4$ & $-2 $ &
$\frac12$  &  $\frac12$ & 1   \\ \hline $(\mathbf v_1\mathbf v_2)$ &
$C_8$ & 1  &   &    & 0 \\ \hline
$(\mathbf v_3\mathbf w_3)$ & $C_5$ & $-4 $ & $1$  &  1 & 2   \\
 \hline 
\end{tabular}
\label{intersections3} 
\end{table}
\par
\subsubsection{The class group}
The class group enters the exact sequence
$$ 0 \to M \xrightarrow{A} \operatorname{Div}{\mathbb T}(\tilde{Y}^{\mathbb{Z}_4}_{[3]}) \to \operatorname{Cl}(\tilde{Y}^{\mathbb{Z}_4}_{[3]})\to 0.$$
So the problem is that of computing the quotient of two free abelian
groups; it may have torsion. The matrix $A$ can be put into a normal
form called the Smith normal form \cite{smith}. This is diagonal,
and the diagonal entries determine the class group: a diagonal block
equal to the identity  corresponds to a free summand of the
appropriate rank, and if a value $m$ appears,   there is a summand
$\mathbb{Z}_m$. For $\tilde{Y}^{\mathbb{Z}_4}_{[3]}$ the Smith
normal form of the matrix $A$ is the identity matrix, so that the
quotient is $\mathbb{Z}$, i.e.,
$$\operatorname{Cl}(\tilde{Y}^{\mathbb{Z}_4}_{[3]}) = \mathbb{Z}.$$
Comparing with Table \ref{divisors3} we see that the morphism $
\operatorname{Pic}(\tilde{Y}^{\mathbb{Z}_4}_{[3]})\to
\operatorname{Cl}(\tilde{Y}^{\mathbb{Z}_4}_{[3]})$ is the
multiplication by $2$ (indeed, $2[D'_{EH}]=[D_4]$).

\subsubsection{$\tilde{Y}^{\mathbb{Z}_4}_{[3]}$ as a line bundle over $\mathbb{WP}_{[1,1,2]}$}
Also $\tilde{Y}^{\mathbb{Z}_4}_{[3]}$ is the total space of a line
bundle, in this case over $\mathbb{WP}_{[1,1,2]}$.
 The fan of $\mathbb{WP}_{[1,1,2]}$  is depicted on the left in Figure \ref{Div3}; it is generated by
the vectors $(1,0)$, $(-1,2)$, $(0,-1)$, corresponding respectively
to the divisors $D_1$, $D_2$, $D_3$. The divisors $D_1$ and $D_2$
are Weil, while $D_3$ is Cartier. In the class group, which is
$\mathbb Z$, they are related by
$$ [D_3] = 2 [D_2] = 2 [D_1].$$
Each of $D_1$ and $D_2$ generates the class group, and $D_3$
generates the Picard group.
\par
We study the line bundle $\cO_{\mathbb{WP}_{[1,1,2]}}(-2D_3) =
\cO_{\mathbb{WP}_{[1,1,2]}}(-4)$. By applying the algorithm in
\cite[\S 7.3]{CoxLS} we see that its fan is that of
$\tilde{Y}^{\mathbb{Z}_4}_{[3]}$, i.e.,
$\tilde{Y}^{\mathbb{Z}_4}_{[3]}=\mbox{tot}(\cO_{\mathbb{WP}_{[1,1,2]}}(-4))$.
Again, since $-2D_3$ is a canonical divisor of
$\tilde{Y}^{\mathbb{Z}_4}_{[3]}$,
$K_{\tilde{Y}^{\mathbb{Z}_4}_{[3]}}$ is trivial. Again the toric
divisors of $\tilde{Y}^{\mathbb{Z}_4}_{[3]}$ may be obtained from
this description: $D_{EH}$ and $D'_{EH}$ are the inverse images of
$D_1$ and $D_2$, while $D_4$ is the inverse image of $D_3$; since
$D_3\cdot D_3= 2$, then $D_4$ is the total space of $\cO(-4)$ on
$\mathbb P^1$. Moreover, $D_c$ is the image of the zero section.
Note that $\tilde{Y}^{\mathbb{Z}_4}_{[3]}$ is obtained by shrinking
$D_{nc}$ to a $\mathbb P^1$; actually $D_{nc}$ is the total space of
the trivial line bundle on the divisor $E$ in $\mathbb F_2$, and
$\mathbb{WP}_{[1,1,2]}$ is indeed obtained by shrinking that divisor
to a point.
\par
Using the generalized Kronheimer construction can calculate the
K\"ahler potential, K\"ahler metric and K\"ahler form of
$\tilde{Y}^{\mathbb{Z}_4}_{[3]}$, verifying that the base of the
bundle is indeed singular, as the periods of the K\"ahler form on
the cycle $C_1$ vanish. This means that $C_1$ shrinks to a point,
and since $C_1$ inside the compact exceptional divisor $\mathbb F_2$
is the exceptional divisor of the blow-down $\mathbb F_2\to \mathbb
P[1,1,2]$, the base variety becomes singular. 
\section{Chamber Structure and the tautological bundles}
\label{camerataccademica}
\begin{figure}
\label{hilton1} \centering
\includegraphics[height=7cm]{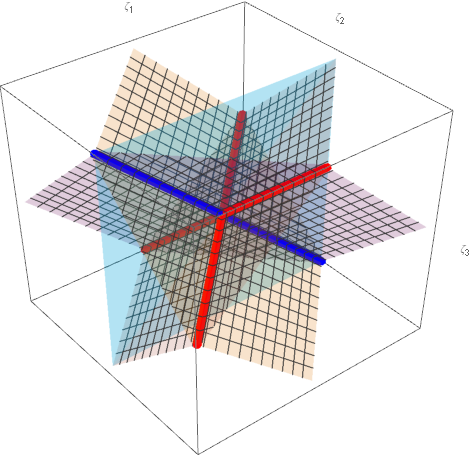}
\caption{\label{hilton1} The structure of the stability chambers.
The space $\mathbb{R}^3$ where the moment map equations always admit
real nonnegative solutions is divided in two halves by the presence
of a wall of type 0, named $\mathcal{W}_0$ which is
  defined by the equation $\zeta_2=0$ and is marked in
the figure as a cyan transparent surface without meshing. In
addition there are other three walls, respectively described by
$\mathcal{W}_1 \Leftrightarrow \zeta=\{x+y,x,y\}$, $\mathcal{W}_2
\Leftrightarrow \zeta=\{x,x+y,y\}$ and $\mathcal{W}_3
\Leftrightarrow \zeta=\{x,y,x+y\}$, where $x,y\in \mathbb{R}$. The
planes $\mathcal{W}_{1,3}$ are of type 0, while $\mathcal{W}_{2}$ is
of type $1$. These three infinite planes provide the partition of
$\mathbb{R}^3$ into eight disjoint chambers that are described in
the main text. The three planes $\mathcal{W}_{1,2,3}$ are marked in
the figure as meshed surfaces of three different colors. On the
three intersections  of two of these planes we find the already
discussed  lines where the moment map equations can be solved by
radicals, corresponding to the Eguchi-Hanson degeneration $
Y_{EH}\times \mathbb{C}$ (blue line) and to the two Cardano
degenerations (red lines).}
\end{figure}
We can now compare the analytical results obtained from the K\"ahler
quotient \`a la Kronheimer  with the general predictions of the
resolution of the singularity provided by toric geometry. This
provides a concrete example of how the chamber structure of the
$\zeta$ parameter space controls the topology of the resolutions of
the singularity \cite{CrawIshii}.
In the following we evaluate the periods of the differential forms
arising from the Kronheimer construction on the cycles given by the
curves $C_1$, $C_2$ that both are contained in $D_c$, namely in the
compact component of the exceptional divisor (actually we are
pulling back the differential forms from $\mathcal{M}_{a,b,c}$ to
$Y$ via the relevant contraction morphism $\gamma\colon Y \to
\mathcal{M}_{a,b,c}$, and we use the fact that the pullback is
injective in cohomology. We succeed evaluating the differential form
periods on the considered curves by restricting the moment map
equations to the relevant exceptional divisor $D_c$ and then to its
relevant sub-loci.
\par
According to the results presented in section \ref{maccaius} for the
Kronheimer construction applied to $\mathbb{C}^3/\mathbb{Z}_4$,
there are three tautological bundles; their first Chern classes are
encoded in the triple  of (1,1)-forms:
\begin{equation}\label{giraldo}
    \omega_{I}^{(1,1)} \, = \, \frac{i}{2\pi} \partial\,{\bar
    \partial} \, \log\left[X^I\right]
\end{equation}
\subsection{The stability chambers}
\label{camerataccademicaWeyl} The result of the  various
calculations subsequently presented  singles out the structure of
the stability  chambers which is summarized in figs.~\ref{hilton1}
and \ref{hilton2}.
\begin{figure}
\label{hilton2} \centering
\includegraphics[height=7cm]{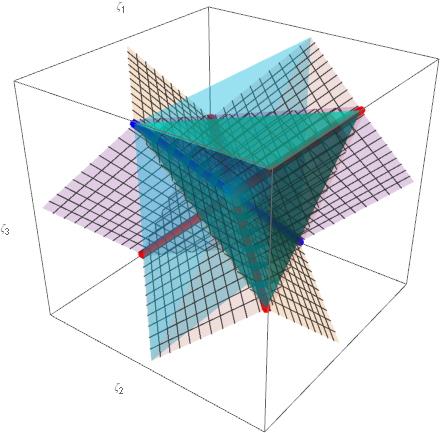}
\caption{\label{hilton2}  In fig.\ref{hilton1} we displayed  only
the walls defining the chamber structure. In the present picture
besides the walls we show also one of the eight chambers, namely
Chamber 1. It is marked as a transparent  greenish-blue colored
portion of three dimensional space. It is a convex cone delimited by
the aforementioned walls. }
\end{figure}
Let us illustrate this structure in detail. Understanding the
geometry of these pictures is a very useful guide through the
subsequent computations. The original data from which we start are
the following ones. To begin with we know that the entire $\zeta$
space is just $\mathbb{R}^3$.

\begin{figure}[htb!]
\centering
\includegraphics[height=9cm]{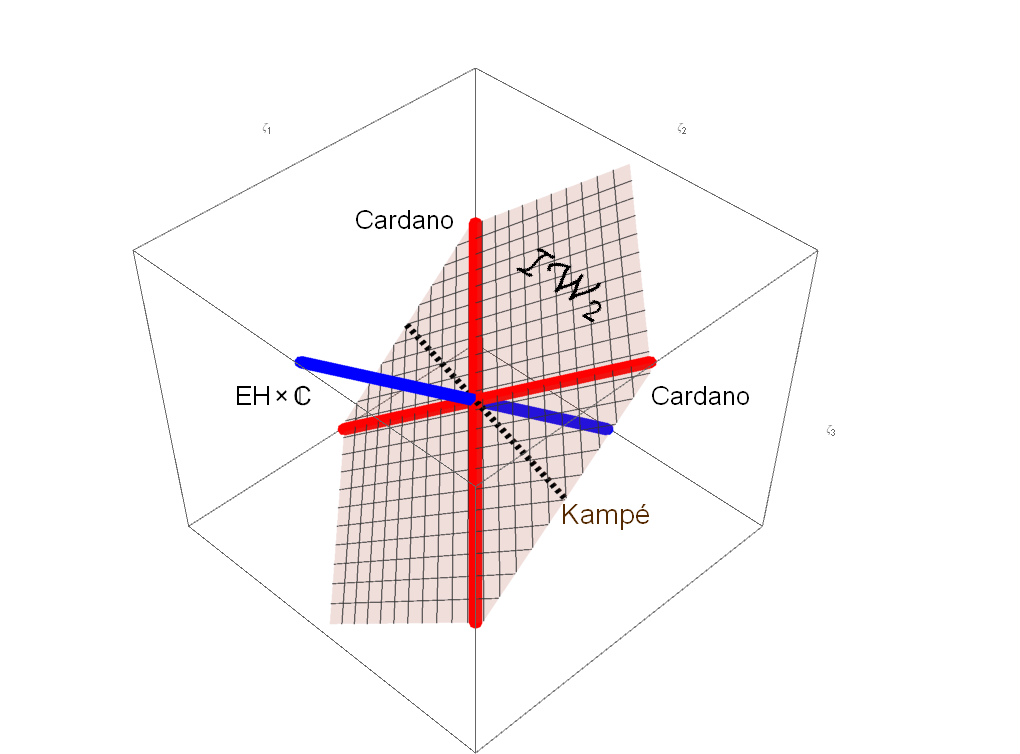}
\caption{ \label{pianoiwdue} This picture shows the type I plane
$\mathcal{W}_2$. The two Cardano manifolds (red lines) and the
Kamp\'{e} manifold (dashed black line) lay all in this plane, whose
generic point corresponds, as we are going to see, to the
degeneration Y3. The Eguchi-Hanson degeneration (blue line) is
instead out of this plane and intersects it only in the origin. }
\end{figure}

In figs.~\ref{hilton1} and \ref{hilton2} we have drawn  3 lines.
These lines correspond to the following 4 instances of degenerate
spaces where the algebraic system of the moment map equations
becomes solvable or partially solvable:
\begin{description}
\item[1)] Eguchi-Hanson case
\begin{equation}\label{ehcasus}
    \zeta_1 \, = \, s \quad ; \quad \zeta_2 \, = \, 0 \quad ; \quad
    \zeta_3 \, = \, s
\end{equation}
In pictures \ref{hilton1}, \ref{hilton2} this line is fat and drawn
in blue color:
\item[2)] Cardano 1
\begin{equation}\label{Carcasus1}
    \zeta_1 \, = \, s \quad ; \quad \zeta_2 \, = \, s \quad ; \quad
    \zeta_3 \, = \, 0
\end{equation}
In pictures \ref{hilton1}, \ref{hilton2} this line is solid fat and
drawn in red color. We remind the reader that the name Cardano is
due to the fact that the solution of the entire moment map equation
system reduces to the solution of a single algebraic equation of the
fourth order (see eqn.~\eqref{baldop}).
\item[3)] Cardano 2
\begin{equation}\label{Carcasus2}
    \zeta_1 \, = \, 0 \quad ; \quad \zeta_2 \, = \, s \quad ; \quad
    \zeta_3 \, = \, s
\end{equation}
 In pictures \ref{hilton1}, \ref{hilton2} this line is solid fat and
drawn in red color.
\end{description}
In addition we have a 4th line that we have not drawn in fig.
\ref{hilton1} and \ref{hilton2}. This line entirely lies on one of
the walls to be described below.
\begin{description}
\item[4)] Kamp\'{e}
\begin{equation}\label{kampus1}
    \zeta_1 \, = \, s \quad ; \quad \zeta_2 \, = \, 2 \,s \quad ; \quad
    \zeta_3 \, = \, s
\end{equation}
In fig.~\ref{pianoiwdue} this line is dashed fat and drawn in black
color. We remind again the reader that the manifold was named
Kamp\'{e} because the solution of the moment map equations reduces
to finding the roots of a single algebraic equation of the sixth
order.
\end{description}
\begin{figure}[htb!]
\centering
\includegraphics[height=9cm]{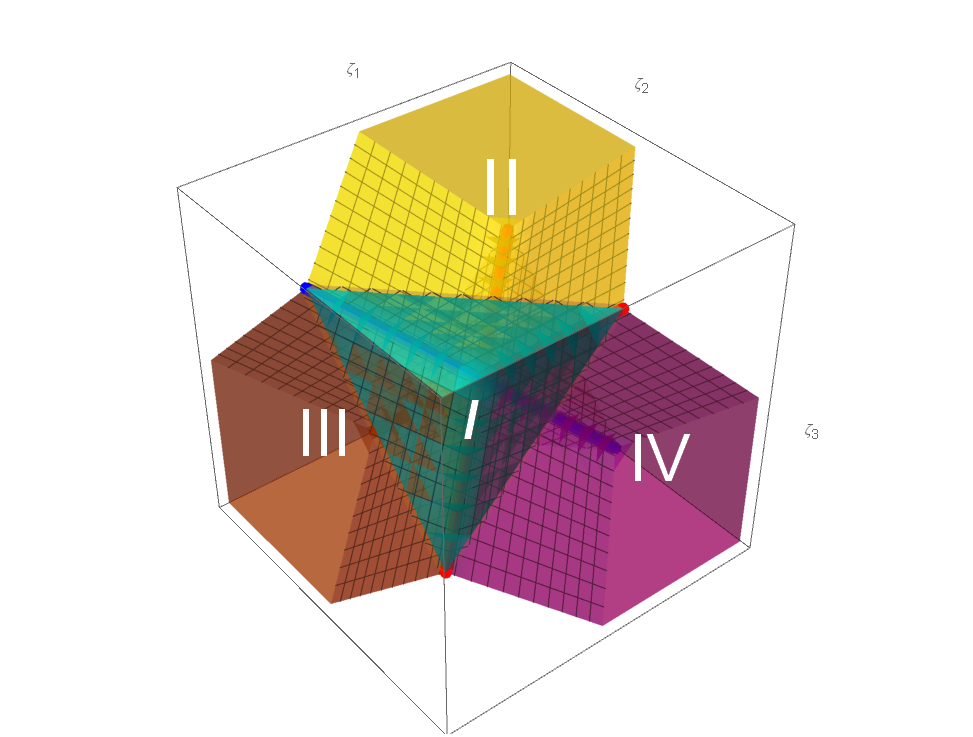}
\caption{ \label{pinetti} The partition of $\mathbb{R}^3$ into 8
convex cones. In the picture, out of the eight regions, we show only
four, marking them in different transparent colors.}
\end{figure}
In all  these cases $s$ is a nonzero real number. Since the
exceptional solvable cases must lie on some walls we have tried to
conjecture which planar cones might partition the infinite cube into
chambers so that the exceptional lines could lie on such planar
cones and possibly be edges at some of their intersections. With
some ingenuity we introduced the following four planar walls (here
$x,y$ are real parameters)
\begin{align}
\mathcal{W}_1 :\qquad &  \{x+y,x,y\}  \label{IW1muro} \\
\mathcal{W}_2  : \qquad  &\{x,x+y,y\}  \label{IW2muro} \\
\mathcal{W}_3 : \qquad  &   \{x,y,x+y\} \label{IW3muro} \\
\mathcal{W}_0  : \qquad  & \{x,0,y\} \label{IW0muro}
\end{align}
that were depicted in fig.~\ref{hilton1}, \ref{hilton2} and split
the space $\mathbb{R}^3$ into eight convex three--dimensional cones.
The list of the eight convex cones that provide as many  stability
chambers is obtained through the following argument. The three
planes $\mathcal{W}_{1,2,3}$ are respectively orthogonal to the
following three vectors:
\begin{eqnarray}
  \pmb{n}_1 &=& \left\{-1,1,1 \right\} \\
  \pmb{n}_2 &=& \left\{1,-1,1 \right\} \\
  \pmb{n}_3 &=& \left\{1,1,-1 \right\} .\label{normalini}
\end{eqnarray}
\begin{table}[htb!]
  \centering
  $$ \begin{array}{|c||c|}
  \hline
  \hline
          \text{Chamber 1} &\begin{array}{c|c|c|c}
\text{cycle} & \int \omega_1 & \int \omega_2 & \int \omega_3 \\
\hline \hline
C_1 & 1& 0 & 1\\
\hline
C_2 & 1& 1 & 1\\
  \end{array}\\
\hline \hline
          \text{Chamber 2} & \begin{array}{c|c|c|c}
          {\text{cycle}} & {\int \omega_1} & {\int \omega_2} & {\int \omega_3} \\
          \hline
C_1 & 1& 0 & 1\\
\hline
C_2 & 1& 1 & 1\\
  \end{array} \\
          \hline
          \hline
         \text{Chamber 3} &  \begin{array}{c|c|c|c}
           {\text{cycle}} &  {\int \omega_1} &  {\int \omega_2} &  {\int \omega_3} \\
           \hline
C_1 & 1& 0 & 1\\
\hline
C_2 & 0 & -1 & 0 \\
  \end{array} \\
         \hline
          \hline
         \text{Chamber 4} & \begin{array}{c|c|c|c}
           {\text{cycle}} &  {\int \omega_1} &  {\int \omega_2} &  {\int \omega_3} \\
           \hline
C_1 & 1& 0 & 1\\
\hline
C_2 & 1& 1 & 1\\
  \end{array} \\
         \hline
          \hline
         \text{Chamber 5} & \begin{array}{c|c|c|c}
           {\text{cycle}} &  {\int \omega_1} &  {\int \omega_2} &  {\int \omega_3} \\
           \hline
C_1 & 1& 0 & 1\\
\hline
C_2 & 0 & -1 & 0 \\
  \end{array} \\
         \hline
          \hline
         \text{Chamber 6} & \begin{array}{c|c|c|c}
           {\text{cycle}} &  {\int \omega_1} &  {\int \omega_2} &  {\int \omega_3} \\
           \hline
C_1 & 1& 0 & 1\\
\hline
C_2 & -1 & -1 & -1 \\
  \end{array} \\
         \hline
          \hline
         \text{Chamber 7} & \begin{array}{c|c|c|c}
           {\text{cycle}} &  {\int \omega_1} &  {\int \omega_2} &  {\int \omega_3} \\
           \hline
C_1 & 1& 0 & 1\\
\hline
C_2 & 0 & -1 & 0 \\
  \end{array} \\
         \hline
          \hline
        \text{Chamber 8} & \begin{array}{c|c|c|c}
           {\text{cycle}} &  {\int \omega_1} &  {\int \omega_2} &  {\int \omega_3} \\
           \hline
C_1 & 1& 0 & 1\\
\hline
C_2 & 0 & -1 & 0 \\
  \end{array} \\
        \hline
          \hline
          \end{array}
   $$
  \caption{The periods of the tautological bundle first Chern classes on the basis of homological cycles
  calculated in the interior points of all the chambers.}\label{periodico}
\end{table}
\begin{table}[htb!]
\renewcommand{\arraystretch}{1.50}
  \centering
  $$ \begin{array}{|c||c|}
  \hline
  \hline
          \text{Wall $\mathcal{W}_0$} &\begin{array}{c|c|c|c}
\text{cycle} & \int \omega_1 & \int \omega_2 & \int \omega_3 \\
\hline \hline
C_1 & 3 & 0 & 3\\
\hline
C_2 & 2 & 0 & -2 \\
  \end{array}\\
\hline \hline
          \text{Wall $\mathcal{W}_1$} & \begin{array}{c|c|c|c}
          {\text{cycle}} & {\int \omega_1} & {\int \omega_2} & {\int \omega_3} \\
          \hline
C_1 & 1 & 0 & 1 \\
\hline
C_2 &0 & -2 & 0 \\
  \end{array} \\
          \hline
          \hline
         \text{Wall $\mathcal{W}_2$} &  \begin{array}{c|c|c|c}
           {\text{cycle}} &  {\int \omega_1} &  {\int \omega_2} &  {\int \omega_3} \\
           \hline
C_1 & 0 & 0 & 0 \\
\hline
C_2 & 1 & 4 & 1 \\
  \end{array} \\
         \hline
          \hline
         \text{Wall $\mathcal{W}_3$} & \begin{array}{c|c|c|c}
           {\text{cycle}} &  {\int \omega_1} &  {\int \omega_2} &  {\int \omega_3} \\
           \hline
C_1 & 1 & 0 & 1 \\
\hline
C_2 &0 & -2 & 0 \\
  \end{array} \\
         \hline
          \hline
          \end{array}
   $$
  \caption{The periods of the tautological bundle first Chern Classes on the basis of homological cycles
  calculated on the 4 walls.
  \label{muraria}}
\end{table}
\begin{table}[htb!]\renewcommand{\arraystretch}{1.50}
  \centering
  $$ \begin{array}{|c||c|}
  \hline
  \hline
          \text{Cardano $\zeta = \{0,s,s\}$} &\begin{array}{c|c|c|c}
\text{cycle} & \int \omega_1 & \int \omega_2 & \int \omega_3 \\
\hline \hline
C_1 & 0 & 0 & 0 \\
\hline
C_2 & -1  & -1 & 0 \\
  \end{array}\\
\hline \hline
          \text{Cardano $\zeta = \{s,s,0\}$} & \begin{array}{c|c|c|c}
          {\text{cycle}} & {\int \omega_1} & {\int \omega_2} & {\int \omega_3} \\
          \hline
C_1 &0 & 0 & 0 \\
\hline
C_2 &0 &  -1  & -1  \\
  \end{array} \\
          \hline
          \hline
         \text{Eguchi-Hanson $\zeta = \{s,0,s\}$} &  \begin{array}{c|c|c|c}
           {\text{cycle}} &  {\int \omega_1} &  {\int \omega_2} &  {\int \omega_3} \\
           \hline
C_1 &  -1  & 0 & \-1  \\
\hline
C_2 & 0 & 0 & 0 \\
  \end{array} \\
         \hline
          \hline
          \end{array}
   $$
  \caption{The periods of the tautological bundle first Chern Classes on the basis of homological cycles
  calculated on the edges.
   }\label{spigolosa}
\end{table}
The eight convex regions are defined by choosing the signs of the
projections $\pmb{n}_{1,2,3}\cdot \zeta$ in all possible ways. In
this way we obtain:
\begin{align}\label{celle8}
   \mbox{Chamber I}
  & \equiv &
  \, \left\{  \zeta_1\, -\,\zeta_2\,-\,\zeta_3\,> \,0
  \quad , \quad
    -\zeta_1\, +\,\zeta_2\,-\,\zeta_3\,> \,0 \quad , \quad
    -\zeta_1\, -\,\zeta_2\,+\,\zeta_3\,> \,0
   \right\} \nonumber \\
    \mbox{Chamber II}
  & \equiv &
  \, \left\{  \zeta_1\, -\,\zeta_2\,-\,\zeta_3\,> \,0
  \quad , \quad
    -\zeta_1\, +\,\zeta_2\,-\,\zeta_3\,> \,0 \quad , \quad
    -\zeta_1\, -\,\zeta_2\,+\,\zeta_3\,< \,0
\right\} \nonumber\\
    \mbox{Chamber III}
  & \equiv &
  \, \left\{  \zeta_1\, -\,\zeta_2\,-\,\zeta_3\,> \,0
  \quad , \quad
    -\zeta_1\, +\,\zeta_2\,-\,\zeta_3\,< \,0 \quad , \quad
    -\zeta_1\, -\,\zeta_2\,+\,\zeta_3\,> \,0
   \right\}
\end{align}
 \begin{align}\label{celle8piu}
 \mbox{Chamber IV}
  & \equiv &
  \, \left\{  \zeta_1\, -\,\zeta_2\,-\,\zeta_3\,< \,0
  \quad , \quad
    -\zeta_1\, +\,\zeta_2\,-\,\zeta_3\,> \,0 \quad , \quad
    -\zeta_1\, -\,\zeta_2\,+\,\zeta_3\,> \,0
   \right\}\nonumber\\
 \mbox{Chamber V}
  & \equiv &
  \, \left\{  \zeta_1\, -\,\zeta_2\,-\,\zeta_3\,< \,0
  \quad , \quad
    -\zeta_1\, +\,\zeta_2\,-\,\zeta_3\,< \,0 \quad , \quad
    -\zeta_1\, -\,\zeta_2\,+\,\zeta_3\,> \,0
    \right\}\nonumber\\
    \mbox{Chamber VI}
  & \equiv &
  \, \left\{  \zeta_1\, -\,\zeta_2\,-\,\zeta_3\,< \,0
  \quad , \quad
    -\zeta_1\, +\,\zeta_2\,-\,\zeta_3\,> \,0 \quad , \quad
    -\zeta_1\, -\,\zeta_2\,+\,\zeta_3\,< \,0
    \right\} \nonumber \\
    \mbox{Chamber VII}
  & \equiv &
  \, \left\{  \zeta_1\, -\,\zeta_2\,-\,\zeta_3\,> \,0
  \quad , \quad
    -\zeta_1\, +\,\zeta_2\,-\,\zeta_3\,< \,0 \quad , \quad
    -\zeta_1\, -\,\zeta_2\,+\,\zeta_3\,< \,0
    \right\}  \nonumber\\
    \mbox{Chamber VIII}
  & \equiv &
  \, \left\{  \zeta_1\, -\,\zeta_2\,-\,\zeta_3\,< \,0
  \quad , \quad
    -\zeta_1\, +\,\zeta_2\,-\,\zeta_3\,< \,0 \quad , \quad
    -\zeta_1\, -\,\zeta_2\,+\,\zeta_3\,< \,0
   \right\}
\end{align}
Four of the above eight chambers are displayed in
fig.~\ref{pinetti}.
\subsection{Edges} The special solvable cases that we have found
all sit at the intersection of two of the three walls
$\mathcal{W}_{1,2,3}$. In particular the Cardano manifolds are edges
at the intersection of the following walls:
\begin{equation}\label{cardanici}
    \text{Cardano 1} \, = \, \mathcal{W}_{1}\bigcap \mathcal{W}_{2} \quad ;
    \quad \text{Cardano 2} \, = \, \mathcal{W}_{2}\bigcap \mathcal{W}_{3}
\end{equation}
while the Eguchi-Hanson case is the intersection:
\begin{equation}\label{ansoniani}
    Y_{EH} \, = \, \mathcal{W}_{1}\bigcap
    \mathcal{W}_{3}
\end{equation}
Note also that  this edge lays entirely on the wall $\mathcal{W}_0$.
From this point of view the Eguchi-Hanson case is similar to the
Kamp\'{e} case, that lays entirely on the wall $\mathcal{W}_2$. The
difference however is that, as we advocate below, the wall
$\mathcal{W}_0$ is of type 0 while $\mathcal{W}_{2}$ is of type 1.
In the first case the Eguchi-Hanson line is the only degeneracy
pertaining to the wall $\mathcal{W}_0$, while in the second case the
Kamp\'{e} line yields an instance of the degeneracy
$\tilde{Y}^{\mathbb{Z}_4}_{[3]}$ as any other point of the same
wall. Actually also the Cardano cases that lay on the same wall
correspond to a different realization of
$\tilde{Y}^{\mathbb{Z}_4}_{[3]}$.
\section{Periods of the Chern classes of the tautological bundles}
\label{Periodstaut} The most appropriate instrument to verify the
degeneracy/non-degeneracy of the singularity resolutions provided by
the K\"ahler quotient with given $\zeta$ parameters is provided by
the calculation  of the period matrix:
\begin{equation}\label{periodare}
  \pmb{\Pi} \, \equiv  \Pi_{i,J} \, = \, \int_{C_i} \omega_{J}
  \quad ; \quad
    i=(1,2), \quad J=(1,2,3)
\end{equation}
where $\omega_J$ are the first Chern classes of the tautological
bundles and the curves $C_i$ provide a basis of     homology
2-cycles. In particular the combination:
\begin{equation}\label{saccius}
    \pmb{Vol}_i \,  = \zeta_I \, \mathfrak{C}^{IJ} \,  \int_{C_i} \omega_J
    =
     \,  \tfrac{i}{2\pi}\, \zeta_I \, \mathfrak{C}^{IJ} \,
    \int_{C_i} \, \partial\bar\partial \log (X_J)^{\alpha_\zeta}\, = \, \int_{C_i}
    \mathbb{K}_\zeta
\end{equation}
is the volume of the cycle $C_i$  in the resolution identified by
the level parameters $\zeta$, having denoted by $\mathbb{K}_\zeta$
the corresponding K\"ahler 2-form.\footnote{Let us remark that in
agreement with eqn.~\eqref{caramboletta} the contribution
$\partial\bar{\partial}\mathcal{K}_0$ to the K\"ahler 2-form  is an
exact form, whose integral on homology cycles therefore vanishes.
Hence the volume of the homology cycles  is  provided by the linear
combination of periods specified in eqn.~\eqref{saccius}. We also
remark that, as the volume of a nonzero cycle is always positive,
this is consistent with the positivity of the so-called Hodge line
bundle $\otimes_{J=1}^3 L_J^{\Theta(\mathcal{D}_J)}$.}  If the
volume of the two homology cycles yielding the homology basis is
nonzero there is no degeneration. Instead in case of degenerations
at least one of such volumes vanishes. This is precisely what
happens on the walls of type III, while in the interior of all
chambers no degeneration appears.
\par
By means of detailed calculations  the authors of
\cite{noietmarcovaldo} succeeded in calculating the periods of the
first Chern forms $\omega_{1,2,3}^{(1,1)}$ on the basis of homology
cycles $C_{1,2}$ both for the interior points of all the chambers
and for all the walls. As far as the interior chamber case is
concerned such results are summarized in Table \ref{periodico}. For
the walls the results are instead summarized in Table \ref{muraria},
while for the edges they are given in Table \ref{spigolosa}.
\par
Let us discuss the results in Table \ref{periodico}. The three
leftmost columns display the degrees of the tautological line
bundles $\mathcal R_I$ restricted to the curves $C_1$, $C_2$; as the
classes of the latter provide a basis of $H_2(Y,\mathbb{Z})$, these
numbers give the first Chern classes of the line bundles over the
integral basis of the Picard group $\operatorname{Pic}(Y)$ given by
the divisors $D_{EH}$, $D_4$, dual to the previously mentioned basis
of $H_2(Y,\mathbb{Z})$. Note that the three columns correspond to
the compact junior conjugacy class, the noncompact junior class, and
the senior class respectively, and the Poincar\'e duality between
$H^2_c(Y) $ and $H^4(Y)$ explains why   in all chambers $\mathcal
R_1$ and $\mathcal R_3$ have the same Chern classes, and are
therefore isomorphic.
\par
We know from the McKay correspondence that the cohomology of
$Y^{\mathbb{Z}_4}_{[3]}$ is generated by algebraic classes which are
in a one-to-one correspondence with the elements of $\mathbb{Z}_4$.
One issue is how these classes are expressible in terms of the Chern
characters of the tautological bundles. In general, this
correspondence is highly nontrivial and is governed by a complicated
combinatorics. This is indeed exemplified by the quoted
calculations.
\newpage
\chapter{The issue of  Ricci-flat metric and the AMSY
action/angle formalism} \label{riccione} The present chapter is
devoted to the issue of the Ricci-flat metric on the resolved
variety $Y^\Gamma_{[3]}$. As we know from previous discussions the
HKLR metric obtained from the K\"ahler quotient \`{a} la Kronheimer
is not Ricci flat in the complex three dimensional case as it
happens to be in the complex two-dimensional one. The reason while
it is Ricci flat in the lower dimensional case is that there we
perform a hyperK\"ahler quotient and all hyperK\"ahler metrics are
Ricci flat by construction. In the higher dimensional case, that is
purely K\"ahler, this facilitation is lost. Hence, in order to find
the Ricci flat metric on $Y^\Gamma_{[3]}$, which is necessary to
derive the classical D3 supergravity solution, one has to work much
harder. The issue was very easily solved for the case
$Y^{\mathbb{Z}_3}_{[3]}= \mathcal{O}_{\mathbb{P}^2}(-3)$ in chapter
\ref{maestro1}, but this is rather exceptional and due to the very
high degree of homogeneity of the exceptional compact divisor that
is a coset manifold for such a choice of $\Gamma$. The next case
discussed in chapter \ref{balengusz4} is already much more
difficult. The several efforts attempted at  the solution of this
problem were discussed in \cite{Bianchi_2021}. In that context the
Ricci flat metric was found for the partial resolution case
$\tilde{Y}^{\mathbb{Z}_4}_{[3]}\,=\,
\operatorname{tot}\left[K\left(\mathbb{WP}_{[1,1,2]}\right)\right]$
as an exact solution of the relevant  Monge Amp\`{e}re equation but
not for the full resolution ${Y}^{\mathbb{Z}_4}_{[3]}\,=\,
\operatorname{tot}\left[K\left(\mathbb{F}_2\right)\right]$. For the
second case it was only shown that the Monge Amp\`{e}re equation can
be iteratively solved in power series of the fibre coordinate but no
systematic for the $n$-th order functional coefficient was found.
Notwithstanding such failure the main positive result following from
the  investigations of \cite{Bianchi_2021} consists in the
following. It became clear that the best approach to any problem
concerning the study of K\"ahler metrics on such manifolds as those
emerging in the context of Kronheimer constructions is provided by
the AMSY symplectic formalism
\cite{abreu,Martelli:2005tp,Bykov:2017mgc} that uses real
action/angle variables. Recasting the problem in that language it
appeared that the Kronheimer induced metrics on the compact
exceptional divisors of both $Y^\Gamma_{[3]}$ and
$\tilde{Y}^{\mathbb{Z}_4}_{[3]}$, namely $\mathbb{F}_2$ and
$\mathbb{WP}_{[1,1,2]}$ are just members of a larger family of
metrics defined by a single function of one variable that are all of
co-homogeneity one. Such family includes other interesting members:
in particular the entire family of so named Calabi extremal metrics
which on its turn includes a subfamily of K\"ahler Einstein metrics.
For the case of the latter ones the Ricci flat metric on the total
space of the canonical bundle of their corresponding manifold was
constructed in \cite{bruzzo2023d3brane}. Such manifolds display a
conical singularity but provide a very interesting alternative
explicit example for  case-study. Indeed taking this different
viewpoint a lot of new questions emerged that are still unanswered
and will be addressed in chapter \ref{aperto}. Indeed the reversed
approach has been brought to attention: given a metric on a line
bundle $\mathcal{L}(\mathcal{M}_B)$ where $\mathcal{M}_B$ is a
compact K\"ahler manifold in real dimension $4$ how can we explore
the possibility that $\mathcal{L}(\mathcal{M}_B)=Y^\Gamma_{[3]}$ is
the crepant resolution of an orbifold singularity $Y^\Gamma_{[3]}\to
\frac{C^3}{\Gamma}$ where $\Gamma$ is a suitable finite subgroup of
$\mathrm{SU(3)}$? In the present chapter I try to report the above
tale in some detail.
\section{Reduction of the moment map equations to the exceptional
divisor and the double fibration} We recall that the moment map
equations for the case $\mathbb{C}^3/\mathbb{Z}_4$ take the
following final form
\begin{equation}\label{rupetto}
    \left(
\begin{array}{c}
 -\frac{\left(X_1^2-X_3^2\right) \left(U
   \left(X_2^2+1\right)+\Sigma  X_1 X_3\right)}{X_1 X_2
   X_3} \\
 \frac{\Sigma  \left(X_2^3+X_2-X_1 X_3
   \left(X_1^2+X_3^2\right)\right)}{X_1 X_2 X_3} \\
 -\frac{\left(X_2^2-1\right) \left(U
   \left(X_1^2+X_3^2\right)+\Sigma  X_2\right)}{X_1 X_2
   X_3} \\
\end{array}
\right) \, = \, \left(
\begin{array}{c}
 \zeta _3-\zeta _1 \\
 -\zeta _1+\zeta _2-\zeta _3 \\
 -\zeta _2 \\
\end{array}
\right)
\end{equation}
where in terms of the complex coordinates $u,v,w$ of the
$Y^{\mathbb{Z}_4}_{[3]}$ three-fold in the first of the four dense
open charts displayed in table \ref{coordinates}\footnote{The dense
open chart $\pmb{\sigma}_1$ is the standard one that I
systematically utilize throughout these lecture notes} we have:
\begin{equation}\label{carlingo}
    \Sigma = \sqrt[4]{\mid w\mid^2} \sqrt{\left(\mid u\mid^2+1\right)^2 \mid
    v\mid^2}\quad ; \quad U \, = \, \sqrt{\mid w \mid^2}
\end{equation}
Furthermore the HKLR K\"ahler potential of the resolved variety is
as follows:
\begin{align}\label{HKLRkallero}
\mathcal{K}_{HKLR} \, = \,& \mathcal{K}_0 \, + \, \zeta _I
\mathfrak{C}^{\text{IJ}} \log \left(X_J\right)& \nonumber\\
\mathcal{K}_0 \,= \, & \frac{U \left(X_2^2+1\right)
   \left(X_1^2+X_3^2\right)+\Sigma  \left(X_2^3+X_2+X_1
   X_3 \left(X_1^2+X_3^2\right)\right)}{X_1 X_2 X_3}&
\end{align}
In a generic point of any chamber, away from the degeneration walls,
the variety $Y^{\mathbb{Z}_4}_{[3]}$ is always the total space of
the canonical bundle of $\mathbb{F}_2$ i.e. the second Hirzebruch
surface. Generically, including the walls and the edges we have:
\begin{equation}\label{seimanigoldi}
 Y^{\mathbb{Z}_4}_{[3]}\,\sim \, \mathcal{M}_6  \, = \,  \text{tot}\left(K\left[\mathcal{M}_B\right]\right)
\end{equation}
where the notation $\mathcal{M}_6$ emphasizes that, at the end of
the day, we are concerned with a 6-dimensional real differentiable
manifold equipped with a K\"ahler metric which is not necessarily
derived in terms of complex coordinates and characterized in terms
of algebraic geometry: indeed we plan to resort to the AMSY
action/angle formalism, to be presented here in later sections, that
has many distinctive advantages and can cover a larger set of
possibilities to be interpreted in the language of algebraic
geometry at a subsequent stage. Similarly the 4-dimensional manifold
$\mathcal{M}_B$ mentioned in eqn. \eqref{seimanigoldi} that is
typically the base manifold of a line-bundle, here the canonical
bundle, is also a K\"ahler manifold, possibly a singular one, whose
description will be provided in the more versatile AMSY action/angle
formalism. For instance on the walls of the chamber for the
$\mathbb{C}^3/\mathbb{Z}_4$ case we have: $\mathcal{M}_B\,
=\,\mathbb{W}\mathbb{P}_{112}$.
\par
A slightly more general framework that can be extracted from the
example at hand of the resolution $\mathcal{M}_6 \longrightarrow
\mathbb{C}^3/\mathbb{Z}_4$ is encoded in the following double
fibration structure:
\begin{equation}\label{doppiafibbia}
    \mathcal{M}_6 \, \stackrel{\pi_1}{\longrightarrow} \, \mathcal{M}_B \,
    \stackrel{\pi_2}{\longrightarrow} \, \mathbb{P}^1 \, \sim \mathbb{S}^2
\end{equation}
Referring to the coordinates $u,v,w$ mentioned in \eqref{carlingo}
$u,v$ are the complex coordinates of the base manifold
$\mathcal{M}_B$ while $w$ is the coordinate along the fibres of the
canonical bundle. The base manifold has, on its turn, the structure
of a fibre bundle with base manifold a 2-sphere. The standard fibre
of the base manifold $\mathcal{M}_B$ is a second 2-sphere in the
smooth case ($\mathcal{M}_B = \mathbb{F}_2$) while it can be a
singular variety possibly homeomorphic, yet not diffeomorphic to a
2-sphere in degenerate cases of the resolution $\mathcal{M}_6
\longrightarrow \mathbb{C}^3/\mathbb{Z}_4$ or in other cases that
cannot be retrieved from the resolution of
$\mathbb{C}^3/\mathbb{Z}_4$, yet adhere to the same double fibration
scheme of eqn. \eqref{doppiafibbia}. This is for instance the case
of the singular  K\"ahler Einstein manifolds $\mathcal{M}^{KE}_4$
discussed later on in these lectures.
\begin{equation}\label{frombolato}
    \forall p \in \mathcal{M}_B \quad : \quad \pi^{-1}_1(p)\, \simeq
    \, \mathbb{C} \quad ; \quad \forall q
    \in \mathbb{S}^2 \quad : \quad \pi_2^{-1} (q) \, \simeq \,
    \left\{\begin{array}{lclcl}
             \mathbb{S}^2 &\null&\null & \text{if} & \mathcal{M}_B \, = \, \mathbb{F}_2\\
             \mathfrak{F} &\underbrace{\sim}_{homeomorphic}& \mathbb{S}^2 & \text{if} &
             \mathcal{M}_B \, =\, \mathcal{M}_B^{KE}\\
             \mathfrak{F} & \simeq & \text{?} & \text{if} &
             \mathcal{M}_B \, = \, \text{other cases}\\
           \end{array}
     \right.
\end{equation}
\begin{figure}[htb!]
\centering
\includegraphics[width=9cm]{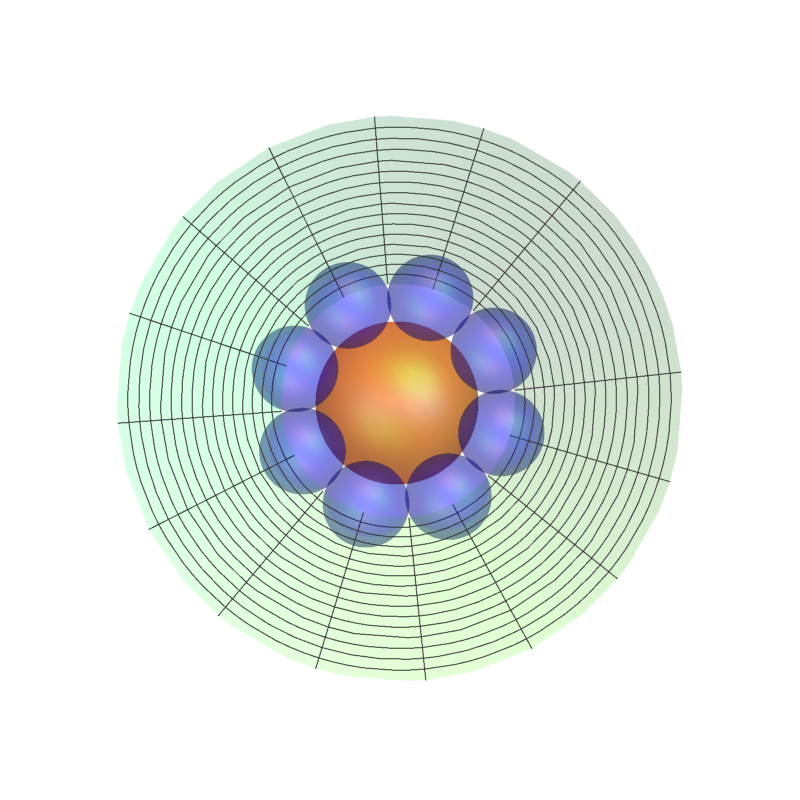}
\caption{\label{Yconcept} A conceptual picture of the resolved
three-fold $\mathcal{M}_6$ displaying its double fibration
structure. The orange sphere in the middle symbolizes the base
manifold of the bundle $\mathcal{M}_B$. A dense complex coordinate
patch for this $\mathbb{P}^1$ is named $u$ in the main body of these
lectures. The blueish spheres around the orange one symbolize the
$\mathbb{P}^1$ fibres of $\mathcal{M}_B$ that can be degenerate and
display conical singularities. A dense complex coordinate patch for
these fibres is named $v$ in the main body of the lectures. Finally
the greenish rays enveloping the base manifold $\mathcal{M}_B$
symbolize the noncompact fibres of the bundle $\mathcal{M}_6$. A
dense coordinate patch for these fibres is named $w$ in the main
body of the lectures.}
\end{figure}
A conceptual picture of the geometric structure encoded in eqn.s
\eqref{doppiafibbia},\eqref{frombolato} is displayed in
fig.\ref{Yconcept}.
\par
\subsection{Explicit reduction of the moment map equations to
$\mathcal{M}_B$} Introducing the definitions:
\begin{equation}\label{lattosio}
    \varpi \, \equiv \,\left(1+ \mid u\mid^2\right)^2 \mid
    v\mid^2 \quad ; \quad \lambda \, \equiv  \, \sqrt{\mid w\mid} \quad ;
    \quad \sigma \,\equiv \,  \mid v\mid \quad ; \quad \delta \,
    \equiv \, \frac{\sqrt{\varpi}}{\sigma}
\end{equation}
we obtain
\begin{equation}\label{rimmel}
  \Sigma \, \to \,  \lambda  \sqrt{\varpi }\quad ; \quad  \, U \, \to \,\lambda ^2
\end{equation}
that can be substituted into the moment map equations
\eqref{rupetto}. These latter can be reduced by means of a limiting
procedure $\lambda\to 0$ to the base manifold $\mathcal{M}_B$ by
introducing suitable rescalings that were determined in
\cite{noietmarcovaldo} and further studied in
\cite{Bianchi_2021,bruzzo2023d3brane}:
\begin{equation}\label{baciodidama}
    X_1\to T_1 \sqrt{\frac{\lambda }{\sigma
   }}\quad ,\quad X_2\to \lambda ^2 T_2\quad ,\quad X_3\to T_3
   \sqrt{\frac{\lambda }{\sigma }}
\end{equation}
Performing the substitutions \eqref{rimmel},\eqref{baciodidama} in
equations \eqref{rupetto} and implementing the limit $\lambda \to 0$
one gets the final result of a new system of algebraic equations for
the \textit{moment maps} $T_{1,2,3}$:
\begin{equation}\label{basemomentmap}
   \left(
\begin{array}{c}
 T_1 \left(\left(\zeta _3-\zeta _1\right) \sigma  T_2
   T_3+T_3^3 \sqrt{\varpi }\right)-\sigma  T_1^2+\sigma
    T_3^2+T_3 T_1^3 \left(-\sqrt{\varpi }\right) \\
 \sigma  T_2 \left(\sigma  \sqrt{\varpi }-\left(\zeta
   _1-\zeta _2+\zeta _3\right) T_1 T_3\right)-T_1 T_3
   \left(T_1^2+T_3^2\right) \sqrt{\varpi } \\
 \sigma  \left(-\zeta _2 T_2 T_3 T_1+\sigma  T_2
   \sqrt{\varpi }+T_1^2+T_3^2\right) \\
\end{array}
\right) \, = \, \left(
\begin{array}{c}
 0 \\
\begin{array}{c}
 0 \\
 0 \\
\end{array}
 \\
\end{array}
\right)
\end{equation}
while the K\"ahler potential of eqn.\eqref{HKLRkallero} in the same
limit $\lambda \to 0$ reduces to an expression for the K\"ahler
potential of the metric on the $\mathcal{M}_B$ base manifold:
\begin{eqnarray}\label{cortisolo}
  \mathcal{K}_4 &=& \frac{\sigma  T_1^2+\sigma
   \left(\sigma  T_2 \sqrt{\varpi }+T_3^2\right)+T_3
   T_1^3 \sqrt{\varpi }+T_3^3 T_1 \sqrt{\varpi
   }}{\sigma  T_1 T_2 T_3}\nonumber\\
   &&+\left(2 \zeta _1-\zeta _2\right) \log
   \left(T_1\right)+\left(-\zeta _1+2 \zeta _2-\zeta
   _3\right) \log \left(T_2\right)+\left(2 \zeta
   _3-\zeta _2\right) \log
   \left(T_3\right)
\end{eqnarray}
\subsubsection{Solution of the reduced moment map equations} The
moment map equations \eqref{basemomentmap} are three for three
unknowns $T_{1,2,3}$ in terms of two variables $\sigma,\varpi$.
There must be a functional relation for one moment map in terms of
the other two. In the $6$-dimensional case such a relation is unique
and reduces on the $4$-dimensional manifold to the following one:
\begin{equation}\label{merluzzo}
    T_3 \, = \, \sqrt{\frac{\zeta _3-\zeta _2}{\zeta _1-\zeta _2}}\, T_1
\end{equation}
Focusing on the Weyl chamber VI as defined in
\eqref{celle8},\eqref{celle8piu}, the branch of the solution to the
moment map equations \eqref{basemomentmap} that is positive for all
the function $T_{1,2,3}$ is given below:
\begin{align}\label{tipositivo}
T_1 & = \,\frac{\sqrt{\frac{\sigma  \left(\sqrt{\left(-\zeta
   _1+\zeta _2-\zeta _3+\varpi \right){}^2+4 \zeta _2
   \varpi }-\zeta _1+\zeta _2-\zeta _3+\varpi
   \right)}{\zeta _2 \sqrt{\frac{\left(\zeta _2-\zeta
   _3\right) \varpi }{\zeta _2-\zeta _1}}}}}{\sqrt{2}}\nonumber\\
T_2 & = \, \frac{\sqrt{\frac{\zeta _3-\zeta _2}{\zeta _1-\zeta
   _2}} \left(\sqrt{\left(-\zeta _1+\zeta _2-\zeta
   _3+\varpi \right){}^2+4 \zeta _2 \varpi }-\zeta _1+3
   \zeta _2-\zeta _3+\varpi \right)}{2 \zeta _2
   \left(\zeta _2-\zeta _3\right)}\nonumber\\
T_3 & = \sqrt{\frac{\zeta _3-\zeta _2}{\zeta _1-\zeta _2}}\, T_1
\end{align}
\paragraph{Choice of a convenient line in the Weyl chamber VI} In
view of the solution \eqref{tipositivo} and of the constraint
\eqref{merluzzo} a convenient choice that in no ways diminishes the
generality of the result is the following one:
\begin{equation}\label{lineagotica}
    \zeta_1 \, = \, \zeta_3 \, = \, s >0 \quad ; \quad \zeta_2 \,=\, s
    \left (2+\alpha\right) \quad ; \quad \alpha >0
\end{equation}
where $s$ is a dimensionful parameter that we can always reabsorb
into a rescaling of the coordinates while $\alpha$ is a
dimensionless positive real number that parameterizes a line in the
VIth Weyl chamber orthogonal to the wall $\mathcal{W}_3$. For
$\alpha=0$ we are on the wall and we get the partial resolution
$\mathcal{M}_6 \, = \,
\operatorname{tot}\left[K\left(\mathbb{WP}_{112}\right)\right]$
while for $\alpha>0$ we have $\mathcal{M}_6 \, = \,
\operatorname{tot}\left[K\left(\mathbb{F}_{2}\right)\right]$. It
follows that the corresponding solutions of the reduced moment maps
\eqref{tipositivo}, upon insertion into eqn. \eqref{cortisolo},
produce the K\"ahler potential of a K\"ahler metric defined either
on $\mathbb{F}_2$, or in the limiting case $\alpha=0$ on
$\mathbb{WP}_{112}$.
\par
On the chosen line the solution of the moment map equations
\eqref{tipositivo} reduces to:
\begin{align}\label{trombetta}
T_3 \, = \,T_1 \, & = \, \frac{\sqrt{\sigma }
\sqrt{\frac{\sqrt{\alpha ^2+6
   \alpha  \varpi +\varpi  (\varpi +8)}+\alpha +\varpi
   }{(\alpha +2) \sqrt{\varpi }}}}{\sqrt{2}}\nonumber\\
T_2 \, & = \, \frac{\sqrt{\alpha ^2+6 \alpha  \varpi +\varpi ^2+8
   \varpi }+3 \alpha +\varpi +4}{2 \alpha ^2+6 \alpha
   +4}
\end{align}
Correspondingly the K\"ahler potential in eqn.\eqref{cortisolo}
becomes
\begin{eqnarray}\label{varpifamiglia}
  \mathcal{K}\left(\varpi\right) &=& \frac{1}{2}
   \left(\sqrt{\alpha ^2+6 \alpha  \varpi +\varpi
   (\varpi +8)}+3 \alpha -\varpi +4\right)\nonumber \\
  &&+ 2 (\alpha +1) \log \left(\frac{\sqrt{\alpha ^2+6 \alpha
    \varpi +\varpi ^2+8 \varpi }+3 \alpha +\varpi +4}{2
   \alpha ^2+6 \alpha +4}\right)\nonumber\\
   &&-2 \alpha
   \log \left(\frac{\sqrt{\frac{\sqrt{\alpha ^2+6
   \alpha  \varpi +\varpi  (\varpi +8)}+\alpha +\varpi
   }{(\alpha +2) \sqrt{\varpi }}}}{\sqrt{2}}\right)
\end{eqnarray}
Notice that dependence on the variable $\sigma$ has disappeared in
the final formula \eqref{varpifamiglia}. In the algebraic part of
the K\"ahler potential (first line of eqn.\eqref{cortisolo}) there
is an exact cancellation between numerator and denominator. In the
second logarithmic part of the K\"ahler potential this happens
because $\log [\sigma] \, = \, \log[v] \, + \, \log[{\bar v}]$ is
the real part of a holomorphic function which does not affect the
metric and  simply corresponds to a K\"ahler gauge transformation.
The ultimate outcome is that the K\"ahler metric induced on the base
manifold $\mathcal{M}_B$ by the Kronheimer construction is encoded
into a K\"ahler potential $\mathcal{K}\left(\varpi\right)$ that
depends only on the coordinate combination $\varpi$ defined in eqn.
(\eqref{lattosio}). This has to do with the isometries of the
resulting metric which, as I explain in the next section, is always
$\mathrm{SU(2) \times U(1)}$ and defines a differentiable manifold
of cohomogeneity one.
\section{The AMSY symplectic action/angle formalism}
\label{amysone} Following the discussions and elaborations of
\cite{Bianchi_2021} based on
\cite{abreu,Martelli:2005tp,Bykov:2017mgc}, given the K\"ahler
potential of a toric complex $n$-dimensional K\"ahler manifold
$\mathcal{K}(|z_1|,..,|z_n|)$,  where {$z_i=e^{x_i+i\Theta_i}$} are
the complex coordinates, introducing the moment variables
\begin{equation}\label{momentini2}
    \mu^i  =
    \partial_{x_i}\mathcal{K}
\end{equation}
we can obtain the so named symplectic potential by means of the
following Legendre transform:
\begin{equation}\label{legendretr}
    {G}\left(\mu_i\right)
    =\sum_{i}^n x_i \,\mu^i \,    - \, \mathcal{K}(|z_1|,..,|z_n|)
\end{equation}
where one assumes that $\mathcal{K}$ only depends on the modules of
the $z$ coordinates to achieve $\mathrm{U}(1)^n$ invariance. The
main issue involved in the use of eqn.~\eqref{legendretr} is the
inversion transformation that expresses the coordinates $x_i$ in
terms of the  moments $\mu^i$. Once this is done one can calculate
the metric in moment variables utilizing the Hessian:
\begin{equation}\label{hessiano}
    {G}_{ij} = \frac{\partial^2}{\partial\mu^i \partial\mu^j} {G}\left(\mu\right)
\end{equation}
and its matrix inverse. Call the $n$ angles by $ \Theta_i $. Complex
coordinates better adapted to the complex structure tensor can be
defined as
\begin{equation}\label{spigolini}
    u_i= e^{z_i} = \exp[x_i \,+\, \mathrm{i} \Theta_i]
\end{equation}
The K\"ahler 2-form has the following universal structure:
\begin{equation}\label{uniKal2}
  \mathbb{K} \,= \,  \sum_{i=1}^{n} \, \mathrm{d}\mu^i\wedge
  \mathrm{d}\Theta_i
\end{equation}
and the metric is expressed as
\begin{equation}\label{sympametra}
    ds^2_{symp} = {\mathbf{G}}_{ij} d\mu^i \, d\mu^j \, + \, {\mathbf{G}}^{-1}_{ij}d\Theta^i \, d\Theta^j
\end{equation}
In the more recent papers pertaining to the overall research project
that the present lectures aim at reviewing, strategic use of this
formalism has been made both in the $4$-dimensional case of the
base-manifold $\mathcal{M}_B$ and in the $6$-dimensional case of the
total bundle manifold $\mathcal{M}_6$.
\section{K\"ahler metrics    with $\mathrm{SU(2)\times U(1)} $ isometry} \label{varposympo}
Inspired by the results of the reduction to the exceptional divisor
of the moment map equations encoded in eqn. \eqref{varpifamiglia} we
are interested, to begin with, in K\"ahler metrics in two complex
dimensions, $n=2$, where the complex coordinates $u,v$ enter the
K\"ahler potential $\mathcal{K}_0(\varpi)$ only through the real
combination $\varpi$ mentioned in eqn.\eqref{lattosio}. This
restriction guarantees invariance of the corresponding metric under
$\mathrm{SU(2)}\times \mathrm{U(1)}$ transformations realized  as
\begin{equation}\begin{array}{rcl}
\text{if}\  \mathbf{g} &=& \left(
          \begin{array}{cc}
            a & b \\
            c & d \\
          \end{array}
        \right) \, \in \, \mathrm{SU(2)}\quad \text{then}  \quad
    \mathbf{g}\left(u,v\right) =\left(\frac{a \, u + b}{c\, u +
    d}, \quad v \, \left(c \,u+d\right)^{2}\right) ;
     \\[3pt]
\text{if}\   \mathbf{g} &=& \exp(i\,\theta_1) \, \in \,
\mathrm{U(1)} \quad  \text{then} \quad \mathbf{g}\left(u,v \right)
=\left(u, \quad \exp(i\,\theta_1)\, v  \right).
\label{ciabattabuona}
\end{array}\end{equation}
The above realization of the isometry  captures the idea that, at
least locally, the manifold  is an $\mathbb{S}^2$ fibration over
$\mathbb{S}^2$ ($u$ being a coordinate on the base and $v$ a fibre
coordinate), although their global topology might be different and
have some kind of singularities.
\par
Two cases are of particular interest within such a framework, namely
\begin{description}
  \item[a)]   the singular weighted projective plane $\mathbb{WP}_{[1,1,2]}$;
  \item[b)]  the  second Hirzebruch surface
  $\mathbb{F}_2$.
\end{description}
In the case of the singular variety $\mathbb{WP}_{[1,1,2]}$ we have
a non K\"ahler--Einstein  metric that emerges from a partial
resolution of the $\mathbb{C}^3/\mathbb{Z}_4$ singularity within the
generalized Kronheimer construction (see
\cite{noietmarcovaldo},\cite{Bianchi_2021}) whose explicit K\"ahler
potential is  the case $\alpha=0$ in eqn. \eqref{varpifamiglia}. On
the other hand the K\"ahler potential of the Kronheimer induced
metric on $\mathbb{F}_2$ corresponds to the case $\alpha>0$ in the
same equation.
\subsection{A family of   4D K\"ahler metrics}
Having mentioned the origin of the idea, now, using the AMSY
approach, we discuss in more general terms  a class of real 4D
K\"ahler manifolds, that we call $\mathcal{M}_B$ and were studied in
\cite{bruzzo2023d3brane}. These are endowed with a metric invariant
under the $\mathrm{SU(2)\times U(1)}$ isometry group acting as in
eqn.~\eqref{ciabattabuona}. As we already said the class of these
manifolds is singled out  by the above assumption that, in the
complex formalism, their K\"ahler potential $\mathcal{K}_0(\varpi)$
is  a function only of the invariant $\varpi$.  The explicit form of
the K\"ahler potential $\mathcal{K}_0(\varpi)$ cannot be worked out
analytically in all cases since the inverse Legendre transform
involves the roots of higher order algebraic equations; yet, using
the $\varpi$-dependence assumption, the K\"ahler metric can be
explicitly worked out in the symplectic coordinates and has a simple
and very elegant form -- actually the metric depends on a single
function of one variable $\mathcal{FK}(\mathfrak{v} )$ which encodes
all the geometric properties and substitutes
$\mathcal{K}_0(\varpi)$. Posing all the questions in this symplectic
language allows  one  to calculate all the geometric properties of
the spaces in the class under consideration and leads also to new
results and to a more systematic overview of the already known
cases. We choose to treat the matter in general, by utilizing a
\textit{local approach} where we discuss the differential equations
in a given open dense coordinate patch $u,v$, and we address the
question of its global topological and algebraic structure  only a
posteriori, once the metric as been found in the considered chart,
just as one typically does in General Relativity.
\par
The symplectic structure of the metric on $\mathcal{M}_B$ is
exhibited in the following way:
\begin{equation}\label{trottolina}
    ds_{\mathcal{M}_B}^2 = \mathbf{g}_{\mathcal{M}_B|\mu\nu}
    \, dq^\mu \, dq^\nu \quad ; \quad q^\mu  =   \left\{\mathfrak{u},\mathfrak{v} ,\phi,\tau\right\} \quad ;
    \quad \mathbf{g}_{\mathcal{M}_B} =\left(\begin{array}{c|c}
    \mathbf{G}_{\mathcal{M}_B} & \mathbf{0}_{2\times 2} \\
    \hline
    \mathbf{0}_{2\times 2} &
    \mathbf{G}^{-1}_{\mathcal{M}_B}
    \end{array}\right)
\end{equation}
where the Hessian $\mathbf{G}_{\mathcal{M}_B}$ is defined by:
\begin{align}\mathbf{G}_{\mathcal{M}_B}  =
    \partial_{\mu^i}\,\partial_{\mu^j} \,
    {G}_{\mathcal{M}_B} \quad ; \quad \mu^i =
    \left\{\mathfrak{u},\mathfrak{v} \right\}\label{reducedhessian}
\end{align}
and
\begin{equation}
{G}_{\mathcal{M}_B}= {G}_0(\mathfrak{u},\mathfrak{v} ) \, + \,
\mathcal{D}(\mathfrak{v} )\label{GBsymplectic} \end{equation}
\begin{equation}
 G_0\left(\mathfrak{u},\mathfrak{v} \right) =  \left(\mathfrak{v} -\frac{\mathfrak{u}}{2}\right) \log (2
   \mathfrak{v} -\mathfrak{u})+\frac{1}{2} \mathfrak{u} \log
   (\mathfrak{u})-\frac{1}{2} \mathfrak{v}  \log (\mathfrak{v} ) \label{lupetto}
\end{equation}
The specific structure \eqref{GBsymplectic},\eqref{lupetto} is the
counterpart  within the symplectic formalism, via Legendre
transform, of the assumption that the K\"ahler potential
$\mathcal{K}_0(\varpi)$ depends only on the $\varpi$ variable.
\par
After noting this important   point, we go back  to  the discussion
of $\mathcal{M}_B$  geometry and   stress that with the given
isometries its Riemannian structure is completely encoded in the
boundary function $ \mathcal{D}(\mathfrak{v} )$. All the other items
in the construction are as follows. For the K\"ahler form we have
\begin{equation}\label{kallerformMB}
    \mathbb{K}^{\mathcal{M}_B} = 2\, \left( \, d\mathfrak{u} \wedge d\phi + d\mathfrak{v}  \wedge d\tau   \right) \,
    = \, \mathbf{K}_{\mu\nu}^{\mathcal{M}_B} \, dq^\mu \wedge
    dq^\nu \quad ; \quad \mathbf{K}^{\mathcal{M}_B} = \left(\begin{array}{c|c}
    \mathbf{0}_{2\times 2} & \mathbf{1}_{2\times 2} \\
    \hline
    - \mathbf{1}_{2\times 2} &
    \mathbf{0}_{2\times 2}
    \end{array}\right)
\end{equation}
and for the complex structure we obtain
\begin{equation}\label{complestruc2}
\mathfrak{J}^{\mathcal{M}_B} = \mathbf{K}^{\mathcal{M}_B}\,
\mathbf{g}^{-1}_{\mathcal{M}_B} = \,\left(\begin{array}{c|c}
\mathbf{0}_{2\times 2} & \mathbf{G}_{\mathcal{M}_B} \\
\hline - \mathbf{G}^{-1}_{\mathcal{M}_B} & \mathbf{0}_{2\times 2}
\end{array}\right)
\end{equation}
Explicitly the $2\times 2 $ Hessian is the following:
\begin{equation}\begin{array}{rcl}\label{GMB}
 \mathbf{G}_{\mathcal{M}_B}& = &    \left(
\begin{array}{cc}
 -\frac{\mathfrak{v} }{\mathfrak{u}^2-2 \mathfrak{u} \mathfrak{v} } &
   \frac{1}{\mathfrak{u}-2 \mathfrak{v} } \\
 \frac{1}{\mathfrak{u}-2 \mathfrak{v} } & \frac{-2 \mathfrak{v}
   (\mathfrak{u}-2 \mathfrak{v} ) \mathcal{D}''(\mathfrak{v} )+\mathfrak{u}+2
   \mathfrak{v} }{2 \mathfrak{v}  (2 \mathfrak{v} -\mathfrak{u})} \\
\end{array}
\right) \\
\mathbf{G}_{\mathcal{M}_B}^{-1}& = &\left(
\begin{array}{cc}
 \frac{\mathfrak{u} \left(-2 \mathfrak{v}  (\mathfrak{u}-2
   \mathfrak{v} ) \mathcal{D}''(\mathfrak{v} )+\mathfrak{u}+2
   \mathfrak{v} \right)}{\mathfrak{v}  \left(2 \mathfrak{v}
   \mathcal{D}''(\mathfrak{v} )+1\right)} & \frac{2 \mathfrak{u}}{2
   \mathfrak{v}  \mathcal{D}''(\mathfrak{v} )+1} \\
 \frac{2 \mathfrak{u}}{2 \mathfrak{v}  \mathcal{D}''(\mathfrak{v} )+1} &
   \frac{2 \mathfrak{v} }{2 \mathfrak{v}  \mathcal{D}''(\mathfrak{v} )+1}
   \\
\end{array}
\right)
\end{array}\end{equation}
The family of metrics \eqref{trottolina} is parameterized by the
choice of a unique one-variable function:
\begin{equation}\label{fungemistero}
    \mathit{f}(\mathfrak{v} )\equiv\mathcal{D}''(\mathfrak{v} )
\end{equation}
and is   worth being considered in its  own  right. The explicit
form of such metrics is the following one:
\begin{equation}\label{metrauniversala}
    ds^2_{\mathcal{M}_B}= \frac{d\mathfrak{v} ^2}{\mathcal{FK}(\mathfrak{v} )}\, + \,\mathcal{FK}(\mathfrak{v} )
     \left[d\phi  (1-\cos \theta )+d\tau
   \right]^2\,+\,\mathfrak{v}
   \underbrace{\left(d\phi^2 \sin ^2\theta +d\theta
   ^2\right)}_{\mathbb{S}^2 \, \text{metric}}
\end{equation}
where we have defined
\begin{equation}\label{FKfunzione}
   \mathcal{FK}(\mathfrak{v} )=\frac{2 \mathfrak{v} }{2 \mathfrak{v}  \mathcal{D}''(\mathfrak{v} )+1}
\end{equation}
The expression \eqref{metrauniversala} for the metric is obtained
performing the following convenient change of variables:
\begin{equation}\label{cambiovariabile}
    \mathfrak{u}\, \rightarrow \, \left(1 -\cos\theta\right)\,\mathfrak{v} \quad ;
    \quad \theta \, \in \, \left[0,\pi\right]
\end{equation}
which automatically takes into account that $\mathfrak{u}\leq
2\,\mathfrak{v} $.
\subsection{The inverse Legendre transform}
\label{reconstruczia} Before proceeding further with the analysis of
the class of metrics \eqref{metrauniversala}, it is convenient to
consider the inverse Legendre transform and see how one reconstructs
the K\"ahler potential on $\mathcal{M}_B$. The inverse Legendre
transform provides the K\"ahler potential through the formula:
\begin{equation}\begin{array}{rcl}\label{K0funzia}
    \mathcal{K}_0 = x_u \, \mathfrak{u} \,+ \,  x_v \, \mathfrak{v}
    \, - \, G_{\mathcal{M}_B}\left(\mathfrak{u},\mathfrak{v} \right)
\end{array}\end{equation}
where $G_{\mathcal{M}_B}\left(\mathfrak{u},\mathfrak{v} \right)$ is
the base manifold symplectic potential defined in
eqn.~\eqref{GBsymplectic}, and
\begin{equation}\label{caramboliere}
    x_u =
    \partial_{\mathfrak{u}}\,G_{\mathcal{M}_B}\left(\mathfrak{u},\mathfrak{v} \right)
    \quad ; \quad x_v =
    \partial_{\mathfrak{v} }\,G_{\mathcal{M}_B}\left(\mathfrak{u},\mathfrak{v} \right)
\end{equation}
which explicitly yields:
\begin{equation}\label{inversioneB}
  x_u=  \frac{1}{2} \left(\log
   \left(\mathfrak{u}\right)-\log \left(2
   \mathfrak{v} -\mathfrak{u}\right)\right)\quad ; \quad x_v=
   \mathcal{D}'\left(\mathfrak{v} \right)+\log \left(2
   \mathfrak{v} -\mathfrak{u}\right)-\frac{1}{2} \log
   \left(\mathfrak{v} \right)+\frac{1}{2}
\end{equation}
Using eqn.~\eqref{inversioneB}) in eqn.~\eqref{K0funzia} we
immediately obtain the explicit form of the base-manifold K\"ahler
potential as a function of the moment $\mathfrak{v} $:
\begin{equation}\label{K0inv0}
    \mathcal{K}_0 = \mathfrak{K}_0\left(\mathfrak{v} \right)=\mathfrak{v}
   \left(\mathcal{D}'\left(\mathfrak{v} \right)+\frac{1}{2}\right)-\mathcal{D}\left(\mathfrak{v} \right)
\end{equation}
The problem is that we need the K\"ahler potential $\mathcal{K}_0$
as a  function of the invariant $\varpi$. Utilizing eqn.
\eqref{inversioneB} it is fairly easy to obtain the expression of
$\varpi$ in terms of the moment $\mathfrak{v} $ for a generic
function $\mathcal{D}(\mathfrak{v} )$ that codifies the geometry of
the base-manifold, obtaining
\begin{equation}\label{omegadiv0}
    \varpi=\left(1+\exp\left[2\,x_u\right]\right)^2
    \,\exp\left[2\,x_u\right] =
    \Omega\left(\mathfrak{v} \right)= 4
    \mathfrak{v}  \exp\left[\,2\,
    \partial_{\mathfrak{v} }\mathcal{D}\left(\mathfrak{v} \right)+1\right]
\end{equation}
If  one is able to invert the function $\Omega\left(\mathfrak{v}
\right)$, the original K\"ahler potential of the base-manifold can
be written as:
\begin{equation}\label{compostadimele}
    \mathcal{K}_0 \left(\varpi\right) =\mathfrak{K}_0\, \circ
    \,
    \Omega^{-1}\left(\varpi\right)
\end{equation}
The inverse function $\Omega^{-1}\left(\varpi\right)$ can be written
explicitly in some simple cases,  but not always, and this inversion
is the main reason why certain K\"ahler metrics can be much more
easily found in the AMSY symplectic formalism which deals only with
real variables than in the complex formalism. Since nothing good
comes without paying a price, the metrics found in the symplectic
approach require that the ranges of the variables $\mathfrak{u}$ and
$\mathfrak{v} $ should be determined, since it is just in those
ranges that the topology and algebraic structure of the underlying
manifold is hidden; indeed the ranges of $\mathfrak{u}$ and
$\mathfrak{v} $ define a convex closed \textit{polytope} in the
$\mathbb{R}^2$ plane that encodes very precious information about
the structure of the underlying manifold.
\section{The Ricci tensor and the Ricci form of the base manifold $\mathcal{M}_B$}  Calculating the Ricci tensor
for the family of metrics \eqref{trottolina} we obtain the following
structure:
\begin{equation}\label{ricciodimare}
\mathrm{Ric}_{\mu\nu}^{\mathcal{M}_B} = \left(\begin{array}{c|c}
\mathbf{P}_\mathrm{U} & \mathbf{0}_{2\times 2} \\
\hline \mathbf{0}_{2\times 2} & \mathbf{P}_\mathrm{D}
\end{array}
\right)
\end{equation}
The expressions for $\mathbf{P}_\mathrm{U}$ and
$\mathbf{P}_\mathrm{D}$ are quite lengthy and we omit them. We
rather consider the Ricci 2-form defined by:
\begin{equation}\label{riccioforma}
\mathbb{R}\text{ic}_{\mathcal{M}_B} =
\mathbf{Ric}_{\mu\nu}^{\mathcal{M}_B} \, dq^\mu
    \wedge dq^\nu
\end{equation}
where:
\begin{equation}\label{riccioformcompo}
    \mathbf{Ric}^{\mathcal{M}_B}=\mathrm{Ric}^{\mathcal{M}_B} \,
    \mathfrak{J}^{\mathcal{M}_B} = \left(\begin{array}{c|c}
    \mathbf{0}_{2\times 2} & \mathbf{R} \\
    \hline
    - \mathbf{R}^T & \mathbf{0}_{2\times 2}
    \end{array}
     \right)
\end{equation}
and\begin{equation}\begin{array}{rcl} \mathbf{R} &=& \left(
\begin{array}{cc}
r_{11} & r_{12} \\
r_{21} & r_{22} \\
\end{array}
\right)
       \\
     r_{11} &=& \frac{2 \mathfrak{v}  \left(\mathfrak{v}  \mathit{f}'(\mathfrak{v} )+2 \mathfrak{v}
   \mathit{f}(\mathfrak{v} )^2+\mathit{f}(\mathfrak{v} )\right)-1}{2 \mathfrak{v}  (2
   \mathfrak{v}  \mathit{f}(\mathfrak{v} )+1)^2}  \\
     r_{12} &=& 0  \\
     r_{21} &=& \displaystyle\frac{2 \mathfrak{u} \mathfrak{v}  \left(\mathfrak{v} ^2 (2 \mathfrak{v}
   \mathit{f}(\mathfrak{v} )+1) \mathit{f}''(\mathfrak{v} )-4 \mathfrak{v} ^3
   \mathit{f}'(\mathfrak{v} )^2+4 \mathfrak{v}  \mathit{f}'(\mathfrak{v} )-4
   \mathfrak{v} ^2 \mathit{f}(\mathfrak{v} )^3-2 \mathfrak{v}
   \mathit{f}(\mathfrak{v} )^2+3 \mathit{f}(\mathfrak{v} )\right)+\mathfrak{u}}{2
   \mathfrak{v} ^2 (2 \mathfrak{v}  \mathit{f}(\mathfrak{v} )+1)^3}  \\[5pt]
     r_{22} &=& \displaystyle \frac{\mathit{f}(\mathfrak{v} ) \left(2 \mathfrak{v} ^2 \left(\mathfrak{v}
   \mathit{f}''(\mathfrak{v} )+\mathit{f}'(\mathfrak{v} )\right)+3\right)+\mathfrak{v}
   \left(\mathfrak{v}  \mathit{f}''(\mathfrak{v} )-4 \mathfrak{v} ^2
   \mathit{f}'(\mathfrak{v} )^2+5 \mathit{f}'(\mathfrak{v} )\right)+2 \mathfrak{v}
   \mathit{f}(\mathfrak{v} )^2}{(2 \mathfrak{v}  \mathit{f}(\mathfrak{v} )+1)^3}  \\
     \mathbf{R}^T&=& \mathbf{P}_\mathrm{D} \,
     \mathbf{G}^{-1}_{\mathcal{M}_B} \label{riccopovero}
   \end{array}\end{equation}
 The last equation is not a definition but rather a consistency
constraint (the Ricci  tensor must be  skew-symmetric).
\par
\paragraph{A two-parameter family of KE metrics for
$\mathcal{M}_B$. }\label{famigliaKE} An interesting and   legitimate
question is whether the family of cohomogeneity one metrics
\eqref{metrauniversala} that we name
$\operatorname{Met}(\mathcal{FV})$ since they are parameterized by
the function $\mathcal{FK}(\mathfrak{v})$, contains K\"ahler
Einstein ones. The answer is positive, and the KE metrics make up a
two-parameter subfamily. As it will be shown in the next section,
such a KE family $\operatorname{Met}(\mathcal{FV})^{KE}$ is a
subfamily of a 4-parameter family of extremal metrics
$\operatorname{Met}(\mathcal{FV})_{ext}$ found many decades ago by
Calabi. Let us first directly retrieve the KE family from the
appropriate differential constraint. A metric is KE if the Ricci
2-form is proportional to the K\"ahler 2-form:
\begin{equation}\label{kallopietra}
     \mathbb{R}\text{ic}^{\mathcal{M}_B} = \frac{\mathit{k}}{4} \, \mathbb{K}^{\mathcal{M}_B}
\end{equation}
where $\mathit{k}$ is a  constant. This amounts to requiring  that
the $2\times 2$ matrix $\mathbf{R}$ displayed in
eqn.~\eqref{riccopovero} be proportional via $\frac{\mathit{k}}{4}$
to the identity matrix $\mathbf{1}_{2\times2}$. This condition
implies differential constraints on the function
$\mathit{f(\mathfrak{v} )}$ that are uniquely solved by the
following function:
\begin{equation}\label{miraculo}
\mathit{f(\mathfrak{v} )} = \frac{-3 \beta +\mathit{k} \mathfrak{v}
^3+3 \mathfrak{v} ^2}{-2 \mathit{k} \mathfrak{v} ^4+6 \mathfrak{v}
^3+6 \beta \mathfrak{v} }
\end{equation}
the parameter $\beta$ being the additional integration constant,
while $k$ is defined by equation \eqref{kallopietra}.  To retrieve
the original symplectic potential $\mathcal{D}(\mathfrak{v} )$ one
has just to perform a double integration in the variable
$\mathfrak{v} $. The explicit calculation of the integral requires a
summation over the three roots $\lambda_{1,2,3}$ of the following
cubic polynomial:
\begin{equation}\label{racete}
    P(x)=x^3\, -\, \frac{3 x^2}{\mathit{k}}\, -\, \frac{3 \beta
    }{\mathit{k}}
\end{equation}
whose main feature is the absence of the linear term. Hence a
beautiful way of parameterizing the family of KE metrics is achieved
by using as parameters two of the three roots of the polynomial
\eqref{racete}. Let us call the independent roots $\lambda_1$ and
$\lambda_2$. The polynomial \eqref{racete} is reproduced by setting:
\begin{equation}\label{rinominopara}
    \mathit{k}\,= \, \frac{3 \left(\lambda _1+\lambda
   _2\right)}{\lambda _1^2+\lambda _2 \lambda _1+\lambda _2^2},\quad \quad \beta
   =  -\frac{\lambda _1^2 \lambda _2^2}{\lambda _1^2+\lambda _2
   \lambda _1+\lambda _2^2},\quad \lambda _3= -\frac{\lambda _1 \lambda
   _2}{\lambda _1+\lambda _2}
\end{equation}
Substituting \eqref{rinominopara} in eqn.~\eqref{miraculo} we obtain
\begin{equation}\begin{array}{rcl}\label{gamellino}
  \mathit{f}(\mathfrak{v} )& = & -\frac{\lambda _1 \mathfrak{v} ^2
   \left(\lambda _2+\mathfrak{v} \right)+\lambda _2 \mathfrak{v} ^2
   \left(\lambda _2+\mathfrak{v} \right)+\lambda _1^2 \left(\lambda
   _2^2+\mathfrak{v} ^2\right)}{2 \mathfrak{v}
   \left(\mathfrak{v} -\lambda _1\right) \left(\mathfrak{v} -\lambda
   _2\right) \left(\lambda _2 \mathfrak{v} +\lambda _1 \left(\lambda
   _2+\mathfrak{v} \right)\right)}
\end{array}\end{equation}
which is completely symmetrical in the exchange of the two
independent roots $\lambda_1,\lambda_2$. Utilizing the expression
\eqref{gamellino} the double integration is easily performed, and we
obtain the following explicit result, where we   omitted
irrelevant linear terms:
\begin{equation}\begin{array}{rcl}\label{sympaKE}
  \mathcal{D}^{KE}(\mathfrak{v} )&=& -\frac{\left(\lambda _1^2+\lambda _2 \lambda _1+\lambda _2^2\right)
   \left(\mathfrak{v} -\lambda _1\right) \log \left(\mathfrak{v} -\lambda
   _1\right)}{\lambda _1^2+\lambda _2 \lambda _1-2 \lambda _2^2} \\[3pt]
   &&  \hskip20mm -\frac{\left(\lambda
   _1^2+\lambda _2 \lambda _1+\lambda _2^2\right) \left(\mathfrak{v} -\lambda
   _2\right) \log \left(\mathfrak{v} -\lambda _2\right)}{-2 \lambda _1^2+\lambda _2
   \lambda _1+\lambda _2^2} \\[3pt]
   &&+\frac{\left(\lambda _1^2+\lambda _2 \lambda _1+\lambda
   _2^2\right) \left(\lambda _2 \mathfrak{v} +\lambda _1 \left(\lambda
   _2+\mathfrak{v} \right)\right) \log \left(\lambda _2 \mathfrak{v} +\lambda _1
   \left(\lambda _2+\mathfrak{v} \right)\right)}{\left(\lambda _1+\lambda _2\right)
   \left(2 \lambda _1^2+5 \lambda _2 \lambda _1+2 \lambda _2^2\right)}-\frac{1}{2}
   \mathfrak{v}  \log (\mathfrak{v} )
\end{array}\end{equation}
Comparing with  the original papers on the AMSY formalism
\cite{abreu,Martelli:2005tp},  we note that the full symplectic
potential for the 4-manifold $\mathcal{M}_B$ has precisely the
structure of what is  there called a \textit{natural symplectic
potential}
\begin{equation}\label{naturalia}
    G_{natural}= \sum_{\ell=1}^r \, c_\ell  \,
    \mathit{p}_\ell\left(\mathfrak{u},\mathfrak{v} \right)\times \log
    \left[\mathit{p}_\ell\left(\mathfrak{u},\mathfrak{v} \right)\right]
    \quad ;\quad
    \mathit{p}_\ell\left(\mathfrak{u},\mathfrak{v} \right) =
    \text{linear functions of the moments}
\end{equation}
{(with $r=4$ in our case).} The only difference is that in
\cite{abreu} the coefficients $c_\ell$ are all equal while here they
differ one from the other in a precise way that depends on the
parameters $\lambda_1$, $\lambda_2$ defining the metric and the
argument of the logarithms. As we are going to discuss later on, the
same thing  happens also for the nonKE metric on the second
Hirzebruch surface $\mathbb{F}_2$ derived from the Kronheimer
construction.
\par
Next we turn to a general discussion of the properties of the
metrics in $\operatorname{Met}(\mathcal{FV})$ and to the
organization of the latter into special subfamilies. This will also
clarify the precise location of the KE metrics in the general
landscape.
\section{Properties of the   family of metrics $\operatorname{Met}(\mathcal{FV})$}
\label{generafam} Let us then perform a study of the considered
class of 4-dimensional metrics.  We do not start from  a given
manifold but rather from the family of metrics
$\operatorname{Met}(\mathcal{FV})$ parameterized by the choice of
the function $\mathcal{FK}(\mathfrak{v} )$ of one variable
$\mathfrak{v} $, given explicitly in coordinate form. The first
tasks we are confronted with are the definition of the maximal
extensions of our coordinates, and the search of possible
singularities in the metric and/or in the Riemannian curvature,
which happens to be the cleanest probing tool. Secondly we might
calculate integrals of the Ricci and K\"ahler 2-forms as it was done
in papers \cite{Bianchi_2021,bruzzo2023d3brane}. All this
information can easily computed since everything reduces to the
evaluation of a few integral-differential functionals of the
function $\mathcal{FK}(\mathfrak{v} )$. Thirdly, one might construct
geodesics relatively to the given metric and  explore their
behavior. This is probably the finest and most accurate tool to
visualize the geometry of a manifold and was explicitly done in
\cite{bruzzo2023d3brane} showing also that the geodesics equation
are fully integrable due to the presence of an additional
hamiltonian in involution with the principal one. In these lectures
I will skip reviewing such results since although interesting they
are somehow aside from the main issue that is that of establishing
holographic dual pairs. The explicit integration of  the complex
structure was also performed in \cite{bruzzo2023d3brane}, but we
skip it here for the same reason as above.
\subsection{Analysis of the family of metrics} We  begin by observing that all the metrics
deriving from the symplectic potential defined by eqn.s
\eqref{GBsymplectic},\eqref{lupetto}\footnote{In this section which
deals only with the base manifold $\mathcal{M}_B$ and where there is
no risk of confusion we drop the suffix $0$ in the moment variables,
in order to make formulas simpler.} admit the general form which we
already displayed in \eqref{metrauniversala}. Furthermore the change
of coordinates performed to arrive at \eqref{metrauniversala}
clearly reveals that all the three dimensional sections of
$\mathcal{M}_B$ obtained by fixing $\mathfrak{v} =\text{const}$ are
$\mathbb{S}^1$ fibrations on $\mathbb{S}^2$ which is consistent with
the isometry $\mathrm{SU(2)} \times \mathrm{U(1)}$. Indeed all the
spaces $\mathcal{M}_B$ have cohomogeneity equal to one and the
moment variable $\mathfrak{v} $ is the only one whose dependence is
not fixed by isometries.
\par
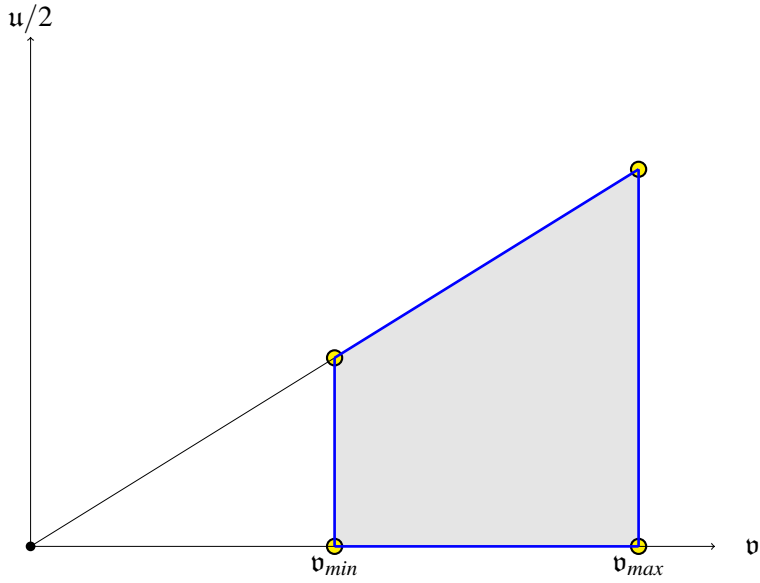
\begin{figure}
\begin{center}
\begin{tikzpicture}[scale=0.50]
\path [fill=gray,opacity=0.2] (-4,-7.5) to (4,-7.5) to  (4,2.5) to
(-4,-2.5) to (-4,-7.5); \node at (4,-8) {$\mathfrak{v} _{max}$};
\node at (-4,-8) {$ \mathfrak{v} _{min}$}; \draw [thick]
[fill=yellow] (-4,-7.5) circle (0.2 cm) ; \draw [thick]
[fill=yellow] (4,-7.5) circle (0.2 cm) ; \draw [thick] [fill=yellow]
(4,2.5) circle (0.2 cm); \draw [thick] [fill=yellow] (-4,-2.5)
circle (0.2 cm); \draw [thick] [fill=black] (-12,-7.5) circle (0.1
cm); \draw [black,line width=0.03] (-12,-7.5)--(4,2.5); \draw
[black,line width=0.03][->] (-12,-7.5)--(6,-7.5); \draw [black,line
width=0.03][->] (-12,-7.5)--(-12,6);
 \draw [blue,line width=1](-4,-7.5) --(4,-7.5);
\draw[blue,line width=1](-4,-7.5)--(-4,-2.5); \draw[blue,line
width=1](-4,-2.5)--(4,2.5); \draw[blue,line
width=1](4,-7.5)--(4,2.5); \node at (7,-7.5) {$\mathfrak{v} $};
\node at (-12,6.5) {$\mathfrak{u}/2$};
\end{tikzpicture}
\caption{\label{politoppo} The universal polytope in the
$\mathfrak{v} ,\frac{\mathfrak{u}}{2}$ plane for all the metrics of
the $\mathcal{M}_B$ manifolds considered in these lectures and
defined in equation \eqref{metrauniversala}}
\end{center}
\end{figure}

The next important point is that the metric \eqref{metrauniversala}
is positive definite only in the interval of the positive
$\mathfrak{v} $-axis where $\mathcal{FK}(\mathfrak{v} )\geq 0$. Let
us name the lower and upper endpoints of such  interval
$\mathfrak{v} _{min}$ and $\mathfrak{v} _{max}$, respectively. If
the interval $\left[\mathfrak{v} _{min},\mathfrak{v} _{max}\right]$
is finite, then the space $\mathcal{M}_{B}$ is compact and the
domain of the coordinates $\mathfrak{u},\mathfrak{v} $ is provided
by the trapezoidal polytope displayed in Figure \ref{politoppo}.
\par
Our two main examples, which both correspond to the same universal
polytope of Figure  \ref{politoppo}, are provided by the case of the
one-parameter family of \textit{Kronheimer metrics} on the
$\mathbb{F}_2$-surface, studied in
\cite{Bianchi_2021,noietmarcovaldo}, whose K\"ahler potential was
given in eqn.~\eqref{varpifamiglia}), and by a family
$\operatorname{Met}(\mathcal{FV})_{ext}$ of \textit{extremal
K\"ahler metrics} due to Calabi that we illustrate below and which
includes the above described K\"ahler Einstein metrics. In addition,
within the first class, we have the degenerate case where the
parameter $\alpha$ goes to zero and the trapezoid degenerates into a
triangle. That case corresponds to the singular space $\mathcal{M}_B
= \mathbb{WP}_{[1,1,2]}$ (a weighted projective plane).
\par
Extremal metrics are defined in the present cohomogeneity one case
by the differential equation (see \cite{abreu2009toric}):
\begin{equation}\label{extremality}
    \frac{\partial^2}{\partial \mathfrak{v}^2} \, \mathcal{R}_s(\mathfrak{v})\, = \,0
\end{equation}
where $\mathcal{R}_s(\mathfrak{v})$ is the scalar curvature.
\begin{table}[htb!]
\begin{equation}
\begin{array}{|l|l|l|l|}
\hline
\null&\null&\null&\null\\
 \mathcal{FK}^{\mathbb{F}_2}_{Kro}(\mathfrak{v} ) \, =\,
\frac{\left(1024 \mathfrak{v} ^2-81 \alpha ^2\right) (32
\mathfrak{v} -9 (3 \alpha +4))}{16 \left(81 \alpha ^2+1024
\mathfrak{v} ^2-576 (3 \alpha +4) \mathfrak{v} \right)} &
\mathfrak{v} _{min} \, =\, \frac{9 \alpha }{32} & \mathfrak{v}
_{max} \,
=\, &\null\\
\null&\null&\frac{9}{32}(3 \alpha +4)
&\alpha >0 \\
\null&\null&\null&\null\\
\hline
\null&\null&\null&\null\\
\mathcal{FK}^{\mathbb{WP}_{[1,1,2]}}_{Kro}(\mathfrak{v} ) \, = \,
\frac{\mathfrak{v}  (8 \mathfrak{v} -9)}{4 \mathfrak{v} -9} &
\mathfrak{v} _{min} \, =\,0 &\mathfrak{v} _{max} \,= \, \frac{9}{8} & \alpha =0\\
\null&\null&\null&\null\\
\hline
\null&\null&\null&\null\\
\mathcal{FK}_{ext}(\mathfrak{v} )\, = \, \frac{-\mathcal{A}+8
\mathcal{C} \mathfrak{v}^3-16 \mathcal{D}
   \mathfrak{v}^4+4 \mathfrak{v}^2-2 \mathfrak{v}
   \mathcal{B}}{4 \mathfrak{v}}&\mathfrak{v} _{min}\, =\, \lambda^r_1 &\mathfrak{v} _{max}
   \,= \,  \lambda^r_2 & 0 < \lambda^r_1 < \lambda^r_2\\
\null&\null&\null&\null\\
\hline
\null&\null&\null&\null\\
\underbrace{\mathcal{FK}^{KE}(\mathfrak{v}
)}_{\mathcal{B}=\mathcal{D}=0} \, = \, -\frac{\left(\mathfrak{v}
-\lambda _1\right) \left(\mathfrak{v} -\lambda
   _2\right) \left(\lambda _2 \mathfrak{v} +\lambda _1 \left(\lambda
   _2+\mathfrak{v} \right)\right)}{\left(\lambda _1^2+\lambda _2 \lambda
   _1+\lambda _2^2\right) \mathfrak{v} } &  \mathfrak{v} _{min} \, =\, \lambda_1 &  \mathfrak{v} _{max} \, = \,
       \lambda_2 &  0 < \lambda_1 < \lambda_2 \\
\null&\null&\null&\null\\
\hline
\null&\null&\null&\null\\
\mathcal{FK}^{KE}_{0}(\mathfrak{v} ) \, = \,
\frac{\mathfrak{v}(\lambda_2-\mathfrak{v})}{\lambda_2}
 &  \mathfrak{v} _{min} \, =\,0 &  \mathfrak{v} _{max} \, = \,
       \lambda_2 &    \lambda_2  >0\\
\null&\null&\null&\null\\ \hline
\null&\null&\null&\null\\
\mathcal{FK}^{cone}(\mathfrak{v} ) \, = \,  \mathfrak{v}   &
\mathfrak{v} _{min} \, =\, 0 & \mathfrak{v}_{max} \,= \,
       \infty &  \null\\
\null&\null&\null&\null\\
\hline
\null&\null&\null&\null\\
 \mathcal{FK}^{\mathbb{F}_2}_{ext}(\mathfrak{v})\, = \,
\frac{(\mathit{a}-\mathfrak{v}) (\mathit{b}-\mathfrak{v})
\left(\mathit{a}^2 (3 \mathit{b}-\mathfrak{v})+\mathit{a}
\left(\mathit{b}^2+4 \mathit{b} \mathfrak{v}+3
\mathfrak{v}^2\right)+\mathit{b} \mathfrak{v}
(\mathit{b}+\mathfrak{v})\right)}{\mathfrak{v} \left(\mathit{a}^3+3
\mathit{a}^2 \mathit{b}-3 \mathit{a}
\mathit{b}^2-\mathit{b}^3\right)}& \mathfrak{v} _{min} \,
=\,\mathit{a} & \mathfrak{v}_{max} \, = \,
       \mathit{b} & \mathit{b}>\mathit{a}>0 \\
\null&\null&\null&\null\\
\hline
     \end{array}
\end{equation}
\caption{\label{casoni} Notable choices for the function
$\mathcal{FK}$.}
\end{table}
Inserting in eqn.\eqref{extremality} the expression of
$\mathcal{R}_s$ calculated later in eqn.\eqref{conturbante} we
obtain the following linear differential equation of order four:
\begin{equation}\label{quartdifeq}
    \frac{\mathfrak{v}^2 \left(-\mathcal{F}\mathcal{K}^{(3)}(\mathfrak{v})\right)+2
   \mathfrak{v} \mathcal{F}\mathcal{K}''(\mathfrak{v})-2
   \mathcal{F}\mathcal{K}'(\mathfrak{v})+2}{\mathfrak{v}^3}-\frac{1}{2}
   \mathcal{F}\mathcal{K}^{(4)}(\mathfrak{v})\, = \, 0
\end{equation}
whose general integral contains four integration constants (we name
them $\mathcal{A},\mathcal{B},\mathcal{C},\mathcal{D}$) and can be
written as follows:
\begin{equation}\label{AbreuCalabfam}
    \mathcal{FK}_{ext}(\mathfrak{v}) \, = \, \frac{-\mathcal{A}+8 \,\mathcal{C} \mathfrak{v}^3-16 \,\mathcal{D} \mathfrak{v}^4+4
   \mathfrak{v}^2-2 \mathfrak{v}\, \mathcal{B}}{4 \mathfrak{v}}
\end{equation}
The explicit expression \eqref{AbreuCalabfam} is very much inspiring
and useful. The function $\mathcal{FK}_{ext}(\mathfrak{v})$ is
rational and it is the quotient of a quartic polynomial with a fixed
coefficient of the quadratic term divided by the linear polynomial
$4\,\mathfrak{v}$. A convenient way of parameterizing the entire
family of metrics is therefore in terms of the four roots
$\lambda_1,\lambda_2,\lambda_3,\lambda_4$, as we did in section
\ref{famigliaKE} for the cubic polynomial of eqn.\eqref{racete} (see
eqn. \eqref{rinominopara}). Combining eqn.\eqref{miraculo} with
eqn.s \eqref{fungemistero} and \eqref{FKfunzione} we obtain the
expression of the function $\mathcal{FK}^{KE}$ corresponding to the
KE metrics:
\begin{equation}\label{pagnatta}
    \mathcal{FK}^{KE}(\mathfrak{v}) \, = \,\frac{3 \beta -k \mathfrak{v}^3+3 \mathfrak{v}^2}{3 \mathfrak{v}}
\end{equation}
Comparing eqn.~\eqref{pagnatta} with eqn.~\eqref{AbreuCalabfam} we
see that the KE metrics belong to the family of extremal metrics and
are singled out by the constraint
\begin{equation}\label{immergokello}
    \mathcal{A}\, = \, -4\,\beta \quad ; \quad \mathcal{B} \, = \, 0
    \quad ; \quad \mathcal{C}\, = \, \frac{k}{6} \quad ; \quad
    \mathcal{D} \, = \, 0
\end{equation}
The most relevant aspect of the above eqn.~\eqref{immergokello} is
the suppression of the quartic and linear terms
($\mathcal{D}\,=\,\mathcal{B}\, = \, 0$), which fixes the number of
free roots to two, as we know, the third being fixed in terms of
$\lambda_1,\lambda_2$.
\par
We have summarized the relevant choices of the function
$\mathcal{FK}(\mathfrak{v})$ in Table \ref{casoni}. Inspecting this
table we see that the functions
$\mathcal{FK}_{Kro}^{\mathbb{F}_2}(\mathfrak{v} )$ and
$\mathcal{FK}^{KE}(\mathfrak{v} )$ show strict similarities but also
a difference which is expected to account for different topologies.
In both cases the function $\mathcal{FK}(\mathfrak{v} )$ is the
ratio of a cubic polynomial having three real roots, two positive
and one negative, and of a denominator that has no zeros in the
$\left[\mathfrak{v} _{min},\mathfrak{v} _{max}\right]$ interval. In
the KE case there is a simple pole at $\mathfrak{v} =0$ while for
$\mathbb{F}_2$ (which is not KE) the denominator has two zeros and
therefore $\mathcal{FK}(\mathfrak{v} )$ has two simple poles  at
\begin{equation}\label{baldrino}
  \mathfrak{v} _{poles} =  \frac{9}{32} \left[\left(3 \alpha +4\right)\pm 2 \sqrt{2} \sqrt{\alpha ^2+3 \alpha
   +2}\right]
\end{equation}
These poles are  out of the interval
$\left[\mathfrak{v}_{min},\mathfrak{v}_{max}\right]$ for any
positive $\alpha
>0$, namely these poles  do not correspond to points of  the
manifold $\mathcal{M}_B$, just as it is the case for the single pole
$\mathfrak{v} =0$ in the KE case. One also notes that the function
$\mathcal{FK}^{\mathbb{F}_2}_{ext}(\mathfrak{v})$ cannot be reduced
to the form \eqref{AbreuCalabfam} by any choice of the parameters
$\mathcal{A},\mathcal{B},\mathcal{C},\mathcal{D}$, so that the
smooth K\"ahler metric induced on the second Hirzebruch surface by
the Kronheimer construction (\cite{Bianchi_2021,noietmarcovaldo}) is
not an extremal metric. A consistency check comes from the
evaluation on $\mathcal{FK}^{\mathbb{F}_2}_{Kro}(\mathfrak{v})$ of
the scalar curvature provided by the later eqn.\eqref{conturbante}.
In this case the scalar curvature is in no way linear in
$\mathfrak{v}$, being a rational function of degree $6$ in the
numerator and of degree $7$ in the denominator. The same is true of
the limiting case $\alpha \to 0$. The last   case
($\mathcal{FK}(\mathfrak v)=\mathfrak v$) corresponds to a metric
cone on the $3$-sphere, i.e., $\mathbb{C}^2/\mathbb{Z}_2$ with a
flat metric. The case $\mathcal{FK}_0^{KE}$ will be discussed in
Section \ref{singularity}.
\par
Finally in table \ref{casoni} we observe the choice of the function
\begin{equation}\label{candela}
\mathcal{FK}^{\mathbb{F}_2}_{ext}(\mathfrak{v})\, = \,
\frac{(\mathit{a}-\mathfrak{v}) (\mathit{b}-\mathfrak{v})
\left(\mathit{a}^2 (3 \mathit{b}-\mathfrak{v})+\mathit{a}
\left(\mathit{b}^2+4 \mathit{b} \mathfrak{v}+3
\mathfrak{v}^2\right)+\mathit{b} \mathfrak{v}
(\mathit{b}+\mathfrak{v})\right)}{\mathfrak{v} \left(\mathit{a}^3+3
\mathit{a}^2 \mathit{b}-3 \mathit{a}
\mathit{b}^2-\mathit{b}^3\right)}
\end{equation}
That above in eqn.\eqref{candela} is a particular case of the
general case $\mathcal{FK}_{ext}(\mathfrak{v} )$, corresponding to
the following choice of the parameters:
\begin{eqnarray}\label{cromotalpa}
 \mathcal{A}& =&  -\frac{4 \mathit{a}^2 \mathit{b}^2 (3
   \mathit{a}+\mathit{b})}{\mathit{a}^3+3 \mathit{a}^2 \mathit{b}-3 \mathit{a}
   \mathit{b}^2-\mathit{b}^3}\quad ; \quad \mathcal{B}\, = \, \frac{8 \mathit{a}^3
   \mathit{b}}{\mathit{a}^3+3 \mathit{a}^2 \mathit{b}-3 \mathit{a}
   \mathit{b}^2-\mathit{b}^3}\nonumber\\
 \mathcal{C}& =& -\frac{2 \mathit{a}^2}{\mathit{a}^3+3
   \mathit{a}^2 \mathit{b}-3 \mathit{a} \mathit{b}^2-\mathit{b}^3}\quad ; \quad
   \mathcal{D}\, = \,
   \frac{3 \mathit{a}+\mathit{b}}{4 \left(-\mathit{a}^3-3 \mathit{a}^2 \mathit{b}+3
   \mathit{a} \mathit{b}^2+\mathit{b}^3\right)}
\end{eqnarray}
where the parameters $\mathit{a},\mathit{b}$ are real, positive and
naturally ordered $\mathit{b}>\mathit{a}>0$. Wherefrom does the
special form \eqref{cromotalpa} originate? We claim that the metric
defined by the function \eqref{candela} is a smooth metric on the
smooth $\mathbb{F}_2$ surface. The algebraic constraints that reduce
the four parameters
$\mathcal{A},\mathcal{B},\mathcal{C},\mathcal{D}$ to the form
\eqref{cromotalpa} are derived from the conditions, already
preliminarily  discussed in \cite{noietmarcovaldo} and specifically
worked out in section 8.2.2 of \cite{Bianchi_2021}, on the periods
of the Ricci two-form localized on the standard toric homology
cycles $C_1$ and $C_2$ of $\mathbb{F}_2$ (see section
\ref{toricresolution} and in particular table \ref{coordinates}).
Utilizing eqn.s \eqref{ricciolone} and \eqref{fantastilione3}
derived in the next section \ref{quattrogambe} we have (see also
eqn.(8.14) of \cite{Bianchi_2021}):
\begin{equation}\label{cracchiato}
    \mathbb{R}\mathrm{ic}\mid_{C_1} \, = \,\mathfrak{A}(\mathfrak{v}) \, \sin[\theta] \,
  d\theta\wedge d\phi \quad ; \quad \mathbb{R}\mathrm{ic}\mid_{C_2} \, = \, \mathfrak{C}(\mathfrak{v}) \,
  d\theta\wedge d\phi
\end{equation}
the relevant functions being given in \eqref{fantastilione3}. The
conditions on the periods are as follows:
\begin{equation}\label{craniodibronzo}
   \frac{1}{2\pi}\, \int_{C_1}\mathbb{R}\mathrm{ic} \, = \, 0 \quad ; \quad
   \frac{1}{2\pi}\,\int_{C_2} \mathbb{R}\mathrm{ic} \, = \, 2
\end{equation}
that, as shown in section 8.2.2 of \cite{Bianchi_2021}) are
automatically verified  by the Kronheimer metric.
\par
The two conditions \eqref{craniodibronzo} become two statements on
the functions $\mathfrak{A}(\mathfrak{v}),\mathcal{D}(\mathfrak{v})$
defined in eqn. \eqref{fantastilione3} and completely determined in
terms of the function $\mathcal{FK}(\mathfrak{v})$ and its
derivatives:
\begin{equation}\label{parcondicio}
    \mathfrak{A}[\mathfrak{v}_{min}]\, = \, 0 \quad ; \quad
    \int_{\mathfrak{v}_{min}}^{\mathfrak{v}_{max}}\mathfrak{D}[\mathfrak{v}]d\mathfrak{v} \, = \, 2
\end{equation}
The result provided in eqn. \eqref{candela} corresponding  to the
parameter choice \eqref{cromotalpa} is deduced in the following way.
First we re-parameterize the function \eqref{AbreuCalabfam} in terms
of the four roots of the quartic polynomial appearing in the
numerator that we name $\mu_1,\mu_2,\mu_3,\mu_4$ obtaining:
\begin{align}\label{torriano}
&\mathcal{FK}_{ext}(\mathfrak{v}) \, = \, \nonumber\\
&\frac{\mu _1 \left(\mathfrak{v} \left(\mu _3+\mathfrak{v}\right)
\left(\mu
   _4+\mathfrak{v}\right)+\mu _2 \left(\mu _3 \left(\mathfrak{v}-\mu
   _4\right)+\mathfrak{v} \left(\mu _4+\mathfrak{v}\right)\right)\right)}{2
   \mathfrak{v} \left(\mathfrak{v}-\mu _1\right) \left(\mathfrak{v}-\mu _2\right)
   \left(\mathfrak{v}-\mu _3\right) \left(\mathfrak{v}-\mu
   _4\right)} + \nonumber\\
&\frac{\mu _2 \left(\mu _3+\mathfrak{v}\right) \left(\mu
   _4+\mathfrak{v}\right)+\mathfrak{v} \left(\mathfrak{v} \left(\mu
   _4-\mathfrak{v}\right)+\mu _3 \left(\mu _4+\mathfrak{v}\right)\right)}{2
   \left(\mathfrak{v}-\mu _1\right) \left(\mathfrak{v}-\mu _2\right)
   \left(\mathfrak{v}-\mu _3\right) \left(\mathfrak{v}-\mu _4\right)}
\end{align}
Secondly we rename $\mu_2 = a$,$\mu_3 =b$ deciding that
$0<a<b<\infty$ and we calculate the two conditions
\eqref{parcondicio} using the function
$\mathcal{FK}_{ext}(\mathfrak{v})$ in eqn.\eqref{torriano} as an
input. We get a system of quadratic algebraic equations for the
remaining roots $\mu_1,\mu_4$ that has the following solutions
\begin{align}
&\mu_2 \, = \, \mathit{a} \quad ; \quad \mu_3 \, = \, \mathit{b}\nonumber \\
&\mu _1 \, = \, \frac{\mathit{a}^2-\left(\mathit{b}^2\pm
\sqrt{\mathit{a}^4-44
   \mathit{a}^3 \mathit{b}-10 \mathit{a}^2 \mathit{b}^2+4 \mathit{a}
   \mathit{b}^3+\mathit{b}^4}\right)-4 \mathit{a} \mathit{b}}{6 \mathit{a}+2
   \mathit{b}} \nonumber\\
&\mu _4 \, = \, \frac{\mathit{a}^2-\left(\mathit{b}^2\mp
   \sqrt{\mathit{a}^4-44 \mathit{a}^3 \mathit{b}-10 \mathit{a}^2 \mathit{b}^2+4
   \mathit{a} \mathit{b}^3+\mathit{b}^4}\right)-4 \mathit{a} \mathit{b}}{6
   \mathit{a}+2 \mathit{b}} \label{coffele}
\end{align}
Substitution of eqn.\eqref{coffele} into eqn. \eqref{torriano}
produces the function
$\mathcal{FK}^{\mathbb{F}_2}_{ext}(\mathfrak{v})$ presented in
\eqref{torriano} and recalled in table \ref{casoni}. Furthermore as
long as the roots $\mu _1,\mu_4$ as given above are complex
conjugate of each other or, being real, do not fall in the interval
$[a,b]$, then the K\"ahler metric generated by the function
$\mathcal{FK}^{F_2}_{ext}(\mathfrak{v})$ is smooth and well defined
on the second Hirzebruch surface $\mathbb{F}_2$. The domain where
this happens in the plane $\mathit{a},\mathit{b}$ can be easily
studied looking at the discriminant under the square root in
\eqref{coffele}.
\subsection{Vielbein formalism and the curvature 2-form of
$\mathcal{M}_B$}\label{quattrogambe} The metric
\eqref{metrauniversala} is in diagonal form so it is easy to write a
set of vierbein 1-forms. Indeed if we set
\begin{equation}\label{vierbeine}
    \mathbf{e}^i = \left\{\frac{d\mathfrak{v} }{\sqrt{\mathcal{FK}(\mathfrak{v} )}},\,\sqrt{\mathcal{FK}(
   \mathfrak{v} )}\, \left[d\phi  (1-\cos \theta )+d\tau
   \right],\,\sqrt{\mathfrak{v} }\,
   d\theta ,\,\sqrt{\mathfrak{v} }\, d\phi \,\sin \theta\right\}
\end{equation}
the line element \eqref{metrauniversala} reads
\begin{equation}\label{vilbone}
    ds^2_B = \sum_{i=1}^4 \, \mathbf{e}^i \otimes \mathbf{e}^i
\end{equation}
Furthermore we can calculate the matrix vielbein and its  inverse
quite  easily, obtaining:
\begin{equation}\begin{array}{rcl}
  \mathbf{e}^i &\ = & E^i_\mu \, dy^\mu \quad ; \quad y^\mu \, = \{\mathfrak{v} ,\theta ,\phi ,\tau \}   \\
  E^i_\mu &=& \left(
\begin{array}{cccc}
 \frac{1}{\sqrt{\mathcal{FK}(\mathfrak{v} )}} & 0 & 0 & 0 \\
 0 & 0 & \sqrt{\mathcal{FK}(\mathfrak{v} )} (1-\cos \theta ) &
   \sqrt{\mathcal{FK}(\mathfrak{v} )} \\
 0 & \sqrt{\mathfrak{v} } & 0 & 0 \\
 0 & 0 & \sqrt{\mathfrak{v} } \sin \theta  & 0 \\
\end{array}
\right)   \\
  E^\nu_j &=& \left(
\begin{array}{cccc}
 \sqrt{\mathcal{FK}(\mathfrak{v} )} & 0 & 0 & 0 \\
 0 & 0 & \frac{1}{\sqrt{\mathfrak{v} }} & 0 \\
 0 & 0 & 0 & \frac{\csc (\theta )}{\sqrt{\mathfrak{v} }} \\
 0 & \frac{1}{\sqrt{\mathcal{FK}(\mathfrak{v} )}} & 0 & \frac{(\cos \theta -1)
   \csc \theta }{\sqrt{\mathfrak{v} }} \\
\end{array}
\right)
\end{array}\end{equation}
By means of the {\sc Mathematica} package {\sc
Vielbgrav23}\footnote{{\sc Vielbgrav23} is a MATHEMATICA  package
for the calculation of the spin connection the curvature 2-form and
the intrinsic components of the Riemann tensor in vielbein
formalism. Constantly updated, it was originally written by the
present author, almost thirty years ago. It can be furnished upon
request and it will be at disposal on the the De Gruyter site for
the readers of a forthcoming book entitled \textit{Discrete Finite
and Lie Groups}.} we can easily calculate the Levi-Civita spin
connection and the curvature 2-form from the definitions
\begin{equation}\label{radicchione}
    0 = \mathfrak{T}^i= d\mathbf{e}^i \, + \, \omega^{ij} \, \wedge \,
    \mathbf{e}^j \quad ; \quad \mathfrak{R}^{ij}=d\omega^{ij}
    \, + \, \omega^{ik} \, \wedge \, \omega^{kj} =
    \mathcal{R}^{ij}_{\phantom{ik}k\ell} \,\mathbf{e}^k \, \wedge \,
    \mathbf{e}^\ell
\end{equation}
obtaining
\begin{equation}\begin{array}{rcl}\label{Rdueforma}
  \mathfrak{R}^{12} &=& -\frac{\mathcal{FK}''(\mathfrak{v} )}{2} \mathbf{e}^{1}\wedge
  \mathbf{e}^{2}
   -\frac{
   \left(\mathfrak{v}
   \mathcal{FK}'(\mathfrak{v} )-\mathcal{FK}(\mathfrak{v} )\right)}{2 \mathfrak{v} ^2}\,\mathbf{e}^{3}\wedge
   \mathbf{e}^{4}  \\[5pt]
  \mathfrak{R}^{13} &=& -\frac{\left(\mathfrak{v}
   \mathcal{FK}'(\mathfrak{v} )-\mathcal{FK}(\mathfrak{v} )\right)}{4
   \mathfrak{v} ^2}\, \mathbf{e}^{1}\wedge \mathbf{e}^{3} \, -\frac{\left(\mathfrak{v}
   \mathcal{FK}'(\mathfrak{v} )-\mathcal{FK}(\mathfrak{v} )\right)}{4 \mathfrak{v} ^2}\, \mathbf{e}^{2}\wedge \mathbf{e}^{4}  \\[5pt]
  \mathfrak{R}^{14} &=& \frac{\left(\mathfrak{v}
   \mathcal{FK}'(\mathfrak{v} )-\mathcal{FK}(\mathfrak{v} )\right)}{4
   \mathfrak{v} ^2}\, \mathbf{e}^{2}\wedge \mathbf{e}^{3}  \, -\, \frac{\left(\mathfrak{v}
   \mathcal{FK}'(\mathfrak{v} )-\mathcal{FK}(\mathfrak{v} )\right)}{4 \mathfrak{v} ^2} \, \mathbf{e}^{1}\wedge \mathbf{e}^{4}  \\[5pt]
  \mathfrak{R}^{23} &=& \frac{ \left(\mathfrak{v}
   \mathcal{FK}'(\mathfrak{v} )-\mathcal{FK}(\mathfrak{v} )\right)}{4
   \mathfrak{v} ^2}\, \mathbf{e}^{1}\wedge \mathbf{e}^{4}\, -\,\frac{\left(\mathfrak{v}
   \mathcal{FK}'(\mathfrak{v} )-\mathcal{FK}(\mathfrak{v} )\right)}{4 \mathfrak{v} ^2}\,\mathbf{e}^{2}\wedge \mathbf{e}^{3}   \\[5pt]
  \mathfrak{R}^{24}&=& -\frac{\left(\mathfrak{v}
   \mathcal{FK}'(\mathfrak{v} )-\mathcal{FK}(\mathfrak{v} )\right)}{4
   \mathfrak{v} ^2}\,\mathbf{e}^{1}\wedge \mathbf{e}^{3} \, -\, \frac{\left(\mathfrak{v}
   \mathcal{FK}'(\mathfrak{v} )-\mathcal{FK}(\mathfrak{v} )\right)}{4 \mathfrak{v} ^2} \, \mathbf{e}^{2}\wedge \mathbf{e}^{4}  \\[5pt]
  \mathfrak{R}^{34} &=& \frac{
   (\mathfrak{v} -\mathcal{FK}(\mathfrak{v} ))}{\mathfrak{v} ^2}\, \mathbf{e}^{3}\wedge \mathbf{e}^{4}\, -\frac{ \left(\mathfrak{v}
   \mathcal{FK}'(\mathfrak{v} )-\mathcal{FK}(\mathfrak{v} )\right)}{2
   \mathfrak{v} ^2} \, \mathbf{e}^{1}
   \wedge \mathbf{e}^{2}
\end{array}\end{equation}
Equation \eqref{Rdueforma} shows that the   Riemann tensor
$\mathcal{R}^{ij}_{\phantom{ik}k\ell}$ is constructed  in terms of
only three functions:
\begin{equation}\label{trefunzie}
    \mathcal{CF}_1(\mathfrak{v} ) = \mathcal{FK}''(\mathfrak{v} ) \quad ; \quad
    \mathcal{CF}_2(\mathfrak{v} )= \frac{\left(\mathfrak{v}
   \mathcal{FK}'(\mathfrak{v} )-\mathcal{FK}(\mathfrak{v} )\right)}{
   \mathfrak{v} ^2} \quad ; \quad \mathcal{CF}_3(\mathfrak{v} )=\frac{(\mathfrak{v} -\mathcal{FK}(\mathfrak{v} ))}{\mathfrak{v} ^2}
\end{equation}
If  these functions  are regular in the interval $\left[\mathfrak{v}
_{min},\mathfrak{v} _{max}\right]$ the Riemann tensor is well
defined and finite in the entire polytope of Figure \ref{politoppo}
and $\mathcal{M}_B$ should be a smooth compact manifold. From the
expression \eqref{Rdueforma} the {\sc Mathematica Code}  immediately
derives the Riemann and Ricci tensors and the curvature scalar. This
latter reads as follows:
\begin{equation}\label{conturbante}
    \mathcal{R}_s \, = \, -\frac{\mathfrak{v} \mathcal{F}\mathcal{K}''(\mathfrak{v})+2
   \mathcal{F}\mathcal{K}'(\mathfrak{v})-2}{2 \mathfrak{v}}
\end{equation}
and its form was used above to define the extremal metrics.
Similarly in the anholonomic vielbein basis, the Ricci tensor takes
the following form: {\scriptsize
\begin{eqnarray}\label{riccetto}
   &&\mathcal{R}_{ij} \, = \, \nonumber\\
   &&\left(
\begin{array}{cccc}
 \frac{\mathcal{F}\mathcal{K}(\mathfrak{v})-\mathfrak{v} \left(\mathfrak{v}
   \mathcal{F}\mathcal{K}''(\mathfrak{v})+\mathcal{F}\mathcal{K}'(\mathfrak{v})\right)}
   {4 \mathfrak{v}^2} & 0 & 0 & 0 \\
 0 & \frac{\mathcal{F}\mathcal{K}(\mathfrak{v})-\mathfrak{v} \left(\mathfrak{v}
   \mathcal{F}\mathcal{K}''(\mathfrak{v})+\mathcal{F}\mathcal{K}'(\mathfrak{v})\right)}
   {4 \mathfrak{v}^2} & 0 & 0 \\
 0 & 0 & -\frac{\mathfrak{v}
   \left(\mathcal{F}\mathcal{K}'(\mathfrak{v})-2\right)
   +\mathcal{F}\mathcal{K}(\mathfrak{v})}{4 \mathfrak{v}^2} & 0 \\
 0 & 0 & 0 & -\frac{\mathfrak{v}
   \left(\mathcal{F}\mathcal{K}'(\mathfrak{v})-2\right)
   +\mathcal{F}\mathcal{K}(\mathfrak{v})}{4 \mathfrak{v}^2} \\
\end{array}
\right)\nonumber\\
\end{eqnarray}}
All the metrics in the considered family are of cohomogeneity one
and have the same isometry, furthermore they are all K\"ahler and
share the same K\"ahler 2-form that can be written as it follows:
\begin{equation}\label{kalleroduef}
    \mathbb{K} = d\mathfrak{u}\wedge d\phi \, + \, d\mathfrak{v}  \wedge
    d\tau = \mathbf{e}^1 \wedge \mathbf{e}^2 \, + \, \mathbf{e}^3 \wedge
    \mathbf{e}^4 = \frac{1}{2} \, \mathfrak{J}_{ij}\, \mathbf{e}^i \, \wedge \,
    \mathbf{e}^j
\end{equation}
where:
\begin{equation}\label{complessostrut}
    \mathfrak{J}^i_{\phantom{i}j} = \left(
\begin{array}{cccc}
 0 & 1 & 0 & 0 \\
 -1 & 0 & 0 & 0 \\
 0 & 0 & 0 & 1 \\
 0 & 0 & -1 & 0 \\
\end{array}
\right)=  \delta^{ik} \, \mathfrak{J}_{kj}
\end{equation}
is the complex structure in flat indices. Utilizing
$\mathfrak{J}^i_{\phantom{i}j}$ the Ricci 2-form is defined by:
\begin{equation}\label{canzellero}
    \mathbb{R}\mathrm{ic} = \mathbb{R}_{ij} \, \mathbf{e}^i \wedge \mathbf{e}^j
    \quad ; \quad \mathbb{R}_{ij}= \mathcal{R}_{i\ell} \, \mathfrak{J}^\ell_{\phantom{\ell}j}
\end{equation}
and explicitly one obtains:
\begin{equation}\label{ricciolone}
  \mathbb{R}\mathrm{ic} \, = \,   \mathfrak{A}(\mathfrak{v}) \, \sin[\theta] \,
  d\theta\wedge d\phi \, + \, \mathfrak{B}(\mathfrak{v}) \,\left(1\, -\,
  \cos[\theta] \right) d\mathfrak{v}\wedge d\phi + \mathfrak{C}(\mathfrak{v}) \, d\mathfrak{v}\wedge
  d\tau
\end{equation}
In eqn. \eqref{ricciolone} the functions of $\mathfrak{v}$ are the
following ones:
\begin{align}
\mathfrak{A}(\mathfrak{v}) &= -\frac{\mathfrak{v}
\left(\mathcal{F}\mathcal{K}'(\mathfrak{v})-2\right)+\mathcal{F}\mathcal{K}(\mathfrak{v})}{2 \mathfrak{v}}\nonumber\\
\mathfrak{B}(\mathfrak{v}) &=
-\frac{\mathcal{F}\mathcal{K}(\mathfrak{v})-\mathfrak{v}
\left(\mathfrak{v}
\mathcal{F}\mathcal{K}''(\mathfrak{v})+\mathcal{F}\mathcal{K}'(\mathfrak{v})\right)}
{2 \mathfrak{v}^2} \nonumber\\
\mathfrak{C}(\mathfrak{v}) &=  \frac{\mathfrak{v}^2
\left(-\mathcal{F}\mathcal{K}''(\mathfrak{v})\right)-\mathfrak{v}
\mathcal{F}\mathcal{K}'(\mathfrak{v})+\mathcal{F}\mathcal{K}(\mathfrak{v})}{2
\mathfrak{v}^2} \label{fantastilione3}
\end{align}
\par
\begin{figure}
\centering
\includegraphics[width=8cm]{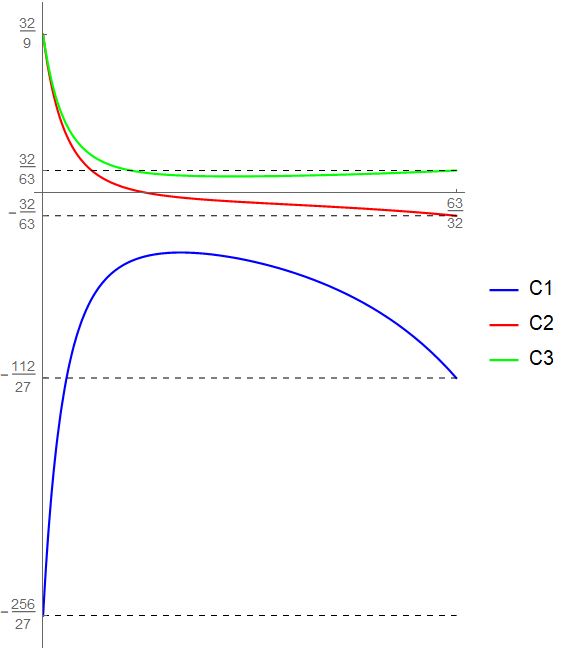}\hskip10mm\includegraphics[width=8cm]{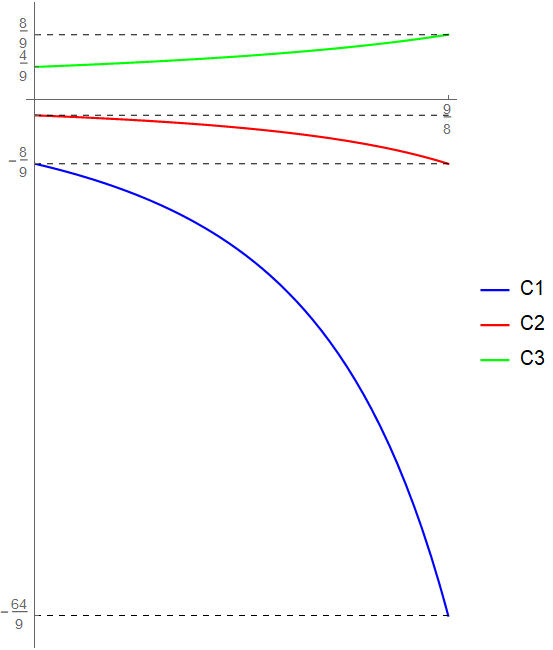}
\caption{\label{plotto123}  A (left):  Plot of the three functions
$\mathcal{CF}^{\mathbb{F}_2}_{1,2,3}(\mathfrak{v} )$ entering the
intrinsic Riemann curvature tensor for the   ``Kronheimer'' metric
on $\mathbb{F}_2$ with the choice of the parameter $\alpha = 1$. B
(right): Plot of the three functions
$\mathcal{CF}^{\mathbb{WW}_{112}}_{1,2,3}(\mathfrak{v} )$ entering
the intrinsic Riemann curvature tensor for the  Kronheimer metric on
$\mathbb{WP}_{[1,1,2]}$ with the choice of the parameter $\alpha =
0$. Comparing this picture with the one on the left we see the
discontinuity. In all smooth cases the functions
$\mathcal{CF}^{\mathbb{F}_2}_{2,3}$ attain the same value in the
lower endpoint of the interval while for the singular case of the
weighted projective space, the initial values of
$\mathcal{CF}^{\mathbb{WP}_{[1,1,2]}}_{2,3}(\mathfrak{v} )$ are
different.}
\end{figure}

\paragraph{The   $\mathbb{F}_2$ Kronheimer case.}
In the   $\mathbb{F}_2$ case with the ``Kronheimer''  metric we
have:
\begin{equation}\begin{array}{rcl}
  \mathcal{CF}^{\mathbb{F}_2}_1(\mathfrak{v} ) &=& \frac{331776 (\alpha +1) (\alpha +2) \left(729 \alpha ^2 (3 \alpha +4)+16384
   \mathfrak{v} ^3-3888 \alpha ^2 \mathfrak{v} \right)}{\left(81 \alpha ^2+1024
   \mathfrak{v} ^2-576 (3 \alpha +4) \mathfrak{v} \right)^3}  \\[5pt]
\mathcal{CF}^{\mathbb{F}_2}_2(\mathfrak{v} )&=& -\frac{9}{32
   \mathfrak{v} ^2 \left(81 \alpha ^2+1024 \mathfrak{v} ^2-576 (3 \alpha +4)
   \mathfrak{v} \right)^2}\, \times \, \
\left(6561 \alpha ^4 (3 \alpha +4)+1048576 (3 \alpha +4)
   \mathfrak{v} ^4\right. \\[5pt]
   &&\left.-1179648 \alpha ^2 \mathfrak{v} ^3
    +497664 \alpha ^2 (3 \alpha
   +4) \mathfrak{v} ^2-93312 \alpha ^2 (3 \alpha +4)^2 \mathfrak{v} \right) \\[5pt]
\mathcal{CF}^{\mathbb{F}_2}_3(\mathfrak{v} ) &=& \displaystyle
\frac{\mathfrak{v} -\displaystyle\frac{\left(1024 \mathfrak{v} ^2-81
\alpha ^2\right) (32
   \mathfrak{v} -9 (3 \alpha +4))}{16 \left(81 \alpha ^2+1024
   \mathfrak{v} ^2-576 (3 \alpha +4) \mathfrak{v} \right)}}{\mathfrak{v} ^2}
\end{array}\end{equation}
The three functions
$\mathcal{CF}^{\mathbb{F}_2}_{1,2,3}(\mathfrak{v} )$ are smooth in
the interval $\left(\frac{9 \alpha }{32},\frac{9}{32} (3 \alpha +4)
\right)$ and they are defined at the endpoints: see for instance
Figure \ref{plotto123}A.
\par
\begin{figure}
\centering
\includegraphics[width=9cm]{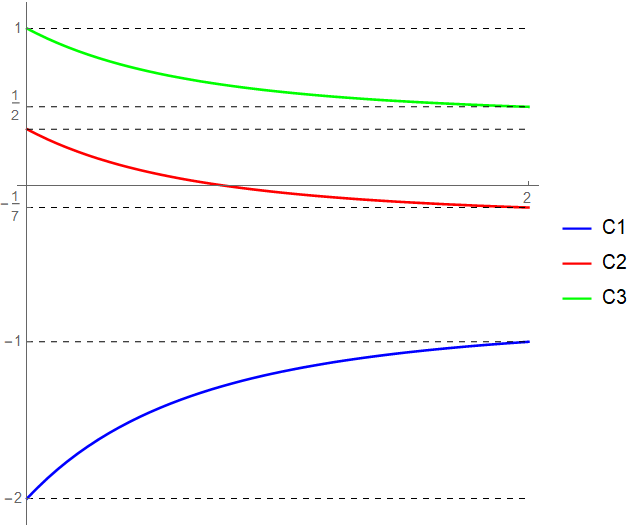}
\caption{\label{KEfunzie} Plot of the three functions
$\mathcal{CF}^{KE}_{1,2,3}(\mathfrak{v} )$ entering the intrinsic
Riemann curvature tensor for the  KE metric  with the choice of the
parameter $\lambda_1 = 1$, $\lambda_2 = 2$. }
\end{figure}
\begin{figure}
\centering
\includegraphics[width=7.3cm]{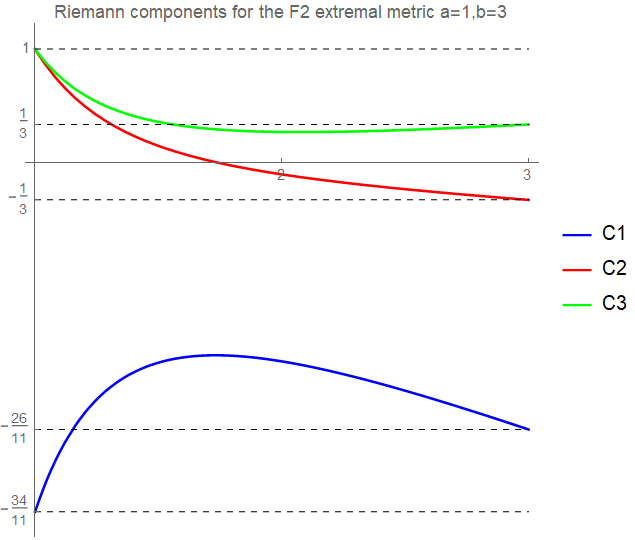}
\includegraphics[width=7cm]{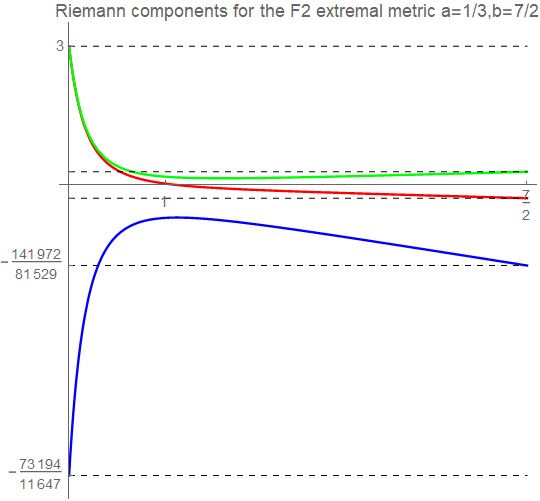}
\caption{\label{RiefunzioF2ext} Plot of the three functions
$\mathcal{CF}^{F2ext}_{1,2,3}(\mathfrak{v} )$ entering the intrinsic
Riemann curvature tensor for the  extremal K\"ahler metrics on
$\mathbb{F}_2$ with with two different  choices of the parameters
$\mathit{a} = 1$, $\mathit{b} = 2$ and $\mathit{a} = 1/3$,
$\mathit{b} = 7/2$. }
\end{figure}

Indeed the values of the three functions at the endpoints   are
\begin{equation}\begin{array}{rcl}\label{parbino}
 \mathcal{CF}^{\mathbb{F}_2}_{1,2,3}\left(\mathfrak{v} _{min}\right)& = &  \left\{-\frac{128 (\alpha +1)}{9 \alpha  (\alpha +2)},
 \frac{32}{9 \alpha
   },\frac{32}{9 \alpha }\right\} \\
   \mathcal{CF}^{\mathbb{F}_2}_{1,2,3}\left(\mathfrak{v} _{max}\right)& = &\left\{-\frac{32 (3 \alpha +4)}{9 \left(\alpha ^2+3 \alpha
   +2\right)},-\frac{32}{27 \alpha +36},\frac{32}{9 (3 \alpha +4)}\right\}
\end{array}
\end{equation}
 The singularity which might be developed by the space
corresponding to the value $\alpha=0$ is evident from
eqns.~\eqref{parbino}. The intrinsic components of the Riemann
curvature seem to have a singularity in the lower endpoint of the
interval, for $\alpha = 0$.
\paragraph{The case of the singular manifold $\mathbb{WP}_{[1,1,2]}$.}
In the previous section we utilized the wording \textit{seem to have
a singularity} for the components of the Riemann curvature in the
case of the space $\mathbb{WP}_{[1,1,2]}$ since  such a singularity
in the curvature actually does not exist. The space
$\mathbb{WP}_{[1,1,2]}$ has indeed a singularity at $\mathfrak{v}
\, = \, 0$ but it is very mild since the intrinsic components of the
Riemann curvature are well-behaved in $\mathfrak{v}  = 0$ and have a
finite limit. It depends on the way one does the limit $\alpha
\rightarrow 0$. If we first compute the value of the curvature
$2$-form at the endpoints for generic $\alpha$ and then we do the
limit $\alpha \to 0$ we see the singularity that is evident from
equations \eqref{parbino}. On the other hand, if we first reduce the
function $\mathcal{FK}(\mathfrak{v} )$ to its $\alpha=0$ form we
obtain:
\begin{equation}\label{wp112FK}
    \mathcal{FK}^{\mathbb{WP}_{[1,1,2]}}(\mathfrak{v} )\, =
    \,\frac{\mathfrak{v}  (8 \mathfrak{v} -9)}{4 \mathfrak{v} -9}
\end{equation}
and the corresponding functions appearing in the curvature are:
\begin{equation}\label{funzie123WP}
\mathcal{CF}^{\mathbb{WP}_{[1,1,2]}}_{1,2,3}\left(\mathfrak{v}
\right)\, = \, \left\{\frac{648}{(4 \mathfrak{v}
-9)^3},-\frac{18}{(9-4
   \mathfrak{v} )^2},\frac{4}{9-4 \mathfrak{v} }\right\}
\end{equation}
which are perfectly regular in the interval $\left[0,9/8\right]$ and
have finite value at the endpoints (see Figure \ref{plotto123}B).
\paragraph{The case of the KE  manifolds.}
In the case of the KE metrics the function
$\mathcal{FK}(\mathfrak{v} )$ is
\begin{equation}\label{felino}
    \mathcal{FK}^{KE}(\mathfrak{v} )= -\frac{\left(\mathfrak{v} -\lambda _1\right) \left(\mathfrak{v} -\lambda
   _2\right) \left(\lambda _1 \lambda _2+\left(\lambda _1+\lambda _2\right)
   \mathfrak{v} \right)}{\left(\lambda _1^2+\lambda _2 \lambda _1+\lambda
   _2^2\right) \mathfrak{v} }
\end{equation}
and the corresponding functions entering the intrinsic components of
the Riemann curvature are
\begin{equation}\label{perdinci}
\mathcal{CF}^{KE}_{1,2,3}\left(\mathfrak{v} \right)= \left\{-\frac{2
\left(\lambda _1^2 \lambda _2^2+\left(\lambda
   _1+\lambda _2\right) \mathfrak{v} ^3\right)}{\left(\lambda
   _1^2+\lambda _2 \lambda _1+\lambda _2^2\right)
   \mathfrak{v} ^3},\frac{2 \lambda _1^2 \lambda
   _2^2-\left(\lambda _1+\lambda _2\right) \mathfrak{v} ^3}{2
   \left(\lambda _1^2+\lambda _2 \lambda _1+\lambda _2^2\right)
   \mathfrak{v} ^3},\frac{\lambda _1^2 \lambda _2^2+\left(\lambda
   _1+\lambda _2\right) \mathfrak{v} ^3}{\left(\lambda
   _1^2+\lambda _2 \lambda _1+\lambda _2^2\right)
   \mathfrak{v} ^3}\right\}
\end{equation}
the interval of variability of the moment coordinate $\mathfrak{v} $
being the following $\mathfrak{v}
\in\left[\lambda_1,\lambda_2\right]$. Correspondingly the  boundary
values are
\begin{equation}\begin{array}{rcl}\label{pergiovetonante}
 \mathcal{CF}^{KE}_{1,2,3}\left(\mathfrak{v} _{min}\right)&=&\left\{-\frac{2}{\lambda _1},\frac{1}{\lambda _1}-\frac{3
   \left(\lambda _1+\lambda _2\right)}{2 \left(\lambda
   _1^2+\lambda _2 \lambda _1+\lambda
   _2^2\right)},\frac{1}{\lambda _1}\right\}  \\
  \mathcal{CF}^{KE}_{1,2,3}\left(\mathfrak{v} _{max}\right) &=&
  \left\{-\frac{2}{\lambda _2},\frac{1}{\lambda _2}-\frac{3
   \left(\lambda _1+\lambda _2\right)}{2 \left(\lambda
   _1^2+\lambda _2 \lambda _1+\lambda
   _2^2\right)},\frac{1}{\lambda _2}\right\} \\
\end{array}\end{equation}
We can use the case $\lambda_1=1,\lambda_2=2$ as a standard example.
In this case the behavior of the three functions is displayed in
Figure \ref{KEfunzie}
\paragraph{The case of the extremal K\"ahler metric on the second Hizebruch surface  $\mathbb{F}_2$.}
Finally we consider the case of the extremal metric on
$\mathbb{F}_2$ discussed in the previous pages and defined by the
function \eqref{torriano}. In this case the three functions
\eqref{trefunzie} parameterizing the curvature 2-form and hence the
intrinsic components of the Riemann tensor are the following ones:
\begin{eqnarray}
\label{piripicchio}
  \mathcal{CF}^{F2ext}_1 &=& \frac{2 \mathit{a}^2 \mathit{b}^2 (3 \mathit{a}+\mathit{b})-8 \mathit{a}^2
   \mathfrak{v}^3+6 \mathfrak{v}^4 (3 \mathit{a}+\mathit{b})}{\mathfrak{v}^3
   (\mathit{a}-\mathit{b}) \left(\mathit{a}^2+4 \mathit{a}
   \mathit{b}+\mathit{b}^2\right)} \\
  \mathcal{CF}^{F2ext}_2 &=& \frac{\mathit{a}^3 \mathit{b} (2 \mathfrak{v}-3 \mathit{b})-\mathit{a}^2
   \left(\mathit{b}^3+2 \mathfrak{v}^3\right)+3 \mathit{a} \mathfrak{v}^4+\mathit{b}
   \mathfrak{v}^4}{\mathfrak{v}^3 (\mathit{a}-\mathit{b}) \left(\mathit{a}^2+4
   \mathit{a} \mathit{b}+\mathit{b}^2\right)}\\
 \mathcal{CF}^{F2ext}_3 &=& \frac{4 \mathit{a}^3 \mathit{b} \mathfrak{v}-\mathit{a}^2 \mathit{b}^2 (3
   \mathit{a}+\mathit{b})+4 \mathit{a}^2 \mathfrak{v}^3+\mathfrak{v}^4 (-(3
   \mathit{a}+\mathit{b}))}{\mathfrak{v}^3 (\mathit{a}-\mathit{b})
   \left(\mathit{a}^2+4 \mathit{a} \mathit{b}+\mathit{b}^2\right)}
\end{eqnarray}
A plot of the three functions for the extremal $\mathbb{F}_2$
metric, to be compared with the analogous plot relative to the
Kronheimer metric (Fig. \ref{plotto123}A) on the same manifold is
shown in Fig.\ref{RiefunzioF2ext}.
\subsection{The structure of $\mathcal{M}_3$ and the conical singularity}
\label{singularity} The two real manifolds defined by the
restriction to the dense chart $\mathfrak{u},\mathfrak{v}
,\phi,\tau$, of the surface $\mathbb{F}_2$ and of  the manifold
$\mathcal{M}_B^{KE}$ are fully analogous. Cutting the compact four
manifold into $\mathfrak{v}  \, = \,\text{const}$ slices we always
obtain the same result, namely a three manifold $\mathcal{M}_3$ with
the structure of a circle fibration on $\mathbb{S}^2$:
\begin{equation}\label{frullatore}
\mathcal{M}_B  \, \supset \,\mathcal{M}_3 \,
\stackrel{\pi}{\longrightarrow} \, \mathbb{S}^2
   \quad ; \quad \forall p \in \mathbb{S}^2 \quad \pi^{-1}(p) \sim
   \mathbb{S}^1
\end{equation}
The metric on $\mathcal{M}_3$ is the standard one for fibrations:
\begin{equation}\label{m3metric}
    ds^2_{\mathcal{M}_3}=\mathfrak{v}
    \, \left(d\phi^2 \sin ^2\theta +d\theta
   ^2\right)\, + \,\mathcal{FK}(\mathfrak{v} ) \left[d\phi  (1-\cos \theta )
    +d\tau\right]^2
\end{equation}
\par
The easiest way to understand $\mathcal{M}_3$ is to study its
intrinsic curvature by using the dreibein formalism. Referring to
equation \eqref{m3metric} we introduce the following dreiben
1-forms:
\begin{equation}\label{trebanni}
    \pmb{\epsilon}^1 = \sqrt{\mathfrak{v} }\,d\theta \quad ;
    \quad \pmb{\epsilon}^2= \sqrt{\mathfrak{v} }\,\sin \theta d\phi
    \quad ; \quad \pmb{\epsilon}^3 = \sqrt{\mathcal{FK}(\mathfrak{v} )} \left[d\phi
   (1-\cos \theta )+d\tau \right]
\end{equation}
The fixed parameter $\mathfrak{v} $ plays the role of the squared
radius of the sphere $\mathbb{S}^2$ while
$\sqrt{\mathcal{FK}(\mathfrak{v} )}$ weights the contribution of the
circle fibre defined over each point $p\in \mathbb{S}^2$. At the
endpoints of the intervals $\mathcal{FK}(\mathfrak{v} _{min})\, =
\,\mathcal{FK}(\mathfrak{v} _{max})\, = 0$  the fibre shrinks to
zero.
\par
\begin{figure}
\centering
\includegraphics[width=10cm]{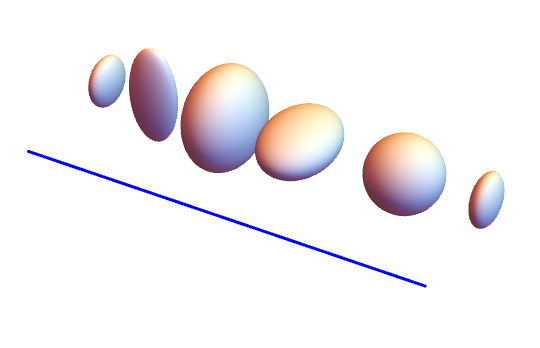}
\caption{\label{concetto}   A conceptual picture of the
$\mathcal{M}_B$ spaces that include also the second Hirzebruch
surface. The finite blue segment represent the $\mathfrak{v}
$-variable varying from its minimum to its maximum value. Over each
point of the line we have a three dimensional space $\mathcal{M}_3$
which is homeomorphic to a 3-sphere but is variously deformed at
each different value $\mathfrak{v} $. At the initial and final
points of the blue segment the three dimensional space degenerates
into an $S^2$ sphere. Graphically we represent the deformed 3-sphere
as an ellipsoid and the 2-sphere as a flat filled circle.}
\end{figure}

Using the standard formulas of differential geometry and once again
the {\sc Mathematica} packgage {\sc Vielbgrav23} we calculate the
spin connection and the curvature 2-form. We obtain:
\begin{equation}\label{gundashapur}
    \mathfrak{R}=\left(
\begin{array}{cc||c}
 0 & \frac{ (4
   \mathfrak{v} -3
   \mathcal{FK}(\mathfrak{v} ))}{4
   \mathfrak{v} ^2}\,\, \pmb{\epsilon}^1\, \wedge \,\pmb{\epsilon}^2 &
   \frac{
   \mathcal{FK}(\mathfrak{v} )}{4
   \mathfrak{v} ^2}\, \,\pmb{\epsilon}^1\, \wedge \,\pmb{\epsilon}^3 \\
-\, \frac{ (4
   \mathfrak{v} -3
   \mathcal{FK}(\mathfrak{v} ))}{4
   \mathfrak{v} ^2}\,\, \pmb{\epsilon}^1\, \wedge \,\pmb{\epsilon}^2 & 0 &
   \frac{
   \mathcal{FK}(\mathfrak{v} )}{4
   \mathfrak{v} ^2} \,\, \pmb{\epsilon}^2\, \wedge \,\pmb{\epsilon}^3 \\
   \hline
 -\frac{
   \mathcal{FK}(\mathfrak{v} )}{4
   \mathfrak{v} ^2}\,\,\pmb{\epsilon}^1\, \wedge \,\pmb{\epsilon}^3 &
    -\frac{
   \mathcal{FK}(\mathfrak{v} )}{4
   \mathfrak{v} ^2} \, \, \pmb{\epsilon}^2\, \wedge \,\pmb{\epsilon}^3 & 0 \\
\end{array}
\right)
\end{equation}
The Riemann curvature 2-form in flat indices has constant components
and if the coefficient $\frac{ (4
   \mathfrak{v} -3
   \mathcal{FK}(\mathfrak{v} ))}{4
   \mathfrak{v} ^2}$ were equal to the coefficient $\frac{
   \mathcal{FK}(\mathfrak{v} )}{4
   \mathfrak{v} ^2}$ the 2-form in eqn.~\eqref{gundashapur} would be the
   standard Riemann 2-curvature of the homogeneous space
   $\mathrm{SO(4)/SO(3)}$, namely the the 3-sphere
   $\mathbb{S}^3$. What we learn from this easy calculation is that
    every section $\mathfrak{v}  = \text{constant}$ of
   $\mathcal{M}_B$ is homeomorphic to a 3-sphere endowed with a
   metric that is not the maximal symmetric one with isometry $\mathrm{SU(2)
   \times SU(2)}$ but a slightly deformed one with isometry $\mathrm{SU(2)
   \times U(1)}$: in other words we deal with a 3-sphere deformed
into the 3-dimensional analogue of an ellipsoid. At the endpoints of
the $\mathfrak{v} $-interval the ellipsoid degenerates into a sphere
since the third dreibein $\pmb{\epsilon}$ vanishes. A conceptual
picture of the full space $\mathcal{M}_B$ is provided in picture
Figure \ref{concetto}.
\paragraph{Global properties of $\mathcal{M}_3$.}
Expanding on the global properties of $\mathcal{M}_3$, we describe
it as a magnetic monopole bundle over $\mathbb{S}^2$,  and prove
that the corresponding monopole strength   is $n=2$. We start from
the definition of the action of the ${\rm SU}(2) $ isometry
\eqref{ciabattabuona} and describe the 2-sphere $\mathbb{S}^2$
spanned by $\theta$ and $\phi$ as $\mathbb{P}^1$ with projective
coordinates $U^0,\,U^1$:
\begin{equation}
    U^0 =   r \sin\left(\frac{\theta}{2}\right)\,e^{i\,\frac{\gamma+\phi}{2}}\,, \quad   U^1 =
      r \cos\left(\frac{\theta}{2}\right)\,e^{i\,\frac{\gamma-\phi}{2}}\label{U0U1}
\end{equation}
where $0\le \theta \le \pi$, $0\le \phi<2\pi$, $0\le \gamma<4\pi$.
In the North patch $\mathcal{U}_N$, $U^1\neq 0$ and the sphere is
spanned by the stereographic coordinate $u_N\ =  U^0/U^1$, while in
the south patch $\mathcal{U}_S$, $U^0\neq 0$ and the stereographic
coordinate is $u_S =  U^1/U^0$. The transformation properties
\eqref{ciabattabuona} define a line bundle whose local
trivializations about the two poles are:
\begin{align}
    \phi^{-1}_N(\mathcal{U}_N)&=\left(u_N,\,v_N\right)= \left(\frac{U^0}{U^1},\,\xi \, (U^1)^2\right)\,,\\
      \phi^{-1}_S(\mathcal{U}_S)&=\left(u_S,\,v_S\right)=\left(\frac{U^1}{U^0},\,\xi \, (U^0)^2\right)\,,
\end{align}
where $\xi$ is a complex number in the fibre not depending on the
patch. As $(U^0,\,U^1)$ transform linearly under the an ${\rm
SU}(2)$-transformation:
\begin{equation}
 \left(\begin{matrix}U^0\cr U^1\end{matrix}\right)\rightarrow \left(\begin{matrix}d & c\cr b & a\end{matrix}\right)
  \left(\begin{matrix}U^0\cr U^1\end{matrix}\right)\,,
\end{equation}
the fibre coordinate $v$ transforms so that $(1+|u_N|^2)^2 |v_N|^2$
and $(1+|u_S|^2)^2 |v_S|^2$, in $\mathcal{U}_N$ and $\mathcal{U}_S$,
respectively, are invariant. The transition function on the fibre
reads, at the equator $\theta=\pi/2$:
\begin{equation}
    t_{NS}=\left(\frac{U^1}{U^0}\right)^2=e^{-2i\phi}=e^{-in\,\phi}\,,
\end{equation}
implying that the ${\rm U}(1)$-bundle associated with the phase of
$v$ (i.e. the  submanifold of the K\"ahler-Einstein space at
constant $|v|$), is a monopole bundle with monopole strength $n=2$.
This has to be contrasted with the Hopf-fibreing description of
$\mathbb{S}^3$, for which the local trivializations have fibre
components $U^1/|U^1|$ and $U^0/|U^0|$ in the two patches,
respectively, and $t_{NS}$ at the equator is $U^1/U^0=e^{-i\phi}$.
In this case,  the monopole strength is $n=1$.
\paragraph{Conical singularities   and regularity of $\mathbb{F}_2$.}
{Let us analyze the exact form of the singularities (when they are
present). The restriction of the metric to a fibre spanned by
$\mathfrak{v} $ and $\tau\in (0,2\pi)$ is }
\begin{equation}
ds^2=\frac{d\mathfrak{v} ^2}{\mathcal{FK}(\mathfrak{v}
)}+\mathcal{FK}(\mathfrak{v} )\,d\tau^2\,.\label{metfib}
\end{equation}
Let  $\lambda$ denote one of the two  roots $\lambda_1,\,\lambda_2$
of $\mathcal{FK}(\mathfrak{v} )$. Close to $\lambda$, to first order
in $\mathfrak{v} $, in the KE case, the metric \eqref{metfib} is
flat and features a deficit angle signalling a conifold singularity.
This singularity is absent in the $\mathbb{F}_2$ cases, as expected.
To show this   let us Taylor expand $\mathcal{FK}(\mathfrak{v} )$
about $\lambda$:
\begin{equation}
    \mathcal{FK}(\mathfrak{v} )= \mathcal{FK}'(\lambda)(\mathfrak{v} -\lambda)+\mathcal{O}((\mathfrak{v} -\lambda)^2).
\end{equation}
We can verify that:
\begin{equation}
\begin{array}{lclcl}
\mbox{KE $\phantom{Kronheimer}$}&:&\mathcal{FK}_{KE}'(\lambda_1)=
 \frac{(\lambda_2-\lambda_1)(\lambda_1+2 \lambda_2)}{\lambda_1^2+
 \lambda_2^2+\lambda_1\lambda_2} &;&
 \mathcal{FK}'_{KE}(\lambda_2)=\frac{(\lambda_1-\lambda_2)(2\lambda_1+ \lambda_2)}{\lambda_1^2+ \lambda_2^2+\lambda_1\lambda_2}\,,
   \\[2pt]
\mbox{$\mathbb{F}_2$
Kronheimer}&:&\mathcal{FK}_{\mathbb{F}_2|Kro}'\left(\frac{9\alpha}{32}\right)=2&
; &
\mathcal{FK}_{\mathbb{F}_2|Kro}'\left(\frac{9(3\alpha+4)}{32}\right)=-2 \\
\mbox{$\mathbb{F}_2$
Extremal}&:&\mathcal{FK}_{\mathbb{F}_2|ext}'(\mathit{a}) = 2 & ;& \mathcal{FK}_{\mathbb{F}_2|ext}'(\mathit{b})=-2 \\
\end{array}
\end{equation}
Next we replace the first-order expansion  of this function in the
fibre metric:
\begin{equation}
ds^2=\frac{d\mathfrak{v} ^2}{\mathcal{FK}'(\lambda)(\mathfrak{v}
-\lambda)}+\mathcal{FK}'(\lambda)(\mathfrak{v}
-\lambda)\,d\tau^2\,,\label{metfib2}
\end{equation}
and write it as a flat metric in polar coordinates:
\begin{equation}
ds^2=dr^2+\beta^2\,r^2\,d\tau^2\,.\label{metfib3}
\end{equation}
One can easily verify that:
\begin{equation}
    r=2\,\sqrt{\frac{\mathfrak{v} -\lambda}{\mathcal{FK}'(\lambda)}}\,\,,\,\,\,\beta=\frac{|\mathcal{FK}'(\lambda)|}{2}\,.
\end{equation}
Defining $\tilde{\varphi}= \beta \,\tau$, we can write the fibre
metric as follows:
$$ds^2=dr^2+r^2\,d\tilde{\varphi}^2\,.$$
Now the polar angle varies in the range: $\tilde{\varphi}\in
[0,\,2\pi\,\beta]$. If $\beta<1$ we have a deficit angle:
$$\Delta\phi=2\pi (1-\beta)\,.$$
\par
Let us see what this implies in the various possible cases of Table
\ref{casoni}.
\begin{enumerate} \item
In the case of the $\mathbb{F}_2$ manifold one has
$|\mathcal{FK}'(\lambda)|=2$ and $\beta=1$, both for the Kronheimer
metric and for the extremal one of the Calabi family, so there is no
conical singularity, as expected.
\item In the case of $\mathbb{WP}_{[1,1,2]}$ we have
$$\mathcal{FK}(\lambda) = \frac{32\lambda^2-144\lambda+81}{(4\lambda-9)^2}.$$
For the limiting value $\lambda=0$ we obtain $\beta=\frac12$, i.e.,
a $\mathbb{C}^2/\mathbb{Z}_2$ singularity, while for
$\lambda=\frac98$ we have $\beta=1$, i.e., no singularity, as we
expected as $\mathbb{WP}_{[1,1,2]}$ is an orbifold
$\mathbb{P}^2/\mathbb{Z}_2$ with one singular point.
\item In the KE manifold case considering $\lambda=\lambda_1$, we have:
\begin{equation}
    \mathcal{FK}'(\lambda_1)=\frac{(\lambda_2-\lambda_1)(\lambda_1+2 \lambda_2)}{\lambda_1^2+
    \lambda_2^2+\lambda_1\lambda_2}=-1+\frac{3\lambda_2^2}{\lambda_1^2+ \lambda_2^2
    +\lambda_1\lambda_2}<-1+\frac{3\lambda_2^2}{ \lambda_2^2}=2\,\Rightarrow\,\,\beta<1\,,
\end{equation}
and $|\mathcal{FK}'(\lambda_2)|<|\mathcal{FK}'(\lambda_1)|$, so that
$\beta<1$ also at $\lambda_2$. The manifold has two conical
singularities, both in the same fibre of the projection to one of
the $\mathbb{S}^2$'s. One of the singularities will be an orbifold
singularity of type $\mathbb{C}^2/\mathbb{Z}_n$ if the corresponding
value of $\beta$ is
\begin{equation}\label{betaint} \beta = 1 - \frac1n. \end{equation}
It is interesting to note that when this happens, the form of the
function $\mathcal{FK}$ does not allow the other singularity to be
of this type as well, as the corresponding integer $m$ should
satisfy
\begin{equation}\label{missionimpossible}
    m = \frac{4n}{2+5n \pm \sqrt{9n^2+12n-12}}
\end{equation}
which is not satisfied by any pair $(m,n)$ where both $m$, $n$ are
integers greater than 1; so the singular fibre can never be a
football or a spindle.
\item
We   discuss the case $\lambda_1=0$. In this case
\begin{equation}
    \mathcal{FK}^0(\mathfrak{v}) =\frac{\mathfrak{v}(\lambda_2-\mathfrak{v})}{\lambda_2}\,.
\end{equation}
If we focus on the fibre metric:
\begin{equation}
ds^2=\frac{d\mathfrak{v} ^2}{\mathcal{FK}_0(\mathfrak{v}
)}+\mathcal{FK}_0(\mathfrak{v} )\,d\tau^2\,.\label{metfib0}
\end{equation}
We can easily verify that in the coordinates $\tilde{\theta}\in
[0,\pi]$ and $\tilde{\varphi}\in [0,\pi)$ defined by
$$
\mathfrak{v}(\tilde{\theta})=R^2\,\sin^2\left(\frac{\tilde{\theta}
}{2}\right) \le
R^2=\lambda_2\,\,,\,\,\,\tilde{\varphi}=\frac{\tau}{2}\,,$$ where $R
= \sqrt{\lambda_2}$, the fibre metric \eqref{metfib0} becomes
 \begin{equation}
     ds^2=R^2\,\left(d\tilde{\theta}^2+\sin^2(\tilde{\theta})\,d\tilde{\varphi}^2\right)\,.
 \end{equation}
 Since $\tilde{\varphi}=\tau/2\in\,[0,\pi)$, the fibre is the splindle $\mathbb{S}^2/\mathbb{Z}_2$.
 {Topologically, the entire 4-manifold is still $\mathbb{S}^2\times \mathbb{S}^2$.}
\item
For $\mathcal{FK}(\mathfrak v)=\mathfrak v$ we   get
$\beta=\frac12$, in accordance with the fact that the variety in
this case is $\mathbb{C}^2/\mathbb{Z}_2$.
 \end{enumerate}
\par
{In the cases 3 and 4 the singular locus is of the form
$\mathbb{S}^2\times p_\pm$, where $p_\pm$ are the ``poles'' of the
fibres of the projection $\mathbb{S}^2\times \mathbb{S}^2\to S$. In
complex geometric terms, it is a pair of divisors, both isomorphic
to $\mathbb{P}^1$.}
\subsubsection{Conclusion on $KE$ manifolds}
As we are going to see in next sections the condition that the base
manifold $\mathcal{M}_{B}$ should be a K\"ahler Einstein manifold is
essential in order to be able to construct the predicted Ricci flat
metric on
$\operatorname{tot}\left[K\left(\mathcal{M}_B\right)\right]$ by
means of a general construction due to Calabi and named the
\textit{Calabi Ansatz}. In want of such a construction the solution
of the Monge Amp\`{e}re equation has so far resisted all attempts at
going beyond a perturbative series. In any case the elaboration of
the \textit{Calabi Ansatz} within the AMSY action/angle formalism
was not known and it was instead  one of the main results of the
recent paper \cite{bruzzo2023d3brane}. Summarizing the previous
discussion, we can say that, at the price of allowing the two above
mentioned conical singularities one can start from a manifold almost
identical and homeomorphic to the resolution of the
$\mathbb{C}^3/\mathbb{Z}_4$ singularity, namely the $\mathbb{F}_2$
surface and subsequently apply the Calabi ansatz construction as I
will report in the sequel. In this way one obtains an exact D3 brane
solution. The conical singularities are one problem in the use of
such an exact solution as a member of a holographic pair, yet there
is another  more intriguing one. The considered $KE$ metrics are not
a degeneration of Kronheimer metrics on the wall of a known chamber
struture, rather they occur in another family, that of the extremal
metrics, and at the moment it is not yet clear, as I will discuss in
chapter \ref{aperto}, how such metrics might appear in a McKay
quiver setup \`{a} la Kronheimer.
\par
\section{The Calabi Ansatz and the AMSY symplectic formalism}\label{trick}
Having studied in some detail the KE base manifolds
$\mathcal{M}_B^{KE}$, we turn now  to the main issue namely to the
construction of a Ricci-flat metric on each of their canonical
bundles. The Monge Amp\`{e}re equations for the determination of
such metrics can be written in general terms and it was studied in
\cite{Bianchi_2021}, yet as shown there, although an iterative
series procedure is always available, exact solutions of such a
partial differential non linear equation are difficult to be found.
For that reason we turn to the method introduced by Calabi, which,
however, works  only for the case of KE base manifolds.
\paragraph{Cautionary remark}
As we anticipated above the Calabi ansatz method produces a
Ricci-flat metric on the canonical bundle
$\operatorname{tot}\left[K\left(\mathcal{M}^{KE}\right)\right]$ that
one might be tempted to consider diffeomorphic to the Ricci-flat
metric on the metric cone over the Sasaki-Einstein manifolds of
\cite{Gauntlett:2004yd}; actually this is not true, notwithstanding
the very close relation of the K\"ahler Einstein manifolds discussed
above with the 5-dimensional Sasaki-Einstein manifolds of
\cite{Gauntlett:2004yd}. We have so far postponed the comparison of
the KE metrics discussed in the above sections with the base
manifolds of the 5-dimensional SE fibrations of
\cite{Gauntlett:2004yd} since such a comparison will be done more
appropriately just in one stroke together with the comparison of the
6-dimensional Ricci-flat metrics.
\paragraph{Calabi's Ansatz} Calabi's paper
\cite{Calabi-Metriques} introduces the following Ansatz for the
local K\"ahler potential $\mathcal{K}(z,\bar{z},w,\bar{w})$ of a
K\"ahler  metric $g_E$ on the total space of a holomorphic vector
bundle $E \to \mathcal{M}$, where $\mathcal{M}$ is a compact
K\"ahler manifold:
 \begin{equation} \label{Ansatz}
    \mathcal{K}(z,\bar{z},w,\bar{w})=\mathcal{K}_0(z,\bar{z})+ U(\lambda)
\end{equation}
where   $\mathcal{K}_0(z,\bar{z})$ is a K\"ahler potential for
$g_{\mathcal{M}}$, ($z^i$, $i=1,\dots \,
\mathrm{dim_\mathbb{C}}\mathcal{M}$ being the complex coordinates of
the base manifold) and $U$ being a function of a real variable
$\lambda$, which we shall identify with the function
\begin{equation}\label{squarnorm}
    \lambda = \mathcal{H}_{\mu\bar\nu}(z,\bar z) \,w^\mu \,  w^{\bar\nu}\, =
    \parallel w \parallel^2
\end{equation}
\textit{i.e} with the square norm   of a section of the bundle with
respect to a fibre metric $\mathcal{H}_{\mu\bar\nu}(z,\bar z)$. If
$\theta$ is the Chern connection on $E$, canonically determined by
the Hermitian structure $\mathcal{H}$ and the holomorphic structure
of $E$, its local connections forms can be written as
\begin{equation}\label{connetto}
\theta^{\phantom{\nu}\lambda}_\nu=\sum_i dz^{i}
\,L^{\phantom{i|\nu}\lambda}_{i|\nu}
\end{equation}
where
\begin{equation}\label{concof}
 L^{\phantom{i|\nu}\lambda}_{i|\nu}=\sum_{\bar\mu}\mathcal{H}^{\lambda\bar{{\mu}}}\frac{\partial}{\partial z^{i}}
 \mathcal{H}_{\nu\bar{\mu}}
 \qquad ; \qquad [\mathcal{H}^{\lambda\bar{\mu}}]=([\mathcal{H}_{\lambda\bar{\mu}}]^{-1})^{T}
\end{equation}
The curvature 2-form $\Theta$ of the connection $\theta$ is given
by:
\begin{equation}\label{ciccio}
    \Theta^{\phantom{\nu}\lambda}_{\nu}=\sum_{i,\bar\ell} \,
    dz^{i}\wedge d\bar{z}^{\bar\ell}
    \,S^{\phantom{i{\bar\ell}|\nu}\lambda}_{i{\bar\ell}|\nu}
    \qquad; \qquad
    S^{\phantom{i{\bar\ell}|\nu}\lambda}_{i{\bar\ell}|\nu} \, =
    \, \frac{\partial}{\partial
    {\bar z}^{\bar\ell}}\, L^{\phantom{i|\nu}\lambda}_{i|\nu}
\end{equation}
The K\"ahler metric $g_{E}$ corresponding to the K\"ahler potential
$\mathcal{K}$ can be written  as follows:
\begin{multline}
\partial\bar{\partial}\mathcal{K}\, =\,
\sum_{i,{\bar\ell}}\Bigl[g_{i{\bar\ell}}+\lambda\, U'(\lambda) \,
\sum_{\lambda,\nu,\bar\mu} \mathcal{H}_{\sigma\bar{\mu}}
S^{\phantom{i{\bar\ell}|\rho}\sigma}_{i{\bar\ell}|\rho} w^{\rho}
\bar{ w}^{\bar\mu}\Bigr]dz^{i}d\bar{z}^{\ell}
+\sum_{\sigma,\bar\mu}\Bigl[U'(\lambda)\,+\, \lambda \,
U''(\lambda)\Bigr] \, \mathcal{H}_{\sigma\bar{\mu}} \, \triangledown
w^{\sigma}\,\triangledown\bar{ w}^{\bar\mu}.
\end{multline}
If $E$ is a line bundle then the above equation reduces to
\begin{equation}\label{pbarppsi}
\partial\bar{\partial}\mathcal{K}\, =\,\sum_{i,\bar{\ell}} \,[g_{i\bar{\ell}}+\lambda U'(\lambda)\,
S_{i\bar{\ell}}]dz^{i}d\bar{z}^{\bar\ell}+[U'(\lambda )\, +\,
\lambda U''(\lambda)]\mathcal{ H}(z,\bar{z})\triangledown
w\triangledown\bar{ w}
\end{equation}
where $\lambda=\mathcal{H}(z,\bar{z})\,w\bar{w}$ is  the nonnegative
real quantity defined in equation \eqref{squarnorm} and
$\triangledown w$ denotes the covariant derivative of the
fibre-coordinate with respect to  the Chern connection $\theta$:
\begin{equation}\label{curlandia}
  \triangledown w  =dw \,+ \, \theta \, w
\end{equation}
\subsection{Ricci-flat metrics on canonical bundles} Now we assume
that $E$ is the canonical bundle  $K\left({\mathcal{M}}\right)$ of a
K\"ahler surface $\mathcal{M}$ ($\dim_\mathbb{C} \mathcal{M} = 2$).
The total space
$\operatorname{tot}\left[K\left(\mathcal{M}\right)\right]$ has
vanishing first Chern class, i.e., it is a noncompact Calabi-Yau
manifold, and we may try to construct explicitly a Ricci-flat metric
on it. Actually, following Calabi, we can reduce the  condition that
$g_{E}$ is Ricci-flat to a differential equation for the function
$U(\lambda)$ introduced in equation \eqref{Ansatz}. Note that under
the present assumptions $S$ is a scalar-valued 2-form on
$\mathcal{M}$.
\par
Since our main target is the construction of a Ricci-flat metric on
the space $\operatorname{tot}\left[\mathcal
K\left(\mathcal{M}^{KE}_B\right)\right]$, where $\mathcal{M}^{KE}_B$
denotes any of the KE manifolds discussed at length in previous
sections, we begin precisely with an analysis of that case: this
 allows  the derivation of a general form of $U(\lambda)$ as a function of the moment $\mathfrak{w}$ associated
with the $\mathrm{U(1)}$ group acting by phase transformations of
the fibre coordinate $w$,  and of certain coefficients $A,B,F$ that
are determined in terms of the K\"ahler potential $\mathcal{K}_0$ of
the base manifold $\mathcal{M}$. Consistency of the Calabi Ansatz
requires that these coefficients should be constant, which happens
in the case of those base manifolds that are equipped  with a
K\"ahler Einstein metric. KE metrics do not exist on Hirzebruch
surfaces and the Calabi Ansatz is not applicable in this case. As we
discuss in the sequel, there exists a Ricci-flat metric on the
canonical bundle of a singular blow-down of $\mathbb{F}_2$, namely
the weighted projective plane $\mathbb{WP}_{[1,1,2]}$, which is
known in the AMSY symplectic toric formalism of \cite{abreu} and
\cite{Martelli:2005tp}. If we were able to do the inverse Legendre
transform we might reconstruct the so far missing K\"ahler potential
and get inspiration on possible generalizations of the Calabi
Ansatz. Hence we are going to pay a lot of attention to both
formulations, the K\"ahler one and the symplectic one.
\subsection{Calabi Ansatz   for 4D K\"ahler  metrics
with $\rm{SU}(2)\times\rm{U}(1)$ isometry} \label{F2Calab}
The Calabi Ansatz can be applied with success or not according to
the structure of the K\"ahler potential $\mathcal{K}_0$  for the
base manifold $\mathcal{M}$ and  the algebraic form of the invariant
combination $\Omega$ of the complex coordinates $u,v$ which is the
only real variable from which the K\"ahler potential
$\mathcal{K}_0=\mathcal{K}_0(\Omega) $ is assumed to depend. On the
other hand   $\Omega$ codifies  the group of isometries which is
imposed on the K\"ahler metric of $\mathcal{M}$.
\par
In the case of the  metrics discussed in section \ref{varposympo},
that have $\rm{SU}(2)\times\rm{U}(1)$ isometry, the invariant is
chosen to be
\begin{equation}\label{OmegaHirze}
   \Omega =  \varpi
\end{equation}
where $\varpi$ was defined in eqn.~\eqref{lattosio}. This choice
guarantees the isometry of the K\"ahler metric $g_\mathcal{M}$
against the group $\mathrm{SU(2)} \times \mathrm{U(1)}$ with the
action described in eqn.~\eqref{ciabattabuona}. Hence we focus on
such manifolds and we consider a K\"ahler potential for
$\mathcal{M}$ that for the time being is a generic function
$\mathcal{K}_0(\varpi)$ of the invariant variable. In this case  the
determinant of the K\"ahler metric $g_\mathcal{M}$ has  an explicit
expression in terms of $\mathcal{K}_0(\varpi)$
\begin{equation}\label{determio}
    \mathrm{det}\left(g_\mathcal{M}\right) \, =
    \, 2 \varpi  {\mathcal{K}_0}'(\varpi ) \left(\varpi  {\mathcal{K}_0}''(\varpi )+{\mathcal{K}_0}'(\varpi
   )\right)
\end{equation}
while the determinant of the Ricci tensor takes the   form
\begin{equation}\begin{array}{rcl}
\mathrm{det}\left(\mathrm{Ric}_\mathcal{M}\right)&=&
\frac{N_{Ric}}{D_{Ric}} \\
N_{Ric}& = & 2 \left(\varpi ^2  {\mathcal{K}_0}''(\varpi )^2+
{\mathcal{K}_0}'(\varpi )^2+\varpi
    {\mathcal{K}_0}'(\varpi ) \left(\varpi   {\mathcal{K}_0}^{(3)}(\varpi )+4  {\mathcal{K}_0}''(\varpi
   )\right)\right) \times  \\
   && \times \left(-\varpi ^3  {\mathcal{K}_0}''(\varpi )^4+\varpi ^2  {\mathcal{K}_0}'(\varpi
   ) \left(\varpi   {\mathcal{K}_0}^{(3)}(\varpi )- {\mathcal{K}_0}''(\varpi )\right)
    {\mathcal{K}_0}''(\varpi )^2 \right.  \\
   &&\left.+\varpi   {\mathcal{K}_0}'(\varpi )^2 \left(-\varpi ^2
    {\mathcal{K}_0}^{(3)}(\varpi )^2- {\mathcal{K}_0}''(\varpi )^2+\varpi  \left(\varpi
    {\mathcal{K}_0}^{(4)}(\varpi )+2  {\mathcal{K}_0}^{(3)}(\varpi )\right)  {\mathcal{K}_0}''(\varpi
   )\right)\right. \\
   &&\left.+ {\mathcal{K}_0}'(\varpi )^3 \left(3  {\mathcal{K}_0}''(\varpi )+\varpi  \left(\varpi
    {\mathcal{K}_0}^{(4)}(\varpi )+5  {\mathcal{K}_0}^{(3)}(\varpi
    )\right)\right)\right) \\
    D_{Ric}& = &  {\mathcal{K}_0}'(\varpi )^3 \left(\varpi   {\mathcal{K}_0}''(\varpi )+ {\mathcal{K}_0}'(\varpi )\right)^3
\end{array}\end{equation}
and the scalar curvature
\begin{equation}\label{Rscalar}
   R_s=\mathrm{Tr}\left(\mathrm{Ric}_\mathcal{M}\, g^{-1}_\mathcal{M}\right)
\end{equation}
is
\begin{equation}\begin{array}{rcl}\label{scalocurvo}
   R_s & = & \frac{N_s}{D_s}  \\
   N_s & = &  {\mathcal{K}_0}'(\varpi )^3+\varpi ^3  {\mathcal{K}_0}''(\varpi )^2 \left(2 \varpi
    {\mathcal{K}_0}^{(3)}(\varpi )+5  {\mathcal{K}_0}''(\varpi )\right) \\
   &&+\varpi ^2  {\mathcal{K}_0}'(\varpi
   ) \left(-\varpi ^2  {\mathcal{K}_0}^{(3)}(\varpi )^2+9  {\mathcal{K}_0}''(\varpi )^2+\varpi
   \left(\varpi   {\mathcal{K}_0}^{(4)}(\varpi )+4  {\mathcal{K}_0}^{(3)}(\varpi )\right)
    {\mathcal{K}_0}''(\varpi )\right) \\
   &&+\varpi   {\mathcal{K}_0}'(\varpi )^2 \left(9
    {\mathcal{K}_0}''(\varpi )+\varpi  \left(\varpi   {\mathcal{K}_0}^{(4)}(\varpi )+6
    {\mathcal{K}_0}^{(3)}(\varpi )\right)\right) \\
    D_s & = & \varpi   {\mathcal{K}_0}'(\varpi ) \left(\varpi   {\mathcal{K}_0}''(\varpi )+ {\mathcal{K}_0}'(\varpi
   )\right)^3
\end{array}\end{equation}
Given these base manifold data,  we introduce a K\"ahler potential
for a metric on the canonical bundle
$\operatorname{tot}\left[K\left(\mathcal{M}\right)\right]$ in
accordance with the Calabi Ansatz, namely
\begin{equation}\label{carabina}
    \mathcal{K}\left(\varpi,\lambda\right)={\mathcal{K}_0}(\varpi )\, +
    \, U(\lambda);  \quad \lambda  =
    \underbrace{\exp\left [\mathcal{P}(\varpi) \right]}_{\text{fibre metric $\mathcal{H}(\varpi)$}}  \mid w\mid^2 =
    \parallel w \parallel^2
\end{equation}
where $\lambda$ is  the  square norm of a section of the canonical
bundle and $\exp\left [\mathcal{P}(\varpi) \right]$ is some   fibre
metric. The determinant of the corresponding K\"ahler metric $g_E$
on the total space of the canonical bundle  is
\begin{equation}\begin{array}{rcl}\label{detGE}
 \mathrm{det}{g_E} & = &   2 \varpi  \Sigma (\lambda ) e^{\mathcal{P}(\varpi )} \Sigma '(\lambda ) \left(\varpi
   \mathcal{K}_0''(\varpi ) \mathcal{P}'(\varpi )+\mathcal{K}_0'(\varpi ) \left(\varpi
   \mathcal{P}''(\varpi )+2 \mathcal{P}'(\varpi )\right)\right)  \\
   &&+2 \varpi
   e^{\mathcal{P}(\varpi )} \Sigma '(\lambda ) \mathcal{K}_0'(\varpi ) \left(\mathcal{K}_0'(\varpi
   )+\varpi  \mathcal{K}_0''(\varpi )\right)+2 \varpi  \Sigma (\lambda )^2 e^{\mathcal{P}(\varpi
   )} \Sigma '(\lambda ) \mathcal{P}'(\varpi ) \left(\varpi  \mathcal{P}''(\varpi
   )+\mathcal{P}'(\varpi )\right) \\
\end{array}\end{equation}
where  we   set
\begin{equation}\label{sigmus}
    \Sigma(\lambda) =\lambda \, U'(\lambda)
\end{equation}
If we impose the Ricci-flatness condition, namely, that the
determinant of the metric $g_E$ is  a constant which we can always
assume to be one since any other number can be reabsorbed into the
normalization of the fibre coordinate $w$, by integration we get
\begin{equation}\label{lambdusingothic}
   \lambda =  \frac{1}{48} \left(A \, \mathfrak{w}^3+2 B\,
   \mathfrak{w}^2+4\, F \, \mathfrak{w}\right)
\end{equation}
where we have set
\begin{equation}\begin{array}{rcl}  \label{coriandolo}
    \Sigma(\lambda)  &= &  2\, \mathfrak{w} \\
  A &=& 4 \varpi  e^{\mathcal{P}(\varpi )} \mathcal{P}'(\varpi ) \left[\varpi
   \mathcal{P}''(\varpi )+\mathcal{P}'(\varpi )\right]  \\
  B &=& 6 \varpi  e^{\mathcal{P}(\varpi )} \left[\varpi  \mathcal{K}_0''(\varpi ) \mathcal{P}'(\varpi
   )+\mathcal{K}_0'(\varpi ) \left(\varpi  \mathcal{P}''(\varpi )+2 \mathcal{P}'(\varpi
   )\right)\right] \\
  F &=& 12 \varpi  e^{\mathcal{P}(\varpi )} \mathcal{K}_0'(\varpi ) \left[\mathcal{K}_0'(\varpi )+\varpi
   \mathcal{K}_0''(\varpi )\right]
\end{array}\end{equation}
In our complex three dimensional case, setting
\begin{equation}\label{qvar}
    x_u = \log \mid u\mid, \quad x_v = \log \mid v\mid, \quad
    \quad  x_w = \log \mid w \mid,
\end{equation}
the corresponding three moments can be named with the corresponding
gothic letters, and we have
\begin{equation}\label{momentacci}
    \mathfrak{u} =
    \partial_{x_u}\mathcal{K}\left(\varpi,\lambda\right),
    \quad   \mathfrak{v}  =
    \partial_{x_v}\mathcal{K}\left(\varpi,\lambda\right),
        \quad  \mathfrak{w} =
    \partial_{x_w}\mathcal{K}\left(\varpi,\lambda\right).
\end{equation}
As the fibre coordinate $w$ appears only in the function
$U(\lambda)$ via the squared norm $\lambda$,  we have
\begin{equation}\label{vugoticona}
    \mathfrak{w} = 2 \, \lambda \, U'(\lambda) =
    \Sigma(\lambda)
\end{equation}
and this justifies the position \eqref{coriandolo}. At this point
the function $U(\lambda)$ can be easily determined by first
observing that, in view of eqn.~\eqref{lambdusingothic} we can also
set
\begin{equation}\label{toppolina}
    U(\lambda) = \mathrm{U}\left(\mathfrak{w}\right)
\end{equation}
and we can use the chain rule
\begin{equation}\label{catenaria}
   \partial_\mathfrak{w}\,\mathrm{U}\left(\mathfrak{w}\right)=
   \frac{\mathfrak{w} \lambda '(\mathfrak{w})}{2 \lambda
   (\mathfrak{w})} =  \frac{3 A \mathfrak{w}^2+4 B \mathfrak{w}+4 F}{2 A \mathfrak{w}^2+4 B \mathfrak{w}+8
   F}
\end{equation}
which by integration yields the universal function
\begin{equation}\label{Udiwgoth}
 \mathrm{U}(\mathfrak{w})=   -\frac{2 \sqrt{4\, A\, F-B^2} \, \arctan
 \,\left(\frac{A\,
   \mathfrak{w}+B}{\sqrt{4 \,A\, F-B^2}}\right)+B \,\log
   \left(A \, \mathfrak{w}^2\, +\, 2\, B \mathfrak{w}\, +\, 4\,
   F\right)-3 \, A \,\mathfrak{w}}{2 \,A}
\end{equation}
The function $U(\lambda)$ appearing in the K\"ahler potential can
  be obtained by substituting for the argument $\mathfrak{w}$ in
\eqref{Udiwgoth} the unique real root of the cubic equation
\eqref{lambdusingothic}, namely:
\begin{equation}\begin{array}{rcl}\label{fischietto}
    \mathfrak{w} & = &  \displaystyle \frac{\sqrt[3]{8 \sqrt{\left(162 A^2 \lambda +9 A B F-2 B^3\right)^2-4 \left(B^2-3 A
   F\right)^3}+1296 A^2 \lambda +72 A B F-16 B^3}}{3 \sqrt[3]{2} A} \\[12pt]
   && \displaystyle +\frac{4 \sqrt[3]{2}
   \left(B^2-3 A F\right)}{3 A \sqrt[3]{8 \sqrt{\left(162 A^2 \lambda +9 A B F-2
   B^3\right)^2-4 \left(B^2-3 A F\right)^3}+1296 A^2 \lambda +72 A B F-16 B^3}}-\frac{2
   B}{3 A} \\
\end{array}\end{equation}
\paragraph{Consistency conditions for the Calabi Ansatz.}
\label{consistentia} In order for the Calabi Ansatz to yield a
 solution of the Ricci-flatness condition it is
necessary that the universal function $\mathrm{U}(\mathfrak{w})$ in
eqn.~\eqref{Udiwgoth} should depend only on $\mathfrak{w}$, which
happens if and only if the coefficients $A,B,F$ are constant. In the
case under consideration, where the invariant combination of the
complex coordinate $u,v$ is the one provided by $\varpi$ as defined
in eqn. \eqref{lattosio}, imposing such a consistency condition
would require the solution of three ordinary differential equations
for two functions $\mathcal{P}(\varpi)$ and $\mathcal{K}_0(\varpi)$,
namely:
\begin{equation}\begin{array}{rcl}\label{equinate}
  k_1 &=& 4 \varpi  e^{\mathcal{P}(\varpi )} \mathcal{P}'(\varpi ) \left[\varpi
   \mathcal{P}''(\varpi )+\mathcal{P}'(\varpi )\right]  \\
  k_2 &=& 6 \varpi  e^{\mathcal{P}(\varpi )} \left[\varpi  \mathcal{K}_0''(\varpi ) \mathcal{P}'(\varpi
   )+\mathcal{K}_0'(\varpi ) \left(\varpi  \mathcal{P}''(\varpi )+2 \mathcal{P}'(\varpi
   )\right)\right] \\
  k_3 &=& 12 \varpi  e^{\mathcal{P}(\varpi )} \mathcal{K}_0'(\varpi ) \left[\mathcal{K}_0'(\varpi )+\varpi
   \mathcal{K}_0''(\varpi )\right]
\end{array}\end{equation}
where $k_{1,2,3}$ are three constants. It is clear from their
structure that the crucial differential equation is the first one.
If we could find a solution for it then it would suffice to identify
the original K\"ahler potential $\mathcal{K}_0(\varpi )$ with a
linear function of $\mathcal{P}(\varpi )$ and we could solve the
three of them. So far we were not able to find any analytical
solution of these equations but if we could find one, we still
should verify that the K\"ahler metric following from such
$\mathcal{K}_0$ is a good metric on the Hirzebruch surface
$\mathbb{F}_2$.
\par
On the contrary for the well known K\"ahler potentials obtained from
the Kronheimer construction that define a one parameter family of
\textit{bona fide} K\"ahler metrics on $\mathbb{F}_2$ and were
discussed in \cite{Bruzzo:2017fwj,noietmarcovaldo}, namely those
presented whose K\"ahler potential is given in
eqn.~\eqref{varpifamiglia}, equations \eqref{equinate} cannot be
satisfied and no Ricci-flat metric on the canonical bundle can be
obtained by means of the Calabi Ansatz.
\paragraph{The general case with the natural fibre metric  $\mathcal{H}=\frac{1}{\mathrm{det}\left(g_\mathcal{M}\right)}$.}
\label{naturalla} If we consider the general case of a toric
two-dimensional compact manifold $\mathcal{M}$ with a K\"ahler
metric $g_\mathcal{M}$ derived from a K\"ahler potential of the
form:
\begin{equation}\label{taralluccio}
    \mathcal{K}_0 = \mathcal{K}_0\left(|u|^2 ,|v|^2\right)
\end{equation}
choosing as fibre metric the natural one for the canonical bundle,
namely setting:
\begin{equation}\label{carneade0}
    \lambda =\mathcal{H} \, |w|^2 \, =
    \,\frac{1}{\mathrm{det}\left(g_\mathcal{M}\right)} \, |w|^2
\end{equation}
and going through the same steps as in section \ref{F2Calab} we
arrive at an identical result for the function
$\mathrm{U}(\mathfrak{w})$ as in equation \eqref{Udiwgoth} but with
the following coefficients:
\begin{equation}\label{cumulone}
    A = 2 \,
    \frac{\mathrm{det}\left(\mathrm{Ric}_\mathcal{M}\right)}{\mathrm{det}\left(g_\mathcal{M}\right)}, \quad B = 3 \,
    \mathrm{Tr}\left(\mathrm{Ric}_\mathcal{M}\, g^{-1}_\mathcal{M}\right), \quad F =6
\end{equation}
It clearly appears why the Calabi Ansatz works perfectly if the
starting metric on the base manifold is KE. In that case the Ricci
tensor is proportional to the metric tensor:
\begin{equation}\label{KalEinst}
    \mathrm{Ric}_{i\bar\ell} = \kappa \, g_{i\bar\ell}
\end{equation}
and we get:
\begin{equation}\label{robilante}
    \mathrm{det}\left(\mathrm{Ric}_\mathcal{M}\right) = \kappa^2
    \, \mathrm{det}\left(g_\mathcal{M}\right) , \quad \mathrm{Tr}\left(\mathrm{Ric}_\mathcal{M}\,
    g^{-1}_\mathcal{M}\right) = 2 \, \kappa
\end{equation}
which implies:
\begin{equation}\label{ABFconst}
    A=2 \, \kappa^2 ,  \quad B = 6   \, \kappa, \quad
    F = 6.
\end{equation}
\subsection{The AMSY symplectic formulation for the Ricci-flat metric
on $\operatorname{tot}\left[K\left(\mathcal{M}_B\right)\right]$}
 According with the discussion
of the AMSY symplectic formalism presented in section \ref{amysone},
given the K\"ahler potential of a toric complex three manifold
$\mathcal{K}(|u|,|v|,|w|)$,   we can define the moments
\begin{equation}\label{momentini}
    \mathfrak{u} =
    \partial_{x_u}\mathcal{K},
    \quad   \mathfrak{v}  =
    \partial_{x_v}\mathcal{K},
    \quad  \mathfrak{w} =
    \partial_{x_w}\mathcal{K}
\end{equation}
and we can obtain the   symplectic potential by means of the
Legendre transform:
\begin{equation}\label{legendretr2}
    {G}\left(\mathfrak{u},\mathfrak{v} ,\mathfrak{w}\right)
    = x_u \,\mathfrak{u} \, + \, x_v \,\mathfrak{v}  + \, x_w
    \,\mathfrak{w}\,  - \, \mathcal{K}(|u|,|v|,|w|)
\end{equation}
The main issue   in the use of eqn.~\eqref{legendretr2} is the
inversion transformation that expresses the coordinates $x_i \, =
\,\{x_u,x_v,x_w\}$ in terms of the three moments $\mu^i\, =
\{\mathfrak{u},\mathfrak{v} ,\mathfrak{w}\}$. Once this is done one
can calculate the metric in moment variables utilizing the Hessian
as explained in section \ref{amysone}. Relying once again on the
results of that section we know that the K\"ahler 2-form has the
following universal structure:
\begin{equation}\label{uniKal3}
  \mathbb{K} \,= \,  \mathrm{d}\mathfrak{u}\wedge \mathrm{d}\phi \, + \, \mathrm{d}\mathfrak{v} \wedge \mathrm{d}\tau \,
  + \, \mathrm{d}\mathfrak{w} \wedge \mathrm{d}\chi
\end{equation}
and the metric is expressed as displayed in eqn.~\eqref{sympametra})
\paragraph{The symplectic potential in the case
with $\mathrm{SU(2)\times U(1) \times U(1)}$ isometries}
\label{varposympoloto} In the case where the K\"ahler potential has
the special structure which guarantees an $\mathrm{SU(2)}\times
\mathrm{U(1)}\times\mathrm{U(1)}$ isometry, namely it depends only
on the two variables $\varpi$ (see eqn.~\eqref{lattosio})) and $
|w|^2$, also the symplectic potential takes a more restricted form.
Indeed we find
\begin{equation}\begin{array}{rcl}\label{specG}
    G\left(\mathfrak{u},\mathfrak{v} ,\mathfrak{w}\right) & = &\underbrace{\left(\mathfrak{v} -\frac{\mathfrak{u}}{2}\right)
   \log (2 \mathfrak{v} -\mathfrak{u})+\frac{1}{2} \mathfrak{u} \log
   (\mathfrak{u})\, -\, \frac{1}{2} \,\mathfrak{v}  \log (\mathfrak{v} )}_{\text{universal part $G_0(\mathfrak{u},\mathfrak{v} )$}}\,
   + \,\underbrace{\mathcal{G}(\mathfrak{v} ,\mathfrak{w})}_{\text{variable
   part}}
\end{array}\end{equation}
where $\mathcal{G}(\mathfrak{v} ,\mathfrak{w})$ is a function of two
variables that encodes the specific structure of the metric. Note
that when we freeze the fibre moment coordinate $\mathfrak{w}$ to
some fixed constant value, for instance $0$, the function
$\mathcal{G}(\mathfrak{v} ,0)=\mathcal{D}(\mathfrak{v} )$ can be
identified with the boundary function that appears in
eqns.~\eqref{GBsymplectic},\eqref{lupetto}, namely in the symplectic
potential for the K\"ahler metric of the base manifold.
\par
With the specific structure \eqref{specG} of the symplectic
potential we obtain the following form for the Hessian
\eqref{hessiano}:
\begin{equation}\label{Gijspec}
    \mathbf{G} = \left(
\begin{array}{ccc}
 -\frac{\mathfrak{v} }{\mathfrak{u}^2-2 \mathfrak{u} \mathfrak{v} } &
   \frac{1}{\mathfrak{u}-2 \mathfrak{v} } & 0 \\
 \frac{1}{\mathfrak{u}-2 \mathfrak{v} } & \frac{-2 \mathfrak{v}
   (\mathfrak{u}-2 \mathfrak{v} )
   \mathcal{G}^{(2,0)}(\mathfrak{v} ,\mathfrak{w})+\mathfrak{u}+2
   \mathfrak{v} }{2 \mathfrak{v}  (2 \mathfrak{v} -\mathfrak{u})} &
   \mathcal{G}^{(1,1)}(\mathfrak{v} ,\mathfrak{w}) \\
 0 & \mathcal{G}^{(1,1)}(\mathfrak{v} ,\mathfrak{w}) &
   \mathcal{G}^{(0,2)}(\mathfrak{v} ,\mathfrak{w}) \\
\end{array}
\right)
\end{equation}
\section{The general form of the symplectic potential for the Ricci
flat metric on
{$\operatorname{tot}\left[K\left(\mathcal{M}^{KE}_B\right)\right]$}}
\label{formullageneralla} Having seen that KE metrics do indeed
exist  in  the form described in
eqs.\eqref{trottolina},\eqref{reducedhessian}, it is natural to
inquire how we can utilize the Calabi Ansatz  to write immediately
the symplectic potential for a Ricci-flat metric on the canonical
bundle of $\mathcal{M}_B^{KE}$ without going through the process of
inverting the Legendre transform. Namely,  we would like to make the
back and forth trip via inverse and direct Legendre transform only
once and in full generality rather than case by case. Our goal is
not only a simplification of the computational steps but it  also
involves a conceptual issue. Indeed, when we introduce intermediate
steps that rely on the variable $\varpi$ whose range is
$[0,+\infty)$ we necessarily have to choose a branch of a cubic
equation whose coefficients are determined by the root parameters
$\lambda_{1,2}$. On the contrary, if we are able to determine
directly the symplectic potential in terms of the symplectic
coordinates, then we can explore the behavior of the metric and of
its curvature on the full available range of variability of these
latter and we learn more on the algebraic and topological structure
of the underlying manifold.
\par
So let us anticipate the final result of our general procedure. As
we did in the previous section we assume that the Ricci form of
$\mathcal{M}_B$ is proportional to the K\"ahler form via a
coefficient
\begin{equation}\label{ricappo}
    \kappa = \frac{k}{4}
\end{equation}
as in eqs.\eqref{kallopietra},\eqref{KalEinst}. The complete
symplectic potential for the Ricci-flat metric on $\mathcal{M}_T =
\operatorname{tot}\left[K\left(\mathcal{M}^{KE}_B\right)\right]$ has
the following structure:
\begin{equation}\begin{array}{rcl}\label{pulcherrima}
  G_{\mathcal{M}_T^{KE}}\left(\mathfrak{u},\mathfrak{v} ,\mathfrak{w}\right) &=& G_0\left(\mathfrak{u},\mathfrak{v} \right)\, +\,
  \mathcal{G}^{KE}\left(\mathfrak{v} ,\mathfrak{w}\right)  \\
  G_0\left(\mathfrak{u},\mathfrak{v} \right) &= & \left(\mathfrak{v} -\frac{\mathfrak{u}}{2}\right) \log [2
   \mathfrak{v} -\mathfrak{u}]+\frac{1}{2} \mathfrak{u} \log
   [\mathfrak{u}]-\frac{1}{2} \mathfrak{v}  \log [\mathfrak{v} ]  \\
\mathcal{G}^{KE}\left(\mathfrak{v} ,\mathfrak{w}\right)
&=&\left(\frac{\kappa \mathfrak{w}}{2}+1\right)
\mathcal{D}^{KE}\left(\frac{2
   \mathfrak{v} }{\kappa  \mathfrak{w}+2}\right)-\frac{1}{2} \mathfrak{v}  \log
   \left(\frac{\kappa  \mathfrak{w}}{2}+1\right)+\frac{1}{2}
   \mathfrak{w} \log (\mathfrak{w}) \\
   &&\displaystyle+\frac{(\kappa  \mathfrak{w}+3) \log
   (\kappa  \mathfrak{w} (\kappa  \mathfrak{w}+6)+12)}{2 \kappa }
   +\frac{\sqrt{3} \arctan\left(\frac{\kappa  \mathfrak{w}+3}{\sqrt{3}}\right)}{\kappa }
  \\
\end{array}\end{equation}
where the second equation is a repetition for the reader's
convenience of  eqn. \eqref{lupetto} and
$\mathcal{D}^{KE}\left(\mathfrak{v} _0\right)$ is the boundary
function 
defined in equation \eqref{sympaKE}; the relation between the two
independent roots $\lambda_{1,2}$ and the parameter $\kappa$ ia
provided by equations \eqref{rinominopara},\eqref{ricappo}. The
reason while we have used the argument
\begin{equation}\label{v0defi}
   \mathfrak{v} _0= \frac{2
   \mathfrak{v} }{\kappa\mathfrak{w}+2}
\end{equation}
is that the symplectic variable $\mathfrak{v} _0$ associated with
the base-manifold metric and the symplectic variable $\mathfrak{v} $
associated with the metric on the canonical bundle
$\mathcal{M}^{KE}_T$ are not the same; their relation is precisely
that in eqn.~\eqref{v0defi} which is a direct consequence of the
Calabi Ansatz as we explain below.
\par
\paragraph{Derivation of the
formula for $\mathcal{G}^{KE}\left(\mathfrak{v}
,\mathfrak{w}\right)$. } The general formula \eqref{pulcherrima} is
a direct yield of the direct Legendre transform after  the Calabi
Ansatz:
\begin{equation}\label{GsymMT}
    G_{\mathcal{M}_T^{KE}}\left(\mathfrak{u},\mathfrak{v} ,\mathfrak{w}\right)
   =x_u \,\mathfrak{u} \, + x_v \, \mathfrak{v}  \, + \, x_w \,  \mathfrak{w} \, - \, \mathcal{K}_0(\mathfrak{v} _0) \, -
    \, U(\lambda)
\end{equation}
where
\begin{equation}\begin{array}{rcl}\label{rollandoforte}
    \lambda & = &  \frac{ w\, \bar{w}}{\text{
    det}g_{\mathcal{M}_B}}=  \text{const}\, \times \, w\, \bar{w}\, \, \exp\left[
     \kappa \, \mathcal{K}_0(\mathfrak{v} _0) \right] \,
    = \, \Lambda(\mathfrak{w})  =\frac{1}{24} \mathfrak{w} \left(\kappa^2
    \mathfrak{w}^2+6\kappa
    \mathfrak{w}+12\right) \\ \displaystyle
    \frac{\mathfrak{w}}{2} & = & \lambda \, U'(\lambda)  \\
   U(\lambda)& = & \mathbb{U}(\mathfrak{w})= \displaystyle \frac{-3 \log \left(2 \left(\kappa ^2 \mathfrak{w}^2+6 \kappa
   \mathfrak{w}+12\right)\right)+3 \kappa  \mathfrak{w}-2 \sqrt{3} \arctan\left(\frac{\kappa  \mathfrak{w}+3}{\sqrt{3}}\right)}{2 \kappa }
    \\
\mathcal{K}_0(\mathfrak{v} _0)&=&\mathfrak{v} _0
   \mathcal{D}'\left(\mathfrak{v} _0\right)-\mathcal{D}\left(\mathfrak{v} _0\right)+\frac
   {\mathfrak{v} _0}{2}
\end{array}\end{equation}
The last two lines in eqns.~\eqref{rollandoforte} were derived
earlier, respectively in eqns.~\eqref{Udiwgoth},\eqref{K0inv0}. The
explicit form of $\mathbb{U}(\mathfrak{w})$ follows from
eqn.~\eqref{Udiwgoth} using   the KE condition, namely
eqn.~\eqref{ABFconst}. From the above relations one easily obtains
the relations
\begin{equation}\begin{array}{rcl}
\label{torneodipallacorda} &&\mathfrak{u}_0 = \displaystyle \frac{2
\mathfrak{u}}{\mathit{k}
   \mathfrak{w}+2},  \quad \mathfrak{v} _0 = \frac{2
   \mathfrak{v} }{\mathit{k} \mathfrak{w}+2} \\[8pt]
&& x_w =\frac{1}{2}\left\{ \log\left[\Lambda(\mathfrak{w})\right] -
\kappa \, \mathcal{K}_0(\mathfrak{v} _0) \right\}
\end{array}\end{equation}
The first two relations can be understood as follows. The momenta
$\mathfrak{u}_0,\mathfrak{v} _0$ are, by definition
\begin{equation}\label{transeat}
    \mathfrak{u}_0= \partial_{x_u}\mathcal{K}_0 \quad ; \quad \mathfrak{v} _0= \partial_{x_vu}\mathcal{K}_0
\end{equation}
while we have
\begin{equation}\label{gloriamundi}
    \mathfrak{u}= \partial_{x_u}\mathcal{K} \quad ; \quad \mathfrak{v}  =
    \partial_{x_v}\mathcal{K}
\end{equation}
By the Calabi Ansatz we get:
\begin{equation}\label{giobattabis}
    \mathfrak{u}= \mathfrak{u}_0 +\partial_{x_u} U(\lambda) \,
    = \, \mathfrak{u}_0 + \partial_{xu}\lambda \, \partial_\lambda U(\lambda) =
    \mathfrak{u}_0 + \kappa \partial_{xu}\mathcal{K}_0 \,\lambda \partial_\lambda
    U(\lambda)\,= \,\mathfrak{u}_0\left( 1+ \frac{\kappa}{2} \,
    \mathfrak{w}\right)
\end{equation}
A completely analogous calculation can be done for the case of
$\mathfrak{v} $. Finally let us note that the coordinate $x_u,x_v$
were already resolved in terms of $u_0,v_0$ in
eqns.~\eqref{inversioneB}:
\begin{equation}\label{riducoinveB}
  x_u=  \frac{1}{2} \left(\log
   \left(\mathfrak{u}_0\right)-\log \left(2
   \mathfrak{v} _0-\mathfrak{u}_0\right)\right)\quad ; \quad x_v=
   \mathcal{D}'\left(\mathfrak{v} _0\right)+\log \left(2
   \mathfrak{v} _0-\mathfrak{u}_0\right)-\frac{1}{2} \log
   \left(\mathfrak{v} _0\right)+\frac{1}{2}
\end{equation}
 The information provided in the above equations \eqref{rollandoforte} - \eqref{riducoinveB} is
sufficient to complete the Legendre transform \eqref{GsymMT} and
retrieve the very simple and elegant result encoded in
eqn.~\eqref{pulcherrima}.
\par
To check the correctness of the general formula \eqref{pulcherrima}
we have explicitly calculated, by means of the {\sc mathematica}
package {\sc metricgrav}\footnote{{\sc metricgrav} just as {\sc
vielbgrav23} is a Mathematica package written by this author almost
thirty years ago and constantly updated. It will be available from
the site of the publisher De Gruyter to the readers of the
forthcoming book \textit{Discrete, Finite and Lie Groups}.}, the
Ricci tensor for a few cases of
$\mathcal{M}_T^{[\lambda_1,\lambda_2]}$, always finding zero.
\paragraph{The example of the metric $[2,1]$.}
Here we present the explicit form in symplectic coordinates of the
Ricci-flat metric on the canonical bundle of the KE manifold
$\mathcal{M}_B^{[1,2]}$, namely that determined by the choice:
$\lambda_1=1$, $\lambda_2=2$. We get:
\begin{equation}\begin{array}{rcl}
  ds^2_{M_T^{[1,2]}} &=& d\phi^2 \left(-\frac{\mathfrak{u}^2 (9 \mathfrak{w}+14)^2}{343
   \mathfrak{v} ^3}-\frac{16464 \mathfrak{u}^2}{(9 \mathfrak{w}+14)^4}+2
   \mathfrak{u}\right) \\[8pt]
   && \displaystyle+\frac{d\mathfrak{v} ^2 \left(\mathfrak{u} \left(2058
   \mathfrak{v} ^3+(9 \mathfrak{w}+14)^3\right)-686 \mathfrak{v} ^3 (9
   \mathfrak{w}+14)\right)}{\mathfrak{v}  (2 \mathfrak{v} -\mathfrak{u}) (7
   \mathfrak{v} -9 \mathfrak{w}-14) (14 \mathfrak{v} -9 \mathfrak{w}-14) (21
   \mathfrak{v} +9 \mathfrak{w}+14)} \\[8pt]
   && \displaystyle+2 d\tau d\phi \left(-\frac{\mathfrak{u} (9
   \mathfrak{w}+14)^2}{343 \mathfrak{v} ^2}-\frac{16464 \mathfrak{u} \mathfrak{v} }{(9
   \mathfrak{w}+14)^4}+\mathfrak{u}\right)+\frac{d\mathfrak{u}
   d\mathfrak{v} }{\mathfrak{u}-2 \mathfrak{v} }+\frac{d\mathfrak{u} (\mathfrak{u}
   d\mathfrak{v} -\mathfrak{v}  d\mathfrak{u})}{\mathfrak{u} (\mathfrak{u}-2
   \mathfrak{v} )} \\[8pt]
   && \displaystyle+\frac{36 \mathfrak{u} \mathfrak{w} \left(27 \mathfrak{w}^2+126
   \mathfrak{w}+196\right) d\chi d\phi}{(9 \mathfrak{w}+14)^3}+d\tau^2
   \left(-\frac{16464 \mathfrak{v} ^2}{(9 \mathfrak{w}+14)^4}-\frac{(9
   \mathfrak{w}+14)^2}{343 \mathfrak{v} }+\mathfrak{v} \right) \\[8pt]
   && \displaystyle+\frac{6174 \mathfrak{v} ^2
   d\mathfrak{v}  d\mathfrak{w}}{(7 \mathfrak{v} -9 \mathfrak{w}-14) (14
   \mathfrak{v} -9 \mathfrak{w}-14) (21 \mathfrak{v} +9
   \mathfrak{w}+14)} \\[8pt]
   && \displaystyle+\frac{d\mathfrak{w}^2 \left(5647152 \mathfrak{v} ^3-343
   \mathfrak{v} ^2 (9 \mathfrak{w}+14)^4+(9 \mathfrak{w}+14)^6\right)}{2 \mathfrak{w} (9
   \mathfrak{w}+14) \left(27 \mathfrak{w}^2+126 \mathfrak{w}+196\right) (7
   \mathfrak{v} -9 \mathfrak{w}-14) (14 \mathfrak{v} -9 \mathfrak{w}-14) (21
   \mathfrak{v} +9 \mathfrak{w}+14)} \\[8pt]
   && \displaystyle+\frac{36 \mathfrak{v}  \mathfrak{w} \left(27
   \mathfrak{w}^2+126 \mathfrak{w}+196\right) d\tau d\chi}{(9
   \mathfrak{w}+14)^3}+\frac{2 \mathfrak{w} \left(27 \mathfrak{w}^2+126
   \mathfrak{w}+196\right) d\chi^2}{(9 \mathfrak{w}+14)^2}
\end{array}\end{equation}
\par
\section{The Ricci-flat metric on $\operatorname{tot}\left[K\left(\mathcal{M}^{KE}_B\right)\right]$ versus the
metric cone on the Sasakian fibrations on
$\mathcal{M}^{KE}_B$}\label{sasacco} The present section  is one of
the most relevant ones in these lectures since it clarifies an
unexpected distinction that opens new directions of investigations.
To develop our argument it is appropriate to begin by reformulating
the geometry of the Ricci-flat metric derived from the Calabi ansatz
in terms of vielbeins. In that language the comparison with the
Sasaki-Einstein metrics of \cite{Gauntlett:2004yd} becomes much more
transparent.
\par
With some straightforward yet cumbersome algebraic analysis one can
verify that, after the change of variables \eqref{cambiovariabile},
the Ricci-flat metric in action-angle coordinates coming  from the
symplectic potential $G_{\mathcal{M}^{KE}_T}$, as displayed in
\eqref{pulcherrima}, can be rewritten as a sum of squares in terms
of a set of six vielbein one-forms $V^i$:
\begin{equation}\label{sommaquadra}
    ds^2_{\mathcal{M}^{KE}_T} \, = \, \sum_{i=1}^6 \, \left(V^i\right)^2 \,
    \, = \, \underbrace{\delta_{ij} \, V^i_\mu \, V^j_\nu \,}_{\text{metric }\mathbf{g}_{\mu\nu}} dy^\mu dy^\nu \quad
    ; \quad \underbrace{y^\mu \, = \,\{\theta ,\mathfrak{v},\mathfrak{w},\phi ,\tau ,\chi
    \}}_{\text{coordinates}}
\end{equation}
The explicit general structure of the sechsbein $V^i$, whose matrix
of components $V^i_\mu$ must be invertible and reproduces the metric
\eqref{sommaquadra} is the following one:
\begin{align}
\label{seigambe}
  V^1 & = \sqrt{\mathfrak{v}}\, d\theta \nonumber\\
  V^2 & = \sqrt{\mathfrak{v}} \, d\phi \, \sin (\theta ) \nonumber \\
  V^3 & = \frac{d\mathfrak{v}}{\mathfrak{A}(\mathfrak{v},\mathfrak{w})} \nonumber \\
  V^4 & = \mathfrak{B}(\mathfrak{v},\mathfrak{w}) \left[ d\mathfrak{w}\, + \,
   \mathfrak{C}(\mathfrak{v},\mathfrak{w})\, d\mathfrak{v} \right] \nonumber \\
  V^5 & = \mathfrak{D}(\mathfrak{v},\mathfrak{w})\, \left[d\tau \, + \,  \left(1-\cos (\theta )\right)\, d\phi \right]\nonumber \\
  V^6 & = \mathfrak{E}(\mathfrak{v},\mathfrak{w})\, \left[d\chi \, + \,\mho (\mathfrak{v},\mathfrak{w})
  \left[\,d\tau\, + \,
    \left(1-\cos (\theta )\right)\,d\phi\,
    \right]\right]
\end{align}
For the metric given by the symplectic potential in
\eqref{pulcherrima} the six functions
$\mathfrak{A}(\mathfrak{v},\mathfrak{w})$,
$\mathfrak{B}(\mathfrak{v},\mathfrak{w})$,
$\mathfrak{C}(\mathfrak{v},\mathfrak{w})$,
$\mathfrak{D}(\mathfrak{v},\mathfrak{w})$,
$\mathfrak{E}(\mathfrak{v},\mathfrak{w})$,
$\mho(\mathfrak{v},\mathfrak{w})$ are explicitly given below, where,
in order to make formulae more easily readable, we have renamed
$\lambda_1 =\alpha$, $\lambda_2 = \beta$ : {\fontsize{6.4}{6.5}
\begin{align}
\label{ciamabuti} \mathfrak{A}(\mathfrak{v},\mathfrak{w}) &=-\frac{i
\sqrt{64 \mathfrak{v}^3 (\alpha +\beta ) \left(\alpha ^2+\alpha
\beta
   +\beta ^2\right)^6-4 \mathfrak{v}^2 \left(\alpha ^2+\alpha  \beta +\beta
   ^2\right)^3 \left(2 \left(\alpha ^2+\alpha  \beta +\beta ^2\right)+3
   \mathfrak{w} (\alpha +\beta )\right)^4+\alpha ^2 \beta ^2 \left(2 \left(\alpha
   ^2+\alpha  \beta +\beta ^2\right)+3 \mathfrak{w} (\alpha +\beta )\right)^6}}{2
   \sqrt{\mathfrak{v}} \left(\alpha ^2+\alpha  \beta +\beta ^2\right)^{3/2}
   \left(2 \left(\alpha ^2+\alpha  \beta +\beta ^2\right)+3 \mathfrak{w} (\alpha
   +\beta )\right)^2}\nonumber\\
\mathfrak{B}(\mathfrak{v},\mathfrak{w}) &=
\frac{\sqrt{N_{\mathfrak{B}}(\mathfrak{v},\mathfrak{w})}}{\sqrt{D_{\mathfrak{B}}(\mathfrak{v},\mathfrak{w})}}\nonumber \\
N_{\mathfrak{B}}(\mathfrak{v},\mathfrak{w}) & = 64 \mathfrak{v}^3
(\alpha +\beta ) \left(\alpha ^2+\alpha  \beta +\beta
   ^2\right)^6-4 \mathfrak{v}^2 \left(\alpha ^2+\alpha  \beta +\beta ^2\right)^3
   \left(2 \left(\alpha ^2+\alpha  \beta +\beta ^2\right)+3 \mathfrak{w} (\alpha
   +\beta )\right)^4+\alpha ^2 \beta ^2 \left(2 \left(\alpha ^2+\alpha  \beta
   +\beta ^2\right)+3 \mathfrak{w} (\alpha +\beta )\right)^6\nonumber\\
D_{\mathfrak{B}}(\mathfrak{v},\mathfrak{w}) & =2 \mathfrak{w}
\left(8 \left(\alpha ^2+\alpha  \beta +\beta ^2\right)^3+9
   \mathfrak{w}^3 (\alpha +\beta )^3+24 \mathfrak{w}^2 \left(\alpha ^2+\alpha
   \beta +\beta ^2\right) (\alpha +\beta )^2+24 \mathfrak{w} \left(\alpha
   ^2+\alpha  \beta +\beta ^2\right)^2 (\alpha +\beta )\right)\times \nonumber\\
   &\quad \times \left(8
   \mathfrak{v}^3 (\alpha +\beta ) \left(\alpha ^2+\alpha  \beta +\beta
   ^2\right)^3-4 \mathfrak{v}^2 \left(\alpha ^2+\alpha  \beta +\beta ^2\right)^3
   \left(2 \left(\alpha ^2+\alpha  \beta +\beta ^2\right)+3 \mathfrak{w} (\alpha
   +\beta )\right)+ \right. \nonumber\\
   &\left. \quad +\alpha ^2 \beta ^2 \left(2 \left(\alpha ^2+\alpha  \beta
   +\beta ^2\right)+3 \mathfrak{w} (\alpha +\beta
   )\right)^3\right)\nonumber\\
\mathfrak{C}(\mathfrak{v},\mathfrak{w})   & = \frac{24
\mathfrak{v}^2 \mathfrak{w} (\alpha +\beta ) \left(\alpha ^2+\alpha
   \beta +\beta ^2\right)^3 \left(8 \left(\alpha ^2+\alpha  \beta +\beta
   ^2\right)^3+9 \mathfrak{w}^3 (\alpha +\beta )^3+24 \mathfrak{w}^2 \left(\alpha
   ^2+\alpha  \beta +\beta ^2\right) (\alpha +\beta )^2+24 \mathfrak{w}
   \left(\alpha ^2+\alpha  \beta +\beta ^2\right)^2 (\alpha +\beta )\right)}{64
   \mathfrak{v}^3 (\alpha +\beta ) \left(\alpha ^2+\alpha  \beta +\beta
   ^2\right)^6-4 \mathfrak{v}^2 \left(\alpha ^2+\alpha  \beta +\beta ^2\right)^3
   \left(2 \left(\alpha ^2+\alpha  \beta +\beta ^2\right)+3 \mathfrak{w} (\alpha
   +\beta )\right)^4+\alpha ^2 \beta ^2 \left(2 \left(\alpha ^2+\alpha  \beta
   +\beta ^2\right)+3 \mathfrak{w} (\alpha +\beta )\right)^6}\nonumber\\
\mathfrak{D}(\mathfrak{v},\mathfrak{w})   & =\frac{\sqrt{-8
\mathfrak{v}^3 (\alpha +\beta ) \left(\alpha ^2+\alpha  \beta
   +\beta ^2\right)^3+4 \mathfrak{v}^2 \left(\alpha ^2+\alpha  \beta +\beta
   ^2\right)^3 \left(2 \left(\alpha ^2+\alpha  \beta +\beta ^2\right)+3
   \mathfrak{w} (\alpha +\beta )\right)-\alpha ^2 \beta ^2 \left(2 \left(\alpha
   ^2+\alpha  \beta +\beta ^2\right)+3 \mathfrak{w} (\alpha +\beta )\right)^3}}{2
   \sqrt{\mathfrak{v} \left(\alpha ^2+\alpha  \beta +\beta ^2\right)^3 \left(2
   \left(\alpha ^2+\alpha  \beta +\beta ^2\right)+3 \mathfrak{w} (\alpha +\beta
   )\right)}}\nonumber\\
\mathfrak{E}(\mathfrak{v},\mathfrak{w})   & =\sqrt{2}
\sqrt{\frac{\mathfrak{w} \left(4 \left(\alpha ^2+\alpha  \beta
+\beta
   ^2\right)^2+3 \mathfrak{w}^2 (\alpha +\beta )^2+6 \mathfrak{w} \left(\alpha
   ^3+2 \alpha ^2 \beta +2 \alpha  \beta ^2+\beta ^3\right)\right)}{\left(2
   \left(\alpha ^2+\alpha  \beta +\beta ^2\right)+3 \mathfrak{w} (\alpha +\beta
   )\right)^2}}\nonumber\\
   \mho(\mathfrak{v},\mathfrak{w}) &= \Omega(\mathfrak{v},\mathfrak{w}) \, \equiv \,\frac{3 \mathfrak{v}
    (\alpha +\beta )}{2 \left(\alpha ^2+\alpha  \beta +\beta
   ^2\right)+3 \mathfrak{w} (\alpha +\beta )}
\end{align}
}
\paragraph{Properties of the sechsbein and comparison with Sasakian 5-manifolds.}
The general structure of the \textit{sechsbein} in \eqref{seigambe}
and \eqref{ciamabuti} is very interesting since it  highlights   the
double fibration structure of the underlying manifold
$\mathcal{M}_{T}$ which is a line-bundle (the canonical one) on the
base manifold $\mathcal{M}_B$ which, in its turn, is a (singular)
$\mathbb{P}^1$ fibration over a base $\mathbb{P}^1$:
\begin{equation}\label{doppiafibbra}
    \mathcal{M}_{T} \, \stackrel{\pi_1}{\longrightarrow} \, \mathcal{M}_{B} \,
    \stackrel{\pi_2}{\longrightarrow} \, \mathbb{P}^1
\end{equation}
The projection onto the base manifold is produced by considering the
limit $\mathfrak{w}\to 0$. Very much informative is the development
in series of the sechsbein for small values of the coordinate
$\mathfrak{w}$. The limit is regular for $\mathfrak{w}=0$; two of
the sechsbein $(V^4,V^6)$ vanish and the remaining four 1-forms
$V^1,V^2,V^3,V^5$ attain the values
$\mathbf{e}^3,\mathbf{e}^4,\mathbf{e}^1,\mathbf{e}^2$ corresponding
to the vierbein of the K\"ahler-Einstein 4-dimensional metrics as
given in eqn. \eqref{vierbeine} with the function
$\mathcal{FK}^{KE}(\mathfrak{v})$ as given in table \ref{casoni}. At
order $\sqrt{\mathfrak{w}}$ there is no deformation of the base
manifold vierbein, but the two vielbein corresponding to the fibre
directions do appear. At order $\mathfrak{w}$ we see the beginning
of the deformation of the base-manifold vierbein. Precisely we find:
\begin{equation}\label{canna}
\begin{array}{rclll}
V^1 &=&\mathbf{e}^3&\null&\null\\
V^2 &=&\mathbf{e}^4&\null&\null\\
V^3 &=&\mathbf{e}^1&\null&+\mathfrak{w}\, \Delta\mathbf{e}^1 \, + \, \mathcal{O}(\mathfrak{w}^2)\\
V^4 &=&0&+\sqrt{\mathfrak{w}}\, \Delta\pmb{\Phi}_{\mathfrak{w}} +\mathcal{O}(\mathfrak{w}^{3/2}) &\null \\
V^5 &=&\mathbf{e}^2 & \null  &+\mathfrak{w}\, \Delta\mathbf{e}^2\,+ \, \mathcal{O}(\mathfrak{w}^2)\\
V^6 &=&0&+ \sqrt{\mathfrak{w}}\, \Delta\pmb{\Phi}_\chi  +\mathcal{O}(\mathfrak{w}^{3/2})\,  & \null\\
\end{array}
\end{equation}
where the deformations of the base manifold vielbein are as it
follows:
\begin{align}\label{cantovello}
\Delta\mathbf{e}^1 &= -\frac{3 (\alpha +\beta )
\sqrt{\frac{\mathfrak{v}}{\alpha
   ^2+\alpha  \beta +\beta ^2}} \left(\alpha ^2 \beta ^2-2 \mathfrak{v}^3 (\alpha
   +\beta )\right)}{2 ((\mathfrak{v}-\alpha ) (\beta -\mathfrak{v}) (\alpha
   \beta +\mathfrak{v} (\alpha +\beta )))^{3/2}}\,  d\mathfrak{v}\\
\Delta\mathbf{e}^2 &=\frac{3 (\alpha +\beta ) \left(\mathfrak{v}^3
(\alpha +\beta )-2 \alpha ^2 \beta
   ^2\right) }{4 \left(\alpha
   ^2+\alpha  \beta +\beta ^2\right)^{3/2} \sqrt{\mathfrak{v}
   (\mathfrak{v}-\alpha ) (\beta -\mathfrak{v}) (\alpha  \beta +\mathfrak{v}
   (\alpha +\beta ))}} \, \left[d\tau+\left(1-  \cos\theta \right)\,d\phi \right]
\end{align}
and the initial fibre-vielbein are instead the following ones:
\begin{align}\label{canberra}
\Delta\pmb{\Phi}_{\mathfrak{w}} &= \frac{d\mathfrak{w}}{\sqrt{2}
\mathfrak{w}} \nonumber\\
\Delta\pmb{\Phi}_\chi  &=\sqrt{2} \left(d\chi
   +\frac{3 \mathfrak{v} (\alpha
+\beta ) }{2 \left(\alpha ^2+\alpha  \beta +\beta
^2\right)}\left[d\tau+\left(1-  \cos\theta \right)\,d\phi
\right]\right)
\end{align}
\paragraph{Comparison with the Sasaki-Einstein metrics.}
It is now the appropriate moment to consider the Sasaki-Einstein
metrics introduced in \cite{Gauntlett:2004yd}. In the coordinates
utilized by those authors we have:
\begin{align}
 ds^2_{SE_5} &= \frac{\text{dy}^2 (1-c y)}{2
\left(a+2 c y^3-3 y^2\right)}+\frac{\left(a+2 c
   y^3-3 y^2\right) (c \text{d$\phi $} \cos (\theta )+\text{d$\psi $})^2}{18
   (1-c y)}+\frac{1}{6} (1-c y) \left(\text{d$\theta $}^2+\text{d$\phi $}^2 \sin
   ^2\theta \right)\\
   &+\frac{\pmb{\Phi} _{\text{SE}}^2}{9}\label{5sasacchi}\\
\pmb{\Phi} _{\text{SE}} & =\left[d\xi\,+\,y \left(d\psi +c d\phi
\cos \theta +d\psi\right)-d\phi  \cos \theta \right]
\end{align}
If in equation \eqref{5sasacchi} we apply the following coordinate
transformation and we raname the  parameters as follows:
\begin{equation}\label{overrule}
   a\,\to \, \frac{1}{4} \left(3 \beta  k^2+4\right),\quad c\, \to \, 1,\quad y\to 1-\frac{k
   \mathfrak{v}}{2},\quad \psi \, \to \, -\tau -\phi
\end{equation}
we find the following interesting result:
\begin{align}
    ds^2_{SE_5} & = \, \frac{k}{12}\,\widehat{ds}^2_{5} \label{omero}\\
    \widehat{ds}^2_{5}&= {ds}^2_{KE_4}+\frac{4}{3 k}\, \pmb{\Phi}_{\text{SE}}^2\label{virgilio}\\
{ds}^2_{KE_4} & =
\frac{d\mathfrak{v}^2}{\mathcal{F}\mathcal{K}_{\text{KE}}(\mathfrak{v})}\,
+ \, \mathcal{F}\mathcal{K}_{\text{KE}}(\mathfrak{v}) \left[d\tau
+\left(1 -\cos \theta \right)\,d\phi\right]^2 \, +\,
\mathfrak{v} \left(d\phi^2 \sin ^2\theta +d\theta^2\right)\label{esiodo}\\[3pt]
\mathcal{K}_{\text{KE}}(\mathfrak{v})&= \frac{3 \beta -k
\mathfrak{v}^3+3 \mathfrak{v}^2}{3 \mathfrak{v}} \label{caramba}
\end{align}
As one sees the line element ${ds}^2_{KE_4}$ in \eqref{esiodo}
exactly coincides with the line element of the K\"ahler Einstein
metrics discussed in previous sections and presented in eqn.
\eqref{metrauniversala}. It remains to be seen what is the
appearance  of the 1-form $\pmb{\Phi} _{\text{SE}}$ after the
transformation \eqref{overrule}. If we add also the coordinate
transformation:
\begin{equation}\label{aggiunta}
    \xi \, \to \, p \,\chi \, +\,\tau \, +\, \phi \quad ; \quad p \in \mathbb{R}
\end{equation}
we find:
\begin{equation}\label{crinologo}
    \pmb{\Phi} _{\text{SE}}\, = \, p \, \left[d\chi\, +\,\frac{k}{2 p} \,
    \mathfrak{v}\, \left(d\tau + \left(1-\cos\theta \right)\,d\phi\right)\right]
\end{equation}
Comparing eqn.\eqref{crinologo} with \eqref{canberra} we see that we
obtain:
\begin{equation}\label{cagnolotto}
    \pmb{\Phi} _{\text{SE}} \, \ltimes \, \Delta\pmb{\Phi}_\chi
\end{equation}
provided we choose the constant $p$ as we do below:
\begin{equation}\label{troglodita}
    p \, = \, k \, \frac{\alpha^{2}+\alpha\,\beta+\beta^{2}}{3 \,
    (\alpha \, + \, \beta)} \, = \, 1
\end{equation}
Hence, using eqn.s
\eqref{overrule},\eqref{aggiunta},\eqref{troglodita},\eqref{rinominopara}
we can conclude that the Sasaki-Einstein metric of
\cite{Gauntlett:2004yd} with the choice of the parameters $a,c$
provided in \eqref{overrule} is proportional through the constant
$\frac{k}{12}$ to the following five dimensional metric:
\begin{equation}\label{culatello}
    \widehat{ds}^2_{\Upsilon^{SE}_5} \, = \,\underbrace{{ds}^2_{KE_4}}_{\text{KE metric on $\mathcal{M}_B$}} \, + \,\frac{4}{3\, k}\,
    \,\underbrace{\left(d\chi \, + \,\Omega(\mathfrak{v},0)
  \left[\,d\tau\, + \,
    \left(1-\cos (\theta )\right)\,d\phi\,
    \right]\right)^2}_{\lim_{\mathfrak{w}\to
    0}\frac{V^6}{\sqrt{2\mathfrak{w}}}} \, \equiv \,
    \widehat{g}^{[5]}_{ij}\, d\mathbf{y}^i \times d\mathbf{y}^j
\end{equation}
As it was explicitly checked in \cite{bruzzo2023d3brane}  the metric
in eqn.\eqref{culatello} is an Einstein metric, since its Ricci
tensor in intrinsic components takes the following form:
\begin{equation}\label{krumiro}
    \mathcal{R}ic[\widehat{g}^{[5]}] \, = \, \frac{k}{6} \, \delta_{ij}
\end{equation}
Hence the 5-dimensional variety $\Upsilon^{SE}_5$ implicitly defined
by eqn.\eqref{culatello} is a Sasaki-Einstein manifold.  Indeed it
is a circle bundle on the K\"ahler Einstein manifold
$\mathcal{M}^{KE}_B$:
\begin{equation}\label{sasaccoimperiale}
    \Upsilon^{SE}_5 \, \stackrel{\pi}{\longrightarrow}
    \,\mathcal{M}^{KE}_B \quad ; \quad \forall p \in
    \mathcal{M}^{KE}_B \quad \pi^{-1}\left(p\right) \, \sim \, \mathbb{S}^1
\end{equation}
On the other hand,  the standard metric of the metric cone on an
Einstein space is certainly Ricci-flat. Hence writing a new
sechsbein:
\begin{equation}\label{Ebeina}
    \mathbb{E}_{cone} \, = \, \{\mathbf{E}^1\, \dots, \mathbf{E}^6\}
\end{equation}
where:
\begin{align}\label{ruminia}
\mathbf{E}^1 & = \,R\, \mathbf{e}_3 &; \quad &
\mathbf{E}^2 & = \,R\, \mathbf{e}_4 \nonumber\\
\mathbf{E}^3 & = \,R\, \mathbf{e}_1 &;\quad &
\mathbf{E}^4 & = \,R\, \mathbf{e}_2 \nonumber\\
\mathbf{E}^5 & = \,R\, \sqrt{\frac{4}{3\, k}}\,
    \,\left(d\chi \, + \,\frac{k}{2} \mathfrak{v}\,
  \left[\,d\tau\, + \,
    \left(1-\cos (\theta )\right)\,d\phi\,
    \right]\right) &;\quad &
\mathbf{E}^6 & =  2\sqrt{\frac{3}{k}} dR
\end{align}
we obtain the Ricci flat metric on the metric cone
$\mathcal{C}\left(\Upsilon^{SE}_5\right)$ by setting:
\begin{equation}\label{cortemilia}
    ds^2_{\mathcal{C}\left(\Upsilon^{SE}_5\right)} \, = \,
    \sum_{I=1}^6 \left(\mathbf{E}^I\right)^2
\end{equation}

  As reported
in \cite{bruzzo2023d3brane} with the help of the Mathematica Code
{\sc Vielbgrav23} the curvature 2-form $\mathfrak{R}^{ab}_{con}$ and
the intrinsic components  of the Riemann tensor
$\mathcal{R}ie_{cone}$ were for the metric provided by the sechsbein
\eqref{ruminia}. As the metric is Ricci-flat, the Riemann tensor
coincides with the Weyl tensor $\mathcal{W}_{cone}(R,\mathfrak{v})$.
A similarly calculation was performed for the Ricci-flat metric
constructed with the Calabi ansatz, using the sechsbein defined in
\eqref{seigambe} with the functions displayed in \eqref{ciamabuti}.
In this way one obtains the curvature 2-form
$\mathfrak{R}^{ab}_{CA}$ and the intrinsic components of the Weyl
tensor associated with the Calabi Ansatz metric
$\mathcal{W}_{CA}(\mathfrak{w},\mathfrak{v})$. In order to make the
comparison more precise the sechsbein \eqref{seigambe} was reordered
in a similar way to the ordering utilized in the metric cone case:
\begin{equation}\label{CAbeina}
    \mathbb{E}_{CA} \, = \, \{V^1,V^2,V^3,V^5,V^6,V^4\}
\end{equation}
The two metrics exactly coincide on the base-manifold
$\mathcal{M}_B$ which is one of the KE manifolds with two conical
singularities discussed at length in the previous sections of the
present lectures and they are both Ricci-flat in 6-dimensions, yet
from all what we said in the present section they seem to be
intrinsically different, since the Sasakian Einstein 5-dimensional
metric as described in \eqref{culatello} cannot be obtained by
fixing the fibre coordinate $\mathfrak{w}$ to some appropriate
constant value. Yet one might deem that there exists some clever
change of coordinates that can map one metric into the other. To
show that this is not the case one resorts to the calculation of
Weyl tensor invariants for the two metrics to compare them.
Furthermore in \cite{bruzzo2023d3brane} the polynomial 2-form
curvature invariants were also calculated, in particular the
following 6-forms:
\begin{equation}
\label{tricurvoni}
\begin{array}{rcl}
\mathrm{Ch} & = &\mathfrak{R}^{ab}\wedge \mathfrak{R}^{bc}\wedge
\mathfrak{R}^{ca} \,  = \,Tr\left(\mathfrak{R}\wedge
\mathfrak{R}\wedge \mathfrak{R}\right) \\
 \mathrm{E} & = &\mathfrak{R}^{ab}\wedge \mathfrak{R}^{cd}\wedge
\mathfrak{R}^{fg} \, \epsilon_{abcdfg} \,\\
\end{array}
\end{equation}
The considered Weyl invariants were instead the following ones:
\begin{equation}
\label{invarianti}
\begin{array}{|rcl|rcl|}
\hline Quad_1 & = & \mathcal{W}[abij]\mathcal{W}[ijab]\\Cub_1
&=&\mathcal{W}[abij]\mathcal{W}[ijpq]\mathcal{W}[pqab]\\
Cub_2 &= &\mathcal{W}[ijpq]\mathcal{W}[rpsq]\mathcal{W}[rsij]\\
Cub_3
&=&\mathcal{W}[ijpq]\mathcal{W}[rpsq]\mathcal{W}[risj]\\
Quart_1
&=&\mathcal{W}[abij]\mathcal{W}[ijpq]\mathcal{W}[pqmn]\mathcal{W}[mnab]\\Quart_2
&=&\mathcal{W}[i, j, p, q]\mathcal{W}[p, r, q, s]\mathcal{W}[r, m,
s, n]\mathcal{W}[m, n, i, j]\\
\hline
\end{array}
\end{equation}
The result of the calculations with one special choice of
$\alpha,\beta$ is displayed in table \ref{poltroni}.
\begin{table}[htb!]
\begin{equation}
\begin{array}{|c|l|l|}
\hline \text{Inv} &\parbox{3cm}{\small \ \hfill\\ metric cone \\
over the Sasaki\\ manifold \\ \ \hfill}&
  \text{Calabi Ansatz for the Ricci-flat metric }\\
\hline
\mathrm{Ch}&0&0\\
\hline \mathrm{E}& 0 & \frac{20736 \left(63780651422208
\mathfrak{v}^6+(9
   \mathfrak{w}+14)^{12}\right)}{343 \mathfrak{v}^6 (9 \mathfrak{w}+14)^{12}} \times \text{Vol}
   \\
\hline Quad_1 & \frac{96}{49 R^4 \mathfrak{v}^6} & \frac{6 (9
\mathfrak{w}+14)^4}{117649 \mathfrak{v}^6}+\frac{8131898880}{(9
   \mathfrak{w}+14)^8}\\
   \hline
   Cub_1 & \frac{384}{343 R^6 \mathfrak{v}^9} & \frac{6 (9 \mathfrak{w}+14)^6}{40353607 \mathfrak{v}^9}+\frac{432}{343
   \mathfrak{v}^6}+\frac{348097316216832}{(9
   \mathfrak{w}+14)^{12}}\\
\hline Cub_2 & \frac{192}{343 R^6 \mathfrak{v}^9}& \frac{3 (9
\mathfrak{w}+14)^6}{40353607 \mathfrak{v}^9}+\frac{216}{343
   \mathfrak{v}^6}+\frac{174048658108416}{(9
   \mathfrak{w}+14)^{12}}\\
\hline Cub_3 &\frac{96}{343 R^6 \mathfrak{v}^9}& \frac{3 (9
\mathfrak{w}+14)^6}{80707214 \mathfrak{v}^9}+\frac{108}{343
   \mathfrak{v}^6}+\frac{87024329054208}{(9 \mathfrak{w}+14)^{12}} \\
   \hline
Quart_1 &\frac{4608}{2401 R^8 \mathfrak{v}^{12}}&\frac{18 (9
\mathfrak{w}+14)^8}{13841287201 \mathfrak{v}^{12}}+\frac{576 (9
   \mathfrak{w}+14)^2}{117649 \mathfrak{v}^9}+\frac{20736}{\mathfrak{v}^6 (9
   \mathfrak{w}+14)^4}+\frac{31631121143724146688}{(9
   \mathfrak{w}+14)^{16}}\\
   Quar_2 &\frac{1152}{2401 R^8 \mathfrak{v}^{12}}&\frac{9 (9 \mathfrak{w}+14)^8}{27682574402 \mathfrak{v}^{12}}+\frac{144 (9
   \mathfrak{w}+14)^2}{117649 \mathfrak{v}^9}+\frac{5184}{\mathfrak{v}^6 (9
   \mathfrak{w}+14)^4}+\frac{7907780285931036672}{(9
   \mathfrak{w}+14)^{16}}\\ \ \hfill  & \ \hfill & \ \hfill \\
   \hline
\end{array}
\end{equation}
\caption{\label{poltroni} Table of the polynomial invariants defined
in \eqref{tricurvoni} and \eqref{invarianti} evaluated in the case
of the metric cone over the Sasaki-Einstein manifold  and in the
Calabi Ansatz case of Ricci-flat metrics. Since the calculations are
very long when the parameters are left symbolic the evaluation of
the invariants was done in the reference case $\alpha=1,\beta=2$.}
\end{table}
Inspecting table \ref{poltroni}   one realizes that the Ricci-flat
metric constructed by means of the Calabi Ansatz translated into
action-angle variables is {different} from that associated with the
Sasaki-Einstein metric recalled in \eqref{5sasacchi}. The strongest
evidence is given by vanishing of the invariant $E$ in one case and
the non-zero and coordinate dependent structure of the same
invariant in the second case. However also the other invariants
corroborate the same evidence. One can calculate the value of the
coordinate $R$ in terms of $\mathfrak{v},\mathfrak{w}$ by equating
the quadratic invariant $Quad_1$ of the two metrics. Substituting
the result in the other invariants the two expressions do not agree
for the other invariants. This being clarified one can try to
understand the reason of the disagreement. In the Sasaki Einstein
approach \eqref{virgilio} corresponds to the standard construction
of the metric on a $\mathrm{U(1)}$ bundle. To the metric of the base
manifold, in the present case $\mathcal{M}_B^{KE}$, one adds the
square of a 1-form $\Phi$ of the type $\Phi \, = \,d\tau\, +
\,\text{$\mathrm{U(1)}$-connection on $\mathcal{M}_B$}$. The
remaining coordinate is the radial one $R$. In the Calabi ansatz
approach, instead, one directly constructs the line bundle on the
base manifold.
\section{The harmonic function and final remarks on the KE case}
As it appears in the most evident way from table \ref{poltroni}, the
Ricci flat metric produced on
$\operatorname{tot}\left[K\left(\mathcal{M}^{KE}_{\alpha\beta}\right)\right]$
by the Calabi ansatz and that of the metric cone on the Sasakian
Einstein Manifold defined by eqn. \eqref{ruminia},\eqref{cortemilia}
are essentially different and it is not even clear if they insist on
the same six dimensional manifold, notwithstanding the fact that the
5-dimensional Sasaki Einstein manifold $\Upsilon^{SE}_5$ and
$K\left(\mathcal{M}^{KE}_{\alpha\beta}\right)$ share the same
base-manifold $\mathcal{M}_B^{KE}$ as it appears from the two
fibrations \eqref{doppiafibbra} and \eqref{sasaccoimperiale} that we
rewrite here below for comparison:
\begin{equation}\label{carlandia}
    \begin{array}{rclclccccc}
       \operatorname{tot}\left[K\left(\mathcal{M}^{KE}_B\right)\right] & \equiv & \mathcal{M}_T & \stackrel{\pi_1}{\longrightarrow}
        & \mathcal{M}^{KE}_B & ; & \forall p \in \mathcal{M}^{KE}_B & \pi_1^{-1} & \sim & \mathbb{C}\\
        \Updownarrow \text{?} & \null & \null & \null
        & \null & ; & \null & \null & \null & \null\\
     \mathcal{C}\left[\Upsilon^{SE}_5\right] & \stackrel{\iota}{\longleftarrow} &
     \Upsilon^{SE}_5 & \stackrel{\tilde{\pi}_1}{\longrightarrow}
        & \mathcal{M}^{KE}_B & ; & \forall p \in \mathcal{M}^{KE}_B & \tilde{\pi}_1^{-1} & \sim & \mathbb{S}^1\\
     \end{array}
\end{equation}
The first fibration in eqn. \eqref{carlandia} is a complex
line-bundle, so that the standard fibre is a copy of $\mathbb{C}$,
while the second fibration is a circle bundle so that the standard
fibre is a circle $\mathbb{S}^1$ whose local coordinate is the angle
$\chi$. The relation between
$\operatorname{tot}\left[K\left(\mathcal{M}^{KE}_B\right)\right]$
and $\mathcal{C}\left[\Upsilon^{SE}_5\right]$ would become clear if
the radial coordinate $R$ of the metric cone might be interpreted as
a monotonic function of $|w|^2$ where $w \in \mathbb{C}$ is the
local fibre coordinate on
$\operatorname{tot}\left[K\left(\mathcal{M}^{KE}_B\right)\right]$
according with the general prescriptions of table \ref{coordinates}
and eqn.\eqref{carlingo} and if the the metric \eqref{culatello} on
$\Upsilon^{SE}$ could be obtained in the limit $R\to 0$ within the
framework of the dimensional transmutation mechanism that involves
the harmonic function $H(R)$. In order to explain this crucial point
it is necessary to go back to the complete $10$-dimensional metric
as displayed in \eqref{ansazzo}. Let us adapt that general formula
to our considered case:
\begin{eqnarray}
\label{decumano}
 \text{ds}^2_{[10]}&=&H\left(R[\mathfrak{w}]\right)^{-\frac{1}{2}}\left
(-\eta_{\mu\nu}\,dx^\mu\otimes dx^\nu \right
)+H\left(R[\mathfrak{w}]\right)^{\frac{1}{2}} \, \left(
\text{ds}^2_{[6]}\right) \, \nonumber\\
 \text{ds}^2_{[6]}&=&\left\{\begin{array}{lcl}
                            \text{ds}^2_{\mathcal{M}^{KE}_T} & = & \sum_{i=1}^6 \left(V^i[\mathfrak{w}]\right)^2  \\
                            \text{ds}^2_{\mathcal{C}\left(\Upsilon ^{KE}_5\right)} & = & \sum_{I=1}^6 \left(\mathbf{E}^I[R]\right)^2
                          \end{array}\right.\nonumber\\
\eta_{\mu\nu}&=&{\rm diag}(+,-,-,-)
\end{eqnarray}
In six dimensions for manifolds that are asymptotically locally
euclidian, namely whose metric tends to the flat euclidian limit:
\begin{equation}\label{ALEssa}
    \text{ds}^2_{[6]}\, \stackrel{R \to \infty}{\longrightarrow} \,
    d R^2 +R^2 ds^2_{\mathcal{C}\left(\Upsilon^{SE}_5\right)}
\end{equation}
the harmonic function that respects all the isometries of the Sasaki
Einstein 5-manifold $\Upsilon^{SE}_5$ has the following general
form:
\begin{equation}\label{craniotriestino}
    H(R) \, = \, 1+\frac{1}{R^4}
\end{equation}
where the integration constant has been fixed in such a way that for
$R\to \infty$ the $10$-dimensional metric approaches the locally
Minkowskian limit:
\begin{equation}\label{maloneinfinito}
   \text{ds}^2_{[10]}\, \stackrel{R \to \infty}{\longrightarrow} \,-\eta_{\mu\nu}\,dx^\mu\otimes
   dx^\nu \, + \,  d R^2 +R^2 \text{ds}^2_{\mathcal{C}\left(\Upsilon^{SE}_5\right)}
\end{equation}
On the contrary in the opposite near brane limit $R\to 0$ one
obtains:
\begin{equation}\label{mutaforma}
     \text{ds}^2_{[10]}\, \stackrel{R \to \infty}{\longrightarrow}
     \, \underbrace{R^2 \left( -\eta_{\mu\nu}\,dx^\mu\otimes  dx^\nu\right) +
     \frac{dR^2}{R^2} }_{\text{AdS}_5 \text{ metric}}\, + \,
     \underbrace{\text{ds}^2_{\mathcal{C}\left(\Upsilon^{SE}_5\right)}}_{\text{ SE
     metric}}
\end{equation}
which is the well known dimensional transmutation mechanism giving
rise to the near--brane structure of $10$-dimensional space:
\begin{equation}\label{cavoliamari}
    \mathcal{M}_{[10]} \, \sim \, \text{AdS}_5 \times \Upsilon^{SE}_5
\end{equation}
Superficially it might seem that all what is needed in order to
match the above familiar dimensional transmutation picture are the
following three conditions:
\begin{description}
  \item[1)] The Harmonic function on
  $\operatorname{tot}\left[K\left(\mathcal{M}^{KE}_B\right)\right]$,
  calculated with respect to its Ricci flat metric
  should be a function $H(\mathfrak{w})$ only of the moment
  $\mathfrak{w}$.
  \item[2)] The radial variable
  \begin{equation}\label{equinodirazza}
    R[\mathfrak{w}] \, \equiv \, \frac{1}{\sqrt[4]{H(\mathfrak{w})-1}}
  \end{equation}
upon suitable choice of the integration constants in the solution of
Laplace equation for $H(\mathfrak{w})$ should be monotonic and map
the interval $[0,\infty]$ of $\mathfrak{w}$ into the same interval
for $[0,\infty]$ of $R$.
\item[3)] The limit $\mathfrak{w}\to 0$ should correspond to the
limit $w\to 0$ for the complex coordinate $w$ along the fibres of
the canonical bundle, so that the vanishing section of the canonical
bundle corresponds to the near-brane geometry.
\end{description}
As we are going to see in a moment the above three conditions can
all be satisfied, yet this is not sufficient because of the
following subtle point. In equation \eqref{decumano} we have
carefully emphasized that the vielbein $V^i(\mathfrak{w})$ of the
Ricci-flat metric derived from the Calabi Ansatz, depend on the
moment variable $\mathfrak{w}$ just as the vielbein
$\mathbf{E}^I(R)$ of the metric cone on the Sasakian
$\Upsilon^{SE}_5$ depend on the radial variable $R$. As for the
first case the dependence on $\mathfrak{w}$ is encoded in the
functions \eqref{ciamabuti} that partly appear as multiplicative
factors in front of the various $V^i(\mathfrak{w})$ partly inside
their inner structure (see \eqref{seigambe}). In the second case,
instead the dependence on $R$ of the vielbein $\mathbf{E}^I(R)$ is
just a multiplicative factor $R$ in front of five of them. In order
for the dimensional transmutational picture to work we need the
three above mentioned conditions, but we also need that in the limit
$\mathfrak{w} \to 0$ five of the six vielbein $V^i(\mathfrak{w})$
should approach as $R[\mathfrak{w}] \times \mathbf{e}^i$ those
$\mathbf{e}^i$ of an Einstein 5-dimensional metric, in particular
that of\cite{Gauntlett:2004yd}, while the sixth should approach
$\mathrm{d}R[\mathfrak{w}]$. As we already know this does not happen
and there is no Sasaki-Einstein $5$-manifold in the near-brane
limit.
\subsection{The harmonic function}
\begin{figure}
\centering
\includegraphics[width=12cm]{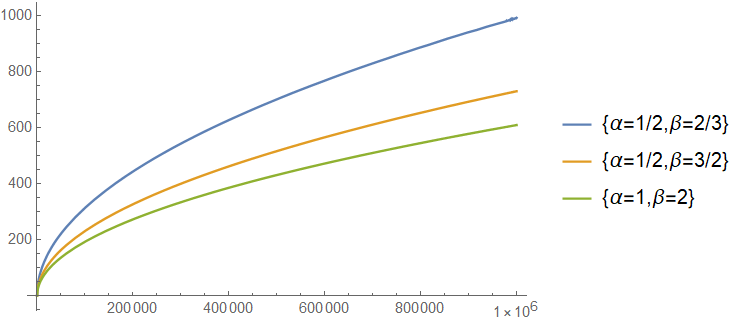}
\caption{\label{monotonale}   Illustration of the monotonic behavior
of the function $R[\mathfrak{w}]$.}
\end{figure}

In any metric, Laplace equation for harmonic functions can be
written as follows:
\begin{equation}\label{laplaccio}
    \triangle \, H(y) \, \equiv \frac{1}{\sqrt{\text{det} \,g}}
    \partial_I \,\left( \sqrt{\text{det} \,g} g^{IJ} \, \partial_J
    \,H(y)\right) \, = \,0
\end{equation}
When we utilize a K\"ahler toric metric in symplectic coordinates,
equation \eqref{laplaccio} simplify considerably because the
determinant of the metric is identically equal to $1$. Furthermore
if we assume that the harmonic function is invariant with respect to
all the toric $\mathrm{U(1)}$ symmetries $H\equiv H(\pmb{\mu})$
depends only on the momentum variables and it does not depend on the
angle variables $\Theta^i$. Because of the structure in eqn.
\eqref{sympametra} eqn. \eqref{laplaccio} becomes:
\begin{equation}\label{laplaccio2}
    \triangle \, H(y) \, \equiv
    \frac{\partial}{\partial \mu^i} \,\left( G^{-1}_{ij} \,  \frac{\partial}{\partial \mu^j}
    \,H(\pmb{\mu})\right) \, = \,0
\end{equation}
Furthermore if we apply eqn.\eqref{laplaccio2} to the case of the
Ricci-flat metric on $\mathcal{M}_T =
\operatorname{tot}\left[K\left(\mathcal{M}^{KE}_B\right)\right]$
defined by the symplectic potential \eqref{pulcherrima} and
furthermore we assume that $H(\pmb{\mu})$ depends only on the moment
$\mathfrak{w}$ we get the following ordinary differential equation:
\begin{align}\label{ommuz}
   & \mathfrak{w} H''(\mathfrak{w}) \left(4 \left(\alpha
   ^2+\alpha  \beta +\beta ^2\right)^2+3 \mathfrak{w}^2
   (\alpha +\beta )^2+6 \mathfrak{w} \left(\alpha ^3+2
   \alpha ^2 \beta +2 \alpha  \beta ^2+\beta
   ^3\right)\right)\nonumber\\
   &+H'(\mathfrak{w}) \left(2
   \left(\alpha ^2+\alpha  \beta +\beta ^2\right)+3
   \mathfrak{w} (\alpha +\beta )\right)^2 \, = \, 0
\end{align}
The general integral to the linear differential eqn.\eqref{ommuz} is
the following one:
\begin{align}\label{grullone}
    H(\mathfrak{w})& = \, c_1-\frac{c_2}{8
   \left(\alpha ^2+\alpha  \beta +\beta ^2\right)^2}\, \left(\log \left[4 \left(\alpha ^2+\alpha
   \beta +\beta ^2\right)^2+3 \mathfrak{w}^2 (\alpha
   +\beta )^2+6 \mathfrak{w} \left(\alpha ^2+\alpha
   \beta +\beta ^2\right) (\alpha +\beta )\right]\right.\nonumber\\
   &\left. +2
   \sqrt{3} \arctan\left[\frac{\sqrt{3} \left(\alpha
   ^2+\alpha  (\beta +\mathfrak{w})+\beta  (\beta
   +\mathfrak{w})\right)}{\alpha ^2+\alpha  \beta
   +\beta ^2}\right]-2 \log [\mathfrak{w}]\right)
\end{align}
where $h,k$ are the integration constants. The condition 2) in the
above list is satisfied fixing:
\begin{equation}\label{corteggioilmare}
    c_1\, = \,1 -\log \left(3 (\alpha +\beta )^2\right)-\sqrt{3} \pi  \quad ; \quad c_2 \, = \,
    -8 \left(\alpha ^2+\alpha  \beta +\beta ^2\right)^2
\end{equation}
In this way one obtains:
\begin{align}\label{corsetto}
R[\mathfrak{w}]\,= \, &  \left\{-\log \left[3 (\alpha +\beta
   )^2\right]+\log \left[4 \left(\alpha ^2+\alpha
   \beta +\beta ^2\right)^2+3 \mathfrak{w}^2 (\alpha
   +\beta )^2+6 \mathfrak{w} \left(\alpha ^2+\alpha
   \beta +\beta ^2\right) (\alpha +\beta )\right]\right.\nonumber\\
   & \left.+2
   \sqrt{3} \arctan\left[\frac{\sqrt{3} \left(\alpha
   ^2+\alpha  (\beta +\mathfrak{w})+\beta  (\beta
   +\mathfrak{w})\right)}{\alpha ^2+\alpha  \beta
   +\beta ^2}\right]-2 \log [\mathfrak{w}]-\sqrt{3} \pi
   \right\}^{-1/4}
\end{align}
Independently from the values of the parameters $\alpha,\beta$ we
can check that:
\begin{equation}\label{limitarsi}
    \lim_{\mathfrak{w} \to 0} R[\mathfrak{w}] \, = \, 0 \quad ;
    \quad \lim_{\mathfrak{w} \to \infty} R[\mathfrak{w}] \, = \,
    \infty
\end{equation}
Furthermore some plots of the function for different allowed values
of the parameters $\alpha,\beta$ shows the $R[\mathfrak{w}]$ is
monotonic (see fig.\ref{monotonale}).

\paragraph{Near brane behavior of the $10$-dimensional metric} In
order to appreciate the distance between the case of the metric cone
and that of the exact $D3$-solution obtained with the Ricci flat
metric generated by the Calabi-Ansatz, it is convenient to consider
the asymptotic near-brane behavior of the $10$-dimensional metric
for $\mathfrak{w}\to 0$ which means $R \to 0$. In the neighborhood
of $R\, =\, 0$ the relation between the momentum $\mathfrak{w}$ and
the radial variable becomes simple and clear. From eqn.
\eqref{corsetto} we find:
\begin{equation}\label{criceto}
    R[w] \, \stackrel{\mathfrak{w}\to 0}{\approx} \,\frac{1}{\sqrt[4]{-2
    \log[\mathfrak{w}]}} \quad ; \quad \mathfrak{w} \, \stackrel{R\to 0}{\approx}
    \,\exp\left[-\frac{1}{2\, R^4}\right]
\end{equation}
Correspondingly we have
\begin{align}\label{casinetto}
    H[\mathfrak{w}]^{-1/2} & \stackrel{\mathfrak{w}\to 0}{\approx}
    \, R^2 \nonumber\\
H[\mathfrak{w}]^{1/2} & \stackrel{\mathfrak{w}\to 0}{\approx}
    \, R^{-2}
\end{align}
so that we find:
\begin{align}\label{senzatopi}
  \text{ds}^2_{[10]} \,\stackrel{\mathfrak{w}\to 0}{\approx}\,& R^2 \left( -\eta_{\mu\nu}\,dx^\mu\otimes  dx^\nu\right)\,
  + \, \frac{2 e^{-\frac{1}{2 R^4}} }{R^{12}}\,\mathrm{d}R^2 \, \nonumber \\
& + \, \frac{1}{R^2} \text{ds}^2_{KE_4}\, +\, \frac{e^{-\frac{1}{2
R^4}}}{R^2} \left(\Delta\pmb{\Phi}_\chi\right)^2
\end{align}
where the initial fibre vielbein $\Delta\pmb{\Phi}_\chi$ was defined
in eqn.\eqref{canberra}. As one sees there is no dimensional
transmutation and no factorization into a product of spaces with
separate metrics. In particular the Einstein metric of
eqn.\eqref{culatello} does not emerge at all in the near-brane limit
since its two terms (the base-manifold metric and the circle-bundle
term) have completely different behavior in the radial variable $R$.
\par
In order to understand the rationale of this result I have
considered the question if any Sasaki Einstein $5$-dimensional
manifold is included in the Riemannian manifold
$\operatorname{tot}\left[K\left(\mathcal{M}^{KE}_B\right)\right]$
equipped with its own Ricci flat metric derived from the Calabi
Ansatz. Having established that the radial coordinate $R$ is a
monotonic function of the moment coordinate $\mathfrak{w}$ the
question about the existence of a Sasaki-Einstein $5$ dimensional
sub-manifold $\Upsilon_5^{SE}$ can be posed in the following way.
Taking the complete 6-dimensional metric defined by the symplectic
potential \eqref{pulcherrima} we reduce it to a $5$-dimensional
slice by setting $\mathrm{d}\mathfrak{w}\, = \, 0$ keeping the value
of $\mathfrak{w}=\text{const} $ as free parameter. In this way one
obtains a family of three-parameter ($\alpha,\beta,\mathfrak{w}$)
metrics of which one can calculate the intrinsic components of the
Riemann and Ricci tensors utilizing the Mathematica Code {\sc
Vielbgrav23}. The result of this calculation is very instructive.
The Ricci tensor in intrinsic components turns out to be diagonal
and we have: {\scriptsize
\begin{align}\label{buttafuoco}
    \mathrm{Ric}_{ii} & = \frac{3 (\alpha +\beta ) \left(4 \left(\alpha
   ^2+\alpha  \beta +\beta ^2\right)^3+9
   \mathfrak{w}^3 (\alpha +\beta )^3+18 \mathfrak{w}^2
   \left(\alpha ^2+\alpha  \beta +\beta ^2\right)
   (\alpha +\beta )^2+12 \mathfrak{w} \left(\alpha
   ^2+\alpha  \beta +\beta ^2\right)^2 (\alpha +\beta
   )\right)}{\left(2 \left(\alpha ^2+\alpha  \beta
   +\beta ^2\right)+3 \mathfrak{w} (\alpha +\beta
   )\right)^4} \quad i=1,2,3,4\nonumber\\
\mathrm{Ric}_{55} & = \frac{9 \mathfrak{w} (\alpha +\beta )^2
\left(4
   \left(\alpha ^2+\alpha  \beta +\beta ^2\right)^2+3
   \mathfrak{w}^2 (\alpha +\beta )^2+6 \mathfrak{w}
   \left(\alpha ^3+2 \alpha ^2 \beta +2 \alpha  \beta
   ^2+\beta ^3\right)\right)}{\left(2 \left(\alpha
   ^2+\alpha  \beta +\beta ^2\right)+3 \mathfrak{w}
   (\alpha +\beta )\right)^4}
\end{align}
} In order to have an Einstein metric we should have that the Ricci
tensor is proportional to the Kronecker delta $\delta_{ij}$; in
other words we should find a value of $\mathfrak{w}$, (finite, null
or infinite) for which the Ricci tensor gets all equal eigenvalues.
Calculating the difference between the two eigenvalues of eqn.
\eqref{buttafuoco} we find:
\begin{equation}\label{rintanato}
    \mathrm{Ric}_{11}-\mathrm{Ric}_{55} \, = \, \frac{12 (\alpha +\beta ) \left(\alpha ^2+\alpha
   \beta +\beta ^2\right)^3}{\left(2 \left(\alpha
   ^2+\alpha  \beta +\beta ^2\right)+3 \mathfrak{w}
   (\alpha +\beta )\right)^4}
\end{equation}
which vanishes only in the limit $\mathfrak{w}\to \infty$. This is
not very useful since the $5$-dimensional metric is asymptotically
flat  and in the limit $\mathfrak{w}\to \infty$ all components of
the Riemann tensor vanish. Indeed we have:
\begin{equation}\label{canaglia}
    \mathrm{Ric}_{ij} \, \stackrel{\mathfrak{w}\to \infty}{\approx}
    \, \frac{1}{3\,\mathfrak{w}} \, \delta_{ij} \, + \,
    \mathcal{O}(\mathfrak{w}^{-2})
\end{equation}
It follows that in the near-brane regime and at finite distance from
the brane no Sasaki-Einstein $5$-dimensional metric emerges from the
Ricci-flat one on the total space of the canonical bundle of the
K\"ahler Einstein manifolds we have been discussing.
\chapter{Open Questions}
\label{aperto} In this chapter I illustrate the several questions
and issues that remain open within the context of the Kronheimer
construction and the analysis of K\"ahler metrics  both on the base
manifold $\mathcal{M}_B$ and on the total space of the line bundles
advocated in eqn. \eqref{doppiafibbia}.
\section{Open questions about the base manifold $\mathcal{M}_B$}
In section \ref{generafam} we have studied the general properties of
the family of $4D$-metrics \eqref{metrauniversala} and, referring to
eqn.\eqref{K0inv0}, we have seen that the K\"ahler potential of such
metrics can be uniquely traced back, by means of the Legendre
transform, to the boundary function $\mathcal{D}(\mathfrak{v})$
appearing in the symplectic potential
\eqref{GBsymplectic},\eqref{lupetto}. Furthermore we have seen that
the same class \eqref{metrauniversala} contains both the
non-extremal K\"ahler metric induced on the second Hirzebruch
surface by the Kronheimer construction and the entire family of
Calabi extremal metrics that includes the KE Einstein metrics
studied in depth in the present lectures. In view of these facts we
want to reconsider the general structure of the
$\mathcal{D}(\mathfrak{v})$ that corresponds to all these cases as
particular ones.
\par
We can write the following general formula:
\begin{align}\label{generalelogo}
    \mathcal{D}(\mathfrak{v}) \, = & \, q \, + \, p\, \mathfrak{v}\, + k_0
    \, \mathfrak{v}\, \log[\mathfrak{v}] + \, \sum_{i=1}^4 \, k_i \, \left(\mathfrak{v}
    -\, \lambda_i\right) \, \log \left[\mathfrak{v}
    -\, \lambda_i\right]
\end{align}
where $\lambda_i$ are the four roots of some quartic polynomial and
$k_i\in \mathbb{R}$ are four real coefficients. The real numbers
$q,p$ are arbitrary since the linear part of the
$\mathcal{D}(\mathfrak{v})$-function does not change the metric and
it can be chosen at will. We can fit into the general formula
\eqref{generalelogo} the case of the non extremal K\"ahler metric on
$\mathbb{F}_2$ generated by the Kronheimer construction by choosing:
\begin{equation}\label{Kronhmetrlog}
    \begin{array}{|rcl|rcl|rcl|rcl|rcl|}
    \hline
    \null&\null&\null&\null&\null&\null&\null&\null&\null&\null&\null&\null&\null&\null&\null\\
      \lambda_0 & = & 0 &  \lambda_1 &=& -\frac{9\alpha}{32}& \lambda_2 &=& \frac{9\alpha}{32}&\lambda_3
      &=&\frac{9(4+3\alpha)}{32}&
      \lambda_4 &=& \text{arbirary}\\
      \null&\null&\null&\null&\null&\null&\null&\null&\null&\null&\null&\null&\null&\null&\null\\
      \hline
      \null&\null&\null&\null&\null&\null&\null&\null&\null&\null&\null&\null&\null&\null&\null\\
      k_0 & = & -\ft 12 &  k_1 &=& \ft 12& k_2 &=& \ft 12 &k_3 &=& \ft 12& k_4
      &=&0\\
      \null&\null&\null&\null&\null&\null&\null&\null&\null&\null&\null&\null&\null&\null&\null\\
      \hline
    \end{array}
\end{equation}
The case of the KE metrics is instead reproduced with the following
choices:
\begin{equation}\label{KElog}
    \begin{array}{|rcl|rcl|rcl|rcl|rcl|}
    \hline
    \null&\null&\null&\null&\null&\null&\null&\null&\null&\null&\null&\null&\null&\null&\null\\
      \lambda_0 & = & 0 &  \lambda_1 &=& \mathit{a}& \lambda_2 &=& \mathit{b}&\lambda_3
      &=&\frac{\mathit{a} \mathit{b}}{\mathit{a}+\mathit{b}}&
      \lambda_4 &=& \text{arbirary}\\
      \null&\null&\null&\null&\null&\null&\null&\null&\null&\null&\null&\null&\null&\null&\null\\
      \hline
      \null&\null&\null&\null&\null&\null&\null&\null&\null&\null&\null&\null&\null&\null&\null\\
      k_0 & = & -\frac{1}{2} &  k_1 &=& -\frac{\mathit{a}^2+\mathit{a}
   \mathit{b}+\mathit{b}^2}{\mathit{a}^2+\mathit{a}
   \mathit{b}-2 \mathit{b}^2}& k_2 &=& -\frac{\mathit{a}^2+\mathit{a}
   \mathit{b}+\mathit{b}^2}{-2 \mathit{a}^2+\mathit{a}
   \mathit{b}+\mathit{b}^2}  &k_3 &=&\frac{\mathit{a}^2+\mathit{a}
   \mathit{b}+\mathit{b}^2}{2 \mathit{a}^2+5 \mathit{a}
   \mathit{b}+2 \mathit{b}^2}& k_4
      &=&0\\
      \null&\null&\null&\null&\null&\null&\null&\null&\null&\null&\null&\null&\null&\null&\null\\
      \hline
    \end{array}
\end{equation}
while the generic  extremal K\"ahler metric in the Calabi family is
characterized by the following  choices:
\begin{equation}\label{extgenlogo}
    \begin{array}{|rcl|rcl|}
    \hline
    \null&\null&\null&\null&\null&\null\\
       \lambda_0  & = & 0 &  k_0 & = & -\frac{1}{2} \\
       \null&\null&\null&\null&\null&\null\\
       \hline
        \null&\null&\null&\null&\null&\null\\
       \lambda_1  & = & \mathit{a} &  k_1 & = & \frac{\mathit{a} (\mathit{a}
   (\mathit{b}+\mathfrak{s}+\mathfrak{t})+\mathit{b}
   (\mathfrak{s}+\mathfrak{t})+\mathfrak{s}
   \mathfrak{t})}{(\mathit{a}-\mathit{b})
   (\mathit{a}-\mathfrak{s}) (\mathit{a}-\mathfrak{t})} \\
   \null&\null&\null&\null&\null&\null\\
       \hline
       \null&\null&\null&\null&\null&\null\\
       \lambda_2  & = & \mathit{b}&  k_2 & = & -\frac{\mathit{b} (\mathit{a}
   (\mathit{b}+\mathfrak{s}+\mathfrak{t})+\mathit{b}
   (\mathfrak{s}+\mathfrak{t})+\mathfrak{s}
   \mathfrak{t})}{(\mathit{a}-\mathit{b})
   (\mathit{b}-\mathfrak{s}) (\mathit{b}-\mathfrak{t})} \\
   \null&\null&\null&\null&\null&\null\\
       \hline
       \null&\null&\null&\null&\null&\null\\
       \lambda_3  & = & \mathfrak{s} &  k_3 & = & -\frac{\mathfrak{s} (\mathit{a}
   (\mathit{b}+\mathfrak{s}+\mathfrak{t})+\mathit{b}
   (\mathfrak{s}+\mathfrak{t})+\mathfrak{s}
   \mathfrak{t})}{(\mathit{a}-\mathfrak{s})
   (\mathfrak{s}-\mathit{b})
   (\mathfrak{s}-\mathfrak{t})} \\
   \null&\null&\null&\null&\null&\null\\
       \hline
       \null&\null&\null&\null&\null&\null\\
       \lambda_4  & = & \mathfrak{t} &  k_4 & = & -\frac{\mathfrak{t} (\mathit{a}
   (\mathit{b}+\mathfrak{s}+\mathfrak{t})+\mathit{b}
   (\mathfrak{s}+\mathfrak{t})+\mathfrak{s}
   \mathfrak{t})}{(\mathit{a}-\mathfrak{t})
   (\mathfrak{t}-\mathit{b})
   (\mathfrak{t}-\mathfrak{s})}\\
   \null&\null&\null&\null&\null&\null\\
   \hline
     \end{array}
\end{equation}
The previous case of KE metrics is retrieved within the general
description of eqn.\eqref{extgenlogo} by means of a limiting
procedure. One chooses the third root $\mathfrak{s}$ in terms of the
first two as follows:
\begin{equation}\label{scelta}
    \mathfrak{s} \, = \, - \, \frac{\mathit{a} \, \mathit{b}}{\mathit{a} \, +
    \,\mathit{b}}
\end{equation}
and performs the limit in which the fourth root goes to infinity:
$\mathfrak{t} \to \infty$. In this limit the coefficient $k_4$ goes
to zero and we get back eqn.\eqref{KElog}. The case of the extremal
K\"ahler metric on $\mathbb{F}_2$ is instead derived from equations
\eqref{extgenlogo} introducing the relation:
\begin{align}\label{F2rulla}
    \mathfrak{s} = &
   \frac{\mathit{a}^2+\sqrt{\mathit{a}^4-44
   \mathit{a}^3 \mathit{b}-10 \mathit{a}^2
   \mathit{b}^2+4 \mathit{a}
   \mathit{b}^3+\mathit{b}^4}-4 \mathit{a}
   \mathit{b}-\mathit{b}^2}{6 \mathit{a}+2
   \mathit{b}}\\
   \mathfrak{t} = &
   \frac{\mathit{a}^2-\sqrt{\mathit{a}^4-44
   \mathit{a}^3 \mathit{b}-10 \mathit{a}^2
   \mathit{b}^2+4 \mathit{a}
   \mathit{b}^3+\mathit{b}^4}-4 \mathit{a}
   \mathit{b}-\mathit{b}^2}{6 \mathit{a}+2
   \mathit{b}}
\end{align}
The main problem in relation with the question whether the extremal
metrics and, among them the KE ones, can be connected with the
Kronheimer construction and the resolution of an orbifold
singularity $\mathbb{C}^3/\Gamma$ becomes evident in the symplectic
action/angle formalism at the level of the Legendre transform: to
this effect we recall the result encoded in equation \eqref{K0inv0}
which we repeat here for convenience:
\begin{equation}\label{K0inv0bis}
    \mathcal{K}\left(\mathfrak{v}\right)\, = \,
    \frac{1}{2}\,\mathfrak{v}\, + \,
   \left(\mathfrak{v}\,\mathcal{D}'\left(\mathfrak{v} \right)-\mathcal{D}\left(\mathfrak{v} \right)\right)
\end{equation}
which expresses the K\"ahler potential of the base manifold
$\mathcal{M}_B$ as a function of the moment variable $\mathfrak{v}$.
Inserting the general expression \eqref{generalelogo} of the
boundary function $\mathcal{D}(\mathfrak{v})$ for the considered
class of metrics into \eqref{K0inv0bis} we obtain the following
general result for the K\"ahler potential as a function of the
moment variable:
\begin{equation}\label{generalekappa}
    \mathcal{K}\left(\mathfrak{v}\right)\, = \,\left(\ft 12 \, + \, k_0 \, + \sum_{i=1}^4 k_i\right) \,
    \mathfrak{v} \, + \, \sum_{i=1}^4 \,k_i \, \lambda_i \, \log
    \left[v-\lambda_i\right]
\end{equation}
Comparing with eqn.\eqref{Kronhmetrlog},\eqref{extgenlogo} we see
that a property shared by the Kronheimer induced metric on
$\mathbb{F}_2$ and all the extremal metrics is:
\begin{equation}\label{cannata}
    k_0 \, = \, - \ft 12
\end{equation}
Hence in view of eqn.\eqref{cannata} the structure of the K\"ahler
potential \eqref{generalekappa} particularizes to the following one:
\begin{equation}\label{cornucopia}
     \mathcal{K}\left(\mathfrak{v}\right)\,= \, \mathfrak{v}\,\sum_{i=1}^4
     k_i \, + \, \sum_{i=1}^4 \,k_i \, \lambda_i \, \log
    \left[v-\lambda_i\right]
\end{equation}
The main structural difference between the Kronheimer induced case
and the extremal metrics, apart from the specific location of the
roots $\lambda_i$, is the following:
\begin{equation}\label{tragocapro}
    \sum_{i=1}^4 \,k_i \, = \, \left\{ \begin{array}{lccl}
                                         3/2 & \neq & 0 & \text{Kronheimer induced metric} \\
                                         \null & \null & \null & \null \\
                                         0 & \null & \null & \text{Calabi extremal metrics}
                                       \end{array}\right.
\end{equation}
Hence in the case of extremal metrics the K\"ahler potential is just
a sum of purely logarithmic terms:
\begin{equation}\label{estremadura}
\mathcal{K}_{ext}\left(\mathfrak{v}\right)\, = \,\sum_{i=1}^4 \,k_i
\, \lambda_i \, \log
    \left[v-\lambda_i\right]
\end{equation}
while in the Kronheimer case it is the sum of a linear term plus the
logarithmic ones:
\begin{equation}\label{kronotipo}
   \mathcal{K}_{Kro}\left(\mathfrak{v}\right)\, = \, \ft 32 \,
   \mathfrak{v} \, +\,\sum_{i=1}^3 \,k_i \, \lambda_i \, \log
    \left[v-\lambda_i\right]
\end{equation}
where we have taken into account that $k_4\, = \, 0$ as it happens
also for the KE metrics. Why the presence or absence of the linear
term is so much relevant? It is relevant in relation with the basic
conjecture \ref{congetto} since the K\"ahler potential on the
exceptional divisor is supposedly obtained in a limiting procedure
from the K\"ahler potential inherent to the Kronheimer construction,
namely that displayed in eqn.\eqref{criceto1} and named HKLR after
Hitchin, Karlhede, Lindstrom and Ro\v{o}cek \cite{HKLR}. Recalling
the structure of the HKLR potential for the reader's convenience:
\begin{equation}\label{ravanatore}
\mathcal{K}_{\mathcal{M}_{\zeta }}= \mathcal{K}_{\mathcal{S}_{\Gamma
}} \mid _{\mathcal{N}_{\zeta }}+
\zeta^I\mathfrak{C}_{\text{IJ}}\text{Log}\left[\text{Det}\left[\mathfrak{H}_{J
}\right]\right]
\end{equation}
we see that indeed it is composed of a non logarithmic term plus a
sum of logarithmic ones, whose coefficients are linear in the Fayet
Iliopoulos parameters (stability parameters in algebraic geometry
parlance) and whose arguments are given by the moment maps
$\mathfrak{H}_{J}$ obtained solving the moment map equations. The
first term $\mathcal{K}_{\mathcal{S}_{\Gamma }} \mid
_{\mathcal{N}_{\zeta }}$ is instead the pull-back on the level
$\zeta$ hypersurface $\mathcal{N}_\zeta$ of the K\"ahler potential
for the flat K\"ahler metric of the $3 |\Gamma|$-dimensional linear
space containing all the scalar multiplets of the gauge theory. This
term is necessarily positive definitive, because of its origin and,
as I shew in eqn.\eqref{HKLRkallero}, it is a rational function of
the moment maps $X_I$ with coefficients that explicitly depend on
the coordinates $u,v,w$. The limiting procedure for the reduction to
the exceptional divisor is  that outlined in eqn.\eqref{baciodidama}
that can be schematized as follows:
\begin{equation}\label{pinocrusca}
    \mathfrak{H}_{J} \, \to \, \ell^{\alpha_J} \, \mathfrak{T}_J
    \quad ; \quad \ell \, \equiv \, |w|
\end{equation}
where $\alpha_J$ are suitable powers (positive or negative) of the
modulus of the fibre coordinate $w$, adopting the convention that
$u,v$ are coordinates of the base manifold $\mathcal{M}_B$
\textit{i.e.} the compact exceptional divisor in the resolution
$Y^{\Gamma} \, \to \, \frac{\mathbb{C}^3}{\Gamma}$. What does it
mean suitable powers $\alpha_J$? It means that the limit $\ell \to
0$ of the moment map equations for the rescaled objects
$\mathfrak{T}_J$ is finite and that in the same limit the K\"ahler
potential is also finite. For the logarithmic addends in the
K\"ahler potential this is guaranteed a priori since terms of the
form $\textit{const} \times \log[\ell]$ can always be disregarded in
$\mathcal{K}$ being the sum of a holomorphic function plus its
complex conjugate. The question remains for the first addend that is
a rational function of the moments $\mathfrak{H}_{J}$. After the
rescaling and in the limit $\ell \to 0$ there are three options.
Either $\mathcal{K}_{\mathcal{S}_{\Gamma }} \mid
_{\mathcal{N}_{\zeta }}$ goes to a finite expression in terms of
moments $\mathfrak{T}_J$, or to zero, or to infinity. The first two
options are both acceptable, while the last has to be discarded,
namely one has to try different powers $\alpha_J$ if they can be
found. In the case of the $\mathbb{C}^3/\mathbb{Z}_4$ quotient, the
limit
\begin{equation}\label{dervisci}
    \lim_{\ell \to 0} \mathcal{K}_{\mathcal{S}_{\Gamma }}\mid_{\mathcal{N}_{\zeta
    }} \, = \, \mathcal{K}_0
\end{equation}
turns out to be finite $\mathcal{K}_0 < \infty$  and one finds the
result discussed at length both above and in previous sections. In
this fully workable case the comparison of the K\"ahler potential as
obtained from the HKLR formula  and as written in the symplectic
formalism \eqref{kronotipo} is very instructive. Upon the
back-transformation from the action/angle variables to the complex
coordinates $u,v$, the linear term in $\mathfrak{v}$ is just the non
logarithmic addend:
\begin{equation}\label{poloviciani}
    \mathcal{K}_0 \, = \, \ft 32 \, \mathfrak{v}
\end{equation}
while the logarithmic terms
\begin{equation}\label{crinetto}
    \sum_{i=1}^3 \,k_i \, \lambda_i \,
\log \left[v-\lambda_i\right]
\end{equation}
reconstruct the logarithmic addends
\begin{equation}\label{credolino}
    \zeta^i\mathfrak{C}_{\text{ij}}\text{Log}\left[\text{Det}\left[\mathfrak{H}_{j
}\right]\right] \, \Leftrightarrow \,\sum_{i=1}^3 \,k_i \, \lambda_i
\, \log \left[v-\lambda_i\right]
\end{equation}
\subsection{Conclusions}
Hence the possibility of connecting the Extremal Calabi Metrics and,
among them the KE ones, to resolutions \`{a} la Kronheimer of
orbifold singularities $\frac{\mathbb{C}^3}{\Gamma}$, obtaining in
this way a full description of the \textbf{holographic dual gauge
theory}, requires to accomplish successfully  the following steps:
\begin{description}
  \item[A)] Find a subgroup $\Gamma \subset \mathrm{SU(3)}$ such
  that the age grading of its conjugacy classes allows for the
  existence of compact components of the exceptional divisor.
  Desirably only one component, which means just one senior class.
  \item[B)] Arrange, if permitted, the scale powers in eqn. \eqref{pinocrusca} so
  that the limit $\ell\to 0$ yields finite moment map equations and a
  K\"ahler potential with vanishing non logarithmic part $\mathcal{K}_0 \, = \,
  0$.
  \item[C)] Investigate whether the logarithms of the moment maps $\log\mathfrak{H}_{j}$
  can be identified, upon Legendre transformation, with suitable
  logarithmic addends \eqref{crinetto}.
  \item[D)] Verify with the help of toric geometry that the compact
  exceptional divisor is indeed the expected manifold for the
  extremal K\"ahler metric, for instance the second Hirzebruch
  surface.
\end{description}
Independently from the extremal Calabi metrics the investigation of
quotients $\frac{\mathbb{C}^3}{\Gamma}$ with $\Gamma$ non abelian
and the features outlined in point A) of the above menu is quite
timely in order to better understand the general pattern encoded in
$\mathcal{D}(\mathfrak{v})$ functions of the form
\eqref{generalelogo} and the issue of cohomogeneity of the
Kronheimer induced K\"ahler metrics.
\section{Non abelian Kronheimer constructions}
The classification of finite subgroups of $\mathrm{SU(3)}$ was
achieved at the beginning of the XXth century in
\cite{blicfeltus,blicfeltus2}. The explicit construction of such
groups and a preliminary discussion of the resolutions of the
corresponding $\mathbb{C}^3/\Gamma$ singularities was discussed in
\cite{roanno}. Similarly to the Kleinian case of $\mathrm{SU(2)}$
finite subgroups, also for $\mathrm{SU(3)}$ there are infinite
series of cyclic and solvable subgroups but there is only a short
list of non abelian groups, somehow the analogues of the three
symmetry groups of Platonic solids. In particular the largest non
abelian finite subgroup of $\mathrm{SU(3)}$ is the simple group
$\mathrm{PSL(2,7)}$ with 168 elements. The resolution of the
singularity $\mathbb{C}^3/\mathrm{PSL(2,7)}$ was studied by
Markushevich in \cite{marcovaldo} using standard blow-up techniques
of algebraic geometry. All the subgroups of $\mathrm{PSL(2,7)}$ have
an $\mathrm{SU(3)}$ embedding via the three dimensional complex
representation of the father group and lead to interesting McKay
quivers that define quite non trivial $D3$-brane gauge theories. The
smallest non abelian subgroup of $\mathrm{SU(3)}$ is the dihedral
group $\mathrm{Dih_3}$ of order six which is isomorphic to the
symmetric group $S_3$. The associated Kronheimer construction for
the resolution of the singularity $\mathbb{C}^3/\mathrm{Dih_3}$ was
described in appendix E of \cite{Bruzzo:2017fwj} and the
corresponding McKay quiver matrix was calculated there.
\section{An Anthology of McKay quivers with  comments} In the present subsection
we provide an anthology of  McKay quiver diagrams for non abelian
finite groups $\Gamma \subset \mathrm{SU(3)}$. Some of them are
obtained as subgroups of the maximal non abelian group
$\mathrm{PSL(2,7)}$ some other are constructed directly introducing
their generators $\gamma_i \in \mathrm{SU(3)}$. Two are the critical
informations we look for. On one side the quiver diagram from which
we read off \textit{the gauge group} of the D3-brane gauge theory
and its \textit{scalar multiplet content}, on the other hand the
\textit{age grading of the conjugacy classes}, which eventually
predicts the number of components of the exceptional divisor. If we
aim at a resolved transverse space that is a line-bundle
$\mathcal{L}\left(\mathcal{M}_B\right)$ over a compact exceptional
divisor $\mathcal{M}_B$ we have to select discrete subgroups
$\Gamma\subset \mathrm{SU(3)}$ that have senior classes. When all
conjugacy classes are junior, the exceptional divisor is the union
of non-compact manifolds and the whole picture that we have
developed in the present lectures does not apply. As the reader can
see from the shown examples, senior classes emerge less frequently
that one might expect and this is a pragmatic criterion to
investigate a subclass of groups more than others.
\subsection{The smallest non
abelian group $\mathrm{Dih_3} \subset\mathrm{PSL(2,7)}$}
\paragraph{Definition of $\mathrm{Dih}_3 \subset\mathrm{SU(3)}$} The
presentation of the dihedral group \(\text{Dih}_m\) is the following
one
\begin{equation}
A^3 = 1 \quad ;\quad  B^2 =1 \quad ;\quad (AB)^2 = \mathbf{1}
\end{equation}
We introduce the following representation of the generators as
matrices acting on \(\mathbb{C}^3\)
\begin{equation}
A \,= \,\left(
\begin{array}{ccc}
 e^{\frac{2\pi }{3}\,i} & 0 & 0 \\
 0 & e^{-\frac{2\pi}{3}\, i} & 0 \\
 0 & 0 & 1 \\
\end{array}
\right)\label{ArepC3}
\end{equation}
\begin{equation}
B \, = \, \left(
\begin{array}{ccc}
 0 & i & 0 \\
 -i & 0 & 0 \\
 0 & 0 & -1 \\
\end{array}
\right)\label{BrepC3}
\end{equation}
\paragraph{Conjugacy classes} The total number of conjugacy classes is
\begin{equation}
\ell  = 3
\end{equation}
that are enumerated as follows:
\begin{eqnarray}
&&\mathcal{C}_1 = \langle E\rangle \nonumber\\
&& \mathcal{C}_2 = \left\langle B,AB,A^2B\right\rangle ,\nonumber\\
&&\mathcal{C}_3 =<A,A^{2}>\label{clasconiugdihm}
\end{eqnarray}
Hence we have one class of population 1, one class of population $3$
and one class of population 2:
\begin{equation}
\left|\text{Dih}_3\right|= 6 = 1+ 3 + 2\label{verif}
\end{equation}
\paragraph{Irreps} Accordingly we expect \(\ell  = 3\)
irreducible representations. They are as follows:
\begin{enumerate}
\item
The one dimensional identity representation \(\mathbb{D}_0\)
\item
The alternating one-dimensional representation \(\mathbb{D}_1\),
obtained by setting { }A $\rightarrow $ 1, { }B $\rightarrow $ -1.
\item
The two-dimensional representation obtained in the following way:
\begin{equation}
\mathbb{D}_{2}[A] = \left(
\begin{array}{cc}
 e^{\frac{2\pi }{3}\,i} & 0 \\
 0 & e^{-\frac{2\pi}{3}\, i} \\
\end{array}
\right)\quad ; \quad \mathbb{D}_{2}[B]=\left(
\begin{array}{cc}
 0 & 1 \\
 1 & 0 \\
\end{array}
\right)
\end{equation}
\end{enumerate}
\paragraph{Characters} In this way we obtain the following character table:
\begin{equation}
\begin{array}{c|cccc}
 0 & \mathcal{C}_1 & \mathcal{C}_2 & \mathcal{C}_3  \\
\hline
 \times  & E & B & A  \\
 \mathbb{D}_1 & 1 & 1 & 1  \\
 \mathbb{D}_2 & 1 & -1 & 1  \\
 \mathbb{D}_3 & 2 & 0 & 2\,\cos\left[\frac{2 \pi }{3}\right] \\
\end{array}
\end{equation}
\paragraph{The quiver matrix of $\mathrm{Dih_3}$ acting on $\mathbb{C}^3$}
Using the above character table  we easily derive the following
decomposition into irreps:
\begin{equation}
 \mathcal{Q}\otimes  \mathbb{D}_i = \underset{j=1}{\overset{3
 }{\oplus }} \text{QC}_{\text{ij}} \mathbb{D}_j
\end{equation}
where $\mathcal{Q}$ is the three-dimensional representation of the
dihedral group defined by eqs.\,(\ref{ArepC3},\ref{BrepC3}) and
\(\mathbb{D}_i\) are the irreducible representations listed above.
The matrix $\text{QC}_{\text{ij}}$ is the following
\begin{equation}
 \text{CQ}\, = \, \left(
\begin{array}{ccc}
 0 & 1 & 1 \\
 1 & 0 & 1 \\
 1 & 1 & 2 \\
\end{array}
\right) \label{dih3quivmat}
\end{equation}
The graphical representation of the quiver is displayed in
fig.\ref{quiverdiedro}. The group has six elements and three
representations, the identity representation, a non trivial
one-dimensional representation and a complex two dimensional
representation. Correspondingly for a stack of $N$-branes at the
resolved orbifold singularity, we expect a gauge theory with the
following gauge group:
\begin{equation}\label{capiscus}
    \mathcal{F}_{\mathrm{Dih_3}} \, = \, \mathrm{S}\left[\mathrm{\mathrm{U_0}(N)\times
    \mathrm{U_1}(N)\times\mathrm{U_2}(2\,N)}\right]
\end{equation}
and a total of $18=3\times \mid \mathrm{Dih_3}\mid$ scalar field
multiplets in the one brane case $N=1$. The symbol $\mathrm{S}$ in
front of the product of $\mathrm{U(m)}$ factors means, as usual that
the overall $\mathrm{U(1)}$ has to be factorized out. Looking at the
diagram we easily see the field content of such a theory:
\begin{equation}\label{festivo}
    \begin{array}{|c|c|c|}
    \hline
    \null & \null & \null\\
       \mbox{Field} & \mathrm{U_0(1)\times U_1(1) \times U_2(2)} & \# \mbox{of components} \\
       \null & \null & \null\\
       \hline
       \Phi_{0,1} & (1,\bar{1},1) & 1 \\
       \Phi_{1,0} & (\bar{1},1,1) & 1\\
       \Phi_{0,2} & (\bar{1},1,2) & 2 \\
       \Phi_{2,1} & (1,1,\bar{2}) & 2 \\
       \Phi_{1,2} & (1,\bar{1},2) & 2 \\
       \Phi_{2,1} & (1,1,\bar{2}) & 2 \\
      \Phi_{2,2} & (1,1,4) & 4\\
       \Phi_{2,2}^\prime & (1,1,4) & 4\\
       \hline
       \null & \null & \null\\
       \null & \null & 18\\
       \null & \null & \null\\
       \hline
     \end{array}
\end{equation}
The adjoint 4-dimensional representations can be given a mass and
one can study the resulting quiver theory.
\begin{figure}
\begin{center}
\begin{tikzpicture}[scale=0.50]
\draw [thick] [fill=yellow] (-5,-5) circle (1.5cm); \node at (-5,-5)
{$\mathrm{U_0(N)}$}; \draw [thick] [fill=yellow] (5,-5) circle
(1.5cm); \node at (5,-5) {$\mathrm{U_1(N)}$}; \node at (0,0)
{$\mathrm{U_1(N)}$}; \draw [thick] [fill=green] (0,0) circle
(1.5cm); \node at (0,0) {$\mathrm{U_2(2N)}$};  \draw [blue, line
width=0.05cm] [-](-2.5,-4) to (-3.5,-5)  ; \draw [blue, line
width=0.05cm] [->] (0,-1.5) to (-2.5,-4)  ; \draw [red, line
width=0.05cm] [->] (-5,-3.5) to (-2.5,-1); \draw [red, line
width=0.05cm] [-] (-2.5,-1) to (-1.5,0);
\draw [black, line width=0.04cm] [-] (0,3) to [out=0,in=40] (1.5,0);
\draw [black, line width=0.04cm] [->] (-1.5,0) to [out=120,in=180]
(0,3);
 \draw [blue, line width=0.05cm] [-] (0,-1.5) to
 (2.5,-4);
 \draw [blue, line width=0.05cm] [->]
 (3.5,-5) to (2.5,-4) ;
\draw [red, line width=0.05cm] [->] (1.5,0) to (4,-2.5); \draw [red,
line width=0.05cm] (4.0,-2.5) to (5,-3.5);
\draw [blue, line width=0.05cm] [->] (-3.5,-5) to (0,-5);
\draw[blue, line width=0.05cm] (0,-5) to (3.5,-5);
\draw [red, line width=0.05cm] [->] (5,-6.5) to (0,-6.5); \draw
[red, line width=0.05cm] (0,-6.5) to (-5,-6.5);
\draw [black, line width=0.04cm] [-] (0,4) to [out=0,in=15] (1.5,0);
\draw [black, line width=0.04cm] [->] (-1.5,0) to [out=160,in=180]
(0,4);
\end{tikzpicture}
\caption{ \label{quiverdiedro} The quiver diagram for the dihedral
group embedding  $\mathrm{Dih}_3 \to \mathrm{SU(3)}$}
\end{center}
\end{figure}
\paragraph{Ages} In a similar easy way we derive the age grading of the
three conjugacy classes and the associated triple of integer numbers
$a_i$ that define the weights in the weighted blowup procedure.
Using the generators to obtain the matrix corresponding to each
group element we find the following result:
\newpage
\begin{enumerate}
\item age = 0 \hskip 0.7cm ; \hskip 0.7cm  $1\{0,0,0\}$
\item
age = 1 \hskip 0.7cm ; \hskip 0.7cm  $\frac{1}{2}\{1,1,0\}$
\item
age = 1 \hskip 0.7cm ; \hskip 0.7cm  $\frac{1}{3}\{0,2,1\}$
\end{enumerate}
Hence apart from the age = 0 class of the identity we find two
junior classes and no senior one. Hence we conclude that the Hodge
numbers of the resolved variety $\mathcal{M}_T$ are as follows
$h^{0,0}=1$, $h^{1,1} \, = \, 2$, $h^{2,2} \, = \, 0$, so that no
(1,1)-class has compact support and the exceptional divisor has non
compact component.
\subsection{The case of the group $\mathrm{PSL(2,7)}$} Starting from
the construction of the irreducible three-dimensional complex
representation discussed in \cite{miol168,Cerchiai:2018shs}, in
\cite{Bruzzo:2017fwj}  the age was computed  of  each of the
conjugacy classes for the holomorphic action of the group
$\mathrm{\mathrm{PSL(2,7)}}$ on $\mathbb{C}^3$.
\par
In order to be able to compare with Markusevich{'}s paper
\cite{marcovaldo}, it is important to note that the form given by
Markusevich of the generators which he calls $\tau $ , $\chi $ and
$\omega $, respectively of order 7, 3 and 2, does not correspond to
the standard generators in the presentation of the group
\(\mathrm{PSL(2,7)}\) utilized by the present author in  paper
\cite{miol168}.  Yet there is no problem since we have a translation
vocabulary.
Setting:\\
\begin{equation}
R = \omega .\chi \quad; \quad S \, = \, \chi .\tau \quad; \quad T \,
=\,\chi ^2. \omega\label{agnaturio}
\end{equation}
these new generators satisfy the standard relations of the
presentation displayed below:
\begin{equation}\label{abstroL168}
\mathrm{\mathrm{PSL(2,7)}} \, = \, \left(R,S,T \,\parallel \, R^2 \,
= \, S^3 \, = \, T^7 \, = \, RST \, = \, \left(TSR\right)^4 \, = \,
\mathbf{e}\right)
\end{equation}
From now on we utilize the abstract notation in terms of $\rho=R$,
$\sigma=S$, $\tau=T$.
\par
We begin by constructing explicitly the group \(\mathrm{PSL(2,7)}\)
in Markushevich's basis substituting the analytic form of the
generators which follows from the identification (\ref{agnaturio}).
We find
\\
\(\epsilon \to  \left(
\begin{array}{ccc}
 1 & 0 & 0 \\
 0 & 1 & 0 \\
 0 & 0 & 1 \\
\end{array}
\right)\\
\\
\rho \to  \left(
\begin{array}{ccc}
 -\frac{2 \text{Cos}\left[\frac{\pi }{14}\right]}{\sqrt{7}} & -\frac{2 \text{Cos}\left[\frac{3 \pi
 }{14}\right]}{\sqrt{7}} & \frac{2 \text{Sin}\left[\frac{\pi
}{7}\right]}{\sqrt{7}} \\
 -\frac{2 \text{Cos}\left[\frac{3 \pi }{14}\right]}{\sqrt{7}} & \frac{2 \text{Sin}\left[\frac{\pi
 }{7}\right]}{\sqrt{7}} & -\frac{2 \text{Cos}\left[\frac{\pi
}{14}\right]}{\sqrt{7}} \\
 \frac{2 \text{Sin}\left[\frac{\pi }{7}\right]}{\sqrt{7}} & -\frac{2 \text{Cos}\left[\frac{\pi
}{14}\right]}{\sqrt{7}} & -\frac{2 \text{Cos}\left[\frac{3
\pi }{14}\right]}{\sqrt{7}} \\
\end{array}
\right)\\
\\
\sigma \to  \left(
\begin{array}{ccc}
 0 & 0 & -(-1)^{1/7} \\
 (-1)^{2/7} & 0 & 0 \\
 0 & (-1)^{4/7} & 0 \\
\end{array}
\right)\\
\\
\tau \to  \left(
\begin{array}{ccc}
 \frac{i+(-1)^{13/14}}{\sqrt{7}} & -\frac{(-1)^{1/14} \left(-1+(-1)^{2/7}\right)}{\sqrt{7}} &
 \frac{(-1)^{9/14} \left(1+(-1)^{1/7}\right)}{\sqrt{7}}
\\
 \frac{(-1)^{11/14} \left(-1+(-1)^{2/7}\right)}{\sqrt{7}} & \frac{i+(-1)^{5/14}}{\sqrt{7}} &
 \frac{(-1)^{3/14} \left(1+(-1)^{3/7}\right)}{\sqrt{7}}
\\
 -\frac{(-1)^{11/14} \left(1+(-1)^{1/7}\right)}{\sqrt{7}} & -\frac{(-1)^{9/14}
 \left(1+(-1)^{3/7}\right)}{\sqrt{7}} & -\frac{-i+(-1)^{3/14}}{\sqrt{7}}
\\
\end{array}
\right)\)\\
\\
\begin{equation}
\label{birillus1}
\end{equation}
We remind the reader that $\rho $,$\sigma $,$\tau $ are the abstract
names for the generators of \(\mathrm{PSL(2,7)}\) whose 168 elements
are written as words in these letters (modulo relations).
Substituting these letters with explicit matrices that satisfy the
defining relation of the group one obtains an explicit
representation of the latter. In the present case the substitution
\ref{birillus1} produces the irreducible 3-dimensional
representation \(\text{DA}_3\).
\par
Utilizing this explicit representation it is straightforward to
calculate the age of each conjugacy class and one obtains the result
displayed in the following table.
\begin{equation}\label{vecchierello168}
\mbox{
\begin{tabular}{||c|c|c|c|c|c|c||}
\hline \hline Conjugacy class of $\mathrm{PSL(2,7)}$
&$\mathcal{C}_1$&$\mathcal{C}_2$&$\mathcal{C}_3$&$\mathcal{C}_4$&$\mathcal{C}_5$&$\mathcal{C}_6$\\
  \hline
  \hline
  representative of the class  & $\mathbf{e}$ & $R$ & $S$ &$TSR$ & $T$ & $SR$ \\
  \hline
  order of the elements in the class & 1 & 2 & 3 & 4 & 7 & 7 \\
  \hline
  age &0 & 1 & 1 & 1 & 1 & 2 \\
  \hline
  number of elements in the class & 1 & 21 & 56 & 42 & 24 & 24  \\
  \hline
  \hline
\end{tabular}}
\end{equation}
According with the general results discussed in section
\ref{anglonipponico} the resolved manifold
$\mathcal{M}_{\zeta|\mathrm{\mathrm{PSL(2,7)}}}$ must have the
following Hodge numbers:
\begin{equation}\label{caramellamu}
    h^{(1,1)}\left(\mathcal{M}_{\zeta|\mathrm{\mathrm{PSL(2,7)}}}\right) \, = \,
    4 \quad ; \quad h^{(2,2)}\left(\mathcal{M}_{\zeta|\mathrm{\mathrm{PSL(2,7)}}}\right)
    \, = \,
    1
\end{equation}
and by Poincar\'e duality one concludes that the exceptional divisor
has four components, three non compact and one compact, whose
geometry has to be derived.
\paragraph{Expectations, Conjectures and Open Problems}
Indeed, from the experience developed in previous worked out
examples, like that extensively discussed in chapter
\ref{balengusz4}, and on the basis of the mathematical theory
developed in papers \cite{giapumckay},\cite{CrawIshii}, we expect
that there should be a chamber structure (see section
\ref{camerataccademica}) and that inside each chamber, away from the
walls, the resolved manifold should be an appropriate line bundle:
\begin{align}\label{otgodai}
 \pmb{\zeta}\, = \, \text{generic} \quad :  \quad &
 \mathcal{M}_{\pmb{\zeta}|\mathrm{\mathrm{PSL(2,7)}}} \, = \, \mathcal{L}\left(
 \mathcal{M}_{B|\mathrm{PSL(2,7)}}\right) \,
 \stackrel{\pi}{\longrightarrow} \,
 \mathcal{M}_{B|\mathrm{PSL(2,7)}} \nonumber\\
 & \forall
 p\in\mathcal{M}_{B|\mathrm{PSL(2,7)}} \, \quad \pi^{-1}(p) \, \sim
 \, \mathbb{C}
\end{align}
Whether such expectations come up true and the appropriate algebraic
geometry identification of the manifold
$\mathcal{M}_{B|\mathrm{PSL(2,7)}} $ is one of the open problems,
probably the most difficult one, in the context of the
investigations reviewed in these lectures. Similarly it would be
quite important to identify the isometry group of the K\"ahler
metric induced on $\mathcal{M}_{B|\mathrm{PSL(2,7)}} $ by the
Kronheimer construction. Last but not least one wonders whether the
conjecture \ref{congetto} applies in the non abelian cases and in
particular in the present one of the maximal non abelian finite
subgroup of $\mathrm{SU(3)}$.
\par
\begin{figure}
\begin{center}
\begin{tikzpicture}[scale=0.50]
\draw [thick] [fill=yellow] (-10,-5) circle (1.5cm); \node at
(-10,-5) {$\mathrm{U(8N)}$}; \draw [thick] [fill=yellow] (10,-5)
circle (1.5cm); \node at (10,-5) {$\mathrm{U(6 N)}$}; \node at (0,0)
{$\mathrm{U(7 N)}$}; \draw [thick] [fill=yellow] (0,0) circle
(1.5cm); \node at (0,0) {$\mathrm{U(7 N)}$};
\draw [thick] [fill=cyan] (-10,-15) circle (1.5cm); \node at
(-10,-15) {$\mathrm{U_A(3N)}$}; \draw [thick] [fill=cyan] (10,-15)
circle (1.5cm); \node at (10,-15) {$\mathrm{U_B(3 N)}$};
\draw [thick] [fill=green] (0,-20) circle (1.5cm); \node at (0,-20)
{$\mathrm{U(N)}$};
\draw [blue, line width=0.05cm] [->](0,-1.5) to (-4.25,-3.225)  ;
\draw [blue, line width=0.05cm] [-] (0,-1.5) to (-8.5,-5)  ;
\draw [blue, line width=0.05cm] [->] (-8.5,-5) to (0,-5);
\draw[blue, line width=0.05cm] (0,-5) to (8.5,-5);
\draw [blue, line width=0.05cm] [-] (0,-1.5) to (8.5,-5);
 \draw [blue, line width=0.05cm] [->]
 (8.5,-5) to (4.25,-3.25) ;
\draw [red, line width=0.05cm] [-] (-10,-3.5) to (-1.5,0); \draw
[red, line width=0.05cm] [->] (-10,-3.5) to (-5.725,-1.75);
\draw [red, line width=0.05cm] [-] (1.5,0) to (10,-3.5); \draw [red,
line width=0.05cm][->] (1.5,0) to (4.25,-1.15); \draw [red, line
width=0.05cm] [->] (10,-6.5) to (0,-6.5); \draw [red, line
width=0.05cm] (0,-6.5) to (-10,-6.5);
\draw [black, line width=0.05cm] [->] (-8.5,-15) to (-4.25,-16.725)
; \draw [black, line width=0.05cm] [-] (0,-18.5) to (-8.5,-15) ;
\draw [black, line width=0.05cm] [->] (0,-18.5) to (4.25,-16.725);
\draw[black, line width=0.05cm] (0,-18.5) to (8.5,-15);
\draw [black, line width=0.05cm] [-] (0,-1.5) to (8.5,-5);
\draw [black, line width=0.05cm] [->] (8.5,-15) to (0,-15);
\draw[black, line width=0.05cm] (0,-15) to (-8.5,-15);
\draw[black, line width=0.05cm] (-10,-13.5) to (-10,-6.5);
\draw[black, line width=0.05cm][->] (-10,-13.5) to (-10,-10);
\draw[black, line width=0.05cm] (10,-13.5) to (10,-6.5);
\draw[black, line width=0.05cm][->] (10,-6.5) to (10,-10);
\draw[black, line width=0.05cm] [-](8.5,-15) to (-10,-6.5);
\draw[black, line width=0.05cm] [->](-10,-6.5) to (-0.75,-10.750);
\draw[black, line width=0.05cm] [-](10,-6.5) to (-8.5,-15);
\draw[black, line width=0.05cm] [->](10,-6.5) to (-0.75,-11.45);
\end{tikzpicture}
\caption{ \label{quivettoPL27} The quiver diagram of the finite
group $\mathrm{\mathrm{PSL(2,7)}}\subset \mathrm{SU(3)}$ }
\end{center}
\end{figure}
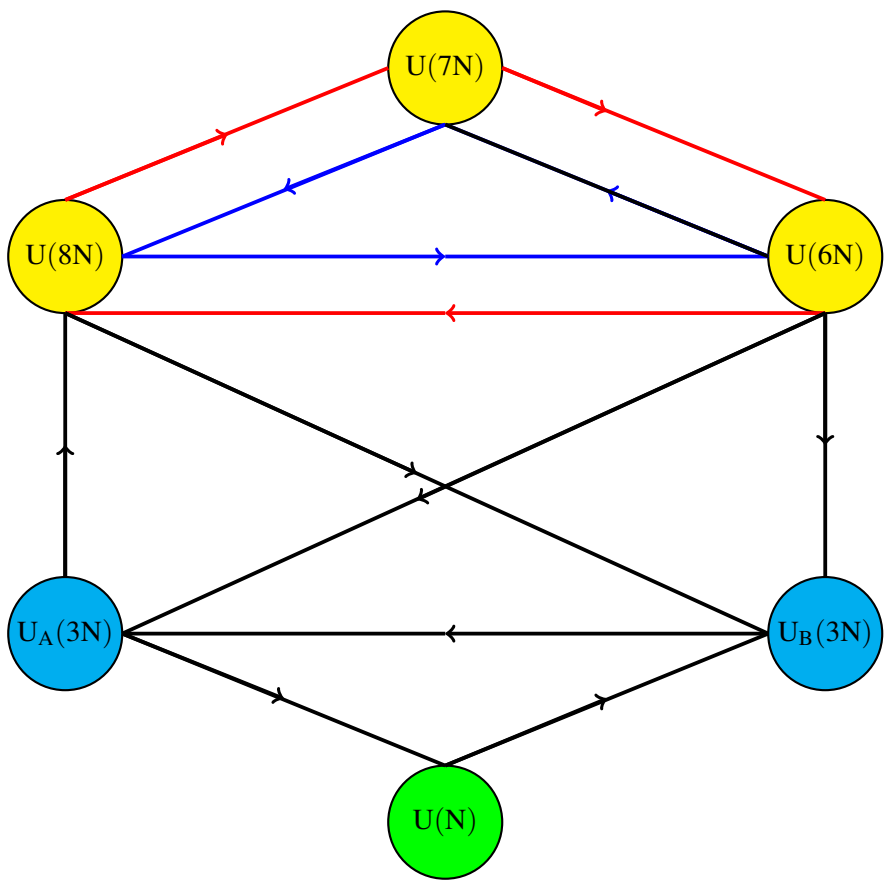

\paragraph{The McKay quiver of $\mathrm{\mathrm{PSL(2,7)}}$} We calculate
the McKay matrix defined by
\begin{equation}\label{quiverro}
    \mathcal{Q}\otimes \mathrm{D}_i \, = \, \bigoplus_{j=1}^6 \, \mathcal{A}_{ij}\,\mathrm{D}_j
\end{equation}
where $\mathcal{Q}$ is the three-dimensional complex representation
defining the action of $\mathrm{\mathrm{PSL(2,7)}}$ on
$\mathbb{C}^3$ while $\mathrm{D}_i$ denote the 6 irreducible
representation ordered in the standard way we have so far adopted,
namely:
\begin{equation}\label{orinnno}
    \mathrm{D}_i \, = \, \left\{\mathrm{D_1,D_6,D_7,D_8,D_3,D_{\bar 3}} \right\}
\end{equation}
We find the following matrix:
\begin{equation}\label{Amatricia}
    \mathcal{A}\, = \, \left(
\begin{array}{cccccc}
 0 & 0 & 0 & 0 & 1 & 0 \\
 0 & 0 & 1 & 1 & 0 & 1 \\
 0 & 1 & 1 & 1 & 0 & 0 \\
 0 & 1 & 1 & 1 & 1 & 0 \\
 0 & 1 & 0 & 0 & 0 & 1 \\
 1 & 0 & 0 & 1 & 0 & 0 \\
\end{array}
\right)
\end{equation}
The matrix $\mathcal{A}$ admits the  graphical representation
displayed in fig.\ref{quivettoPL27}, named the \textit{McKay quiver}
of the quotient $\mathbb{C}^3/\mathrm{\mathrm{PSL(2,7)}}$
Therefore for a stack of $N$ branes at the resolved singular point,
the gauge-group has the following structure:
\begin{equation}\label{cunegonda}
    \mathcal{F}_{\mathrm{PSL(2,7)}}\,= \, \mathrm{S}\left[\mathrm{U(8N)\times U(7N)\times U(6N)\times U_A(3N)\times U_B(3N)\times
    U(N)}\right]
\end{equation}
The quiver diagram provides the representation assignments of the $3
\times 168$ scalar multiplets of the theory. Each line of the
diagram is a scalar multiplet in the bi-fundamental representation
of the two gauge groups, that in the start node and that in the
arrival node.
\subsection{ The case of the maximal subgroup $\Gamma =
\mathrm{G_{21}} \subset \mathrm{\mathrm{PSL(2,7)}}$} 
\begin{figure}
\begin{center}
\begin{tikzpicture}[scale=0.50]
\draw [thick] [fill=yellow] (-5,0) circle (1.5cm); \node at (-5,0)
{$\mathrm{U_A(N)}$}; \draw [thick] [fill=yellow] (5,0) circle
(1.5cm); \node at (5,0) {$\mathrm{U_B(N)}$}; \draw [thick]
[fill=yellow] (15,0) circle (1.5cm);\node at (15,0)
{$\mathrm{U_C(N)}$};
\draw [thick] [fill=green] (0,5) circle (1.5cm); \node at (0,5)
{$\mathrm{U_D(3N)}$}; \draw [thick] [fill=green] (0,-5) circle
(1.5cm); \node at (0,-5) {$\mathrm{U_F(3N)}$};
\draw [black, line width=0.07cm] [->] (-1.5,-5) to (-5,-1.5);
\draw [black, line width=0.07cm] [->] (-5,1.5) to (-1.5,5);
\draw [blue, line width=0.07cm] [->] (1.5,-5) to (5,-1.5);
\draw [blue, line width=0.07cm] [->] (5,1.5) to (1.5,5);
\draw [blue, line width=0.08cm] [->] (1.5,5) to (1.5,-5); 
\draw [black, line width=0.08cm] [->] (-1.5,5) to (-1.5,-5);
\draw [red, line width=0.08cm] [->] (0,-3.5) to (0,3.5); 
\draw [brown, line width=0.06cm] [->] (0,-6.5) to [out=-90,in=-90](15,-1.5); 
\draw [brown, line width=0.06cm] [->] (15,1.5) to
[out=90,in=90](0,6.5); 
\draw [red, line width=0.04cm] [->] (-1.5,-5) to
[out=-180,in=90](-4,-8); \draw [red, line width=0.04cm] [->] (-4,-8)
to [out=-60,in=-90](0,-6.5); 
\draw [red, line width=0.04cm] [->] (0,6.5) to [out=90,in=90](-4,8);
\draw [red, line width=0.04cm] [->] (-4,8) to
[out=-90,in=180](-1.5,5);
\end{tikzpicture}
\caption{\label{quivetto21} The quiver diagram of the finite group
$\mathrm{G_{21}}\subset \mathrm{\mathrm{PSL(2,7)}}$. In this
non-abelian case the corresponding gauge theory has three
$\mathrm{U(N)}$ factors and two $\mathrm{U(3N)}$ factors in its
gauge group. }
\end{center}
\end{figure}
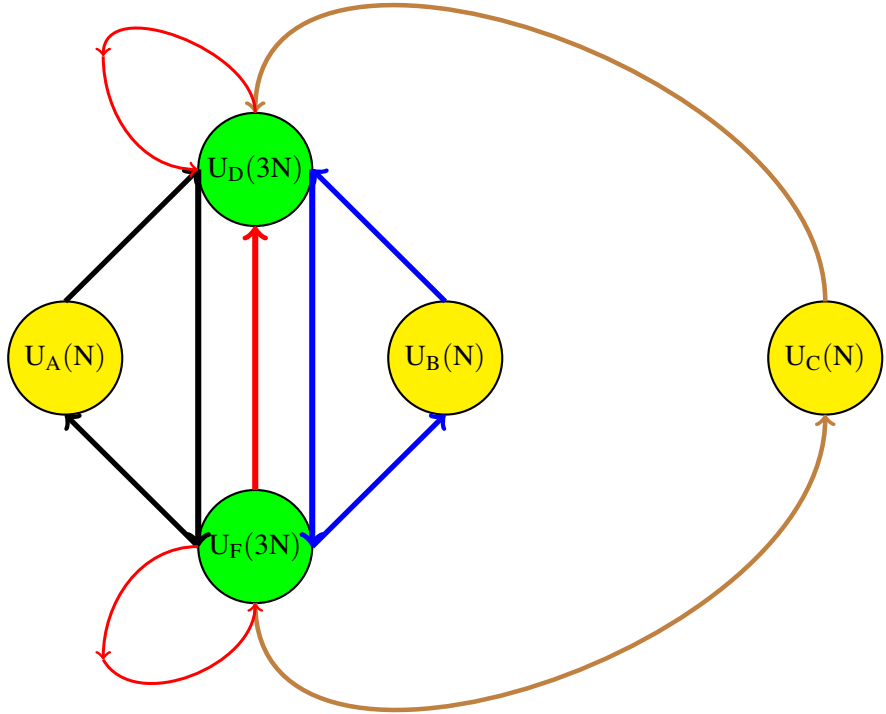

Another
interesting example, with a somewhat reduced dimensionality which
offers the possibility of investigating with more ease the same
basic questions outlined above, is provided by the maximal subgroup
$\mathrm{G_{21}}\subset\mathrm{\mathrm{PSL(2,7)}}$. In order to
obtain the ages for the conjugacy classes of the maximal subgroup
$\mathrm{G_{21}}$, we just need to obtain the explicit
three-dimensional form of its generators $\mathcal{X}$ and
$\mathcal{Y}$ satisfying the defining relations:
\begin{equation}
\mathcal{X}^3 \, = \, \mathcal{Y}^7 \, = \,\mathbf{ 1} \quad ; \quad
\mathcal{X}\mathcal{Y}  = \mathcal{Y}^2 \mathcal{X} \label{faloluna}
\end{equation}
This latter is determined by the above explicit form of the
$\mathrm{\mathrm{PSL(2,7)}}$ generators, by recalling the embedding
relations:
\begin{equation}\label{guftollo}
    \mathcal{Y} \, = \, \rho\,\sigma\,\tau^3\, \sigma\, \rho \quad ;
    \quad  \mathcal{X} \, = \, \sigma \, \rho\, \sigma\,
    \rho\, \tau^2
\end{equation}
In this way we obtain the following explicit result:
\begin{eqnarray}
 \mathcal{Y} \, \to \, \mathrm{Y} &=& \left(
\begin{array}{ccc}
 -(-1)^{3/7} & 0 & 0 \\
 0 & (-1)^{6/7} & 0 \\
 0 & 0 & -(-1)^{5/7} \\
\end{array}
\right)\nonumber \\
 \mathcal{Y} \, \to \,  \mathrm{X} &=& \left(
\begin{array}{ccc}
 0 & 1 & 0 \\
 0 & 0 & 1 \\
 1 & 0 & 0 \\
\end{array}
\right)\label{pesciYX}
\end{eqnarray}
Hence, for the action on $\mathbb{C}^3$ of the maximal subgroup
$\mathrm{G_{21}}\subset \mathrm{\mathrm{PSL(2,7)}}$  we obtain the
following ages of its conjugacy classes:
\begin{equation}\label{vecchioG21}
\begin{array}{||c|c|c|c|c|c||}
\hline \hline \mbox{Conjugacy} \mbox{ Class of $\mathrm{G_{21}}$} &
C_1 & C_2 & C_3 & C_4 & C_5 \\ \hline \mbox{representative of the
class} & e & \mathcal{Y} & \mathcal{X}^2
\mathcal{Y}\mathcal{X}\mathcal{Y}^2 & \mathcal{Y}\mathcal{X}^2 &
\mathcal{X}
\\
\hline \mbox{order of the elements in the class} & 1 & 7 & 7 & 3& 3 \\
\hline \mbox{age} & 0 & 2 & 1 & 1 & 1 \\
\hline \mbox{number of elements in the class} & 1 & 3 & 3 & 7 & 7 \\
\hline
\end{array}
\end{equation}
According with the general results discussed in
\cite{Bruzzo:2017fwj}, reviewed in section \ref{anglonipponico}, the
resolved manifold $\mathcal{M}_{\zeta|\mathrm{G_{21}}}$ must have
the following Hodge numbers:
\begin{equation}\label{caramellamu}
    h^{(1,1)}\left(\mathcal{M}_{\zeta|\mathrm{G_{21}}}\right) \, = \,
    3 \quad ; \quad h^{(2,2)}\left(\mathcal{M}_{\zeta|\mathrm{G_{21}}}\right)
    \, = \,
    1
\end{equation}
and by Poincar\'e duality we conclude that the exceptional divisor
has three components, two non compact and one compact, whose
geometry has to be derived. We are therefore in the same situation
as that of the maximal subgroup $\mathrm{PSL(2,7)}$, namely we
expect that in generic points of the chambers, away from the walls
we should have:
\begin{align}\label{otgodaiG21}
 \pmb{\zeta}\, = \, \text{generic} \quad :  \quad &
 \mathcal{M}_{\pmb{\zeta}|\mathrm{\mathrm{G_{21}}}} \, = \, \mathcal{L}\left(
 \mathcal{M}_{B|\mathrm{G_{21}}}\right) \,
 \stackrel{\pi}{\longrightarrow} \,
 \mathcal{M}_{B|\mathrm{G_{21}}} \nonumber\\
 & \forall
 p\in\mathcal{M}_{B|\mathrm{G_{21}}} \, \quad \pi^{-1}(p) \, \sim
 \, \mathbb{C}
\end{align}
Once again all the questions raised in relation with the manifold
$\mathcal{M}_{B|\mathrm{PSL(2,7)}}$ apply also to the exceptional
divisor $\mathcal{M}_{B|\mathrm{G_{21}}}$ and are equally
unanswered.
\paragraph{The McKay quiver of $\mathrm{G_{21}}$}
We calculate the McKay matrix defined by
\begin{equation}\label{quiverro21}
    \mathcal{Q}\otimes \mathrm{D}_i \, = \, \bigoplus_{j=1}^5 \,
    \mathcal{A}_{ij}\,\mathrm{D}_j
\end{equation}
where $\mathcal{Q}$ is the three-dimensional complex representation
defining the action of $\mathrm{G_{21}}$ on $\mathbb{C}^3$ while
$\mathrm{D}_i$ denote the 5 irreducible representations ordered in
the standard way we have  adopted, namely:
\begin{equation}\label{orinnno21}
    \mathrm{D}_i \, = \, \left\{\mathrm{D_{0},D_1,D_{\bar 1},D_3,D_{\bar 3}}
    \right\}
\end{equation}
We find the following matrix:
\begin{equation}\label{Amatricia21}
    \mathcal{A}\, = \, \left(
\begin{array}{ccccc}
 0 & 0 & 0 & 1 & 0 \\
 0 & 0 & 0 & 1 & 0 \\
 0 & 0 & 0 & 1 & 0 \\
 0 & 0 & 0 & 1 & 2 \\
 1 & 1 & 1 & 1 & 1 \\
\end{array}
\right)
\end{equation}
The matrix $\mathcal{A}$ admits the  graphical representation
presented in fig. \ref{quivetto21}, named the \textit{McKay quiver}
of the quotient $\mathbb{C}^3/\mathrm{G_{21}}$. From the diagram we
immediately read off the gauge group of the corresponding gauge
theory and all the representation assignments of the scalar
multiplets represented by the various directed lines, whose total
number is $N \times 63 \, = \, N\times 3 \times \mid \mathrm{G_{21}}
\mid$

\subsection{The case of the two maximal octahedral subgroups} In order to appreciate the
relevance of the two previous cases I review also the case of the
other two maximal subgroups of $\mathrm{PSL(2,7)}$, namely
$\mathrm{O_{24A}}$ and $\mathrm{O_{24B}}$ that are both isomorphic
to the symmetric group $S_4$ but not conjugate to each other inside
the father group $\mathrm{PSL(2,7)}$. The age grading of their
conjugacy classes was calculated in \cite{Bruzzo:2017fwj} starting
from the two embedding conditions of the generators $S$ and $T$,
satisfying the defining relations:
\begin{equation}
S^2 \, = \, T^3\,  = \,(ST)^4 \, = \,  \mathbf{1}
\end{equation}
\paragraph{Subgroup $\mathrm{O_{24A}}$}
\begin{equation} \label{O24Aembed}
  T \, = \, \rho\,\sigma \, \rho \, \tau^2 \, \sigma\, \rho \, \tau \quad ; \quad S =
  \tau^2 \, \sigma \, \rho \, \tau \, \sigma^2
\end{equation}
\paragraph{Subgroup $\mathrm{O_{24B}}$}
\begin{equation} \label{O24Bembed}
  T \, = \, \rho\, \tau\, \sigma\, \rho\, \tau^2\, \sigma\, \rho\, \tau \quad ; \quad S =
  \sigma\, \rho\, \tau\, \sigma\, \rho\, \tau
\end{equation}
In this way in \cite{Bruzzo:2017fwj} the following result was
obtained:
\begin{equation}\label{o24Avecchione}
  \begin{array}{|c|c|c|c|c|c|} \hline
 \mbox{Conjugacy Class of the $\mathrm{O_{24A}}$} & C_1 & C_2 & C_3 & C_4 & C_5 \\
\hline
 \mbox{representative of the class} & e & T & STST  & S &  ST  \\
\hline
 \mbox{order of the elements in the class} & 1 & 3 & 2 & 2 & 4 \\
\hline  \mbox{age} & 0 & 1 & 1 & 1 & 1 \\
\hline \mbox{number of elements in the class} & 1 & 8 & 3 & 6 & 6 \\
\hline
\end{array}
\end{equation}
and
\begin{equation}\label{o24Bvecchione}
  \begin{array}{|c|c|c|c|c|c|} \hline
 \mbox{Conjugacy Class of the $\mathrm{O_{24B}}$} & C_1 & C_2 & C_3 & C_4 & C_5 \\
\hline
 \mbox{representative of the class} & e & T & STST  & S &  ST  \\
\hline
 \mbox{order of the elements in the class} & 1 & 3 & 2 & 2 & 4 \\
\hline  \mbox{age} & 0 & 1 & 1 & 1 & 1 \\
\hline \mbox{number of elements in the class} & 1 & 8 & 3 & 6 & 6 \\
\hline
\end{array}
\end{equation}
Hence in both cases the exceptional divisor of the resolution has no
compact components, so that the picture of eqn.
\eqref{otgodai},\eqref{otgodaiG21} does not apply. A feeling of what
actually happens is obtained by inspecting the simpler case of the
subgroup $\mathbb{Z}_3 \subset \mathrm{PSL(2,7)}$ also analyzed in
\cite{Bruzzo:2017fwj}.
\subsection{The case of the cyclic subgroups $\mathbb{Z}_3$ and $\mathbb{Z}_7$}
We consider the age grading for the quotient singularities
$\mathbb{C}^3/\mathbb{Z}_3$ and $\mathbb{C}^3/\mathbb{Z}_7$. As
generators of the two cyclic groups we respectively choose the
matrices $\mathrm{X}$ and $\mathrm{Y}$ displayed in
eqn.\,(\ref{pesciYX}). In other words we utilize either one of the
two generators of the maximal subgroup $\mathrm{G_{21}}\subset
\mathrm{PSL(2,7)}$.
\paragraph{The $\Gamma =\mathbb{Z}_3$ case.} The first step consists of diagonalizing the action of the
generator $\mathrm{X}$. Introducing the unitary matrix:
\begin{equation}\label{qumatta}
    \mathfrak{q}\, = \, \left(
\begin{array}{ccc}
 \frac{1}{\sqrt{3}} & \frac{1}{\sqrt{3}} & \frac{1}{\sqrt{3}} \\
 \frac{-1+i \sqrt{3}}{2 \sqrt{3}} & \frac{-1-i \sqrt{3}}{2 \sqrt{3}} & \frac{1}{\sqrt{3}} \\
 \frac{-1-i \sqrt{3}}{2 \sqrt{3}} & \frac{-1+i \sqrt{3}}{2 \sqrt{3}} & \frac{1}{\sqrt{3}} \\
\end{array}
\right)
\end{equation}
we obtain:
\begin{equation}\label{xtilda}
    \widetilde{\mathrm{X}} \, \equiv \,\mathfrak{q}^\dagger \, \mathrm{X} \, \mathfrak{q} \, = \, \left(
\begin{array}{ccc}
 e^{\frac{2 i \pi }{3}} & 0 & 0 \\
 0 & e^{-\frac{2 i \pi }{3}} & 0 \\
 0 & 0 & 1 \\
\end{array}
\right)
\end{equation}
This shows that the quotient singularity $\mathbb{C}^3/\mathbb{Z}_3$
is actually of the form $\mathbb{C}^2/\mathbb{Z}_3\times \mathbb{C}$
since it suffices to change basis of $\mathbb{C}^3$ by introducing
the new complex coordinates:
\begin{equation}\label{ciangiotto}
    \tilde{z}_a \, = \, \mathfrak{q}_{a}^{\phantom{a}b} \, z_b
\end{equation}
It follows that in the resolution of the singularity we will obtain:
\begin{equation}\label{ghinotucco}
    ALE_{\mathbb{Z}_3} \times \mathbb{C} \, \rightarrow \, \frac{\mathbb{C}^3}{\mathbb{Z}_3}
\end{equation}
Yet, the starting setup $\mathbb{C}^3/\Gamma$  produces a special
type of ALE-manifold where all the holomorphic moment map levels are
frozen to zero and only the K\"ahler quotient parameters are
switched on. As I claimed above this particularly simple situation
illustrates the general situation of what happens in the case that
no compact component of the exceptional divisor is present. Each non
compact component is probably reducible to a copy of $\mathrm{C}$
times the base manifold of a line-bundle over a complex surface
$\Sigma$ like it is the case of ALE manifolds. For instance the
simplest ALE, namely the Eguchi Hanson manifold is the total space
of the cotangent bundle of the sphere $\mathbb{P}^1$.
\par
Eq.\,(\ref{xtilda}) corresponds also to the decomposition of the
three-dimensional representation of $Z_3$ into irreducible
representations of $\mathbb{Z}_3$. From the diagonalized form
(\ref{xtilda}) of the generator we immediately obtain the ages of
the conjugacy classes:
\begin{equation}\label{z3vecchione}
  \begin{array}{|c|c|c|c|} \hline
 \mbox{Conjugacy Class of $\mathbb{Z}_3$} & C_1 & C_2 & C_3  \\
\hline
 \mbox{representative of the class} & e & \mathrm{X} & \mathrm{X}^2   \\
\hline
 \mbox{order of the elements in the class} & 1 & 3 & 3  \\
\hline  \mbox{age} & 0 & 1 & 1  \\
\hline \mbox{number of elements in the class} & 1 & 1 & 1  \\
\hline
\end{array}
\end{equation}
The above result shows that we have only junior classes and hence no
compact component of the exceptional divisor.
\paragraph{The $\Gamma =\mathbb{Z}_7$ case.} In the $\mathbb{Z}_7$ case, the generator $Y$ is already diagonal and,
as we see,  none of the three complex coordinates is invariant under
the action of the group. Hence differently from the previous case we
obtain:
\begin{equation}\label{ghinotuccone}
    \mathcal{M}_{\mathbb{Z}_7} \, \rightarrow \, \frac{\mathbb{C}^3}{\mathbb{Z}_7}
\end{equation}
where the resolved smooth manifold is not the direct product of
$\mathbb{C}$ with an ALE-manifold:
\begin{equation}\label{cominato}
 \mathcal{M}_{\mathbb{Z}_7} \, \neq \,  ALE_{\mathbb{Z}_7} \times \mathbb{C}
\end{equation}
 From the explicit diagonal  form (\ref{pesciYX}) of the
generator we immediately obtain the ages of the conjugacy classes:
\begin{equation}\label{z7vecchione}
  \begin{array}{|c|c|c|c|c|c|c|c|} \hline
 \mbox{Conjugacy Class of $\mathbb{Z}_7$} & C_1 & C_2 & C_3 & C_4 & C_5 & C_6 & C_7  \\
\hline
 \mbox{representative of the class} & e & \mathrm{Y} & \mathrm{Y}^2 & \mathrm{Y}^3 & \mathrm{Y}^4 & \mathrm{Y}^5 & \mathrm{Y}^6\\
\hline
 \mbox{order of the elements in the class} & 1 & 7 & 7 & 7 & 7 & 7 & 7 \\
\hline  \mbox{age} & 0 & 2 & 2 & 1 & 2 & 1 & 1  \\
\hline \mbox{number of elements in the class} & 1 & 1 & 1 & 1 & 1 & 1 & 1 \\
\hline
\end{array}
\end{equation}
In this case since we have three junior classes and three senior
classes, the exceptional divisor is made out of three compact
components,  non compact components being absent.
\paragraph{The
McKay quiver for $\mathbb{Z}_7$}
\begin{figure}
\centering
\includegraphics[height=12cm]{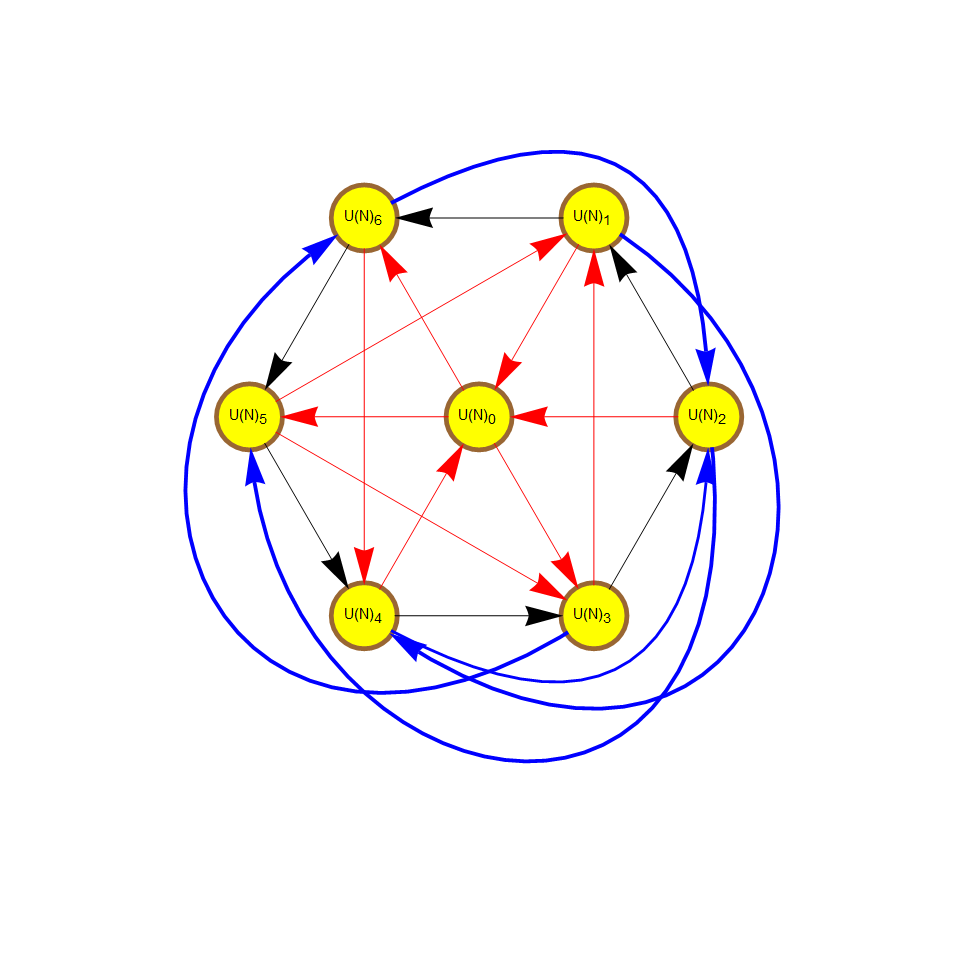}
\caption{ \label{quivettoz7} The quiver diagram of the finite group
$\mathbb{Z}_7\subset \mathrm{PSL(2,7)}$. The seven vertices
correspond to the seven irreducible representations, $\mathbf{1}$,
marked $0$, and $\psi,\dots,\psi^6$, marked $1,2,\dots,6$, where
$\psi$ is a primitive seventh root of the unity.  In each vertex
converge three lines and three lines depart from each vertex. }
\end{figure}In \cite{Bruzzo:2017fwj}  the McKay
matrix for the case where $\mathcal{Q}$ is the three-dimensional
representation of the group $\mathbb{Z}_7$ generated by $\mathrm{Y}$
as given  in eqn.\,(\ref{pesciYX}) and $D_{i}$ are the seven
irreducible one-dimensional representations of $\mathbb{Z}_7$ was
calculated. The result is displayed below.
\begin{eqnarray}
  \mathcal{Q}\otimes \mathrm{D}_i &=& \bigoplus_{j=1}^7 \mathcal{A}_{ij} \, \mathrm{D}_j \\
  \mathcal{A}_{ij} &=& \left(
\begin{array}{ccccccc}
 0 & 0 & 0 & 1 & 0 & 1 & 1 \\
 1 & 0 & 0 & 0 & 1 & 0 & 1 \\
 1 & 1 & 0 & 0 & 0 & 1 & 0 \\
 0 & 1 & 1 & 0 & 0 & 0 & 1 \\
 1 & 0 & 1 & 1 & 0 & 0 & 0 \\
 0 & 1 & 0 & 1 & 1 & 0 & 0 \\
 0 & 0 & 1 & 0 & 1 & 1 & 0 \\
\end{array}
\right) \label{komsomolt7}
\end{eqnarray}
The McKay matrix in eqn.\,(\ref{komsomolt7}) admits the graphical
representation of fig.\ref{quivettoz7}.
\paragraph{Open questions and comments}. The explicit geometry of
the compact exceptional divisor is not clear at the moment. An in
depth study of this question is certainly necessary in order to
understand how the three compact components make up a single compact
manifold of complex dimension two or, equivalently of real dimension
4. What one can speculate a priori is that being composed of three
separate parts the compact exceptional divisor
$\mathcal{M}_{B|\mathbb{Z}_7}$, although toric, inherits from the
Kronheimer construction a K\"ahler metric with a limited number of
isometries, probably only two. Deriving both the algebraic geometry
characterization of this manifold and its Kronheimer metric (if
possible) is an important question.
\subsection{$\mathrm{G_{48}}$: a challenging alternative for the group $\Gamma$}
So far I have considered finite subgroups $\Gamma$ of
$\mathrm{SU(3)}$ that are in the embedding chain:
\begin{equation}\label{kriviccio}
    \Gamma \, \subset \,  \mathrm{PSL(2,7)} \,  \subset \, \mathrm{SU(3)}
\end{equation}
yet there are others that are not contained in the maximal subgroup
$\mathrm{PSL(2,7)}$. One of them has recently attracted my attention
since it contains two $\mathbb{Z}_4$ subgroups and since it displays
just one senior class. These facts point in the direction of  the
existence of a unique interesting compact exceptional divisor that
might be connected with the Hirzebruch surfaces and which might also
inherit a K\"ahler metric endowed with  a number of isometries
larger than the minimal two. This group of order $48$ that we name
$\mathrm{GS_{48}}$ is generated by the following two matrices:
\begin{equation}\label{generatoriG48}
   A \, = \, \left(
\begin{array}{ccc}
 i & 0 & 0 \\
 0 & -i & 0 \\
 0 & 0 & 1 \\
\end{array}
\right) \quad ; \quad S \, = \,\left(
\begin{array}{ccc}
 0 & 1 & 0 \\
 0 & 0 & 1 \\
 1 & 0 & 0 \\
\end{array}
\right)
\end{equation}
That satisfy the following relations:
\begin{equation}\label{gamin}
    A^4 \, = \, S^3 \, = \, \left(A\cdot S\right)^3 \, = \,  \left(S\cdot
    A\right)^3 \, = \, \mathbf{e}
\end{equation}
where $e$ is the neutral element. The generator $A$ is the same as
in eqn. \eqref{generatoreAZ4} and therefore the $\mathbb{Z}_4$ group
studied in chapter \ref{balengusz4} is a subgroup of the here
considered $\mathrm{G_{48}}$. The number of inequivalent words
written with the two letters $A,S$ turns out to be indeed $48$ and
the resulting group is organized into $8$ conjugacy classes
$\mathcal{C}_i$ ($i=0,\dots ,7$) whose populations and order are as
follows:
\begin{equation}\label{tablizza}
\begin{array}{||c|c|c|c|c|c|c|c|c||}\hline
\null & \mathcal{C}_0 & \mathcal{C}_1 & \mathcal{C}_2 & \mathcal{C}_3 &
\mathcal{C}_4 & \mathcal{C}_5 &\mathcal{C}_6 & \mathcal{C}_7 \\
\hline
\text{$\#$ of elements in the class}& 1 & 3 & 3 & 3 & 3 & 3 & 16 & 16 \\
\hline
\text{order} & 1 & 2 & 4 & 4 & 4 & 4 & 3 & 3\\
\hline
\end{array}
\end{equation}
The group $\mathrm{G_{48}}$ contains a normal subgroup
$\mathrm{N_{16}}$ that is isomorphic to $\mathbb{Z}_4 \times
\mathbb{Z}_4$:
\begin{equation}\label{forchettone}
    \mathbb{Z}_4 \times
\mathbb{Z}_4 \, \sim \, \mathrm{N_{16}}\, \lhd \,
\mathrm{G_{48}}\quad ; \quad \frac{\mathrm{G_{48}}}{\mathrm{N_{16}}}
\, \sim \, \mathbb{Z}_3
\end{equation}
Hence following the general procedure for the construction of
irreducible representations of solvable groups (see for instance
\cite{nuovogruppo}) we conclude that the only irreducible
representations of this group $\mathrm{G_{48}}$ can be either
1-dimensional or three dimensional. Hence from numerology we
immediately deduce that the 8 irreducible representations are
necessarily so distributed: three one-dimensional representations
and five three dimensional ones. The construction is straightforward
yet laborious, so it has not yet been done. Consequently I cannot
write the McKay matrix and diagram at this stage, yet I have
calculated the age grading and I have found what follows:
\begin{equation}\label{ospizio}
\begin{array}{|c|ccc|c|c|}
\hline \mathcal{C} &\text{age}&\null&\text{grading}&\null& \text{age vector}\\
\hline
0 &\text{age} & = & 0 & \text{identity} & 1\, \{0, 0, 0\}\\
\hline
1 &\text{age} & = & 1  & \text{junior} &  \frac 12\, \{1, 1, 0\}\\
\hline
2 &\text{age} & = & 2  & \text{senior} &  \frac 14\, \{2, 3, 3\}\\
\hline
3 &\text{age} & = & 1  & \text{junior}& \frac 14\, \{2, 1, 1\}\\
\hline
4 &\text{age} & = & 1  &\text{junior} & \frac 14\, \{1, 3, 0\}\\
\hline
5 &\text{age} & = & 1  & \text{junior} & \frac 14\, \{1, 3, 0\}\\
\hline
6 &\text{age} & = & 1  & \text{junior} &  \frac 13\, \{0, 2, 1\}\\
\hline
7 &\text{age} & = & 1  &\text{junior} &  \frac 13\, \{0, 2, 1\}\\
\hline
\end{array}
\end{equation}
\paragraph{Speculations}
Clearly there is only one compact component of the exceptional
divisor and it is that associated with junior third class. The age
vector suggests that just as in the case of $\mathbb{Z}_4$ quotient
there should be a partial resolution in the form of the canonical
bundle on the weighted projective space $\mathbb{WP}_{[1,1,2]}$. In
the case of the $\mathbb{C}^3/\mathbb{Z}_4$ quotient introducing the
resolution associated with the remaining junior class that generates
a non compact component of the exceptional divisor, turns the
compact divisor $\mathbb{WP}_{[1,1,2]}$ into the second Hirzebruch
surface $\mathbb{F}_2$. Here the remaining junior classes are five
and the question is what will emerge in place of $\mathbb{F}_2$?
Furthermore what will be the K\"ahler metric on this compact
exceptional divisor? Such questions have to be studied in order to
enlarge the spectrum of case studies and unveil the hidden
mechanisms behind the resolutions and the Ricci flat metrics on the
resolved 6-dimensional spaces transverse to the D3 brane.
\newpage

\end{document}